\documentclass[paperpaper,superscriptaddress,english,floatfix, showpacs, amsfonts, amssymb]{revtex4}
\usepackage{epsfig,graphicx,psfrag,amsmath,amssymb,float}
\usepackage[T1]{fontenc}
\usepackage[latin9]{inputenc}
\usepackage{amsmath}
\usepackage{amssymb}
\usepackage{bbold}
\usepackage{amscd}
\usepackage{bm}
\usepackage{psfrag}
\usepackage{epsfig}% Include figure files
\usepackage{graphicx}% Include figure files
\usepackage{dcolumn}% Align table columns on decimal point
\usepackage{bm}% bold math
\usepackage{subfigure}
\DeclareMathOperator{\arccot}{arccot} 
\DeclareMathOperator{\arcsinh}{arcsinh} 
\usepackage{babel}

\begin{document}
\title{One-electron singular spectral features of the 1D Hubbard model at finite magnetic field}
\author{J. M. P. Carmelo}
\affiliation{Department of Physics, University of Minho, Campus Gualtar, P-4710-057 Braga, Portugal}
\affiliation{Center of Physics of University of Minho and University of Porto, P-4169-007 Oporto, Portugal}
\affiliation{Beijing Computational Science Research Center, Beijing 100193, China}
\affiliation{University of Gothenburg, Department of Physics, SE-41296 Gothenburg, Sweden}
\author{T. \v{C}ade\v{z}}
\affiliation{Beijing Computational Science Research Center, Beijing 100193, China}
\affiliation{Center of Physics of University of Minho and University of Porto, P-4169-007 Oporto, Portugal}

\date{25 May 2016}

\begin{abstract}
The momentum, electronic density, spin density, and interaction dependences of the exponents
that control the $(k,\omega)$-plane singular features of the $\sigma =\uparrow,\downarrow$ one-electron 
spectral functions of the 1D Hubbard model at finite magnetic field are studied. The usual half-filling
concepts of one-electron lower Hubbard band and upper Hubbard band are defined
for all electronic density and spin density values and the whole finite repulsion range in terms 
of the rotated electrons associated with the model Bethe-ansatz solution. Such rotated electrons 
are the link of the non-perturbative relation between the electrons and the pseudofermions. Our results
further clarify the microscopic processes through which the pseudofermion dynamical theory 
accounts for the $\sigma $ one-electron matrix elements between the ground state and
excited energy eigenstates.
\end{abstract}


\maketitle

\section{Introduction}
\label{Introduction}

The one-dimensional (1D) Hubbard model with nearest-neighbor hopping integral $t$ and on-site repulsion $U$ is an important correlated 
electronic system whose Bethe anstaz (BA) solution was first derived by the coordinate BA \cite{Lieb,Lieb-03}, following
a similar solution for a related continuous model with repulsive $\delta$-function interaction \cite{Yang-67}. 
For the 1D Hubbard model such a solution is also reachable by the BA inverse-scattering method \cite{Martins}.
In the thermodynamic limit (TL) the imaginary part of its BA complex rapidities simplifies \cite{Takahashi}. 
The Hubbard model was originally introduced as a toy model to study d-electrons in transition metals \cite{Gutzwiller,Hubbard}. 
It is possibly the most studied lattice system of correlated electrons. Static properties such as the charge and spin stiffnesses 
of the 1D Hubbard model under periodic boundary conditions can be determined from the use of the response of the energy 
eigenvalues to an external flux piercing the ring \cite{Shastry-90,Carmelo-97-00}. 

On the other hand, one of the main challenges in the study of the 1D Hubbard model properties is the 
calculation of dynamical correlation functions. Its BA solution provides the exact spectrum of the energy eigenstates, 
yet it has been difficult to apply to the derivation of high-energy dynamical correlation functions. (In this paper we 
use the designation {\it high energy} for all energy scales larger than the model low-energy 
limit associated with the Tomonaga-Luttinger-liquid regime \cite{Tomonaga-50,Luttinger-63,Solyom-79,Voit,Woy-89,Lederer}.) 
The high-energy dynamical correlation functions of both some integrable models with spectral gap 
\cite{Karowski,Smirnov,Mussardo,Zamolodchikov,Tsvelik,Altshuler,Konik} and spin lattice systems
\cite{Jimbo,Bougourzi,Biegel,Kitanine,Caux,Konno} can be studied by the form-factor approach.
However, form factors of the 1D Hubbard model $\sigma =\uparrow,\downarrow$ electron creation and 
annihilation operators involved in the spectral functions studied in this paper remains an unsolved problem. 

The low-energy behavior of the correlation functions of the 1D Hubbard model at finite magnetic field was addressed 
in Refs. \cite{Woy-89,Frahm,Frahm-91,Ogata-91}. On the other hand, in what high-energy behavior of 
dynamical correlation functions is concerned the method used in Refs. \cite{Karlo,Karlo-97} has been a breakthrough 
to address it for one-electron removal and addition spectral functions at zero magnetic field 
in the $u\rightarrow\infty$ limit, which have been derived for the whole $(k,\omega)$ plane. 
That method relies on the spinless-fermion phase shifts imposed by Heisenberg spins $1/2$. Such 
elementary objects naturally arise from the zero spin density and $u\rightarrow\infty$ electron wave-function factorization 
\cite{Woy,Woy-82,Ogata}.

A related pseudofermion dynamical theory (PDT) relying on a representation of the model BA solution in terms of
the pseudofermions generated by a unitary transformation from the corresponding pseudoparticles considered
in Ref. \cite{Carmelo-04} was introduced in Refs. \cite{V-1,LE}. It is an extension of the $u\rightarrow\infty$ method
of Refs. \cite{Karlo,Karlo-97} to the whole $u\equiv U/4t>0$ range of the 1D Hubbard model. A key property is that the
pseudofermions are inherently constructed to their energy spectrum having no interaction terms. This allows the expression of the
one-electron spectral functions in terms of convolutions of pseudofermion spectral functions. The price to pay for the lack of
pseudofermion energy spectrum interaction terms is that creation or annihilation
of pseudofermions under transitions to excited states imposes phase shifts to the remaining pseudofermions.
Within the PDT such phase shifts fully control the one- and two-electron spectral-weight distributions over the
$(k,\omega)$ plane. That approach has been the first breakthrough for the derivation of analytical expressions of the 
zero-magnetic-field 1D Hubbard model high-energy dynamical correlation functions for the whole finite $u>0$ range. 
Recently a modified form of the PDT was used to study the high-energy spin dynamical correlation functions of
the 1D Hubbard model electronic density $n_e=1$ Mott-Hubbard insulator phase \cite{CarCadez}.

After the PDT of the 1D Hubbard model was introduced in Refs. \cite{V-1,LE}, a set of novel methods have been developed to also
tackle the high-energy physics of 1D correlated quantum problems, beyond the low-energy Tomonaga-Luttinger-liquid limit \cite{Glazman}.
In the case of the 1D Hubbard model at zero magnetic field such methods reach the same results as the PDT. For instance,
the momentum, electronic density, and on-site repulsion $u=U/4t>0$ dependence of the exponents that control 
the line shape of the one-electron spectral function of the model at zero magnetic field
calculated in Refs. \cite{Essler,Essler-14} in the framework of a mobile impurity model using input from the BA 
solution is exactly the same as that obtained previously by the use of the PDT.

However, the applications to the study of the repulsive 1D Hubbard model one-electron spectral functions of
both such methods \cite{Essler,Essler-14}, those of the PDT \cite{TTF,spectral0,spectral,spectral-06}, 
and the time-dependent density-matrix renormalization group (tDMRG) method \cite{Kohno-10,Benthien-04}
have been limited to zero magnetic field. The tDMRG studies of Ref. \cite{Feiguin-09} studied the one-electron 
spectral-weight distributions of the attractive 1D Hubbard model at finite magnetic field. Under the canonical transformation
that maps that model into the repulsive 1D Hubbard model, the one-electron spectral-weight distributions plotted in 
Figs. 1 (c) and Fig. 2 of that reference correspond to electronic densities $n_e=1$ and $n_e=0.9$, respectively,
and spin density $m=1/2$. The results refer to a finite system with $40$ electrons. While
they provide some information on the one-electron spectral-weight distributions, it is not possible
to extract from them the momentum dependence of the exponents that {\it in the TL} control the
line shapes near the $\sigma$ one-electron spectral functions singularities.

The main goal of this paper is to extend the PDT applications to the study of the $\sigma $ one-electron spectral functions of the
repulsive 1D Hubbard model at finite magnetic field $h$ in the TL near their singularities. In the TL the corresponding line shapes are
controlled by exponents whose momentum, on-site repulsion $u=U/4t$, electronic density $n$, and spin density
$m$ dependences we study for $u>0$, $n\in [0,1[$, and $m \in [0,n_e]$. In addition, the issue of how the
$\sigma$ one-electron creation and annihilation operators matrix elements between the ground state and
excited energy eigenstates are accounted for by the PDT introduced in Refs. \cite{V-1,LE} is further clarified
in this paper. Beyond the preliminary analysis of these references, the corresponding microscopic processes 
are shown to involve the rotated electrons as a needed link of the non-perturbative relation between
the electrons and PDT pseudofermions.

Our studies refer to the TL of the Hubbard model under periodic boundary conditions on 
a 1D lattice with an even number $L\rightarrow\infty$ of sites and in a chemical potential $\mu$ and magnetic field $h$,
\begin{eqnarray}
{\hat{H}} & = & {\hat{H}}_u + 2\mu\,{\hat{S}}_{\eta}^{z} + 2\mu_B h\,{\hat{S}}_s^{z} \, ,
\nonumber \\
{\hat{H}}_u & = & -t\sum_{\sigma=\uparrow,\downarrow }\sum_{j=1}^{L}\left(c_{j,\sigma}^{\dag}\,
c_{j+1,\sigma} + c_{j+1,\sigma}^{\dag}\,c_{j,\sigma}\right) 
+ U\sum_{j=1}^{L}\left(c_{j,\uparrow}^{\dag}\,c_{j,\sigma} -1/2\right)
\left(c_{j,\downarrow}^{\dag}\,c_{j,\sigma} -1/2\right) \, ,
\nonumber \\
{\hat{S}}_{\eta}^{z} & = & -{1\over 2}(L-\hat{N})  \, ; \hspace{0.50cm} 
{\hat{S}}_s^{z} = -{1\over 2}({\hat{N}}_{\uparrow}-{\hat{N}}_{\downarrow}) \, .
\label{H}
\end{eqnarray}
Here the first and second terms of ${\hat{H}}_u$ are the kinetic-energy operator and the electron on-site repulsion operator,
respectively, the operator $c_{j,\sigma}^{\dagger}$ (and $c_{j,\sigma}$)
creates (and annihilates) one spin-projection $\sigma$ electron at lattice site
$j=1,...,L$, and the electron number operators read
${\hat{N}}=\sum_{\sigma=\uparrow ,\downarrow }\,\hat{N}_{\sigma}$ and
${\hat{N}}_{\sigma}=\sum_{j=1}^{L}\hat{n}_{j,\sigma}= \sum_{j=1}^{L}c_{j,\sigma}^{\dag}\,c_{j,\sigma}$.
Moreover, $\mu_B$ is the Bohr magneton and ${\hat{S}}_{\eta}^{z}$ and ${\hat{S}}_s^{z}$
are the diagonal generators of the Hamiltonian ${\hat{H}}_u$ global $\eta$-spin and spin $SU(2)$ symmetry algebras, respectively.
In this paper we use in general units of lattice constant one, so that the number of lattice sites $N_a$ equals
the lattice length $L$. The model properties depend on the ratio $U/t$ and in this paper the 
corresponding parameter $u= U/4t$ is often used.

The lowest weight states (LWSs) and highest weight states (HWSs) of the $\eta$-spin and spin $SU(2)$ 
symmetry algebras have numbers $S_{\alpha} = - S_{\alpha}^{z}$ and $S_{\alpha} = S_{\alpha}^{z}$, respectively,
for $\alpha = \eta,s$. Here $S_{\eta}$ is the states $\eta$-spin, $S_{s}$ their spin, and $S_{\eta}^{z}$ and $S_{s}^{z}$
are the corresponding projections, respectively, which are the eigenvalues of the spin operators given in
Eq. (\ref{H}). Let $\{\vert l_{\rm r},l_{\eta s},u\rangle\}$ be the complete set of $4^{L}$ energy eigenstates of the
Hamiltonian $\hat{H}$, Eq. (\ref{H}), associated with the BA solution for $u>0$. The LWSs of both $SU(2)$ symmetry algebras
are here denoted by $\vert l_{\rm r},l_{\eta s}^0,u\rangle$ where the $u$-independent label $l_{\eta s}$ is a short notation for 
the set of quantum numbers, 
\begin{equation}
l_{\eta s} = S_{\eta},S_{s},n_{\eta},n_s \, ; \hspace{0.50cm} n_{\alpha} = 
S_{\alpha}+S_{\alpha}^{z} = 0,1,..., 2S_{\alpha} \, , \hspace{0.50cm} \alpha = \eta, s \, .
\label{etas-states-ll}
\end{equation}
Furthermore, the label $l_{\rm r}$ refers to the set of all remaining $u$-independent quantum numbers needed
to uniquely specify an energy eigenstate $\vert l_{\rm r},l_{\eta s},u\rangle$. The latter $u$-independent 
quantum numbers naturally emerge from the BA solution and are given below in Section \ref{pseudoRelect}. 

We call a {\it Bethe state} an energy eigenstate that is a LWS of both $SU(2)$ symmetry algebras.
For a Bethe state one then has that $n_{\eta}= n_s=0$ in Eq. (\ref{etas-states-ll}), so that $l_{\eta s}^0$ stands for $S_{\eta},S_{s},0,0$. 
The non-LWSs $\vert l_{\rm r},l_{\eta s},u\rangle$ can be generated from the corresponding 
Bethe states $\vert l_{\rm r},l_{\eta s}^0,u\rangle$ as \cite{Completeness},
\begin{eqnarray}
\vert l_{\rm r},l_{\eta s},u\rangle & = & \prod_{\alpha=\eta, s}\left(\frac{1}{
\sqrt{{\cal{C}}_{\alpha}}}({\hat{S}}^{+}_{\alpha})^{n_{\alpha}}\right)\vert l_{\rm r},l_{\eta s}^0,u\rangle \, ;
\hspace{0.50cm} {\cal{C}}_{\alpha} = (n_{\alpha}!)\prod_{j=1}^{n_{\alpha}}(\,2S_{\alpha}+1-j\,) 
\, , \hspace{0.50cm} n_{\alpha}=1,...,2S_{\alpha} \, ,
\nonumber \\
{\hat{S}}^{+}_{\eta} & = & \sum_{j=1}^{L}(-1)^j\,c_{j,\downarrow}^{\dag}\,c_{j,\uparrow}^{\dag} \, ;
\hspace{0.50cm}
{\hat{S}}^{+}_{s} = \sum_{j=1}^{L}c_{j,\downarrow}^{\dag}\,c_{j,\uparrow} \, .
\label{Gstate-BAstate}
\end{eqnarray}
Here ${\cal{C}}_{\alpha}$ are normalization constants and $\alpha =\eta,s$. 
The model in its full Hilbert space can be described either directly within the BA solution \cite{Woy-82,Braak} 
or by application onto the Bethe states of the $\eta$-spin and spin $SU(2)$ symmetry algebras off-diagonal 
generators, as given in Eq. (\ref{Gstate-BAstate}).

Relying on the model symmetries, for simplicity and without loss in generality the
studies of this paper refer to electronic densities and spin densities in the ranges $n_e \in [0,1[$ and
$m\in [0,n_e]$, respectively. For such electronic densities and spin densities the model ground states 
are LWSs of both the $\eta$-spin and spin $SU(2)$ symmetry algebras so that in the studies of this 
paper we use the LWS formulation of 1D Hubbard model BA solution. 

The PDT is used in it to clarify one of the unresolved questions concerning  
the physics of the 1D Hubbard model at finite magnetic field, Eq. (\ref{H}), by deriving the momentum, repulsive interaction $u=U/4t$, 
electron-density $n_e$, and spin-density $m$ dependences of the exponents that control the singularities at 
the $\sigma$ one-electron spectral functions. These exponents control the line shape near the
singularities of the following $\sigma$ one-electron spectral function $B_{\sigma,\gamma} (k,\,\omega)$ such that
$\gamma=-1$ (and $\gamma=+1$) for one-electron removal (and addition),
\begin{eqnarray}
B_{\sigma,-1} (k,\,\omega) & = & \sum_{\nu^-}
\vert\langle\nu^-\vert\, c_{k,\sigma} \vert \,GS\rangle\vert^2 \,\delta (\omega
+ (E_{\nu^-}^{N_{\sigma}-1}-E_{GS}^{N_{\sigma}})) \hspace{0.5cm} \omega \leq 0 \, ,
\nonumber \\
B_{\sigma,+1} (k,\,\omega) & = & \sum_{\nu^+}
\vert\langle\nu^+\vert\, c^{\dagger}_{k,\sigma} \vert
\,GS\rangle\vert^2 \,\delta (\omega - (E_{\nu^+}^{N_{\sigma}+1}-E_{GS}^{N_{\sigma}}))  \hspace{0.5cm} \omega \geq 0 \, .
\label{Bkomega}
\end{eqnarray}
Here $c_{k,\sigma}$ and $c^{\dagger}_{k,\sigma}$ are electron
annihilation and creation operators, respectively, of momentum $k$ and $\vert GS\rangle$ denotes the
initial $N_{\sigma}$-electron ground state of energy $E_{GS}^{N_{\sigma}}$. The $\nu^-$ and $\nu^+$
summations run over the $N_{\sigma}-1$ and $N_{\sigma}+1$-electron excited 
energy eigenstates, respectively, and $E_{\nu^-}^{N_{\sigma}-1}$ and 
$E_{\nu^+}^{N_{\sigma}+1}$ are the corresponding energies.

The remainder of the paper is organized as follows. In Section \ref{relation} 
the $\sigma$ one-electron lower-Hubbard band (LHB) and upper-Hubbard band (UHB)
are defined for $u>0$ and all densities in terms of quantum numbers associated with
the $\sigma$ rotated-electron energy eigenstates occupancies. Moreover, the relation of the $\beta $ pseudoparticle representation 
to such $\sigma$ rotated electrons, which are uniquely defined in terms of the matrix elements of the
electron - rotated-electron unitary operator between all model $4^L$ energy
and momentum eigenstates, is an issue also addressed in that section. 
The PDT suitable for the study of the $\sigma$ 
one-electron spectral weights and further information beyond that provided in Refs. \cite{V-1,LE}
on how that dynamical theory accounts for the matrix elements of the $\sigma$
electron operators between the ground state and the excited energy eigenstates
are the issues revisited and studied in Section \ref{PDT}. In Section \ref{DSGzzxx} the $(k,\omega)$-plane
line shape near the singular spectral features of the $\sigma$ one-electron spectral functions, Eq. (\ref{Bkomega}),
is studied. Finally, the concluding remarks are presented in Section \ref{concluding}.

\section{Lower- and upper-Hubbard bands and the pseudoparticle representation 
emerging from the rotated electrons associated with the BA solution}
\label{relation}

Concerning the $\sigma $ one-electron addition processes
that contribute to the $\gamma=1$ spectral function, Eq. (\ref{Bkomega}), important concepts for 
our study are those of a LHB and a UHB. Those are defined for $u>0$ and all
densities in Section \ref{QPTLUHB} by the rotated-electron quantum numbers that
define the $\sigma $ one-electron addition excited energy eigenstates.
The corresponding unique definition of the electron - rotated-electron unitary transformation associated with
the BA solution and the separation of the rotated-electron occupancy configurations that
generate the exact $u>0$ energy eigenstates into occupancy configurations of three types 
of fractionalized particles, specifically the spinless $c$ pseudoparticles, the rotated spins $1/2$, 
and the rotated $\eta$-spins $1/2$, are the issues addressed in Section
\ref{pseudoRelect}. Such a relation allows the introduction and expression in Section \ref{pseuOpRel} 
of operators for the $c$ pseudoparticles, rotated spins $1/2$, and rotated $\eta$-spins $1/2$ in terms 
of the $\sigma$ rotated-electron creation and annihilation operators. In Section \ref {quantum-liquid} 
the pseudoparticle energy dispersions and other quantities that emerge from the pseudoparticle quantum liquid and
determine and control the $(k,\omega)$-plane line shape near the singular spectral features 
of the $\sigma =\uparrow,\downarrow$ one-electron spectral functions, Eq. (\ref{Bkomega}),
are introduced.

\subsection{Definition of $\sigma$ one-electron lower- and upper-Hubbard bands}
\label{QPTLUHB}

The concept of $\sigma$ one-electron UHB addition is well established at electronic density $n_e=1$ for $u>0$ 
\cite{Lieb,Lieb-03,DSF-n1}. Below we define the concepts of a LHB and a UHB for $n_e\neq 1$ and $u>0$ such that
due to a quantum phase transition at $n_e=1$ there is only $\sigma$ one-electron UHB addition
whereas for $n_e\neq 1$ there is both $\sigma$ one-electron LHB and UHB addition.
The Hamiltonian $\hat{H}$, Eq. (\ref{H}), quantum phases are associated with different ranges of electronic density $n_e$ and spin
density $m$ and are marked by important energy scales. Those correspond to limiting values
of the charge energy scale $2\mu = 2\mu (n_e)$ and magnetic energy scale $2\mu_B\,h=2\mu_B\,h (m)$
involving the chemical potential $\mu$ and and magnetic field $h$, respectively.

The energy scales $2\mu = 2\mu (n_e)$ and $2\mu_B\,h=2\mu_B\,h (m)$
are odd functions of the hole concentration $(1-n_e)$ and spin density $m$, respectively. One may
then consider for instance the ranges $n_e\in [0,1[$ and $m\in [0,n_e]$. The interval $n_e\in ]0,1[$
refers for $m<n_e$ to a metallic quantum phase for which $2\mu = 2\mu (n_e)$ is a continuous function 
of $n_e$. It smoothly decreases from $2\mu = (U+4t)$
for $n_e\rightarrow 0$ to $2\mu = 2\mu_{u}$ for $n_e\rightarrow 1$ where $2\mu_{u} <(U+4t)$
is the Mott-Hubbard gap. On the other hand, at $n_e=1$ the chemical potential varies in the range
$\mu \in [-\mu_{u},\mu_{u}]$ in spite of the electronic density remaining constant, which is
a property of the corresponding $n_e=1$ and $u>0$ Mott-Hubbard insulator quantum phase. 

The $n_e=1$ Mott-Hubbard gap $2\mu_{u}$ is the energy scale associated with the phase
transition between the two above mentioned quantum phases. For $u>0$ it 
remains finite for all spin densities, $m\in [0,1[$. For instance, in the limits $m\rightarrow 0$ \cite{Lieb,Lieb-03,Ovchi} 
and $m\rightarrow 1$ it reads,
\begin{eqnarray}
2\mu_{u} & = & U - 4t + 8t\int_0^{\infty}d\omega {J_1 (\omega)\over\omega\,(1+e^{2\omega u})} 
= {16\,t^2\over U}\int_1^{\infty}d\omega {\sqrt{\omega^2-1}\over\sinh\left({2\pi t\omega\over U}\right)} 
\, , \hspace{0.50cm} m\rightarrow 0 \, ,
\nonumber \\
& = & \sqrt{(4t)^2+U^2} - 4t \, , \hspace{0.50cm} m\rightarrow 1 \, ,
\label{2mu0}
\end{eqnarray}
respectively. Its $u\ll 1$ limiting behaviors 
are $2\mu_{u} \approx (8/\pi)\,\sqrt{t\,U}\,e^{-2\pi \left({t\over U}\right)}$ for $m\rightarrow 0$
and $2\mu_{u} \approx U^2/8t$ for $m\rightarrow 1$ and the $u\gg 1$ behavior
is $2\mu_{u} \approx (U - 4t)$ for $m \in [0,1]$.
 
On the other hand, for the metallic quantum phase corresponding to the
spin density interval $m\in [0,n_e[$ for $n_e\in [0,1[$ the magnetic energy scale $2\mu_B\,h$ 
is a continuous function of $m$. It smoothly increases from zero at $m=0$
to $2\mu_B\,h_c$ for $m\rightarrow n_e$. Here $h_c$ is the critical field for the
fully polarized ferromagnetism quantum phase transition. Indeed, for $h>h_c$ there is no electron 
double occupancy, so that the on-site repulsive interaction term in the Hamiltonian, Eq. (\ref{H}), has no effects 
and the system is driven into a non-interactive quantum phase. 

The magnetic energy scale $2\mu_B\,h_c$ associated with such a quantum phase transition is an even function 
of the hole concentration $(1-n_e)$. For the ranges $n_e\in [0,1[$ and $m\in [0,n_e]$ 
it has the following closed-form expression in terms of $u=U/4t$ and the electronic density $n_e$ \cite{Carmelo-91-92},
\begin{eqnarray}
2\mu_B\,h_c & = & 2t\Bigl[\sqrt{1+u^2}\left(1 - {2\over\pi}\arccot\left({\sqrt{1+u^2}\over u}\tan (\pi n_e)\right)\right) 
\nonumber \\
& - & 2u\,n_e - {2\over\pi}\cos (\pi n_e)\arctan\left({\sin (\pi n_e)\over u}\right)\Bigr] \, .
\label{hc}
\end{eqnarray}
In the $n_e\rightarrow 0$ and $n_e\rightarrow 1$ limits this gives, 
\begin{eqnarray}
2\mu_B\,h_c & = & 0 \, , \hspace{0.50cm} n_e\rightarrow 0 \, ,
\nonumber \\
& = & \sqrt{(4t)^2+U^2} - U \, , \hspace{0.50cm} n_e\rightarrow 1 \, ,
\label{mBHc-n0-n1}
\end{eqnarray} 
respectively. For the density range $n_e \in [0,1]$ it behaves as $2\mu_B\,h_c = 4t\sin^2 (\pi\,n_e/2)$
for $u\rightarrow 0$ and as $2\mu_B\,h_c =  (2t\,n_e/u)[1 - \sin (2\pi n_e)/(2\pi n_e)]$ for $u\gg 1$.

The definition of the $\sigma$ one-electron LHB and UHB addition for the
whole $u>0$ range, electronic densities $n_e \in [0,1]$, and spin densities
$m\in [0,n_e]$ relies on the occupancy configurations of uniquely defined {\it rotated electrons}. This involves
selecting out of the many choices of $u\rightarrow\infty$ degenerate $4^{L}$ energy eigenstates, 
those obtained from the $u>0$ Bethe states and corresponding non-LWSs, Eq. (\ref{Gstate-BAstate}),
as $\vert  l_{\rm r},l_{\eta s},\infty \rangle = \lim_{u\rightarrow\infty}\vert  l_{\rm r},l_{\eta s},u\rangle$.

The important point is that for the $u\rightarrow\infty$ energy eigenstates $\vert  l_{\rm r},l_{\eta s},\infty \rangle$, 
$\sigma$ electron single occupancy, double occupancy, and non-occupancy are good
quantum numbers. We call {\it $V$ tower} the set of energy eigenstates $\vert  l_{\rm r},l_{\eta s},u\rangle$ with exactly the
same $u$-independent quantum numbers $l_{\rm r}$ and $l_{\eta s}$ and different $u$ values in the range $u>0$. 
$\sigma$ electron single occupancy, electron double occupancy, 
and electron non-occupancy are not good quantum numbers for the finite-$u$ energy eigenstates $\vert  l_{\rm r},l_{\eta s},u\rangle$ belonging to the same
$V$ tower. For instance, upon decreasing $u$ there emerges for ground states a finite electron double occupancy
expectation value, which vanishes for $u\rightarrow\infty$ \cite{Carmelo-03}. 

Since for any $u>0$ value the set of energy eigenstates $\vert  l_{\rm r},l_{\eta s},u\rangle$ that belong to the same $V$ tower are
generated by exactly the same occupancy configurations of the $u$-independent quantum
numbers $l_{\rm r}$ and $l_{\eta s}$ given in Eq. (\ref{etas-states-ll}) and below in Section \ref{pseudoRelect}, 
respectively, the Hilbert space is the same for the whole $u>0$ range. Hence for any $u>0$ there is a 
uniquely defined unitary operator ${\hat{V}}={\hat{V}} (u)$ such that 
$\vert  l_{\rm r},l_{\eta s},u\rangle={\hat{V}}^{\dag}\vert  l_{\rm r},l_{\eta s},\infty \rangle$. 
This operator ${\hat{V}}$ is the $\sigma $ electron - rotated-electron unitary operator such that,
\begin{equation}
{\tilde{c}}_{j,\sigma}^{\dag} =
{\hat{V}}^{\dag}\,c_{j,\sigma}^{\dag}\,{\hat{V}}
\, ; \hspace{0.50cm}
{\tilde{c}}_{j,\sigma} =
{\hat{V}}^{\dag}\,c_{j,\sigma}\,{\hat{V}} 
\, ; \hspace{0.50cm}
{\tilde{n}}_{j,\sigma} = {\tilde{c}}_{j,\sigma}^{\dag}\,{\tilde{c}}_{j,\sigma} \, , \hspace{0.5cm} j = 1,...,L
\, , \hspace{0.5cm} \sigma = \uparrow, \downarrow \, ,
\label{rotated-operators}
\end{equation} 
are the operators that create and annihilate, respectively, the $\sigma$ rotated electrons as defined here. Moreover, 
$\vert  l_{\rm r},l_{\eta s},\infty \rangle=\hat{G}^{\dag}_{l_{\rm r},l_{\eta s}}\,\vert 0\rangle$
where $\vert 0\rangle$ is the electron and rotated-electron vacuum and $\hat{G}^{\dag}_{l_{\rm r},l_{\eta s}}$ a uniquely defined operator.
It then follows that $\vert  l_{\rm r},l_{\eta s},u\rangle=\tilde{G}^{\dag}_{l_{\rm r},l_{\eta s}}\,\vert 0\rangle$
where the generator $\tilde{G}^{\dag}_{l_{\rm r},l_{\eta s}}={\hat{V}}^{\dag}\,\hat{G}^{\dag}_{l_{\rm r},l_{\eta s}}\,{\hat{V}}$ has 
the same expression in terms of the $\sigma$ rotated-electron creation and annihilation operators
as $\hat{G}^{\dag}_{l_{\rm r},l_{\eta s}}$ in terms of $\sigma$ electron creation and annihilation operators, respectively.
The $\sigma$ electron - $\sigma$ rotated-electron unitary operator ${\hat{V}}$ in Eq. (\ref{rotated-operators}) is uniquely defined 
in Section \ref{pseudoRelect} for $u>0$ by its matrix elements between all $4^L$ energy and momentum eigenstates,
Eq. (\ref{Gstate-BAstate}).

That $\sigma$ electron single occupancy, electron double occupancy, and electron non-occupancy 
are good quantum numbers for a $u\rightarrow\infty$ energy eigenstate $\vert  l_{\rm r},l_{\eta s},\infty \rangle$
then implies that for all the finite-$u$ energy eigenstates $\vert  l_{\rm r},l_{\eta s},u\rangle$ belonging to the same
$V$ tower $\sigma$ rotated-electron single occupancy, rotated-electron double occupancy, and 
rotated-electron non-occupancy are good quantum numbers for $u>0$. For $u>0$ this applies to all 
$4^L$ energy and momentum eigenstates.

Fortunately and as confirmed in Section \ref{pseudoRelect}, the BA quantum numbers are directly related to the
numbers of sites singly occupied, doubly occupied, and unoccupied by $\sigma$ rotated electrons. 
From the use of that relation it is found that for electronic densities $n_e \in [0,1[$ and spin densities $m\in [0,n_e]$ the model ground 
states have zero rotated-electron double occupancy. The
$\sigma$ one-electron LHB addition spectral function $B^{\rm LHB}_{\sigma,+1} (k,\,\omega)$
and UHB addition spectral function $B^{\rm UHB}_{\sigma,+1} (k,\,\omega)$ are then uniquely defined
for $u>0$ as follows,
\begin{eqnarray}
B_{\sigma,+1} (k,\,\omega) & = & B^{\rm LHB}_{\sigma,+1} (k,\,\omega) + B^{\rm UHB}_{\sigma,+1} (k,\,\omega) \, ,
\nonumber \\
B^{\rm LHB}_{\sigma,+1} (k,\,\omega) & = & \sum_{\nu^+_0}
\vert\langle\nu^+_0\vert\, c^{\dagger}_{k,\sigma} \vert
\,GS\rangle\vert^2 \,\delta (\omega - (E_{\nu^+_0}^{N_{\sigma}+1}-E_{GS}^{N_{\sigma}}))  \hspace{0.5cm} \omega \geq 0 \, , 
\nonumber \\
B^{\rm UHB}_{\sigma,+1} (k,\,\omega) & = & \sum_{\nu^+_D}
\vert\langle\nu^+_D\vert\, c^{\dagger}_{k,\sigma} \vert
\,GS\rangle\vert^2 \,\delta (\omega - (E_{\nu^+_D}^{N_{\sigma}+1}-E_{GS}^{N_{\sigma}}))  \hspace{0.5cm} \omega \geq 0 \, , 
\label{LUHB}
\end{eqnarray}
where the $\nu^+_0$ and $\nu^+_D$ summations run over the $N_{\sigma}+1$-electron excited 
energy eigenstates with zero and $D>0$, respectively, rotated-electron double occupancy 
and $E_{\nu^+_0}^{N_{\sigma}-1}$ and $E_{\nu_D^+}^{N_{\sigma}+1}$ are the corresponding energies.

The $\sigma$ one-electron spectral functions obey the following sum rules,
\begin{eqnarray}
\sum_k\int_{-\infty}^{\infty}d\omega\,B_{\sigma,-1} (k,\,\omega) & = & N_{\sigma} \, ; \hspace{0.75cm}
\sum_k\int_{-\infty}^{\infty}d\omega\,B_{\sigma,+1} (k,\,\omega) = L - N_{\sigma} \, ,
\nonumber \\
\sum_k\int_{-\infty}^{\infty}d\omega\,B^{\rm LHB}_{\sigma,+1} (k,\,\omega) & = & L - N \, ; \hspace{0.55cm}
\sum_k\int_{-\infty}^{\infty}d\omega\,B^{\rm UHB}_{\sigma,+1} (k,\,\omega) = N - N_{\sigma} \, .
\label{sumrules}
\end{eqnarray}
The first two sum rules are well known and exact for all $u$ values. The $B^{\rm LHB}_{\sigma,+1} (k,\,\omega)$
and $B^{\rm UHB}_{\sigma,+1} (k,\,\omega)$ sum rules are found to be exact both in the
limits $n_e\rightarrow 0$ and $n_e\rightarrow 1$ for $u>0$. Both in the $u\ll 1$ and $u\gg 1$ limits
they are exact as well for electronic densities $n_e \in [0,1[$ and spin densities $m\in [0,n_e]$.
They are likely exact also for intermediate $u$ values yet we could not prove it. If otherwise,
they are a very good approximation. Fortunately, clarification of this issue is not needed for our
studies, as it focuses only on the line shapes in the vicinity of the singularities of the 
$\sigma$ one-electron spectral functions and not on the detailed weight distribution over the whole
$(k,\omega)$ plane. The line shape near the singularities is that observed in experiments on
actual condensed matter systems and spin $1/2$ ultra-cold atom systems.
The important point for the present study is rather the definition of $\sigma$ one-electron LHB and UHB for $u>0$,
$n_e \in [0,1]$, and $m\in [0,n_e]$, Eq. (\ref{LUHB}), which follows from the corresponding unique
definition of rotated electrons in Sec. \ref{pseudoRelect} in terms of quantities of the exact BA solution. 

The present definition for $u>0$ and all densities of the concepts of a LHB and a UHB is directly 
associated with a global lattice $U(1)$ symmetry of the Hamiltonian ${\hat{H}}_u$, Eq. (\ref{H}),
beyond its well-known $SO(4) = [SU(2)\otimes SU(2)]/Z_2$ symmetry, which contains the 
$\eta$-spin and spin $SU(2)$ symmetries \cite{Yang,Yang-90,Lieb-89}. Such a global lattice $U(1)$ symmetry exists
for the model on the 1D and any other bipartite lattice \cite{bipartite} and is behind its global symmetry being 
actually larger than $SO(4)$ and given by $[SO(4)\otimes U(1)]/Z_2=[SU(2)\otimes SU(2)\otimes U(1)]/Z_2^2$, 
which is equivalent to $SO(3)\otimes SO(3)\otimes U(1)$. (The factor $1/Z_2^2$ follows from the total number $4^{L}$ of independent 
representations of the group $[SU(2)\otimes SU(2)\otimes U(1)]/Z_2^2$ being four times smaller than the
dimension $4^{L+1}$ of the group $SU(2)\otimes SU(2)\otimes U(1)$.) 

That the Hamiltonian ${\hat{H}}_u$, Eq. (\ref{H}), global symmetry is $[SO(4)\otimes U(1)]/Z_2$ has
direct effects on the $4^{L}$ energy and momentum eigenstates of the Hamiltonian ${\hat{H}}$ in the presence
of a chemical potential and magnetic field also given in Eq. (\ref{H}), as these states
refer to $4^{L}$ state representations of the group $[SO(4)\otimes U(1)]/Z_2$ in the model full Hilbert space. 
In the present 1D case the occurrence
of the global lattice $U(1)$ symmetry justifies for instance that the spin and charge monodromy matrices of the BA inverse-scattering method
have different ABCD and ABCDF forms associated with the spin $SU(2)$ and charge $U(2)=SU(2)\otimes U(1)$ 
symmetries, respectively. Consistently, the latter matrix is larger than the former and involves more fields \cite{Martins}. 
If the model global symmetry was $SO(4)=[SU(2)\otimes SU(2)]/Z_2$, the charge and a spin monodromy matrices 
would have the same traditional ABCD form, which is that of the spin-$1/2$ $XXX$ Heisenberg chain \cite{Faddeev}. 

The relation of the global lattice $U(1)$ symmetry beyond $SO(4)$ to the LHB and UHB 
as defined here for $u>0$ and all densities results from its generator being the operator that counts the
number $N_s^R$ of rotated-electron singly occupied sites or, alternatively, the
number $N_{\eta}^R = L -N_s^R$ of rotated-electron 
unoccupied sites plus doubly occupied sites. Indeed, for the
electronic density ranges (i) $n_e \in [0,1]$ and (ii) $n_e \in [1,2]$ the UHB exactly originates from 
transitions to energy eigenstates with a finite number of (i) rotated-electron doubly
occupied sites and (ii) rotated-electron unoccupied sites, respectively. 

\subsection{Rotated-electron separation in terms of $c$ pseudoparticles,
rotated spins $1/2$, and rotated $\eta$-spins $1/2$}
\label{pseudoRelect}

The charge-only and spin-only fractionalized particles that emerge in 1D correlated electronic systems
are usually identified with holons and spinons, respectively \cite{Essler-94}. 
Such holons and spinons are in 1D integrable correlated electronic models associated with specific quantum 
numbers of the exact solutions. The use of holon and spinon representations provides a suitable description of 
these models low-energy physics. Some of such quantum liquids exotic properties survive at higher energies 
yet the exponents characterizing the dynamical correlation functions singularities are functions 
of the momentum and differ significantly from the predictions of the linear Tomonaga-Luttinger liquid 
theory \cite{V-1,Glazman,Essler,Essler-14}. This applies to the 1D Hubbard model. 

Furthermore, a careful analysis of the high-energy dynamical correlation functions reveals that their spectral weights 
are controlled by the scattering of both spinless fractionalized particles and neutral composite objects whose constituents 
are spin-$1/2$ or $\eta$-spin $1/2$ fractionalized particles. Both such spinless fractionalized particles
and composite elementary objects refer to the pseudofermions of the PDT representation used in this paper 
to study the $\sigma$ one-electron spectral functions, Eq. (\ref{Bkomega}).
Such pseudofermions are identical to the pseudoparticles of Ref. \cite{Carmelo-04} except that their momentum values
are slightly shifted by a well defined unitary transformation. The direct relation of the corresponding spinless $c$ pseudoparticles 
and spin-$1/2$ or $\eta$-spin $1/2$ fractionalized particles within the neutral composite pseudoparticles 
to the rotated electrons whose operators are given in Eq. (\ref{rotated-operators}) refers to the
above mentioned needed link of the corresponding non-perturbative relation between
the electrons and PDT pseudofermions. 

For the 1D Hubbard model there is an infinite number of transformations that generate $\sigma$ rotated 
electrons from the $\sigma$ electrons such that $\sigma$ rotated-electron single occupancy is a good quantum number 
for $u>0$ \cite{bipartite}. The pseudoparticle representation and corresponding pseudofermion representation 
refer though to a specific choice of $\sigma$ rotated electrons. Those are generated from the $\sigma$ electrons
by a unitary transformation uniquely defined by the BA. Actually, the BA solution performs such a 
transformation. The corresponding electron - rotated-electron unitary operator ${\hat{V}}$ in Eq. (\ref{rotated-operators})
can be defined by its matrix elements between the model $4^L$ energy and momentum eigenstates. Fortunately, such
matrix elements can be expressed in terms of the following known BA amplitudes of the Bethe states 
$\vert l_{\rm r},l_{\eta s}^0,u\rangle$ \cite{Woy,Woy-82}, 
\begin{equation}
f_{l_{\rm r},l_{\eta s}^0,u} (x_1\sigma_1,...,x_{N^0}\sigma_{N^0}) = 
\langle x_1\sigma_1,...,x_{N^0}\sigma_{N^0}\vert l_{\rm r},l_{\eta s}^0,u \rangle \, .
\label{f-U-0}
\end{equation}
Such amplitudes are uniquely defined in Eqs. (2.5)-(2.10) of Ref. \cite{Woy} in terms of BA solution quantities. 
In them, $\vert x_1\sigma_1,...,x_{N^0}\sigma_{N^0}\rangle$ denotes a local state in which the $N^0 = L - 2S_{\eta}$ electrons with spin projection $\sigma_1,...,\sigma_{N^0}$ are located at sites of spatial coordinates $x_1,...,x_{N^0}$,
respectively. For a LWS their numbers are $N_{\uparrow}^0 = L/2 -S_{\eta} + S_s$ and $N_{\downarrow}^0 = L/2 -S_{\eta} - S_s$.
Due to symmetry, the amplitudes of the non-LWSs $\vert l_{\rm r},l_{\eta s},u\rangle$ generated 
from each Bethe state as given in Eq. (\ref{Gstate-BAstate}) differ from
it by the trivial phase factor $(-1)^{n_{\eta}}$. Here $n_{\eta}=S_{\eta}+S_{\eta}^z$ is the non-LWS
number given in Eq. (\ref{etas-states-ll}).

For the set of Bethe states corresponding to different finite $u>0$ values and belonging to the same $V$ tower 
the amplitudes, Eq. (\ref{f-U-0}), smoothly and continuously behave as a function of $u$.
That the amplitudes $\langle n_{\eta};n_s;x_1\sigma_1,...,x_{N^0}\sigma_{N^0}\vert l_{\rm r},l_{\eta s},u \rangle$
of a non-LWS involving the states $\vert n_{\eta};n_s;x_1\sigma_1,...,x_{N^0}\sigma_{N^0}\rangle$
are given in terms of those of the corresponding Bethe state 
merely by $(-1)^{n_{\eta}}\langle x_1\sigma_1,...,x_{N^0}\sigma_{N^0}\vert l_{\rm r},l_{\eta s}^0,u \rangle$
and thus by $(-1)^{n_{\eta}}f_{l_{\rm r},l_{\eta s}^0,u} (x_1\sigma_1,...,x_{N^0}\sigma_{N^0})$
follows from except for the phase factor $(-1)^{n_{\eta}}$ the non-LWS amplitudes
being insensitive to the $n_{\eta}$ created electrons pairs and their spatial coordinates. These electrons pairs emerge
as a result of the application onto the Bethe state of the $\eta$-spin off-diagonal generator ${\hat{S}}^{+}_{\eta}$ 
a number of times $n_{\eta}$, as given in Eq. (\ref{Gstate-BAstate}). Moreover, 
such amplitudes are insensitive to the spatial coordinates of the $n_{s}$ electrons whose spin has been flipped 
by the $n_{s}$ spin off-diagonal generators $({\hat{S}}^{+}_{s})^{n_{s}}$, Eq. (\ref{Gstate-BAstate}). 
Such insensitivities are behind denoting the local states 
$\vert x_1'\sigma_{1'},...,x_{N^0+2n_{\eta}}'\sigma_{(N^0+2n_{\eta})'}\rangle$
in which the $N^0+2n_{\eta}$ electrons with spin projection $\sigma_{1'},...,\sigma_{(N^0+2n_{\eta})'}$ 
are located at sites of spatial coordinates $x_1',...,x_{N^0+2n_{\eta}}'$ by
$\vert n_{\eta};n_s;x_1\sigma_1,...,x_{N^0}\sigma_{N^0}\rangle$.
They also imply that, as for the Bethe states, for the set of any energy eigenstates corresponding to different 
finite $u$ values and belonging to the same $V$ tower 
the general amplitudes $f_{l_{\rm r},l_{\eta s},u} (x_1\sigma_1,...,x_{N^0}\sigma_{N^0}) 
= \langle n_{\eta};n_s;x_1\sigma_1,...,x_{N^0}\sigma_{N^0}\vert l_{\rm r},l_{\eta s},u \rangle$ smoothly and 
continuously behave as a function of $u$.

It then follows from basic quantum mechanics arguments that the electron - rotated-electron unitary operator ${\hat{V}}$ 
in Eq. (\ref{rotated-operators}) is uniquely defined by the set of the following matrix elements between the energy eigenstates,
\begin{eqnarray}
\langle l_{\rm r},l_{\eta s},u\vert {\hat{V}}\vert  l_{\rm r}',l_{\eta s}',u \rangle & = &
\delta_{l_{\eta s},l_{\eta s}'}\,\sum_{x = 1}^{L}... \sum_{x_{N^0} = 1}^{L}
f_{l_{\rm r},l^0_{\eta s},u}^* (x_1\sigma_1,...,x_{N^0}\sigma_{N^0})\,f_{l_{\rm r}',l^0_{\eta s},\infty} (x_1\sigma_1,...,x_{N^0}\sigma_{N^0}) \, .
\label{ME-Vll}
\end{eqnarray}
Here and throughout this paper $\delta_{l,l'}$ is the usual Kronecker symbol such that $\delta_{l,l'}=1$ for $l=l'=0,1,2,...$
and $\delta_{l,l'}=0$ for $l\neq l'$ and $f_{l_{\rm r},l^0_{\eta s},u} (x_1\sigma_1,...,x_{N^0}\sigma_{N^0})$ and 
$f_{l_{\rm r}',l^0_{\eta s},\infty} (x_1\sigma_1,...,x_{N^0}\sigma_{N^0})$ are the amplitudes defined by Eqs. (2.5)-(2.10) of 
Ref. \cite{Woy} for $u>0$ and Eq. (2.23) of Ref. \cite{Woy-82}
for $u\rightarrow\infty$, respectively. The factor $\delta_{l_{\eta s},l_{\eta s}'}$ implies that the phase factors $(-1)^{n_{\eta}}$ 
always occur in pairs, which gives rise to an overall phase factor $(-1)^{2n_{\eta}}=1$. Since the set of
$4^L\times 4^L = 4^{2L}$ matrix elements of form, Eq. (\ref{ME-Vll}), are 
between all $4^L$ energy and momentum eigenstates that span the model full Hilbert space they uniquely define the electron - rotated-electron 
unitary operator ${\hat{V}}$. 

That because of symmetries behind the factor $\delta_{l_{\eta s},l_{\eta s}'}$ many of the matrix 
elements, Eq. (\ref{ME-Vll}), vanish simplifies the quantum problem under consideration. 
Indeed, the electron - rotated-electron unitary operator ${\hat{V}}$ commutes
with the three generators of both the global $\eta$-spin and spin $SU(2)$ symmetry algebras and the charge density
operator. As a result, the $\sigma$ rotated electrons have the same charge, spin $1/2$, and $\eta$-spin $1/2$ degrees of freedom as the 
$\sigma$ electrons. Application of the operator ${\hat{V}}$ onto the $\sigma$ electron operators merely changes the 
$\sigma$ electrons lattice spatial occupancy configurations.
On the other hand, from analysis of the relation between (i) the BA quantum numbers and (ii) rotated-electron 
occupancy configurations, respectively, that generate the finite-$u$ exact energy eigenstates 
$\vert  l_{\rm r},l_{\eta s},u\rangle={\hat{V}}^{\dag}\vert  l_{\rm r},l_{\eta s},\infty \rangle$
of any $V$ tower one reaches important physical information. 

First, the $\sigma$ rotated-electron spatial occupancy configurations that generate the finite-$u$ energy 
eigenstates $\vert  l_{\rm r},l_{\eta s},u\rangle={\hat{V}}^{\dag}\vert  l_{\rm r},l_{\eta s},\infty \rangle$ of any 
$V$ tower are exactly the same as the $\sigma$ electron spatial occupancy configurations of the tower
$u\rightarrow\infty$ energy eigenstate $\vert  l_{\rm r},l_{\eta s},\infty \rangle$. Hence for $u>0$ the number 
$N^R_{s,\pm 1/2}$ of spin-projection $\pm 1/2$ rotated-electron singly occupied sites, $N^R_{\eta,+1/2}$
of rotated-electron unoccupied sites, and $N^R_{\eta,-1/2}$ of rotated-electron doubly occupied sites are conserved. 
Such numbers obey the sum rules $N^R_{s,\pm 1/2}+N^R_{\eta,-1/2}=N_{\pm 1/2}$, $N^R_{s}+2N^R_{\eta,-1/2}=N$,
and $N^R_{s}+N^R_{\eta}=L$. The $\sigma $ rotated-electron numbers values equal those of the $\sigma $ electrons,
so that here $N_{\pm 1/2}$ denotes the number of electrons and rotated electrons of
spin projection $\pm 1/2$. On the other hand, for finite $u$ values the numbers $N^R_{s}=N^R_{s,+1/2}+N^R_{s,-1/2}$ 
of rotated-electron singly occupied sites and $N^R_{\eta}=N^R_{\eta,+1/2}+N^R_{\eta,-1/2}$ of rotated-electron doubly 
occupied plus unoccupied sites are only conserved for rotated electrons.

Second, for $u>0$ a non-perturbative three degrees of freedom lattice - $\eta$-spin - spin separation occurs at all energy scales. Here the 
lattice - $\eta$-spin degrees of freedom separation may be considered as a separation of the charge degrees of freedom. At 
energy scales lower than $2\vert\mu\vert$ one has that $D=N^R_{\eta,-1/2}=0$ (and $N^R_{\eta,+1/2}=0$) for $n_e \in [0,1[$ (and
$n_e \in ]1,2]$), so that the $\eta$-spin degrees of freedom are hidden and the three degrees of freedom
non-perturbative lattice - $\eta$-spin - spin separation is seen as the usual two degrees of freedom charge - spin 
separation. Within the former general separation the (i) lattice global $U(1)$ symmetry, (ii) $\eta$-spin global $SU(2)$
symmetry, and (iii) spin global $SU(2)$ symmetry state representations are in each fixed number $N^R_{s}$
of rotated-electron singly occupied sites subspace generated by the occupancy configurations of (i) 
$N_c=N^R_{s}$ spinless {\it $c$ pseudoparticles} and corresponding $N_c^h=N^R_{\eta}$ $c$ pseudoparticle holes
whose $c$ effective lattice is identical to the original lattice and thus has $N^R_{s}+N^R_{\eta}=L$ sites,
(ii) $M_{s,\pm 1/2} = N^R_{s,\pm 1/2}$ spin-$1/2$ fractionalized particles of spin projection $\pm 1/2$
that we call {\it rotated spins $1/2$}, and (iii) $M_{\eta,\pm 1/2} = N^R_{\eta,+1/2}$ $\eta$-spin-$1/2$ fractionalized 
particles of $\eta$-spin projection $\pm 1/2$ that we call {\it rotated $\eta$-spins $1/2$},
respectively. ($+1/2$ and $-1/2$ $\eta$-spin projections refer to $\eta$-spin degrees of freedom
of rotated-electron unoccupied and doubly occupied sites, respectively.) It then follows that these numbers are such that,
\begin{eqnarray}
&& M_s = M_{s,+1/2} + M_{s,-1/2} = N_c \, ,
\nonumber \\
&& M_{\eta} = M_{\eta,+1/2} + M_{\eta,-1/2} = L - N_c = N_c^h \, ,
\nonumber \\
&& M_{s,+1/2} - M_{s,-1/2} = -2S_s^z = N_{\uparrow} - N_{\downarrow} \, ,
\nonumber \\
&& M_{\eta,+1/2} - M_{\eta,-1/2} = -2S_{\eta}^z = L - N \, ,
\label{severalM}
\end{eqnarray}
where $M_s$ denotes the number of rotated spins and $M_{\eta}$ that of rotated $\eta$-spins, which
equal those $N_c$ of $c$ pseudoparticles and $N_c^h = L - N_c$ of $c$ pseudoparticle holes, respectively.
Consistently with the $N_c$ $c$ pseudoparticles, $M_{\eta,\pm 1/2}$ rotated $\eta$-spins of $\eta$-spin projection $\pm 1/2$, 
and $M_{s,\pm 1/2}$ rotated spins of spin projection $\pm 1/2$ under consideration stemming from rotated-electron occupancy 
configurations degrees of freedom separation, their numbers are fully controlled by those of rotated electrons as follows, 
\begin{eqnarray}
&& N_ c = N_{R}^{s} \, ; \hspace{0.5cm} N_ c^h = N_{R}^{\eta} \, ; \hspace{0.5cm}
N_ c + N_ c^h = N_{R}^{s} + N_{R}^{\eta} = L \, ,
\nonumber \\
&& M_{\alpha,\pm 1/2} = N_{R,\pm 1/2}^{\alpha} \, ; \hspace{0.5cm}
M_{\alpha} = M_{\alpha,+1/2} + M_{\alpha,-1/2} = N_{R}^{\alpha} \, , \hspace{0.50cm} \alpha=\eta, s \, .
\label{NRpm}
\end{eqnarray}
Indeed the degrees of freedom of each rotated-electron occupied site decouple into one spinless $c$ pseudoparticle that 
carries the rotated-electron charge and one rotated spin $1/2$ that carries its spin. On the other hand, the degrees of freedom of 
each rotated-electron unoccupied and doubly occupied site decouple into one $c$ pseudoparticle
hole and one rotated $\eta$-spin $1/2$ of projection $+1/2$ and $-1/2$, respectively. Hence the rotated-electron
on-site separation refers to two degrees of freedom associated with the lattice global $U(1)$ symmetry and
one of the two global $SU(2)$ symmetries, respectively. That the rotated-electron occupancy configurations give rise to the independent 
occupancy configurations of the $c$ pseudoparticles, rotated spins $1/2$, and rotated $\eta$-spins $1/2$ is behind the 
exotic properties of the corresponding quantum liquid. 

Third, from the further analysis of the relation between the BA quantum numbers and the three degrees of freedom 
separation of the rotated-electron occupancy configurations one finds that such quantum numbers are
directly associated with the occupancy configurations of the three types of fractionalized particles that generate
all $4^L$ energy eigenstates, Eq. (\ref{Gstate-BAstate}). For the densities ranges $n_e\in [0,1]$ and $m\in [0,n_e]$ one 
has that $N^R_{s,+1/2}\geq N^R_{s,-1/2}$ and $N^R_{\eta,+1/2}\geq N^R_{\eta,-1/2}$. For the corresponding exact 
Bethe states, there is a number $M_{s\,{\rm sp}}=N^R_{s,-1/2}$ of spin-singlet pairs $(\alpha =s)$ and $M_{\eta\,{\rm sp}}=N^R_{\eta,-1/2}$ 
of $\eta$-spin-singlet pairs $(\alpha =\eta)$ within which all $N^R_{s,-1/2}$ rotated spins of projection $-1/2$ are paired
with an equal number of rotated spins of projection $+1/2$ and all $N^R_{\eta,-1/2}$ rotated $\eta$-spins of projection $-1/2$ are paired
with an equal number of rotated $\eta$-spins of projection $+1/2$, respectively. Such $M_{\alpha\,{\rm sp}}$ spin-singlet $(\alpha =s)$ 
and $\eta$-spin-singlet $(\alpha =\eta)$ pairs are found to be contained in a set of composite $\alpha n$ pseudoparticles. 
Here $n=1,...,\infty$ gives the number of pairs that refer to their internal structure. One denotes by $N_{\alpha n}$ the number of such 
$\alpha n$ pseudoparticles in each energy and momentum eigenstate. 
The sum rule $M_{\alpha\,{\rm sp}} = \sum_{n=1}^{\infty}n\,N_{\alpha n}$ is then obeyed. 

The remaining $M^{un}_{\alpha}=N^R_{\alpha,+1/2}- N^R_{\alpha,-1/2}=2S_{\alpha}$ unpaired rotated spins $(\alpha =s)$
and rotated $\eta$-spins $(\alpha =\eta)$ have for a Bethe state spin and $\eta$-spin projection $+1/2$. For general
energy eigenstates the configurations of these $2S_s$ unpaired rotated spins and $2S_{\eta}$ unpaired rotated $\eta$-spins 
generate the spin and $\eta$-spin, respectively, towers of non-LWSs. Specifically, the $2S_s$ unpaired rotated spins and 
$2S_{\eta}$ unpaired rotated $\eta$-spins of the Bethe states are flipped upon the application of the corresponding $SU(2)$ 
algebras off-diagonal generators, as given in Eq. (\ref{Gstate-BAstate}). Application of such generators leaves the spin 
$(\alpha =s)$ and $\eta$-spin $(\alpha =\eta)$ singlet configurations of the $M_{\alpha\,{\rm sp}} = \sum_{n=1}^{\infty}n\,N_{\alpha n}$
pairs contained in $\alpha n$ pseudoparticles unchanged. Hence for general $u>0$ energy eigenstates one finds that the 
number $M^{un}_{s,\pm 1/2}$ of unpaired rotated spins of projection $\pm 1/2$ and $M^{un}_{\eta,\pm 1/2}$
of unpaired rotated $\eta$-spins of projection $\pm 1/2$ are good quantum numbers, which read,
\begin{equation}
M^{un}_{\alpha,\pm 1/2} = (S_{\alpha}\mp S_{\alpha}^{z}) \, ;
\hspace{0.50cm}
M^{un}_{\alpha} = M^{un}_{\alpha,-1/2}+M^{un}_{\alpha,+1/2} = 2S_{\alpha} 
\, , \hspace{0.50cm} \alpha = \eta,s \, .
\label{L-L}
\end{equation}
For the $\alpha =\eta,s$ LWSs one has that $M^{un}_{\alpha,+1/2} = M^{un}_{\alpha} = 2S_{\alpha}$ and 
$M^{un}_{\alpha,-1/2} = 0$ for both $\alpha=\eta, s$. The set of $M_{\eta\,{\rm sp}}$ $\eta$-spin-singlet pairs
and $M_{s\,{\rm sp}}$ spin-singlet pairs of an energy eigenstate contains an equal number of rotated
$\eta$-spins and rotated spins, respectively, of opposite projection. Hence the total numbers $M_{\eta,\pm 1/2}$
of rotated $\eta$-spins of projection $\pm 1/2$ and $M_{s,\pm 1/2}$ of rotated spins of projection $\pm 1/2$ read,
\begin{equation}
M_{\alpha,\pm 1/2} = M_{\alpha\,{\rm sp}} + M^{un}_{\alpha,\pm 1/2} \, , \hspace{0.50cm} \alpha = \eta,s \, .
\label{Mtotal}
\end{equation}

Moreover, by combining the above equations one finds that the set of numbers $\{N_{\alpha n}\}$ of composite $\alpha n$ 
pseudoparticles of any $u>0$ energy eigenstate obey the following exact sum rules concerning the number of 
$M_{\alpha\,{\rm sp}}$ of spin $(\alpha =s)$ and $\eta$-spin $(\alpha =\eta)$ singlet pairs of any $u>0$ energy eigenstate,
\begin{eqnarray}
M_{\alpha\,{\rm sp}} & = & \sum_{n=1}^{\infty}n\,N_{\alpha n} = {1\over 2}(L_{\alpha} - 2S_{\alpha})
\, , \hspace{0.50cm} \alpha = s, \eta \, ,
\nonumber \\
M_{\rm sp}^{SU(2)} & \equiv & \sum_{\alpha =\eta,s}\sum_{n=1}^{\infty}n\,N_{\alpha n} = {1\over 2}(L - 2S_s - 2S_{\eta}) \, ,
\label{sum-Nseta}
\end{eqnarray}
where $M_{\rm sp}^{SU(2)}$ denotes the total number of both rotated spins and rotated $\eta$-spins pairs. 

The BA solution contains different types of quantum numbers whose occupancy configurations are
within the pseudoparticle representation described by corresponding occupancy configurations
of spinless $c$ pseudoparticles with no internal structure and composite $\alpha n$ pseudoparticles. 
Complete information on the microscopic details of the latter pseudoparticles internal $\eta$-spin 
$(\alpha =\eta)$ and spin $(\alpha =s)$ $n$-pair configurations is encoded within the BA solution
and is not needed for the goals and studies of this paper. Indeed, within the present TL the problem 
concerning a $\alpha n$ pseudoparticle internal degrees of freedom and that associated with its 
translational degrees of freedom center of mass motion separate. 
Here we merely provide some general information on the internal degrees of freedom issue, which
as further discussed below involves the imaginary part of the BA complex rapidities \cite{Takahashi},
\begin{equation}
\Lambda^{\alpha n,l}(q_{j}) = \Lambda^{\alpha n} (q_{j}) + i\,(n + 1 - 2l)\,u \, , \hspace{0.50cm} l = 1,...,n \, ,
\label{complex-rap}
\end{equation}
where $\alpha = \eta, s$ and $n =1,...,\infty$. The corresponding number $L_{\alpha n}$ of the set $j = 1,...,L_{\alpha n}$ of 
the $\alpha n$ branch BA quantum numbers $\{q_j\}$ and that $L_c$ of the related set $j = 1,...,L_c$ of the $c$ branch 
BA quantum numbers $\{q_j\}$ are given by,
\begin{eqnarray}
L_{\alpha n} & = & N_{\alpha n} + N^h_{\alpha n} \, ; \hspace{0.50cm}
N^h_{\alpha n} = 2S_{\alpha}+\sum_{n'=n+1}^{\infty}2(n'-n)N_{\alpha n'} \, , 
\hspace{0.50cm} \alpha=\eta, s \, , \hspace{0.50cm} n =1,...,\infty \, ,
\nonumber \\
L_{c} & = & N_{c} + N^h_{c} = N^R_{s} + N^R_{\eta} = L \, ,
\label{N-h-an}
\end{eqnarray}
respectively. The real part $\Lambda^{\alpha n}(q_{j})$ of the complex rapidities, Eq. (\ref{complex-rap}), is a function
of the $j = 1,...,L_{\alpha n}$ quantum numbers $q_j$ that has the same value
for the whole set $l = 1,...,n$ of $\alpha n$ rapidities. It is the rapidity function which for each $u>0$ energy eigenstate is
the solution of the BA equations introduced in Ref. \cite{Takahashi} for the TL. Within the pseudoparticle momentum 
distribution functional notation \cite{Carmelo-04}, these equations have the form given in Eqs. (\ref{Tapco1}) and (\ref{Tapco2})
of Appendix \ref{Ele2PsPhaShi} where the sets of $j = 1,...,L_c$ and $j = 1,...,L_{\alpha n}$ of quantum numbers $q_j$, respectively, 
read,
\begin{equation}
q_j = {2\pi\over L}\,I^{\beta}_j  \, , \hspace{0.50cm} j=1,...,L_{\beta} 
\, , \hspace{0.50cm}  \beta = c,\eta n,sn \, , \hspace{0.50cm}  n =1,...,\infty \, .
\label{q-j}
\end{equation} 
These play the role of $\beta = c,\alpha n$ band microscopic momentum values of 
the $\beta = c,\alpha n$ pseudoparticle branches. For a given energy and momentum eigenstate, the $j=1,...,L_{\beta}$ quantum numbers 
$I^{\beta}_j$ on the right-hand side of Eq. (\ref{q-j}) are either integers or half-odd integers according to the following boundary 
conditions \cite{Takahashi},
\begin{eqnarray}
I_j^{\beta} & = & 0,\pm 1,\pm 2,... \hspace{0.50cm}{\rm for}\hspace{0.15cm}I_{\beta}\hspace{0.15cm}{\rm even} \, ,
\nonumber \\
& = & \pm 1/2,\pm 3/2,\pm 5/2,... \hspace{0.50cm}{\rm for}\hspace{0.15cm}I_{\beta}\hspace{0.15cm}{\rm odd} \, .
\label{Ic-an}
\end{eqnarray}
Here the numbers $I_{\beta}$ are given by,
\begin{eqnarray}
I_c & = & N_{\rm ps}^{SU(2)} \equiv \sum_{\alpha =\eta,s}\sum_{n=1}^{\infty}N_{\alpha n} \, ,
\nonumber \\
I_{\alpha n} & = & L_{\alpha n} -1 \, , \hspace{0.50cm}
\alpha = \eta, s \, , \hspace{0.50cm} n=1,...,\infty \, .
\label{F-beta}
\end{eqnarray}

From analysis of the BA quantum numbers, one finds that the set of numbers $\{N_{\alpha n}\}$ of composite $\alpha n$ 
pseudoparticles obey a second exact sum rule in addition to the spin $(\alpha =s)$ and $\eta$-spin $(\alpha =\eta)$ singlet 
pairs sum rule given in Eq. (\ref{sum-Nseta}). It is associated with the
value of the total number $N_{\alpha\,{\rm ps}} = \sum_{n=1}^{\infty}N_{\alpha n}$ of composite $\alpha n$ pseudoparticles of all
$n=1,...,\infty$ branches of a $u>0$ energy eigenstate and reads,
\begin{eqnarray}
N_{s\,{\rm ps}} & = & \sum_{n=1}^{\infty}N_{s n} = {1\over 2}(N_c - N_{s 1}^h) \, ,
\nonumber \\
N_{\eta\,{\rm ps}} & = & \sum_{n=1}^{\infty}N_{\eta n} = {1\over 2}(N_c^h - N_{\eta 1}^h) \, ,
\nonumber \\
N_{\rm ps}^{SU(2)} & = & \sum_{\alpha =\eta,s}\sum_{n=1}^{\infty}N_{\alpha n} = {1\over 2}(L - N_{s 1}^h - N_{\eta 1}^h) \, . 
\label{NpsNapsSR}
\end{eqnarray}
Here $N_{\rm ps}^{SU(2)}$ is the number of both $\alpha =\eta$ and $\alpha =s$ composite $\alpha n$ pseudoparticles 
of all $n=1,...,\infty$ branches also appearing in Eq. (\ref{F-beta}) and $N_{\alpha 1}^h$ is that of $\alpha 1$-band holes, 
Eq. (\ref{N-h-an}) for $\alpha = \eta,s$ and $n =1$. The interesting point is that for given fixed $N_c$ and
thus $N_c^h=L-N_c$ values that of $N_{\alpha\,{\rm ps}}$ is fully determined by the corresponding value of 
the number $N_{\alpha 1}^h$ of $\alpha 1$-band holes.

The $\beta = c,\alpha n$ band successive set $j = 1,...,L_{\beta}$ of momentum values $q_j$, Eq. (\ref{q-j}), 
have only $\beta$ pseudoparticle occupancies zero and one and the usual separation, $q_{j+1}-q_{j}=2\pi/L$. That 
they play the role of $\beta = c,\alpha n$ band momentum values is consistent with within our functional
representation the momentum eigenvalues of all $u>0$ energy and momentum eigenstates reading,
\begin{equation}
P =\sum_{j=1}^{L} q_j\, N_c (q_j)
+ \sum_{n =1}^{\infty}\sum_{j=1}^{L_{s n}}
q_{j}\, N_{sn} (q_{j}) 
+ \sum_{n =1}^{\infty}\sum_{j=1}^{L_{\eta n}}
(\pi -q_{j})\, N_{\eta n} (q_{j}) + \pi M_{\eta,-1/2} \, ,
\label{P}
\end{equation}
being thus additive in $q_j$. Within that representation, the $\beta$-band momentum distribution 
functions $N_{\beta} (q_j)$ in Eq. (\ref{P}) and BA equations, Eqs. (\ref{Tapco1}) and (\ref{Tapco2}) of Appendix \ref{Ele2PsPhaShi}, 
read $N_{\beta} (q_j)=1$ and $N_{\beta} (q_j)=0$ for occupied and unoccupied discrete momentum values, 
respectively. The momentum contribution $\pi M_{\eta,-1/2}$, which from the use of Eq. (\ref{Mtotal}) can be written as
$\pi (M_{\eta\,{\rm sp}} + M^{un}_{\eta,-1/2})$, follows from both the paired and unpaired rotated spins $1/2$ and 
rotated $\eta$-spins $1/2$ of projection $\pm 1/2$ having a momentum given by,
\begin{equation}
q_{s,\pm 1/2} = q_{\eta,+1/2} = 0  \, ; \hspace{0.75cm} q_{\eta,-1/2} = \pi \, .
\label{q-eta-s}
\end{equation}
On the other hand, the $\eta n$ pseudoparticle contribution $(\pi -q_{j})$ to the momentum eigenvalue, Eq. (\ref{P}), refers to
its translational degrees of freedom. It is associated with the center of mass motion of that composite $n$-pair object 
as a whole. That such a contribution to the momentum eigenvalue reads $(\pi -q_{j})$ rather than $q_j$, as
for the $c$ and $sn$ pseudoparticles, follows from the $2n$-rotated-$\eta$-spin configuration of a 
composite $\eta n$ pseudoparticle having an anti-bounding character, as confirmed below in Section \ref{BANTIB}.

The $c$ band is populated by $N_c=N^R_{s}$ $c$ pseudoparticles. They occupy $N_c$ discrete momentum
values out of the $c$ band $j=1,...,L_c$ such momentum values, where $L_c = L$. Hence the number of $c$ pseudoparticle 
holes indeed reads $N_c^h=N^R_{\eta}=L-N^R_{s}$. On the other hand, the number $L_{\alpha n}$ in Eq. (\ref{N-h-an})
refers to that of $\alpha n$ band $j=1,...,L_{\alpha n}$ momentum values $q_j$ in Eq. (\ref{q-j}).
For an energy and momentum eigenstate each such bands is populated by a well defined number
$N_{\alpha n}$ of $\alpha n$ pseudoparticles, so that the corresponding number $N_{\alpha n}^h$ of  
$\alpha n$ pseudoparticle holes is that given in Eq. (\ref{N-h-an}). 

The set $j=1,...,L_{\beta}$ of $\beta =c, \alpha n$ bands discrete momentum values $q_j$ whose different occupancy
configurations generate the energy and momentum eigenstates and determine the corresponding
momentum eigenvalues, Eq. (\ref{P}), belong to well-defined domains, $q_j\in [q_{\beta}^-,q_{\beta}^+]$, where,
\begin{eqnarray}
q_{c}^{\pm} & =  & \pm {\pi\over L}(L-1) \approx \pm\pi \hspace{0.15cm}{\rm for}\hspace{0.15cm}N_{\rm ps}^{SU(2)}\hspace{0.15cm}{\rm odd} \, ;
\hspace{0.35cm}
q_{c}^{\pm} = \pm {\pi\over L}(L-1\pm 1) \approx \pm\pi \hspace{0.15cm}{\rm for}\hspace{0.15cm}N_{\rm ps}^{SU(2)}\hspace{0.15cm}{\rm even} \, ,
\nonumber \\
q_{\alpha n}^{\pm} & = & \pm {\pi\over L}(L_{\alpha n}-1) \, .
\label{qcan-range}
\end{eqnarray}

The label $l_{\rm r}$ in the energy eigenstates $\{\vert l_{\rm r},l_{\eta s},u\rangle\}$, Eq. (\ref{Gstate-BAstate}),
can now be defined. It corresponds to a short notation for the following set of BA quantum numbers, 
\begin{equation}
l_{\rm r} = \{I_j^{\beta}\}\,\,{\rm such}\,\,{\rm that}\,\,N_{\beta} (q_j) =1\,\,{\rm where}\,\,
q_j = {2\pi\over L}I_j^{\beta}\,\,{\rm for}\hspace{0.15cm} j = 1,...,L_{\beta} \, , \hspace{0.15cm} \beta = c, \eta n, sn
\, , \hspace{0.15cm} n = 1,...,\infty \, ,
\label{states-ll}
\end{equation}

Ground states are neither populated by composite $sn$ pseudoparticles with $n>1$ spin-singlet pairs
nor by composite $\eta n$ pseudoparticles with any number $n=1,...,\infty$ of $\eta$-spin-singlet pairs.
For electronic densities $n_e \in [0,1]$ and spin densities $m \in [0,n_e]$, ground states have 
no unpaired rotated spins of projection $-1/2$ and no unpaired rotated $\eta$-spins of projection $-1/2$. 
For them the number $M^{un}_{s}=N^R_{s}=2S_{s}$ of unpaired rotated spins of projection $+1/2$ 
and the number $M^{un}_{\eta}=N^R_{\eta}=2S_{\eta}$ of unpaired rotated $\eta$-spins of projection $+1/2$
equal those $N_{s1}^h =N^R_{s}=2S_{s}$ of $s1$ pseudoparticle holes and $N_c^h=N^R_{\eta}=2S_{\eta}$
of $c$ pseudoparticle holes, respectively. Indeed, within the pseudoparticle representation the 
unpaired rotated spins and unpaired rotated $\eta$-spins play the role of empty sites of the
$c$ effective lattice and squeezed $s1$ effective lattice, respectively, considered in Section \ref{pseuOpRel}.
Hence they are implicitly accounted for by the pseudoparticle representation.

The ground-state $\beta$ band pseudoparticle momentum distribution functions are given by,
\begin{equation}
N_c^0 (q_j) = \theta (q_j - q_{Fc}^{-})\,\theta (q_{Fc}^{+} - q_j)  
\, ; \hspace{0.75cm} 
N_{s 1}^0 (q_j) = \theta (q_j - q_{Fs1}^{-})\,\theta (q_{Fs1}^{+} - q_j)  
\, ; \hspace{0.75cm} 
N_{\alpha n}^0 (q_j) = 0 \, , \hspace{0.50cm} \alpha n \neq s1 \, ,
\label{N0q1DHm}
\end{equation}
where the distribution $\theta (x)$ reads $\theta (x)=1$ for $x> 0$ and 
$\theta (x)=0$ for $x\leq 0$. For the $c$ and $s1$ bands the momentum distribution functions,
Eq. (\ref{N0q1DHm}), refer to compact and symmetrical occupancy configurations. The corresponding
$\beta =c,s1$ Fermi points are associated with the Fermi momentum values $q_{F\beta}^{\pm}$ in Eq. (\ref{N0q1DHm}).
Accounting for ${\cal{O}} (1/L)$ corrections, they are given in Eqs. (C.4)-(C.11) of Ref. \cite{Carmelo-04}. 
If within the TL we ignore such corrections, one finds that $N_{\beta}^0 (q_j) = \theta (q_{F\beta} - \vert q_j\vert)$ 
for $\beta =c,s1$ where the Fermi momentum values are given by,
\begin{equation}
q_{Fc} = 2k_F = \pi\,n_e \, ; \hspace{0.75cm} q_{Fs1} = k_{F\downarrow} = \pi\,n_{e\downarrow} \, . 
\label{q0Fcs}
\end{equation}

The $c$ pseudoparticles have no internal structure. On the other hand, the imaginary part $i\,(n + 1 - 2l)\,u$ 
of the set of $l = 1,...,n$ complex rapidities, Eq. (\ref{complex-rap}), with the same real part 
$\Lambda^{\alpha n} (q_{j})$ refers to the internal degrees of freedom of one
composite $\alpha n$ pseudoparticle with $n>1$ pairs whose center of mass carries $\alpha n$ band
momentum $q_j$. Specifically, for $\alpha =s$ the imaginary part of such $l = 1,...,n$ rapidities 
is associated with a corresponding set $l = 1,...,n$ of spin-singlet pairs of rotated spins $1/2$ and the
binding of these pairs within the composite $sn$ pseudoparticle. For $\alpha =\eta$ it is rather associated 
with a set $l = 1,...,n$ of $\eta$-spin-singlet pairs of rotated $\eta$-spins $1/2$ and the
anti-binding of these pairs within the composite $\eta n$ pseudoparticle. Each such $l = 1,...,n$ rapidities thus refers to one of
the $l = 1,...,n$ singlet pairs bound and anti-bound within the composite $sn$ and $\eta n$ pseudoparticle, respectively. 
For $n =1$ the rapidity imaginary part vanishes. Indeed, the $\alpha 1$ pseudoparticle internal degrees of freedom 
refer to a single singlet pair of rotated spins $1/2$ ($\alpha =s$) or rotated $\eta$-spins $1/2$ ($\alpha =\eta$). 

Below in Section \ref{BANTIB} the form of the composite $s n$ and $\eta n$ pseudoparticle energy dispersions is used
to extract valuable information on the bounding and anti-bounding character of their 
$2n=2,4,...$ paired rotated spins and paired rotated $\eta$-spins configuration, respectively.

\subsection{The $c$ pseudoparticle, rotated spin, and rotated $\eta$-spin operators in terms
of $\sigma$ rotated-electron operators}
\label{pseuOpRel}

That the $c$ pseudoparticles, rotated spins $1/2$, and rotated $\eta$-spins $1/2$ naturally emerge
from the $\sigma$ rotated-electron onsite occupancy configurations separation allows the introduction
of local operators for these fractionalized particles in terms of the local rotated-electron creation
and annihilation operators, Eq. (\ref{rotated-operators}). 

The simplest case refers to the $l = z, \pm$ local operators associated with the rotated spins $1/2$ ($\alpha =s$) and
rotated $\eta$-spins $1/2$ ($\alpha =\eta$), which read,
\begin{eqnarray}
{\tilde{S}}^{l}_{j,\alpha} & = & {\hat{V}}^{\dag}\,{\hat{S}}^{l}_{j,\alpha}\,{\hat{V}} \, , \hspace{0.50cm} l = z, \pm 
\, , \hspace{0.50cm} \alpha = \eta, s \, ,
\nonumber \\
{\tilde{S}}^{\pm}_{j,\alpha} & = & {\tilde{S}}^{x}_{j,\alpha}\pm i\,{\tilde{S}}^{y}_{j,\alpha} \, , \hspace{0.50cm} \alpha = \eta, s \, ,
\label{inf-exp-OS}
\end{eqnarray}
where ${\hat{S}}^{l}_{j,\alpha}$ are the usual unrotated $l = z, \pm$ local spin ($\alpha =s$) and 
$\eta$-spin ($\alpha =\eta$) operators. The $l = z, \pm$ local operators ${\tilde{S}}^{l}_{j,\alpha}$, Eq. (\ref{inf-exp-OS}),
have in terms of creation and annihilation $\sigma$ rotated-electron operators, Eq. (\ref{rotated-operators}), 
exactly the same expressions as the corresponding unrotated $l = z, \pm$ local operators ${\hat{S}}^{l}_{j,\alpha}$ in terms of creation 
and annihilation $\sigma$ electron operators. 

Specifically, the spin operators ${\tilde{S}}^{l}_{j,s}$, which act onto sites singly occupied by $\sigma$ rotated electrons, read
${\tilde{S}}^-_{j,s} = ({\tilde{S}}^+_{j,s})^{\dag} = {\tilde{c}}_{j,\uparrow}^{\dag}{\tilde{c}}_{j,\downarrow}$
and ${\tilde{S}}^z_{j,s} = ({\tilde{n}}_{j,\downarrow} - 1/2)$. Similarly, the $\eta$-spin operators 
${\tilde{S}}^{l}_{j,\eta}$, which act onto sites unoccupied by rotated electrons and sites doubly occupied by rotated electrons, are
given by ${\tilde{S}}^-_{j,\eta} = ({\tilde{S}}^+_{j,\eta})^{\dag} = (-1)^j\,{\tilde{c}}_{j,\uparrow}{\tilde{c}}_{j,\downarrow}$
and ${\tilde{S}}^z_{j,\eta} = ({\tilde{n}}_{j,\downarrow} - 1/2)$. 

Below the $c$ pseudoparticle creation operator $f_{j,c}^{\dag}$ and annihilation operator $f_{j,c}$ on the lattice site $j=1,...,L$ are 
uniquely defined in terms of the local rotated-electron creation and annihilation operators, Eq. (\ref{rotated-operators}). 
(The $c$ effective lattice considered below is identical to the original lattice.)
The $c$ pseudoparticles have inherently emerged from the $\sigma$ rotated electrons to the sites singly occupied by the latter
being occupied by $c$ pseudoparticles and those unoccupied and doubly occupied by rotated electrons 
being unoccupied by $c$ pseudoparticles. Hence the $c$ pseudoparticle local density operator
${\tilde{n}}_{j,c} \equiv f_{j,c}^{\dag}\,f_{j,c}$ and the corresponding operator $(1-{\tilde{n}}_{j,c})$ are the natural projectors 
onto the subset of $N_{R}^{s}=N_c$ original-lattice sites singly occupied by rotated electrons and
onto the subset of $N_{R}^{\eta}=N_c^h=L-N_c$ original-lattice sites unoccupied and doubly occupied by rotated electrons,
respectively. It then follows that the $\alpha =s,\eta$ and $l = z, \pm$ local operators ${\tilde{S}}^{l}_{j,\alpha}$,
Eq. (\ref{inf-exp-OS}), can be written as,
\begin{equation}
{\tilde{S}}^l_{j,s} = {\tilde{n}}_{j,c}\,{\tilde{q}}^l_{j} 
\, ; \hspace{0.75cm}
{\tilde{S}}^l_{j,\eta} = (1-{\tilde{n}}_{j,c})\,{\tilde{q}}^l_{j}
\, , \hspace{0.50cm} l = z, \pm \, ,
\label{sir-pirG}
\end{equation}
respectively, where the $l = z, \pm$ local {\it $\eta s$ quasi-spin} operators,
\begin{equation}
{\tilde{q}}^l_{j} = {\tilde{S}}^l_{j,s} + {\tilde{S}}^l_{j,\eta} \, ,
\hspace{0.50cm} l=\pm,z \, ,
\label{q-operG}
\end{equation}
such that ${\tilde{q}}^{\pm}_{j}= {\tilde{q}}^{x}_{j}\pm i\,{\tilde{q}}^{y}_{j}$, 
have the following expression in terms
of $\sigma$ rotated-electron creation and annihilation operators,
\begin{equation}
{\tilde{q}}^-_{j} = ({\tilde{q}}^+_{j})^{\dag} = 
({\tilde{c}}_{j,\uparrow}^{\dag}
+ (-1)^j\,{\tilde{c}}_{j,\uparrow})\,
{\tilde{c}}_{j,\downarrow} 
\, ; \hspace{0.75cm}
{\tilde{q}}^{z}_{j} = ({\tilde{n}}_{j,\downarrow} - 1/2) \, .
\label{rotated-quasi-spinG}
\end{equation}

The $N_c$ $c$ pseudoparticles live on the $N_{R}^{s}=N_c$ sites singly occupied by the rotated 
electrons, so that their occupancy configurations refer to the lattice degrees of freedom associated
with the relative positions of the $M_s=N_{R}^{s}=N_c$ sites occupied by rotated spins $1/2$ and
$M_{\eta}=N_{R}^{\eta} =N_c^h = L-N_c$ sites occupied by rotated $\eta$-spins $1/2$. The corresponding
three degrees of freedom separation of the $\sigma$ rotated-electron occupancy configurations
then implies that their operators, Eq. (\ref{rotated-operators}), can be written as,
\begin{eqnarray}
{\tilde{c}}_{j,\uparrow}^{\dag} & = &
\left({1\over 2} - {\tilde{S}}^{z}_{j,s} - {\tilde{S}}^{z}_{j,\eta}\right)f_{j,c}^{\dag} + (-1)^j
\left({1\over 2} +{\tilde{S}}^{z}_{j,s} + {\tilde{S}}^{z}_{j,\eta}\right)f_{j,c} \, ;
\hspace{0.5cm} {\tilde{c}}_{j,\uparrow} = ({\tilde{c}}_{j,\uparrow}^{\dag})^{\dag} \, ,
\nonumber \\
{\tilde{c}}_{j,\downarrow}^{\dag} & = &
({\tilde{S}}^{+}_{j,s} + {\tilde{S}}^{+}_{j,\eta})(f_{j,c}^{\dag} + (-1)^j\,f_{j,c}) \, ,
\hspace{0.5cm}
{\tilde{c}}_{j,\downarrow} = ({\tilde{c}}_{j,\downarrow}^{\dag})^{\dag} \, .
\label{c-up-c-downG}
\end{eqnarray}

The local $c$ pseudoparticle operators $f_{j,c}^{\dag}$ and $f_{j,c}$ appearing here 
are then {\it uniquely} defined for $u>0$ in terms of $\sigma$ rotated-electron creation and 
annilihation operators, Eq. (\ref{rotated-operators}), by combining the inversion of the relations, Eq. (\ref{c-up-c-downG}), 
with the expressions of the $l = z, \pm$ local operators ${\tilde{S}}^{l}_{j,\alpha}$ 
associated with the rotated spins $1/2$ ($\alpha =s$) and rotated $\eta$-spins $1/2$ ($\alpha =\eta$), Eq. (\ref{inf-exp-OS}),
provided in Eqs. (\ref{sir-pirG})-(\ref{rotated-quasi-spinG}), which gives, 
\begin{equation}
f_{j,c}^{\dag} = (f_{j,c})^{\dag} = {\tilde{c}}_{j,\uparrow}^{\dag}\,
(1-{\tilde{n}}_{j,\downarrow}) + (-1)^j\,{\tilde{c}}_{j,\uparrow}\,{\tilde{n}}_{j,\downarrow} 
\, ; \hspace{0.50cm} {\tilde{n}}_{j,c} = f_{j,c}^{\dag}\,f_{j,c} \, , \hspace{0.50cm} j = 1,...,L \, ,
\label{fc+G}
\end{equation}
where ${\tilde{n}}_{j,\sigma}$ is the $\sigma$ rotated-electron local density operator
given in Eq. (\ref{rotated-operators}). 

The unitarity of the electron - rotated-electron transformation implies that the rotated-electron operators 
${\tilde{c}}_{j,\sigma}^{\dag}$ and ${\tilde{c}}_{j,\sigma}$, Eqs. (\ref{rotated-operators}) and (\ref{c-up-c-downG}),
have the same anticommutation relations as the corresponding electron 
operators $c_{j,\sigma}^{\dag}$ and $c_{j,\sigma}$, respectively. 
Straightforward manipulations based on Eqs. (\ref{inf-exp-OS})-(\ref{fc+G}) then lead
to the following algebra for the local $c$ pseudoparticle creation and annihilation operators,
\begin{equation}
\{f^{\dag}_{j,c}\, ,f_{j',c}\} = \delta_{j,j'} ;
\hspace{0.75cm}
\{f_{j,c}^{\dag}\, ,f_{j',c}^{\dag}\} =
\{f_{j,c}\, ,f_{j',c}\} = 0 \, .
\label{albegra-cf}
\end{equation}
Furthermore, the local $c$ pseudoparticle operators and the $l = z, \pm$ local rotated quasi-spin operators ${\tilde{q}}^{l}_{j}$, 
Eq. (\ref{rotated-quasi-spinG}), commute with each other and the latter $l = z, \pm$ operators and corresponding rotated 
$\eta$-spin ($\alpha =\eta$) and rotated spin ($\alpha =s$) operators ${\tilde{S}}^{l}_{j,\alpha}$, Eqs. (\ref{inf-exp-OS}) and (\ref{sir-pirG}), 
obey the usual $SU(2)$ operator algebra.

The $c$ pseudoparticle and $\eta s$ quasi-spin operator algebras refer to the whole Hilbert space.
On the other hand, those of the rotated $\eta$-spin and rotated spin operators correspond to well-defined 
subspaces spanned by energy eigenstates whose value of the number $N^R_{s}=N_c$ of rotated-electron singly occupied sites 
and thus of $c$ pseudoparticles is fixed. This ensures that the value of the corresponding rotated $\eta$-spin number 
$M_{\eta}=N^R_{\eta}=L-N_c$ and rotated spin number $M_{s}=N^R_{s}=N_c$ is fixed as well. 

The degrees of freedom separation, Eq. (\ref{c-up-c-downG}), is such that the 
$c$ pseudoparticle operators, Eq. (\ref{fc+G}), rotated-spin $1/2$ and rotated-$\eta$-spin $1/2$ operators, 
Eq. (\ref{sir-pirG}), and the related $\eta s$ quasi-spin operators, Eqs. (\ref{q-operG}) and (\ref{rotated-quasi-spinG}), 
emerge from the $\sigma$ rotated-electron operators by 
an exact local transformation that {\it does not} introduce constraints. 

That as given in Eq. (\ref{N0q1DHm}) ground states are neither populated 
by composite $\eta n$ pseudoparticles nor by composite $sn$ pseudoparticles with $n>1$ spin-singlet pairs 
plays an important role in the PDT. The $s1$ pseudoparticle translational degrees of freedom are associated with its center of 
mass motion and corresponding $s1$ band momentum $q_{j}$. The PDT involves $s1$ pseudoparticle creation and
annihilation operators associated with such translational degrees of freedom. 

As mentioned above, for $u>0$ the $c$ pseudoparticles live on a $c$ effective lattice identical to the original lattice that has $j=1,...,L$ sites and length $L$.
On the other hand, the $s1$ pseudoparticles live in the TL on a squeezed $s1$ effective lattice \cite{Ogata,Karlo-97,Zaanen}
whose $j=1,...,L_{s1}$ sites number $L_{s1}$ equals that of $s1$ band discrete momentum values, Eq. (\ref{N-h-an}) for
$\alpha n=s1$. The $s1$ effective lattice has length $L$. Its spacing is in general larger than $a$ and given by,
\begin{equation}
a_{s1} = {N_a\over L_{s1}}\,a \, ,
\label{a-a-nu}
\end{equation}
which ensures that indeed $L = L_{s1}\,a_{s1}$. (Except in Eq. (\ref{a-a-nu}), in this paper we use units of lattice spacing $a$ one, 
so that the lattice length $L$ equals the number of lattice sites $N_a$.) 

As for the local creation and annihilation $c$ pseudoparticle operators, Eq. (\ref{albegra-cf}), the $s1$ pseudoparticle 
translational degrees of freedom center of mass motion are described by operators $f^{\dag}_{j,s1}$ (and 
$f_{j,s1}$) that create (and annihilate) one $s1$ pseudoparticle at the $s1$ effective lattice site $x_{j}=a_{s1}\,j$ 
where $j = 1,...,L_{s1}$. Such local $s1$ pseudoparticle creation and annihilation operators obey a fermionic algebra, 
consistently with the $\beta =c,s1$ band momentum value $q_j$ having only occupancies zero and one. 

The $s1$ pseudoparticle operator representation is valid for the 1D Hubbard
model in subspaces spanned by energy eigenstates with fixed $L_{s1}$ value, Eq. (\ref{N-h-an}) for $\alpha n=s1$.
That in such subspaces the local $s1$ pseudoparticle operators obey a fermionic algebra, can be confirmed 
in terms of their statistical interactions \cite{Haldane-91}. This is a problem that we address here very briefly. 
The local $s1$ pseudoparticle creation and annihilation operators may be written as,
\begin{equation}
f^{\dag}_{j,s1} = e^{i\phi_{j,s1}}\,g^{\dag}_{j,s1} \, ; \hspace{0.75cm} 
f_{j,s1} = (f^{\dag}_{j,s1})^{\dag} \, , \hspace{0.50cm} j = 1,...,L_{s1} \, ,
\label{ffs1j}
\end{equation}
where $\phi_{j,s1} = \sum_{j'\neq j}f^{\dag}_{j',s1}$ and the operator $g^{\dag}_{j,s1}$ obeys a hard-core bosonic algebra. 
This algebra is justified by the corresponding statistical interaction vanishing for the model in subspaces spanned 
by energy eigenstates with fixed $L_{s1}$ value, Eq. (\ref{N-h-an}) for $\alpha n=s1$. The $s1$ effective lattice has 
been constructed inherently to that algebra being of hard-core type for the operators $g^{\dag}_{j,s1}$ and $g_{j,s1}$. 
Therefore, through a Jordan-Wigner 
transformation, $f^{\dag}_{j,s1} = e^{i\phi_{j,s1}}\,g^{\dag}_{j,s1}$ \cite{Wang-92}, the operators
$f^{\dag}_{j,s1}$ and $f_{j,s1} = (f^{\dag}_{j,s1})^{\dag}$ in Eq. (\ref{ffs1j}) obey indeed a fermionic algebra,
\begin{equation}
\{f^{\dag}_{j,s1}\, ,f_{j',s1}\} = \delta_{j,j'} ;
\hspace{0.75cm}
\{f_{j,s1}^{\dag}\, ,f_{j',s1}^{\dag}\} =
\{f_{j,s1}\, ,f_{j',s1}\} = 0 \, .
\label{ffs1}
\end{equation}

Each of the $N_{s1}$ occupied $s1$ effective lattice sites corresponds to a spin-singlet pair that
involves two original lattice sites occupied by rotated spins $1/2$ of opposite spin projection.
For the densities $n_e \in [0,1[$ and $m\in [0,n_e]$ the line shape in the vicinity of the singular features of the $\sigma$ one-electron
spectral functions, Eq. (\ref{Bkomega}), studied in Sections \ref{PDT} and \ref{DSGzzxx} 
involves ground state transitions to excited energy eigenstates for which 
$N_{sn}=0$ for $n>1$. For both the ground states and such excited states the number of $N^h_{s1}$ unoccupied $s1$ 
effective lattice sites, Eq. (\ref{N-h-an}) for $\alpha n=s1$, reads $N^h_{s1}=2S_s$. Indeed for
such states the $s1$ effective lattice unoccupied sites refer to the $M^{un}_s=M^{un}_{s,+1/2}=2S_s$ sites occupied in the original 
lattice by the unpaired rotated spins $1/2$. Such unpaired rotated spins $1/2$ are used within the 
$s1$ pseudoparticle motion as unoccupied sites with which they interchange position. Hence
they are implicitly accounted for by the pseudoparticle representation.

The $\beta =c,s1$ pseudoparticle operators labelled by the $\beta =c,s1$ band momentum values
defined in Eqs. (\ref{q-j}) and (\ref{Ic-an}), which are the quantum numbers of the exact BA solution
whose occupancy configurations generate the energy eigenstates considered in the studies of
this paper, play a key role in these studies and read, 
\begin{equation}
f^{\dag}_{q_j,\beta} = {1\over \sqrt{L}}\sum_{j'=1}^{L_{\beta}}e^{i\,q_j\,x_{j'}}f^{\dag}_{j',\beta}
\, ; \hspace{0.50cm} 
f_{q_j,\beta} = (f^{\dag}_{q_j,\beta})^{\dag}
\, , \hspace{0.50cm} j = 1,...,L_{\beta} \, , \hspace{0.5cm} \beta = c,s1 \, .
\label{f-f-FT}
\end{equation}
Besides acting within subspaces spanned by energy eigenstates with fixed $L_{s1}$ values, the $s1$ pseudofermion operators
labelled by momentum $q_j$ also appear in the 
expressions of the shake-up effects generators that transform such subspaces quantum number values 
into each other.

The $(k,\omega)$-plane line shapes near the singular features of the $\sigma$ one-electron 
LHB and UHB addition spectral functions, Eq. (\ref{LUHB}), studied in Sections \ref{PDT} and \ref{DSGzzxx} 
for $u>0$ and densities $n_e \in [0,1[$ and $m \in [0,n_e]$ are determined by transitions to excited energy 
and momentum eigenstates with $N_{\eta 1}=0$ and $N_{\eta 1}=1$,
respectively. Such states are not populated by composite $\alpha n$ pseudoparticles with $n>1$ pairs and
have no unpaired rotated spins of projection $-1/2$ and no unpaired rotated $\eta$-spins of projection $-1/2$.

Hence and as further discussed in Section \ref{PDT}, only the $c$ and $s1$ pseudofermion operator representation generated 
from the $\beta =c,s1$ pseudoparticle operators, Eq. (\ref{f-f-FT}), is needed to study such $(k,\omega)$-plane line shapes. The 
unpaired rotated spins of projection $+1/2$ and unpaired rotated $\eta$-spins of projection $+1/2$
are accounted for within both the pseudoparticle and pseudofermion representations as unoccupied sites of the $s1$ and
$c$ effective lattices, respectively. On the other hand, the effects under $\sigma$ one-electron UHB addition of
the creation of one $\eta 1$ pseudofermion are simpler to be accounted for 
than those stemming from the $c$ and $s1$ pseudofermion processes and fortunately do not require the 
explicit use of the $\eta 1$ pseudofermion operator representation.

\subsection{Needed quantities associated with the $\beta $ pseudoparticle quantum liquid}
\label{quantum-liquid}

For the densities $n_e \in [0,1[$ and $m\in [0,n_e]$ considered in this paper
for which ground states are LWSs of both the spin and $\eta$-spin $SU(2)$ symmetry algebras,
a {\it particle subspace} (PS) is spanned by one such ground states and the set of excited energy eigenstates 
generated from it by a finite number of $\beta $ pseudoparticle processes. For these 
excited energy eigenstates, the deviation densities $\delta N_{\beta}/L$, $\delta S_{s}/L$,
and $\delta S_{\eta/L}$ vanish as $L\rightarrow\infty$. For a PS there are though no restrictions on the value
of the excitation energy and excitation momentum. 

It is often convenient within the TL to replace the $\beta =c,\alpha n$ band discrete 
momentum values $q_j$, Eq. (\ref{q-j}), such that $q_{j+1}-q_j=2\pi/L$, by a 
corresponding continuous momentum variable, $q$. It belongs to a domain $q\in [q_{\beta}^-,q_{\beta}^+]$ whose limiting 
momentum values $q_{\beta}^{\pm }$ are given in Eq. (\ref{qcan-range}).  
For the $\beta =\alpha n$ bands the relation $q_{\alpha n}^{-}=-q_{\alpha n}^{+}$
is exact, as given in that equation. Ignoring $1/L$ corrections as $L\rightarrow\infty$, one finds $q_{\beta}^{\pm }\approx \pm q_{\beta}$ where
for all $\beta =c,\alpha n$ bands $q_{\beta}$ has simple expressions for the ground states and 
their PS excited energy eigenstates. For the present densities ranges 
they read \cite{Carmelo-04},
\begin{equation}
q_{c} = \pi \, ; \hspace{0.5cm} q_{s1} = k_{F\uparrow} \, ; \hspace{0.5cm} q_{sn} =
(k_{F\uparrow}-k_{F\downarrow}) = \pi\,m \, , \hspace{0.3cm} n >1  \, ; \hspace{0.5cm}
q_{\eta n} = (\pi -2k_F) = \pi\,(1-n_e) \, .
\label{qcanGS}
\end{equation}

The $\beta =c,\alpha n$ momentum band distribution functions of the PS excited energy eigenstates
are of the general form $N_{\beta} (q_j) = N^{0}_{\beta} (q_j) +  \delta N_{\beta} (q_j)$
where the ground-state $\beta$ band pseudoparticle momentum distribution functions $N_{\beta}^0 (q_j)$
are given in Eq. (\ref{N0q1DHm}). Several physical quantities are then expressed as functionals of the corresponding  
$\beta =c,\alpha n$ momentum band distribution function deviations,
\begin{equation}
\delta N_{\beta} (q_j)  = N_{\beta} (q_j) - N^0_{\beta} (q_j) \, , \hspace{0.50cm}
j = 1,...,L_{\beta}  \, , \hspace{0.50cm} \beta = c, \alpha n \, , \hspace{0.50cm} n =1,...,\infty \, .
\label{DNq}
\end{equation}

Under transitions from a ground state to a PS excited energy eigenstate, there may occur a shakeup effect
associated with the overall $\beta$-band discrete momentum shifts, $q_j\rightarrow q_j + 2\pi\,\Phi_{\beta}^0/L$, 
that follow from the boundary conditions change in Eq. (\ref{Ic-an}). Here $\Phi_{\beta}^0$ reads,
\begin{eqnarray}
\Phi_{c}^0 & = & 0 \, ; \hspace{0.5cm} \delta N_{\rm ps}^{SU(2)} \hspace{0.50cm} {\rm even} 
\, ;  \hspace{0.5cm} \Phi_{c}^0=\pm{1\over 2} \, ;
\hspace{0.50cm}\delta N_{\rm ps}^{SU(2)} \hspace{0.50cm} {\rm odd} \, ; \nonumber \\
\Phi_{\alpha n}^0 & = & 0 \, ; \hspace{0.50cm} \delta N_{c}+\delta N_{\alpha n}
\hspace{0.50cm} {\rm even} \, ; \hspace{0.50cm} \Phi_{\alpha n}^0=\pm {1\over 2} \, ; \hspace{0.50cm}
\delta N_{c}+\delta N_{\alpha n} \hspace{0.50cm} {\rm odd} \, , \hspace{0.50cm} \alpha = \eta,s \, , \hspace{0.50cm} n =1,...,\infty \, ,
\label{pican}
\end{eqnarray}
where $\delta N_{\rm ps}^{SU(2)}$ is the deviation in the number $N_{\rm ps}^{SU(2)}$ in Eq. (\ref{NpsNapsSR}).

Within the continuum $q$ representation, the deviation values $\delta N_{\beta} (q_j)=-1$ and $\delta N_{\beta} (q_j)=+1$,
Eq. (\ref{DNq}), become $\delta N_{\beta} (q)=-(2\pi/L)\delta (q-q_j)$ and $\delta N_{\beta} (q)=+(2\pi/L)\delta (q-q_j)$, respectively. 
Here and throughout this paper, $\delta (x)$ denotes the usual Dirac delta-function distribution.
According to Eqs. (\ref{q-j}) and (\ref{Ic-an}), under a
transition to an excited energy eigenstate the $\beta$ band discrete momentum values $q_j = (2\pi/L)\,I_j^{\beta}$ 
may undergo a collective shift, $(2\pi/L)\,\Phi_{\beta}^0 =\pm \pi/L$. For $q$ at the $\beta =c,s1$ and $\iota =\pm 1$
Fermi points, $\iota\,q_{F\beta}$, such an effect is captured within the continuum representation by 
additional deviations, $\pm (\pi/L)\delta (q-\iota\,q_{F\beta})$. 
For transitions to an excited energy eigenstate for which $\delta L_{\alpha n}\neq 0$, 
the removal or addition of BA $\alpha n$ band discrete momentum values
occurs in the vicinity of the band edges $q_{\alpha n}^-=-q_{\alpha n}^+$, Eq. (\ref{qcan-range}). 
Those are zero-momentum and zero-energy processes. 

The PS energy functionals are derived from the use in the TBA equations,
Eqs. (\ref{Tapco1})-(\ref{Tapco2}) of Appendix \ref{Ele2PsPhaShi}, and general energy spectra, Eq. (\ref{E})
of that Appendix, of distribution functions of general form $N_{\beta} (q_j) = N^{0}_{\beta} (q_j) +  \delta N_{\beta} (q_j)$
for the excited energy eigenstates. The combined and consistent solution of such equations and spectra 
up to second order in the deviations, Eq. (\ref{DNq}), leads to \cite{Carmelo-91-92},
\begin{equation}
\delta E = \sum_{\beta}\sum_{j=1}^{L_{\beta}}\varepsilon_{\beta} (q_j)\delta N_{\beta} (q_j) 
+ {1\over L}\sum_{\beta}\sum_{\beta'}\sum_{j=1}^{L_{\beta}}\sum_{j'=1}^{L_{\beta'}}
{1\over 2}\,f_{\beta\,\beta'} (q_j,q_{j'})\,\delta N_{\beta} (q_j)\delta N_{\beta'} (q_{j'})
+ \sum_{\alpha =\eta,s}\varepsilon_{\alpha,-1/2}\,M^{un}_{\alpha,-1/2} \, ,
\label{DE-fermions}
\end{equation}
where for the present densities ranges the unpaired rotated $\eta$-spin
$(\alpha =\eta)$ and unpaired rotated spin $(\alpha =s)$ energies relative to the ground state zero-energy level read,
\begin{equation}
\varepsilon_{\alpha,-1/2} = 2\mu_{\alpha} \, ; \hspace{0.75cm} \varepsilon_{\alpha,+1/2} = 0
\, , \hspace{0.50cm}  \alpha =\eta,s \, ,
\label{energy-eta}
\end{equation}
and the energy scales $2\mu_{\alpha}$ are given by,
\begin{equation}
2\mu_{\eta} = 2\vert\mu\vert \, ; \hspace{0.75cm} 2\mu_{s} = 2\mu_B\,\vert h\vert \, ,
\label{2mu-eta-s}
\end{equation}
for general electronic and spin densities and by $2\mu_{\eta} = 2\mu$ and $2\mu_{s} = 2\mu_B\,h$ for the densities 
ranges $n_e\in [0,1[$ and $m\in [0,n_e]$ for which Eq. (\ref{energy-eta}) applies. For the $n_e=1$ Mott-Hubbard 
insulator phase the unpaired rotated $\eta$-spin energy rather reads $\varepsilon_{\eta,\mp 1/2} =  (\mu_{u}\pm\mu)$ 
for $\mu \in [-\mu_{u},\mu_{u}]$. The $n_e=1$ Mott-Hubbard gap $2\mu_{u}$ appearing here whose limiting behaviors are
given in Eq. (\ref{2mu0}) is behind the spectra of the one-electron 
and charge excitations of the half-filled 1D Hubbard model being 
gapped \cite{Lieb,Lieb-03,Ovchi}. 

Furthermore, in Eq. (\ref{DE-fermions}) the $\beta =c,\alpha n$ band energy dispersions $\varepsilon_{\beta} (q_j)$ 
are given by,
\begin{equation}
\varepsilon_{\beta} (q_j) = E_{\beta} (q_j) + 
{t\over \pi}\int_{-Q}^{Q}dk\,2\pi\,\bar{\Phi }_{c,\beta}
\left({\sin k\over u}, {\Lambda_{0}^{\beta} (q_j)\over u}\right)\sin k  \, , \hspace{0.50cm}
j = 1,...,L_{\beta} \, .
\label{epsilon-q}
\end{equation} 
Here $\Lambda_{0}^{\beta} (q_j)$ is a ground-state rapidity function
and $E_{\beta} (q_j)$ is for $\beta = c, \eta n, s n$ the energy spectrum,
Eq. (\ref{spectra-E-an-c-0}) of Appendix \ref{Ele2PsPhaShi}, with the rapidity functions in their expressions
given by the ground-state rapidity functions $k^{c}_{0} (q_j)$ and $\Lambda_{0}^{\beta} (q_j)$.
These functions are the solution of Eqs. (\ref{Tapco1}) and (\ref{Tapco2}) of that Appendix for the $\beta$-band ground-state 
distribution function distributions, Eq. (\ref{N0q1DHm}). The parameter $Q$ also
appearing in Eq. (\ref{epsilon-q}) and related parameters $B$, $r_c^0$, and $r^s_0$ read,
\begin{equation}
Q \equiv k^{0}_c (2k_F) \, ; \hspace{0.5cm} B \equiv
\Lambda_{0}^{s1}(k_{F\downarrow})  \, ; \hspace{0.5cm}
r_c^0 = {\sin Q \over u}  \, ; \hspace{0.5cm} r_s^0 = {B\over u} \, .
\label{QB-r0rs}
\end{equation}

Furthermore, the rapidity dressed phase shift $2\pi\,\bar{\Phi }_{c,\beta} (r,r')$ in Eq. (\ref{epsilon-q})
is a particular case of the more general rapidity dressed phase shifts $2\pi\,\bar{\Phi }_{\beta,\beta'} (r,r')$ 
uniquely defined by the set of integral equations given in Eqs. (\ref{Phis1c-m})-(\ref{Phisnsn-m})
of Appendix \ref{Ele2PsPhaShi}. The general expression of the $f$ functions in the second-order terms 
of the energy functional, Eq. (\ref{DE-fermions}), is provided in Eq. (\ref{ff}) of that Appendix and involves 
the related momentum dressed phase shifts $2\pi\,\Phi_{\beta,\beta'}(q_j,q_{j'})$,
\begin{equation}
2\pi\,\Phi_{\beta,\beta'}(q_j,q_{j'}) = 2\pi\,\bar{\Phi }_{\beta,\beta'} \left(r,r'\right) 
\, ; \hspace{0.35cm} r = \Lambda_{0}^{\beta}(q_j)/u
\, ; \hspace{0.35cm} r' = \Lambda_{0}^{\beta'}(q_{j'})/u \, .
\label{Phi-barPhi}
\end{equation}
Such $f$ function expression also involves the $\beta$ band group velocities 
$v_{\beta} (q_j)$ that within the TL continuum $q$ representation are given by,
\begin{equation}
v_{\beta} (q) = {\partial\varepsilon_{\beta} (q)\over \partial q} 
\, , \hspace{0.50cm} \beta = c, \eta n, sn \, , \hspace{0.50cm} n = 1,...,\infty 
\, ; \hspace{0.50cm}
v_{\beta} \equiv v_{\beta} (q_{F\beta}) \, , \hspace{0.50cm} \beta = c,s1 \, ,
\label{vel-beta}
\end{equation}
where the $\beta$ band energy dispersions are given in Eq. (\ref{epsilon-q}).

An overall dressed phase shift functional involving the momentum dressed phase
shifts, Eq. (\ref{Phi-barPhi}), that within the PDT plays an active role in the control of the
$(k,\omega)$-plane $\sigma$ one-electron spectral weight distributions is given by,
\begin{equation}
2\pi\,\Phi_{\beta} (q_j) = \sum_{\beta'}\,\sum_{j'=1}^{N_{a_{\beta'}}}\,2\pi\,\Phi_{\beta,\beta'}(q_j,q_{j'})\, \delta N_{\beta'}(q_{j'}) 
\, , \hspace{0.50cm} j=1,...,L_{\beta} \, , \hspace{0.50cm} \beta = c,s1 \, , 
\label{Phibetaq}
\end{equation}
where the summation $\sum_{\beta'}$ refers to $\beta' =c,s1$ for $\sigma$ one-electron removal and LHB addition
and to $\beta' =c,s1,\eta 1$ for $\sigma$ one-electron UHB addition
and the deviation $\delta N_{\beta'}(q_{j'})$ is defined in Eq. (\ref{DNq}).

The functional energy spectrum, Eq. (\ref{DE-fermions}), describes the 
1D Hubbard model as a quantum liquid of $c$, $\eta n$, and $sn$ pseudoparticles that have 
residual interactions associated with the $f$ functions,
Eqs. (\ref{ff}). While the general energy spectrum, Eq. (\ref{E}) of Appendix \ref{Ele2PsPhaShi}, 
gives the energy eigenvalues, that given in Eq. (\ref{DE-fermions})
rather provides the excited-state energy eigenvalues minus the ground state energy.
The second term of the energy dispersion, Eq. (\ref{epsilon-q}), and the $f$-function
terms in Eq. (\ref{DE-fermions}) are absent from Eq. (\ref{E}) of Appendix \ref{Ele2PsPhaShi}
and stem from such energies difference. This is why that energy dispersion term and the $f$-function expressions
involve dressed phase shifts, Eq. (\ref{Phi-barPhi}). Indeed those emerge under the transitions
from the ground state to energy eigenstates of excitation energy, Eq. (\ref{DE-fermions}). 

As found in Sections \ref{PDT} and \ref{DSGzzxx}, the spectra of the $\sigma$ one-electron spectral functions 
near their singular features are expressed in terms of the $c$ and $s1$ band energy dispersions, Eq. (\ref{epsilon-q}) for $\beta=c,s1$,
the definition of a particular type of such features called a boundary line involves $\beta$ pseudoparticle group velocities, Eq. (\ref{vel-beta}), 
and the exponents that control the line shape in the vicinity of another type of singular features are expressed
in terms of momentum dressed phase shifts, Eq. (\ref{Phi-barPhi}). Hence in Appendix \ref{LimitBV} useful limiting behaviors 
of all such quantities are provided. 

\subsection{Bounding and anti-bounding character of the composite $\alpha n$ pseudoparticle
$2n=2,4,...$ rotated spins $1/2$ ($\alpha =s$) and rotated $\eta$-spins $1/2$ ($\alpha =\eta$) configuration}
\label{BANTIB}

Analysis of the form of the composite $\alpha n$ pseudoparticle energy dispersions, Eq. (\ref{epsilon-q}) for $\beta =\alpha n$,
provides valuable information on the bounding and anti-bounding character of its
$2n=2,4,...$ paired rotated spins $1/2$ ($\alpha =s$) and paired rotated $\eta$-spins $1/2$
($\alpha =\eta$) singlet configuration, respectively. 

Consistently with Eq. (\ref{energy-eta}) for the particular case of densities $n_e\in [0,1[$ and $m\in [0,n_e]$,
for general electronic densities $n_e\neq 1$ and all corresponding spin densities $m$ the energy of two unpaired rotated $\eta$-spins
$(\alpha =\eta)$ and of two unpaired rotated spins $(\alpha =s)$ of opposite projection reads,
\begin{equation}
2\mu_{\alpha} = \varepsilon_{\alpha,-1/2} + \varepsilon_{\alpha,+1/2} \, , \hspace{0.5cm} \alpha = \eta, s \, ,
\label{varep}
\end{equation}
where the energy scale $2\mu_{\alpha}$ is given in Eq. (\ref{2mu-eta-s}).
For $n_e =1$ and  $m\in [-1,1]$ this expression remains being valid for $\alpha =s$ yet rather involves
the Mott-Hubbard gap, Eq. (\ref{2mu0}), and is replaced by
$2\mu_u = \varepsilon_{\eta,-1/2} + \varepsilon_{\eta,+1/2}$ for $\alpha =\eta$. The bare $\eta$-spin-triplet $(\alpha =\eta)$ and 
spin-triplet $(\alpha =s)$ pair energy, Eq. (\ref{varep}), also applies to a $\eta$-spin-singlet $(\alpha =\eta)$ and 
spin-singlet $(\alpha =s)$ pair in case that the corresponding configuration has no bounding or anti-bounding 
character.
  
The $\alpha n$ pseudoparticle energy dispersion, Eq. (\ref{epsilon-q}) for $\beta =\alpha n$, may be written as,
\begin{equation}
\varepsilon_{\alpha n} (q_j) = \varepsilon_{\alpha n}^0 (q_j) + n\,2\mu_{\alpha}
\, , \hspace{0.50cm} \alpha = \eta,s \, , \hspace{0.50cm} n = 1,..., \infty \, .
\label{e-0-bands}
\end{equation}
The term $n\,2\mu_{\alpha}$ in this energy dispersion is merely additive in the 
bare energy $2\mu_{\alpha}$, Eq. (\ref{varep}). On the other hand, $\varepsilon_{\alpha n}^0 (q_j)$ is a 
bounding or anti-bounding energy if $\varepsilon_{\alpha n}^0 (q_j)<0$ or $\varepsilon_{\alpha n}^0 (q_j)>0$, respectively.
The use of such a criterion reveals that the $sn$ pseudoparticles $2n$ rotated spins configuration has a bounding
character, since $\varepsilon_{s 1}^0 (q_j)<0$ for $\vert q_j\vert <q_{s n}$. The $\eta n$ pseudoparticles $2n$ 
rotated $\eta$-spins configuration has in turn an anti-bounding
character because $\varepsilon_{\eta n}^0 (q_j)>0$ for $\vert q_j\vert <q_{\eta n}$. 

Interestingly, $\varepsilon_{\alpha n}^0 (\pm q_{\alpha n})=0$ so that at the $\alpha n$
band limiting values $q_j = \pm q_{\alpha n}$ given in Eq. (\ref{qcanGS}) one has that the energy, Eq. (\ref{e-0-bands}), 
becomes additive in the bare energy $2\mu_{\alpha}$ of two unpaired rotated $\eta$-spins
$(\alpha =\eta)$ and of two unpaired rotated spins $(\alpha =s)$ of opposite projection,
$\varepsilon_{\alpha n} (\pm q_{\alpha n}) = n\,2\mu_{\alpha}$. As discussed
below in Sec. \ref{upUHBs}, this is due to a symmetry that is behind the $\sigma$ one-electron UHB addition singular spectral features 
being for $n_e \in [0,1[$ and under the transformations $k\rightarrow \pi -k$ and $\omega\rightarrow 2\mu -\omega$
similar to those of the  corresponding $\bar{\sigma}$ one-electron removal singular spectral features.

On the other hand, the $c$ pseudoparticle energy dispersion, Eq. (\ref{epsilon-q}) for $\beta =c$,
can be written as,
\begin{equation}
\varepsilon_{c} (q_j) = \varepsilon_{c}^0 (q_j) + \mu_{\eta} - \mu_{s} \, .
\label{e-c-band}
\end{equation}

The magnetic-field energy scale $2\mu_B\,h = 2\mu_B\,h (m)$ dependence on the spin density 
$m\in [0,n_e]$ and the energy scale $2\mu = 2\mu (n_e)$ associated with the
chemical potential $\mu $ dependence on the electronic density 
$n_e\in [0,1[$ are fully determined by the $s1$ band energy dispersion $\varepsilon_{s1}^0 (q_j)$ at 
$q_j = q_{Fs1} = k_{F\downarrow}$ in Eq. (\ref{e-0-bands}) for $\alpha n=s1$ 
and the $c$ band energy dispersion $\varepsilon_{c}^0 (q_j)$ at $q_j = q_{Fc} = 2k_{F}$ 
in Eq. (\ref{e-c-band}), respectively, as follows \cite{Carmelo-91-92},
\begin{eqnarray}
2\mu_B\,h (m) & = & - \varepsilon_{s1}^0 (q_{Fs1}) \in [0,2\mu_B\,h_c]
\nonumber \\
& & {\rm for}\hspace{0.1cm} q_{F{s1}} = k_{F\downarrow} = 
{\pi\over 2}(n_e - m)\hspace{0.1cm}{\rm where}\hspace{0.1cm}m \in [0,n_e]\hspace{0.1cm}{\rm at}\hspace{0.1cm}{\rm fixed}\hspace{0.1cm}n_e \, ,
\nonumber \\
2\mu (n_e) & = & - 2\varepsilon_{c}^0 (q_{Fc}) - \varepsilon_{s1}^0 (q_{F{s1}}) \in [2\mu_u,(U+4t)]
\nonumber \\
& & {\rm for}\hspace{0.1cm}q_{Fc} =  2k_{F} = 
\pi\,n_e\hspace{0.1cm}{\rm and}\hspace{0.1cm} q_{F{s1}} = {\pi\over 2}(n_e - m)\hspace{0.1cm}
{\rm where}\hspace{0.1cm}n_e \in [0,1[\hspace{0.1cm}{\rm at}\hspace{0.1cm}{\rm fixed}\hspace{0.1cm}m < n_e \, ,
\label{mu-muBH}
\end{eqnarray}
where $2\mu_B\,h (0) =0$, $2\mu_B\,h (n_e) =2\mu_B\,h_c$ is the magnetic energy scale, Eq. (\ref{hc}),
$2\mu (0) = (U+4t)$, and $\lim_{n_e\rightarrow 1}2\mu (n_e) = 2\mu_u$ is the Mott-Hubbard gap, Eq. (\ref{2mu0}).

\section{The pseudofermion dynamical theory microscopic processes that account for the $\sigma$ one-electron spectral weights}
\label{PDT} 

The main goal of this section is to provide information beyond that of Refs. \cite{V-1,LE} on the
microscopic processes that control the $\sigma$ one-electron spectral weights at finite magnetic
field. This includes how the PDT accounts through such processes for the matrix elements of the $\sigma $ electron creation or annihilation operators 
between the initial ground state and the excited energy eigenstates. To accomplish that aim, we start by briefly introducing in
Section \ref{matrixelem} the pseudofermion representation to be used for these matrix elements. In Section \ref{leading}
the $\sigma$ one-electron problem is expressed in terms of pseudofermion operators.
The matrix elements of the $\sigma $ electron creation or annihilation operators and the expression of the corresponding 
$\sigma$ one-electron spectral functions in terms of $\beta =c,s1$ pseudofermion spectral functions are the issues
addressed in Section \ref{matrixOnel}. In Section \ref{hocontribu} the effects of
the small higher-order pseudofermion contributions to the $\sigma$ one-electron spectral weight
are discussed. Section \ref{statesumm2} addresses the involved state summations problem and
the analytical expressions obtainable near $\sigma$ one-electron singular spectral features. Finally,
the validity of the expressions for the line shape near such features is the subject of Section \ref{validity}.

\subsection{Pseudofermion representation to be used for the $\sigma $ electron operators matrix elements}
\label{matrixelem} 

For the 1D Hubbard model at a finite magnetic field in a PS as defined in Section \ref{quantum-liquid}, 
the $c$ and $s1$ rapidity functions of the excited energy eigenstates can be expressed in terms of 
those of the corresponding initial ground state
as given in Eq. (\ref{FL}) of Appendix \ref{Ele2PsPhaShi}. The set of $j=1,...,L_{\beta}$ values ${\bar{q}}_j = {\bar{q}} (q_j)$ 
in such excited energy eigenstates rapidity 
expressions $\Lambda^{c}(q_j) = \Lambda_0^{c} ({\bar{q}} (q_j))$ and $\Lambda^{s1}(q_j) = \Lambda^{s1}_0 ({\bar{q}} (q_j))$
are the $\beta = c,s1$ band discrete {\it canonical momentum} values. They are given by,
\begin{equation}
{\bar{q}}_j = {\bar{q}} (q_j) = q_j + {2\pi\,\Phi_{\beta} (q_j)\over L} = {2\pi\over
L}\left(I^{\beta}_j + \Phi_{\beta} (q_j)\right) \, , \hspace{0.50cm} j=1,...,L_{\beta} \, , \hspace{0.50cm} \beta = c,s1 \, . 
\label{barqan}
\end{equation}
Here $2\pi\,\Phi_{\beta} (q_j)$ stands for the dressed phase-shift functional, Eq. (\ref{Phibetaq}), in units of $2\pi$.
The discrete canonical momentum values, Eq. (\ref{barqan}), have spacing ${\bar{q}}_{j+1}-{\bar{q}}_{j}= 2\pi/L + {\rm h.o.}$,
where h.o. stands for contributions of second order in $1/L$. 

We call a {\it $\beta =c,s1$ pseudofermion} each of the $N_{\beta}$ occupied $\beta$-band discrete canonical momentum values ${\bar{q}}_j$
\cite{V-1,LE}. We call a {\it $\beta$ pseudofermion hole} the remaining $N_{\beta}^h$ unoccupied 
$\beta$-band discrete canonical momentum values ${\bar{q}}_j$ of a PS energy eigenstate. There is a pseudofermion representation 
for each ground state and its PS. This holds for all electronic and spin densities. 

The $\beta =c,s1$ pseudofermion creation and annihilation operators are generated from
the corresponding $\beta =c,s1$ pseudoparticle creation and annihilation operators, Eq. (\ref{f-f-FT}), as follows,
\begin{eqnarray}
{\bar{f}}^{\dag}_{{\bar{q}}_j,\beta} & = & f^{\dag}_{q_j + 2\pi\,\Phi_{\beta} (q_j)/L,\beta} =
\left({\hat{S}}^{\Phi}_{\beta} \right)^{\dag}f^{\dag}_{q_j,\beta}\,{\hat{S}}^{\Phi}_{\beta} 
\, ; \hspace{0.75cm} {\bar{f}}_{{\bar{q}}_j,\beta} = ({\bar{f}}^{\dag}_{{\bar{q}}_j,\beta})^{\dag} \, ,
\nonumber \\
{\hat{S}}^{\Phi}_{\beta} & = & 
e^{\sum_{j=1}^{L_{\beta}}f^{\dag}_{q_{j} + 2\pi\,\Phi_{\beta} (q_j)/L,\beta}f_{q_{j},\beta}} 
\, ; \hspace{0.75cm}
\left({\hat{S}}^{\Phi}_{\beta} \right)^{\dag} = 
e^{\sum_{j=1}^{L_{\beta}}f^{\dag}_{q_{j} - 2\pi\,\Phi_{\beta} (q_j)/L,\beta}f_{q_{j},\beta}} \, ,
\label{f-f-Q}
\end{eqnarray}
where and ${\hat{S}}^{\Phi}_{\beta}$ is the $\beta$ pseudoparticle - $\beta$ pseudofermion unitary operator. 
By combining Eq. (\ref{fc+G}) with Eq. (\ref{f-f-FT-Q}) for $\beta =c$, the $c$ pseudofermion operator
given here can be formally expressed in terms of rotated-electron operators as, 
\begin{equation}
{\bar{f}}^{\dag}_{{\bar{q}}_j,c} = {1\over{\sqrt{L}}}\sum_{j'=1}^{L}\,e^{+i{\bar{q}}_j j'}\,
\Bigl({\tilde{c}}_{j',\uparrow}^{\dag}\,
(1-{\tilde{n}}_{j',\downarrow}) + (-1)^{j'}\,{\tilde{c}}_{j',\uparrow}\,{\tilde{n}}_{j',\downarrow}\Bigr) 
\, ; \hspace{0.5cm} {\bar{f}}_{{\bar{q}}_j,c} = ({\bar{f}}^{\dag}_{{\bar{q}}_j,c})^{\dag} \, .
\label{f-f-Q-cG}
\end{equation}

As in the case of the corresponding $\beta =c,s1$ pseudoparticle operators,
the canonical-momentum $\beta =c,s1$ pseudofermion operators, Eq. (\ref{f-f-Q}), are related 
to local $\beta =c,s1$ pseudofermion operators ${\bar{f}}^{\dag}_{j',\beta}$ and ${\bar{f}}_{j',\beta}$ that create and annihilate, respectively, one
$\beta =c,s1$ pseudofermion at the $\beta =c,s1$ effective lattice site $x_{j'}=a_{\beta}\,j'$ where $ j' = 1,...,L_{\beta}$.
The relation reads,
\begin{equation}
{\bar{f}}^{\dag}_{{\bar{q}}_j,\beta} = {1\over \sqrt{L}}\sum_{j'=1}^{L_{s1}}e^{i\,{\bar{q}}_j\,x_{j'}}{\bar{f}}^{\dag}_{j',\beta}
\, ; \hspace{0.75cm} 
{\bar{f}}_{{\bar{q}}_j,\beta} = {1\over \sqrt{L}}\sum_{j'=1}^{L_{s1}}e^{-i\,{\bar{q}}_j\,x_{j'}}{\bar{f}}_{j',\beta} 
\, , \hspace{0.50cm} j = 1,...,L_{\beta} \, , \hspace{0.50cm} \beta = c,s1 \, .
\label{f-f-FT-Q}
\end{equation}
Indeed, the $c$ and $s1$ pseudofermions also live in the $c$ effective lattice, which is identical to
the original lattice, and in the squeezed $s1$ effective lattice, respectively.
As the $c$ pseudoparticles, the $c$ pseudofermions have no internal structure, whereas the $s1$ pseudofermions 
have the same internal structure as the corresponding $s1$ pseudoparticles.
They only differ in their discrete momentum values, which rather refer to the translational degrees of
freedom associated with their center of mass motion.

In the present pseudofermion operator representation a PS ground state has the simple form,
\begin{equation}
\vert GS\rangle = \prod_{{\bar{q}}=-k_{F\downarrow}}^{k_{F\downarrow}}\prod_{{\bar{q}}'=-\pi}^{\pi}
{\bar{f}}^{\dag }_{{\bar{q}},\,s1}\,{\bar{f}}^{\dag }_{{\bar{q}}',\,c}\vert 0\rangle 
= \prod_{j=1}^{N_{\downarrow}}\prod_{j'=1}^{L}
{\bar{f}}^{\dag }_{{\bar{q}}_j,\,s1}\,{\bar{f}}^{\dag }_{{\bar{q}}_{j'},\,c}\vert 0\rangle \, .
\label{GS}
\end{equation}
That representation has been inherently constructed to ${\bar{q}} = q$ for a PS ground state,
so that here the $s1$ and $c$ band momentum values ${\bar{q}} = q = {\bar{q}}_j = q_j$ 
and ${\bar{q}}' = q' = {\bar{q}}_{j'} = q_{j'}$, respectively, are those of the corresponding 
$s1$ and $c$ pseudoparticle occupied ground-state Fermi seas. Moreover,
$\vert 0\rangle$ stands in Eq. (\ref{GS}) for the electron and rotated-electron vacuum and
the ground-state generator has been written in terms of $s1$ and $c$ pseudofermion 
creation operators, Eqs. (\ref{f-f-Q}) and (\ref{f-f-FT-Q}).

The $c$ pseudofermions as defined here refer to an extension to finite $u$ of the usual
$u\rightarrow\infty$ spinless fermions \cite{Karlo,Karlo-97}. Indeed, in the $u\rightarrow\infty$ limit
the momentum rapidity function of the ground state $k^{c}_{0} (q_j)$ simplifies to $k^{c}_{0} (q_j)  = q_j$. 
Hence, according to Eq. (\ref{FL}) of Appendix \ref{Ele2PsPhaShi}, for the PS excited energy eigenstates 
associated with the initial ground state under consideration such a function reads, $k^c (q_j) = {\bar{q}}_j$. 
The $u\rightarrow\infty$ spinless fermions of Refs. \cite{Karlo,Karlo-97} have been 
constructed inherently to carry the momentum rapidity $k_j = k^c (q_j) = {\bar{q}}_j$. This reveals that
such spinless fermions are the $c$ pseudofermions as defined here in the $u\rightarrow\infty$ limit. 
Indeed, the relations ${\bar{f}}^{\dag }_{{\bar{q}}_j,c} = {\hat{V}}^{\dag}\,b^{\dag}_{k_j}\,{\hat{V}}$ and 
${\bar{f}}_{{\bar{q}}_j,c} ={\hat{V}}^{\dag}\,b_{k_j}\,{\hat{V}}$ hold where ${\hat{V}}$ is the electron - rotated-electron unitary operator defined in terms 
of its matrix elements in Eq. (\ref{ME-Vll}) and $b^{\dag}_{k_j}$ and $b_{k_j}$ stand for the $u\rightarrow\infty$ spinless 
fermions creation and annihilation operators that appear in the anti-commutators given in the first equation of 
Section IV of Ref. \cite{Karlo-97}. 

The one-to-one correspondence between a canonical momentum value ${\bar{q}}_j$ and the
corresponding bare momentum value $q_j$ as defined in Eq. (\ref{barqan}) enables the
expression of several ${\bar{q}}_j$-dependent pseudofermion quantities in terms of the
corresponding bare momentum $q_j$. This applies to the dressed phase shift 
$2\pi\,\Phi_{\beta,\beta'}(q_j,q_{j'})$, Eq. (\ref{Phi-barPhi}).
Within the pseudofermion representation it has a precise physical meaning:
$2\pi\,\Phi_{\beta,\beta'}(q_j,q_{j'})$ (and $-2\pi\,\Phi_{\beta,\beta'}(q_j,q_{j'})$) 
is the phase shift acquired by a $\beta$ pseudofermion or
$\beta$ pseudofermion hole of canonical momentum ${\bar{q}}_j={\bar{q}} (q_j)$
upon scattering off a $\beta'$ pseudofermion (and $\beta'$ pseudofermion hole) 
of canonical momentum value ${\bar{q}}_{j'}={\bar{q}} (q_{j'})$
created under a transition from the ground state to a PS excited energy eigenstate. 
Hence the important functional $2\pi\,\Phi_{\beta} (q_j)$, Eq. (\ref{Phibetaq}), in the $\beta =c,s1$ canonical momentum
expression ${\bar{q}}_j = q_j + {2\pi\over L}\Phi_{\beta} (q_j)$, Eq. (\ref{barqan}), is the phase
shift acquired by a $\beta$ pseudofermion or $\beta$ pseudofermion hole of
canonical momentum value ${\bar{q}}_j={\bar{q}} (q_j)$ upon scattering off the set of $\beta'$ pseudofermions 
and $\beta'$ pseudofermion holes created under a transition from the ground state to 
a PS excited energy eigenstate. Hence the $\beta$ pseudofermion phase shift 
$2\pi\,\Phi_{\beta} (q_j)$ has a specific value for each ground-state - excited-state
transition.

The line shape near the $\sigma$ one-electron UHB addition spectral function singular features involves
the creation of a single $\eta 1$ pseudoparticle at one of the $\eta 1$ band limiting momentum values 
$q_j = \pm q_{\eta 1} = \pm (\pi -2k_F)$, Eq. (\ref{qcanGS}). $\eta 1$ band canonical momentum values 
${\bar{q}}_j =q_j + 2\pi\,\Phi_{\eta 1} (q_j)/L$ can be introduced, as in Eq. (\ref{barqan}) for the $\beta =c,s1$ 
bands. Interestingly, one finds that $2\pi\,\Phi_{\eta 1} (q_j)=0$
at the $\eta 1$ band limiting momentum values $q_j = \pm (\pi -2k_F)$, so that
${\bar{q}}_j =q_j$. This reveals that a $\eta 1$ pseudoparticle and a $\eta 1$ pseudofermion
of momenta $\pm (\pi -2k_F)$ are the same quantum object.
Such an invariance under the $\eta 1$ pseudoparticle - $\eta 1$ pseudofermion unitary
transformation follows from symmetries related to the anti-bounding energy $\varepsilon_{\eta 1}^0 (q_j)$
on the right-hand side of Eq. (\ref{e-0-bands}) for $\alpha n = \eta 1$
vanishing at $q_j = \pm q_{\eta 1}= \pm (\pi -2k_F)$. As the unpaired rotated spins and unpaired
rotated $\eta$-spins, the $\eta 1$ pseudofermions of momentum $\pm q_{\eta 1} = \pm (\pi -2k_F)$
do not acquire a phase shift under the transitions from the ground state to the PS excited energy eigenstates.

One can introduce a creation operator $f^{\dag}_{q_j,\eta 1}$ for the $\eta 1$ pseudoparticles that
at $q_j = \iota(\pi -2k_F)$ is identical to the corresponding $\eta 1$ pseudofermion creation operator,
\begin{equation}
{\bar{f}}^{\dag}_{{\bar{q}}_j,\eta 1} = f^{\dag}_{q_j,\eta 1} \hspace{0.1cm}{\rm at}\hspace{0.1cm}
{\bar{q}}_j = q_j = \iota(\pi -2k_F) 
\, , \hspace{0.50cm} \iota = \pm 1 \, ,
\label{feta1}
\end{equation}
where in the present case ${\bar{f}}^{\dag}_{{\bar{q}}_j,\eta 1}$ creates one $\eta 1$ pseudofermion at the
canonical momentum values ${\bar{q}}_j = \pm (\pi -2k_F)$. Although such a $\eta 1$ pseudofermion
does not acquire phase shifts of its own, under its creation within a transition from the ground state
to an excited energy eigenstate the $\beta =c,s1$ pseudofermions of canonical momentum
${\bar{q}}_j$ acquire a phase shift $2\pi\,\Phi_{\beta,\eta 1}(q_j,\pm (\pi -2k_F))$,
Eq. (\ref{Phi-barPhi}) for $\beta' = \eta 1$ and $q_{j'}=\pm (\pi -2k_F)$. After some manipulations relying
on the use of Eqs. (\ref{Phis1cn-m}) and (\ref {Phiccn-m}) 
of Appendix \ref{Ele2PsPhaShi} for $\eta n = \eta 1$, one finds that it can be written as,
\begin{equation}
2\pi\,\Phi_{\beta,\eta 1}(q_j,\pm (\pi -2k_F)) = \pm {1\over 2}\left(\delta_{\beta,c}2\pi  + 2\pi\,\Phi_{\beta,c}(q_j,2k_F)
- 2\pi\,\Phi_{\beta,c}(q_j,-2k_F)\right) \, , \hspace{0.50cm} \beta = c,s1 \, , \hspace{0.50cm} \iota = \pm 1 \, .
\label{Phibetaeta1}
\end{equation}
Hence except for the factor $1/2$ creation of one $\eta 1$ pseudofermion at the canonical momentum values $\pm (\pi -2k_F)$
is felt by a $\beta =c,s1$ pseudofermion as the creation and annihilation of two $c$ pseudofermions at opposite
Fermi points.

The exponents that control the $\sigma$ one-electron spectral weight
in the $(k,\omega)$-plane vicinity of a type of singular features called branch lines are found below in Section
\ref{matrixOnel} to involve both the two-pseudofermion phase shifts $2\pi\,\Phi_{c,\beta}(\pm 2k_F,q_{j})$
and $2\pi\,\Phi_{s1,\beta}(\pm k_{F\downarrow},q_{j})$ where $\beta =c,s1$ 
and the following related $j=0,1$ parameters, 
\begin{equation}
\xi^{j}_{\beta\,\beta'} = \delta_{\beta,\beta'} 
+ \sum_{\iota=\pm 1} (\iota)^j\,\Phi_{\beta,\beta'}\left(q_{F\beta},\iota q_{F\beta'}\right)
\, , \hspace{0.50cm} \beta, \beta' = c, s1 \, , \hspace{0.50cm} j = 0, 1 \, .
\label{x-aa}
\end{equation}
For the particular case of $\beta =\beta'$ and $\iota=1$ in Eq. (\ref{x-aa}), the present 
notation assumes that the two $\beta =c,s1$ Fermi momenta in the argument of the $\beta$ pseudofermion
phase shift, $2\pi\,\Phi_{\beta,\beta}\left(q_{F\beta},q_{F\beta}\right)$, differ by $2\pi/L$.
(For identical momentum values one has that $2\pi\,\Phi_{\beta,\beta}(q_j,q_j)=0$.)

The two-pseudofermion phase-shift related anti-symmetrical $\xi^{1}_{\beta\,\beta'}$ and symmetrical $\xi^{0}_{\beta\,\beta'}$
parameters, Eq. (\ref{x-aa}), that naturally emerge from the pseudofermion representation are actually the entries of the 
low-energy conformal-field theory dressed-charge matrix and of the transposition of its inverse matrix \cite{Woy-89,Frahm,LE,Carmelo-91-92},
\begin{equation}
Z^1 = \left[\begin{array}{cc}
\xi^{1}_{c\,c} & \xi^{1}_{c\,s1}  \\
\xi^{1}_{s1\,c}   & \xi^{1}_{s1\,s1}  
\end{array}\right]
\, ; \hspace{0.5cm}
Z^0 = ((Z^1)^{-1})^T = \left[\begin{array}{cc}
\xi^{0}_{c\,c} & \xi^{0}_{c\,s1}  \\
\xi^{0}_{s1\,c}   & \xi^{0}_{s1\,s1} 
\end{array}\right] \, ,  
\label{ZZ-gen}
\end{equation}
respectively. (Here the dressed-charge matrix definition
of Ref. \cite{Woy-89} has been used, which is the transposition of that of Ref. \cite{Frahm}.)
The limiting behaviors of the parameters, Eq. (\ref{x-aa}), which are the entries of the
matrices, Eq. (\ref{ZZ-gen}), are given in Appendix \ref{LimitBV}.

Moreover, from the combined use of Eqs. (\ref{Phibetaeta1}) and (\ref{x-aa}) one finds,
\begin{equation}
\Phi_{\beta,\eta 1}(\iota q_{F\beta},\iota' (\pi -2k_F)) = \iota' {\xi^{1}_{\beta\,c}\over 2} 
\, , \hspace{0.50cm} \beta = c,s1 \, , \hspace{0.50cm} \iota, \iota' = \pm 1 \, .
\label{Phibetaeta1Fbeta}
\end{equation}

For the PS excited energy eigenstates with densities $n_e\in [0,1[$ and $m\in [0,n_e]$ associated
with the line shape near the $\sigma$ one-electron spectral functions singularities the $\alpha n$ pseudofermion
numbers have values given by $N_{\alpha n}=0$ for $n>1$ and $N_{\eta 1}=0,1$ where when $N_{\eta 1}=1$ the $\eta 1$ 
pseudofermion has canonical momentum $\pm (\pi -2k_F)$.
For the PSs spanned by these excited energy eigenstates and corresponding ground states the pseudoparticle
representation general PS energy functional, Eq. (\ref{DE-fermions}), simplifies to,
\begin{equation}
\delta E = \sum_{\beta=c,s1}\sum_{j=1}^{L_{\beta}}\varepsilon_{\beta} (q_j)\delta N_{\beta} (q_j) 
+ {1\over L}\sum_{\beta =c,s1}\sum_{\beta'=c,s1,\eta 1}\sum_{j=1}^{L_{\beta}}\sum_{j'=1}^{L_{\beta'}}
{1\over 2}\,f_{\beta\,\beta'} (q_j,q_{j'})\,\delta N_{\beta} (q_j)\delta N_{\beta'} (q_{j'})
+ 2\mu\,N_{\eta 1} \, .
\label{DE-fermions0}
\end{equation}
Upon expressing this functional in the pseudofermion representation, which involves
the $\beta =c,s1$ bands discrete canonical momentum values 
${\bar{q}}_j = {\bar{q}} (q_j)$, Eq. (\ref{barqan}), one finds after some algebra
that it reads up to ${\cal{O}}(1/L)$ order,
\begin{equation}
\delta E = \sum_{\beta=c,s1}\sum_{j=1}^{L_{\beta}}\varepsilon_{\beta} ({\bar{q}}_j)\,\delta {\cal{N}}_{\beta}({\bar{q}}_j)
+ 2\mu\,N_{\eta 1} \, . 
\label{DE}
\end{equation}
Here the $\beta =c,s1$ pseudofermion energy dispersions $\varepsilon_{\beta} ({\bar{q}}_j)$
have exactly the same form as those given in Eq. (\ref{epsilon-q})
with the momentum $q_j$ replaced by the corresponding canonical momentum, ${\bar{q}}_j= {\bar{q}} (q_j)$.

If in Eq. (\ref{DE}) one expands the $\beta =c,s1$ band canonical momentum ${\bar{q}}_j=q_j + 2\pi\,\Phi_{\beta} (q_j)/L$
around $q_j$ and considers all energy contributions up to ${\cal{O}}(1/L)$ order, one arrives after some algebra
to the energy functional, Eq. (\ref{DE-fermions0}), which includes terms 
of second order in the deviations $\delta N_{\beta}(q_j)$. Their absence from the corresponding 
energy spectrum, Eq. (\ref{DE}), follows from the functional $2\pi\,\Phi_{\beta} (q_j)$, Eq. (\ref{Phibetaq}), 
being incorporated in the $\beta =c,s1$ band canonical momentum, Eq. (\ref{barqan}).

In contrast to the equivalent energy functional, Eq. (\ref{DE-fermions0}), that in Eq. (\ref{DE}) has no energy 
interaction terms of second-order in the deviations $\delta {\cal{N}}_{\beta}({\bar{q}}_j)$. Indeed
the $\beta =c,s1$ pseudofermions have no such interactions up to ${\cal{O}}(1/L)$ order. Within the present TL, only
finite-size corrections up to that order are relevant. The property that the excitation energy spectrum, Eq. (\ref{DE}), 
has no pseudofermion energy interactions is found below to simplify the expression of the $\sigma$ one-electron
spectral functions in terms of a sum of convolutions of $c$ and $s1$ pseudofermion spectral functions whose 
spectral weights are expressed as Slater determinants of pseudofermion operators.

\subsection{The $\sigma$ one-electron problem expressed in terms of pseudofermion operators}
\label{leading} 

Within the PDT of Refs. \cite{V-1,LE} the $\beta =c,s1$ pseudofermion phase shifts determine 
the dynamical correlation functions spectral-weight distributions.
Here we provide information beyond that given in these references about how that dynamical 
theory accounts for the matrix elements $\langle\nu^-\vert\, c_{k,\sigma} \vert \,GS\rangle$ and
$\langle\nu^+\vert\, c^{\dagger}_{k,\sigma} \vert\,GS\rangle$ in the spectral functions, Eq. (\ref{Bkomega}).
For such spectral functions the elementary processes
that generate the excited energy eigenstates from ground states with densities in the ranges $n_e \in [0,1[$ and
$m\in [0,n_e]$ can be classified into three (A)-(C) classes: 
\vspace{0.50cm}

(A) High-energy and finite-momentum elementary $\beta =c,s1$ pseudofermion processes. Specifically, creation or 
annihilation of one or a finite number of $\beta =c,s1$ pseudofermions 
with canonical momentum values ${\bar{q}}_j\neq \pm {\bar{q}}_{F\beta}$;
\vspace{0.50cm}

(B) Finite-momentum processes of excitation energy zero or $2\mu$ that change the
number of $\beta =c,s1$ pseudofermions at the $\iota=+1$ right and 
$\iota=-1$ left $\beta =c,s1$ Fermi points. The processes contributing
to the line shape near the $\sigma$ one-electron UHB spectral function singular features involve 
creation of one $\eta 1$ pseudofermion at a $\eta 1$ band limiting canonical momentum 
$q_{\eta 1}^{\pm}=\pm (\pi -2k_F)$, Eq. (\ref{qcan-range}) for $\alpha n = \eta 1$, which 
involves a finite-energy $2\mu$. This is the minimal energy for creation of one rotated-electron 
doubly occupied site and stems from the first term of the spectrum $E_{\eta 1} (q_j)$,
Eq. (\ref{spectra-E-an-c-0}) of Appendix \ref{Ele2PsPhaShi} for $\alpha n = \eta 1$, in 
the $\eta 1$ energy dispersion $\varepsilon_{\eta 1} (q_j)$, Eqs. (\ref{epsilon-q}) 
and (\ref{e-0-bands}) for $\beta = \eta 1$;
\vspace{0.50cm}

(C) Low-energy and small-momentum elementary pseudofermion particle-hole processes in the vicinity of the
$\beta =c,s1$ bands right ($\iota=+1$) and left ($\iota=+1$) Fermi points, relative to the excited-state 
$\beta = c,s1$ pseudofermion momentum occupancy configurations generated by the above
elementary processes (A) and (B).
\vspace{0.50cm} \\ 

The creation of one $\eta 1$ pseudofermion associated with the $\sigma$ one-electron UHB addition singular spectral features refers
to transitions from ground states with densities $n_e<1$. At $n_e=1$ the $\sigma$ one-electron UHB involves
instead ground-state transitions to excited energy eigenstates populated by one unpaired rotated $\eta$-spin
$1/2$ of $\eta$-spin projection $-1/2$. This also amounts for creation of one rotated-electron doubly occupied site.

The first two steps to express in the pseudofermion representation the matrix elements $\langle\nu^-\vert\, c_{k,\sigma} \vert \,GS\rangle$
and $\langle\nu^+\vert\, c^{\dagger}_{k,\sigma} \vert\,GS\rangle$ in the spectral functions, Eq. (\ref{Bkomega}), of a $\sigma $ 
electron operator between the ground state and the excited energy eigenstates 
are (i) to express the $\sigma$ electron creation or annihilation operator in terms
of $\sigma $ rotated electron creation and annihilation operators, Eq. (\ref{rotated-operators}),
and (ii) to express the latter operators in terms of rotated spin $1/2$ operators, rotated $\eta$-spin $1/2$ operators,
and $c$ pseudofermion operators. This is accomplished by use of the 
$\sigma $ rotated electron creation and annihilation operators expressions 
in terms of rotated spin $1/2$ operators, rotated $\eta$-spin $1/2$ operators, and $c$ pseudoparticle operators,
Eqs. (\ref{c-up-c-downG}) and (\ref{rota-cksigma}), accounting for the relation between the 
$c$ pseudoparticle and $c$ pseudofermion operators, Eq. (\ref{f-f-Q}) for $\beta =c$.

The momentum $k$ dependent $\sigma $ 
electron operators in the spectral functions Lehmann representation, Eq. (\ref{Bkomega}), 
are related to the corresponding local operators as,
\begin{equation}
c_{k,\sigma} =  {1\over{\sqrt{L}}}\sum_{j=1}^{L}e^{i\,k\,x_{j}}c_{j,\sigma} \, ; \hspace{0.50cm} 
c_{k,\sigma}^{\dag} = (c_{k,\sigma})^{\dag}
\, , \hspace{0.50cm} \sigma = \uparrow,\downarrow \, .
\label{cksigma}
\end{equation}
To write the operators $c_{k,\sigma}$ and $c^{\dagger}_{k,\sigma}$ in terms of $\sigma $ rotated electron 
creation and annihilation operators, Eq. (\ref{rotated-operators}), we use of the Baker-Campbell-Hausdorff formula
to rewrite the relation, Eq. (\ref{rotated-operators}), as follows,
\begin{eqnarray}
c_{k,\sigma} & = & \sum_{i =0}^{\infty}c_{k,\sigma,i} =
{\tilde{c}}_{k,\sigma} + {1\over 1!}\,[{\tilde{c}}_{k,\sigma},{\tilde{S}}\,] + {1\over 2!}\,[[{\tilde{c}}_{k,\sigma},{\tilde{S}}\,],{\tilde{S}}\,] + ... 
\, ; \hspace{0.50cm} c_{k,\sigma}^{\dag} = (c_{k,\sigma})^{\dag}
\, , \hspace{0.50cm} \sigma = \uparrow,\downarrow \, ,
\nonumber \\
c_{k,\sigma,i} & = & [{\tilde{c}}_{k,\sigma},{\tilde{S}}\,]_{i} = {1\over i!}[[{\tilde{c}}_{k,\sigma},{\tilde{S}}\,]_{i-1},{\tilde{S}}\,] \, , \hspace{0.50cm} 
i = 1,...,\infty \, ; \hspace{0.75cm} [{\tilde{c}}_{k,\sigma},{\tilde{S}}\,]_0 = {\tilde{c}}_{k,\sigma} = {\hat{V}}^{\dag}\,c_{k,\sigma}\,{\hat{V}} \, ,
\nonumber \\
{\hat{V}} & = & e^{\hat{S}} = e^{\tilde{S}} \, .
\label{Sr-rot}
\end{eqnarray}
Here the operator $\tilde{S}=\hat{S}$ commutes with ${\hat{V}}$ and thus has the same 
expression in terms of creation and annihilation $\sigma $ rotated-electron operators and $\sigma $ electron operators,
respectively, and the momentum operators ${\tilde{c}}_{k,\sigma}^{\dag} = {\hat{V}}^{\dag}\,c_{k,\sigma}^{\dag}\,{\hat{V}}$ and
${\tilde{c}}_{k,\sigma} = {\hat{V}}^{\dag}\,c_{k,\sigma}\,{\hat{V}}$ 
can be written in terms of the local operators ${\tilde{c}}_{j,\sigma}^{\dag}$ and ${\tilde{c}}_{j,\sigma}$,
respectively, in Eqs. (\ref{rotated-operators}) and (\ref{c-up-c-downG}) as,
\begin{equation}
{\tilde{c}}_{k,\sigma}^{\dag} =  {1\over \sqrt{L}}\sum_{j=1}^{L}e^{i\,k\,x_{j}}{\tilde{c}}_{j,\sigma}^{\dag} \, ; \hspace{0.50cm} 
{\tilde{c}}_{k,\sigma} = ({\tilde{c}}_{k,\sigma}^{\dag})^{\dag}
\, , \hspace{0.50cm} \sigma = \uparrow,\downarrow \, .
\label{rota-cksigma}
\end{equation}

The next step of our program consists in rewriting the rotated-electron expression $c_{k,\sigma} = \sum_{i =0}^{\infty}c_{k,\sigma,i}$
within a related uniquely defined $\beta $ pseudofermion representation as,
\begin{equation}
c_{k,\sigma} = \sum_{i' =0}^{\infty}{\hat{g}}_{i'} (k)\,{\hat{c}}_{\odot} \, .
\label{Oodotpse}
\end{equation}
The new index $i'=0,1,...,\infty$ refers here to $\beta $ pseudofermions processes and
${\hat{c}}_{\odot}$ is a generator that transforms the initial ground state 
$\vert GS\rangle$ into a state with the same electron and rotated-electron numbers
$N_{\uparrow}$ and $N_{\downarrow}$ and compact symmetrical $c$ and $s1$ bands momentum
occupancies as the ground state of the final PS, which we call $\vert GS_f\rangle$. The 
only difference between the states ${\hat{c}}_{\odot}\vert GS\rangle$ and 
$\vert GS_f\rangle$ is their $c$ and $s1$ band discrete momentum values being those of the
initial ground state, $\bar{q}'=q'$, and of the excited-energy eigenstate
$\sum_{i' =0}^{\infty}{\hat{g}}_{i'} (k)\vert GS_f\rangle$, $\bar{q}\neq q$, respectively. 

Each term of index $i'=0,1,...,\infty$ in Eq. (\ref{Oodotpse}) may have contributions from
several terms of different index $i=0,1,...,\infty$ in $c_{k,\sigma} = \sum_{i =0}^{\infty}c_{k,\sigma,i}$, Eq. (\ref{Sr-rot}).
Fortunately, one can compute the operational form in terms of $\beta$ pseudofermion
operators of the leading $i'=0,1,...,\infty$ orders of
$c_{k,\sigma} = \sum_{i' =0}^{\infty}{\hat{g}}_{i'} (k)\,{\hat{c}}_{\odot}$ from the transformation laws 
of the ground state $\vert GS\rangle$, Eq. (\ref{GS}), upon acting onto it the related operators $c_{k,\sigma,i}$ in
the expression $c_{k,\sigma} = \sum_{i =0}^{\infty}c_{k,\sigma,i}$.

The 1D Hubbard model is a non-perturbative quantum problem in terms of $\sigma$ electron processes.
This is behind the computation of the $\sigma$ one-electron spectral functions, Eq. (\ref{Bkomega}), being a 
very complex many-electron problem. On the other hand, a property that plays key role in our
study follows from expressing the $\sigma $ electron operator $c_{k,\sigma}$ in the terms of 
pseudofermion operators as $c_{k,\sigma} = \sum_{i' =0}^{\infty}{\hat{g}}_{i'} (k)\,{\hat{c}}_{\odot}$, Eq. (\ref{Oodotpse}),
rendering the computation of the $\sigma$ one-electron spectral functions,
Eq. (\ref{Bkomega}), a perturbative problem. 

Note that both the expressions $c_{k,\sigma} = \sum_{i =0}^{\infty}c_{k,\sigma,i}$ 
and $c_{k,\sigma} = \sum_{i' =0}^{\infty}{\hat{g}}_{i'} (k)\,{\hat{c}}_{\odot}$ are not 
small-parameter expansions. Consistently, the perturbative character of the
$\beta$ pseudofermions processes refers to the spectral weight contributing to
the spectral functions being dramatically suppressed upon increasing the 
number of corresponding elementary processes
of classes (A) and (B). Those are generated by application onto the
ground state, Eq. (\ref{GS}), of operators in $\sum_{i' =0}^{\infty}{\hat{g}}_{i'} (k)\,{\hat{c}}_{\odot}$ 
with an increasingly large value of the index $i'=0,1,...,\infty$. 

The perturbative character of the 1D Hubbard model upon expressing the $\sigma $ electron creation or annihilation 
operators in the spectral functions, Eq. (\ref{Bkomega}), in terms of 
$c$ pseudofermion operators, rotated spins $1/2$ operators and corresponding $sn$ pseudofermion operators,
and rotated $\eta$-spins $1/2$ operators and corresponding $\eta n$ pseudofermion operators,
follows from the exact energy eigenstates being generated by occupancy configurations of these elementary objects.
The non-perturbative character of the problem in terms of electrons results from their relation to the above
elementary objects having as well a non-perturbative nature, qualitatively different from that of the electrons to 
the quasiparticles of a Fermi liquid.

For simplicity, in the following we denote the $i'=0$ operator ${\hat{g}}_0 (k)$ associated with the
$\sigma $ one-electron operator $c_{k,\sigma}$ (or $c_{k,\sigma}^{\dagger}$)
by ${\hat{g}} (k)$. Such a $i'=0$ leading-order operator term in the one- or two-electron operator expression,
\begin{equation}
c_{k,\sigma} = \left({\hat{g}} (k) + \sum_{i' =1}^{\infty}{\hat{g}}_{i'} (k)\right)\,{\hat{c}}_{\odot} \, ,
\label{Oodot0kGO}
\end{equation}
plays a key role in our study.

The leading-order operators ${\hat{g}} (k)\,{\hat{c}}_{\odot}$ are selected inherently to
all the singular spectral features in the $\sigma$ one-electron 
spectral functions, Eq. (\ref{Bkomega}), being produced by their application onto the ground state. 
The corresponding leading-order pseudofermion processes (A) and (B) that after being dressed by 
low-energy and small-momentum elementary $\beta =c,s1$
pseudofermion particle-hole processes (C) in the vicinity of their
right ($\iota=+1$) and left ($\iota=+1$) Fermi points control the line shape near the singular
features of the $\sigma$ one-electron spectral functions, Eq. (\ref{Bkomega}), are the following:
\vspace{0.50cm}

(1) Removal of one $\uparrow$ electron and thus of one $\uparrow$ rotated electron is a process
that involves annihilation of one $c$ pseudofermion and one unpaired rotated spin $1/2$ of projection $\uparrow$, so that $\delta N_c=-1$. 
That unpaired rotated spin $1/2$ recombines with the annihilated $c$ pseudofermion within the removed 
$\uparrow$ rotated electron. The annihilation of the unpaired rotated spin $1/2$ leaves the number $N_{s1}$ 
$s1$ pseudofermions unchanged and leads to a deviation $\delta N_{s1}^h=-1$ in the number of $s1$ band holes. 
\vspace{0.50cm}

(2) LHB addition of one $\uparrow$ electron and thus of one $\uparrow$ rotated electron
is a process that involves creation of one $c$ pseudofermion and one unpaired rotated spin $1/2$ of projection $\uparrow$, so 
that $\delta N_c=1$. The creation of the unpaired rotated spin $1/2$ leaves the number $N_{s1}$ $s1$ pseudofermions 
unchanged and gives rise to a deviation $\delta N_{s1}^h=1$ in the number of $s1$ band holes.
\vspace{0.50cm}

(3) UHB addition of one $\uparrow$ electron and thus of one $\uparrow$ rotated electron
is a process that involves annihilation of one $c$ pseudofermion and one $s1$ pseudofermion and creation 
of one $\eta 1$ pseudofermion and one unpaired rotated spin $1/2$ of projection $\uparrow$, so that 
$\delta N_c=-1$, $\delta N_{s1}=-1$, and $\delta N_{\eta 1}=1$. The $s1$ pseudofermion annihilation 
occurs through its spin-singlet pair breaking. The rotated spin $1/2$ of projection $\downarrow$ emerging from such a pair breaking
recombines with the annihilated $c$ pseudofermion within one $\downarrow$ rotated electron.
Such a $\downarrow$ rotated electron then pairs with the created $\uparrow$ rotated electron onto a doubly 
occupied site. The rotated $\eta$-spin $1/2$ of projection $-1/2$ that describes the $\eta$-spin degrees
of freedom of such a doubly occupied site
combines with one ground-state unpaired rotated $\eta$-spin $1/2$ of projection $+1/2$ within the
$\eta 1$ pseudofermion $\eta$-spin singlet pair. The creation of one unpaired rotated spin $1/2$ is accounted for by the deviation 
$\delta N_{s1}^h=1$ in the number of $s1$ band holes.
\vspace{0.50cm}

(4) Removal of one $\downarrow$ electron and thus of one $\downarrow$ rotated electron
is a process that involves annihilation of one $c$ pseudofermion and one $s1$ pseudofermion and creation
of one unpaired rotated spin $1/2$ of projection $\uparrow$, so that $\delta N_c=-1$ and $\delta N_{s1}=-1$. The $s1$ pseudofermion annihilation 
spin-singlet pair breaking gives rise to one rotated spin $1/2$ of projection $\downarrow$ that
recombines with the annihilated $c$ pseudofermion within the removed $\downarrow$ rotated electron.
The created rotated spin $1/2$ of projection $\uparrow$ is that left over by the pair breaking. Its creation is
accounted for by the deviation $\delta N_{s1}^h=1$ in the number of $s1$ band holes. 
\vspace{0.50cm}

(5) LHB addition of one $\downarrow$ electron and thus of one $\downarrow$ rotated electron is a process that 
involves the creation of one $c$ pseudofermion and one $s1$ pseudofermion and annihilation of one unpaired rotated 
spin $1/2$ of projection $\uparrow$, so that $\delta N_c=1$ and $\delta N_{s1}=1$. The $s1$ pseudofermion creation involves
a spin-singlet pair formation. The annihilated unpaired rotated spin $1/2$ of projection $\uparrow$
combines with the rotated spin $1/2$ of projection $\downarrow$ of the created $\downarrow$ rotated electron
within such a $s1$ pseudofermion spin-singlet pair. The annihilation of the unpaired rotated spin $1/2$ of projection
$\uparrow$ is accounted for by the deviation $\delta N_{s1}^h=-1$ in the number of $s1$ band holes. 
\vspace{0.50cm}

(6) UHB addition of one $\downarrow$ electron and thus of one $\downarrow$ rotated electron is a process that 
involves the annihilation of one $c$ pseudofermion and one unpaired rotated spin $1/2$ of projection $\uparrow$ 
and creation of one $\eta 1$ pseudofermion, so that $\delta N_c=-1$ and $\delta N_{\eta 1}=1$. The annihilated unpaired rotated 
spin $1/2$ recombines with the annihilated $c$ pseudofermion within one $\uparrow$ rotated electron.
Such a $\uparrow$ rotated electron then pairs with the created $\downarrow$ rotated electron onto a doubly 
occupied site. The rotated $\eta$-spin $1/2$ of projection $-1/2$ that describes the $\eta$-spin degrees
of freedom of such a doubly occupied site combines with one ground-state unpaired rotated $\eta$-spin $1/2$ of projection $+1/2$ within the
$\eta 1$ pseudofermion $\eta$-spin singlet pair. 
The annihilation of one unpaired rotated spin $1/2$ leaves the number $N_{s1}$ $s1$ pseudofermions 
unchanged and gives rise to a deviation $\delta N_{s1}^h=-1$ in the number of $s1$ band holes.
\vspace{0.50cm}

The above elementary processes involving $s1$ pseudofermion annihilation pair breaking and 
$s1$ pseudofermion creation pair formation are behind the squeezed $s1$ effective lattice and corresponding
$s1$ momentum band being exotic, since their number of sites and discrete momentum values,
respectively, which both are given by $L_{s1} = N_{s1} + N_{s1}^h$, has different values for different subspaces. Hence
within the $s1$ pseudofermion operator algebra, one distinguishes the $s1$-band
holes created and annihilated under processes within which one $s1$ pseudofermion
is annihilated and created, respectively, from the $s1$-band
holes created and annihilated upon changing the number $L_{s1} = N_{s1} + N_{s1}^h$ 
of squeezed $s1$ effective lattice sites, which equals that of $s1$-band discrete momentum values. 
(For $S_s>0$ states such exotic $L_{s1}$ variations only lead to $N_{s1}^h$ variations.)

The former processes are described by application of the operators
${\bar{f}}_{\bar{q},s1}$ and ${\bar{f}}^{\dag}_{\bar{q},s1}$, respectively,
onto the initial state. On the other hand, the latter $N_{s1}^h$ variations
that do not conserve $L_{s1} = N_{s1} + N_{s1}^h$ result from vanishing energy and
vanishing momentum processes within which discrete momentum values
are added to and removed from one of the $s1$ band limiting momentum values
$q_{s1}^{\pm}$, Eq. (\ref{qcan-range}) for $\alpha n = s1$. Whether such an addition
or remotion occurs at the left limiting momentum $q_{s1}^{-}$ or at right limiting momentum $q_{s1}^{+}$
is uniquely defined, since the process must leave invariant the $s1$ band symmetrical relation
$q_{s1}^{+}=-q_{s1}^{-}$ for the final state. 

Specifically, in the case of the (i) $\uparrow$ one-electron removal processes (1) and
$\downarrow$ one-electron UHB addition processes (6) and (ii)
$\uparrow$ one-electron LHB addition processes (1) a single discrete momentum value is
(i) removed from and (ii) added to, respectively, the $s1$ band limiting momentum values. Such vanishing
energy and vanishing momentum processes are implicitly
accounted for by the pseudofermion representation through the $s1$ band discrete momentum values
of the final states, which are uniquely defined. 

In the following we use the transformation laws of the ground state, Eq. (\ref{GS}), upon acting onto it with
the $i=0,1,...,\infty$ operators on the right-hand side of the equation, $c_{k,\sigma} = \sum_{i =0}^{\infty}c_{k,\sigma,i}$
(and $c_{k,\sigma}^{\dag} = \sum_{i =0}^{\infty}c_{k,\sigma,i}^{\dag}$), for the $\sigma$ electron annihilation 
(and creation) operators whose first terms are given in Eq. (\ref{Sr-rot}) to derive the expression of the corresponding leading-order operators 
${\hat{g}} (k)\,{\hat{c}}_{\odot}$, Eq. (\ref{Oodot0kGO}), in terms of $c$ and $s1$ pseudofermion operators
for the processes (1), (2), (4), and (5) and in terms of $c$, $s1$, and $\eta 1$ pseudofermion operators
for the $\sigma$ one-electron UHB addition processes (3) and (6). 

Within our study of the line shape near the $\sigma$ one-electron spectral 
weight singular features the expression of the $\sigma$ electron creation and annihilation operators 
in terms of pseudofermion operators can be approximated by the corresponding 
leading-order term, ${\hat{g}} (k)\,{\hat{c}}_{\odot}$. 
In the case of the $\uparrow$ one-electron removal processes (1) one finds the following
leading-order expression,
\begin{eqnarray}
c_{k,\uparrow} & \approx & {\hat{g}}_{\iota} (k)\,{\hat{c}}_{\odot} \, ,
\nonumber \\
{\hat{c}}_{\odot} & = & {\bar{f}}_{\pm 2k_F,c} \, ; \hspace{0.5cm}
\Phi_{c}^0 = 0 \, ; \hspace{0.5cm} \Phi_{s1}^0= \iota/2 \, , \hspace{0.50cm} \iota = \pm 1 \, ,
\nonumber \\
{\hat{g}}_{\iota} (k) & = & {\bar{f}}^{\dag}_{\bar{q}(\pm 2k_F),c}\,{\bar{f}}_{\bar{q}(\iota k_{F\downarrow}),s1}  
\sum_{q = -2k_F}^{2k_F}
\Theta (k_{F\downarrow} - \vert k + q\vert)\, 
{\bar{f}}_{\bar{q}(q),c}\,{\bar{f}}^{\dag}_{\bar{q}(k+q),s1} \, ,
\label{upElremo}
\end{eqnarray}
where the shift parameters $\Phi_{\beta}^0$ whose value results from the ground-state transition to the 
excited energy eigenstates are those in Eq. (\ref{pican}) for $\beta = c,s1$, ${\bar{q}} (q) = q + 2\pi\,\Phi_{\beta} (q)/L$,
and the capital-$\Theta$ distribution $\Theta (x)$ is given here and in the following by $\Theta (x)=1$ for $x\geq 0$ and 
$\Theta (x)=0$ for $x<0$. The momentum $\mp k_{F\downarrow}$
resulting from the $s1$ pseudofermion annihilation at $\bar{q}(\pm k_{F\downarrow})$ exactly
cancels the momentum $\pm k_{F\downarrow}$ stemming from the overall $s1$ band
momentum shift $q_j \rightarrow q_j \pm \pi/L$ associated with $\Phi_{s1}^0=\pm 1/2$. 

Within a $k$ extended zone scheme, the $\omega <0$ spectrum generated by application of the $\uparrow$ one-electron removal
leading-order generator, Eq. (\ref{upElremo}), onto the ground state reads $-\omega = -\varepsilon_c (q) +\varepsilon_{s1} (k+q)$
and has the following two branches, 
\begin{eqnarray}
-\omega (k) & = & -\varepsilon_c (q) +\varepsilon_{s1} (q') \, ; \hspace{0.75cm} k = - q + q' \, ,
\nonumber \\
k & \in & [-k_{F\uparrow}, (2k_F + k_{F\uparrow})] \, ; \hspace{0.5cm} q \in [-2k_F,2k_F]
 \, ; \hspace{0.50cm} q' \in [k_{F\downarrow},k_{F\uparrow}] \, , \hspace{0.4cm}{\rm branch}\hspace{0.1cm}A \, ,
\nonumber \\
k & \in & [-(2k_F + k_{F\uparrow}),k_{F\uparrow}] \, ; \hspace{0.5cm} q \in [-2k_F,2k_F]
 \, ; \hspace{0.50cm} q' \in [-k_{F\uparrow},-k_{F\downarrow}] \, , \hspace{0.4cm}{\rm branch}\hspace{0.1cm}B \, .
\label{SpupElremo}
\end{eqnarray}

In the case of the $\uparrow$ one-electron LHB addition processes (2) the leading-order operator is given by,
\begin{eqnarray}
c^{\dag}_{k,\uparrow} & \approx & {\hat{g}}_{\iota}  (k)\,{\hat{c}}_{\odot} \, ,
\nonumber \\
{\hat{c}}_{\odot} & = & {\bar{f}}^{\dag}_{\pm 2k_F,c} \, ; \hspace{0.5cm}
\Phi_{c}^0 = 0 \, ; \hspace{0.5cm} \Phi_{s1}^0= \iota/2 \, , \hspace{0.50cm} \iota = \pm 1 \, ,
\nonumber \\
{\hat{g}}_{\iota} (k) & = & {\bar{f}}_{\bar{q}(\pm 2k_F),c}\,{\bar{f}}^{\dag}_{\bar{q}(-\iota k_{F\downarrow}),s1}  
(\sum_{q = -\pi}^{-2k_F} + \sum_{q = 2k_F}^{\pi})\,
\Theta (k_{F\downarrow} - \vert k - q\vert)\, 
{\bar{f}}^{\dag}_{\bar{q}(q),c}\,{\bar{f}}_{\bar{q}(-k + q),s1} \, ,
\label{upElLHBadd}
\end{eqnarray}
where the momentum $\mp k_{F\downarrow}$ resulting from the $s1$ pseudofermion creation at $\bar{q}(\mp k_{F\downarrow})$ exactly
cancels again the momentum $\pm k_{F\downarrow}$ stemming from an overall $s1$ band
momentum shift $q_j \rightarrow q_j \pm \pi/L$ that occurs under the ground-state
transition to the excited energy eigenstates. 

The $\omega >0$ spectrum generated by application of the $\uparrow$ one-electron LHB addition
leading-order generator, Eq. (\ref{upElLHBadd}), onto the ground state reads $\omega = \varepsilon_c (q) -\varepsilon_{s1} (k-q)$
and has within a $k$ extended zone scheme again two branches, 
\begin{eqnarray}
\omega (k) & = & \varepsilon_c (q) - \varepsilon_{s1} (q') \, ; \hspace{0.75cm} k = q - q' \, ,
\nonumber \\
k & \in & [k_{F\uparrow}, (\pi + k_{F\downarrow})] \, ; \hspace{0.5cm} q \in [2k_F,\pi]
 \, ; \hspace{0.50cm} q' \in [-k_{F\downarrow},k_{F\downarrow}] \, , \hspace{0.4cm}{\rm branch}\hspace{0.1cm}A \, ,
\nonumber \\
k & \in & [-(\pi + k_{F\downarrow}),-k_{F\uparrow}] \, ; \hspace{0.5cm} q \in [-\pi,-2k_F]
 \, ; \hspace{0.50cm} q' \in [-k_{F\downarrow},k_{F\downarrow}] \, , \hspace{0.4cm}{\rm branch}\hspace{0.1cm}B \, .
\label{SpupElLHBadd}
\end{eqnarray}

In the case of the $\uparrow$ one-electron UHB addition processes (3) the leading-order operator reads,
\begin{eqnarray}
c^{\dag}_{k,\uparrow} & \approx & {\hat{g}}_{\iota} (k)\,{\hat{c}}_{\odot} \, ,
\nonumber \\
{\hat{c}}_{\odot} & = & {\bar{f}}_{\iota 2k_F,c}\,{\bar{f}}_{\pm k_{F\downarrow},s1}\,{\bar{f}}^{\dag}_{-\iota (\pi - 2k_F),\eta 1} 
\, ; \hspace{0.5cm} \Phi_{c}^0 = \Phi_{s1}^0 = 0 \, , \hspace{0.50cm} \iota = \pm 1 \, ,
\nonumber \\
{\hat{g}}_{\iota} (k) & = & {\bar{f}}^{\dag}_{\bar{q}(\iota 2k_F),c}\,{\bar{f}}^{\dag}_{\bar{q}(\pm k_{F\downarrow}),s1}  
\sum_{q = -2k_F}^{2k_F} 
\Theta (k_{F\downarrow} -\vert k - \iota\,(\pi -2k_F) + q\vert)\, 
{\bar{f}}_{\bar{q}(q),c}\,{\bar{f}}_{\bar{q}(-k + \iota\,(\pi -2k_F)-q),s1} \, .
\label{upElUHBadd}
\end{eqnarray}
In this case one has $N_{\eta 1} (q_j)=1$ where $q_j = -\iota (\pi - 2k_F)$ and $M_{\eta,-1/2}=1$ for 
the excited energy eigenstates in the general momentum expression, Eq. (\ref{P}),
so that the momentum $\pi\,M_{\eta,-1/2}=\pi$ combines with $(\pi -q_j)\, N_{\eta 1} (q_j) 
= \pi - q_j$ to give $2\pi - q_j = -q_j =\iota (\pi - 2k_F)$.

Within a $k$ extended zone scheme, the $\omega >0$ spectrum generated by application of the $\uparrow$ one-electron UHB addition
leading-order generator, Eq. (\ref{upElUHBadd}), onto the ground state reads $\omega = 2\mu -\varepsilon_c (q) -\varepsilon_{s1} (k - \iota\,(\pi -2k_F)+q)$
and has two branches corresponding to $\iota = \pm 1$, 
\begin{eqnarray}
\omega (k) & = & 2\mu -\varepsilon_c (q) - \varepsilon_{s1} (q') \, ; \hspace{0.75cm} k = \iota (\pi-2k_F) - q - q' 
\, ; \hspace{0.5cm} q \in [-2k_F,2k_F]
 \, ; \hspace{0.50cm} q' \in [-k_{F\downarrow},k_{F\downarrow}] \, ,
\nonumber \\
k & = & (\pi-2k_F) - q - q' \in [(\pi - 4k_F -k_{F\downarrow}), (\pi + k_{F\uparrow})] \, , \hspace{0.4cm}{\rm branch}\hspace{0.1cm}A \, ,
\nonumber \\
k & = & - (\pi-2k_F) - q - q' \in [-(\pi + k_{F\uparrow}),-(\pi - 4k_F -k_{F\downarrow})] \, , \hspace{0.4cm}{\rm branch}\hspace{0.1cm}B  \, .
\label{SpupElUHBadd}
\end{eqnarray}

In the case of the $\downarrow$ one-electron removal processes (4) the leading-order operator is given by,
\begin{eqnarray}
c_{k,\downarrow} & \approx & {\hat{g}}_{\iota} (k)\,{\hat{c}}_{\odot} \, ,
\nonumber \\
{\hat{c}}_{\odot} & = & {\bar{f}}_{\iota\,2k_F,c}\,{\bar{f}}_{-\iota k_{F\downarrow},s1} \, ; \hspace{0.5cm}
\Phi_{c}^0 = \iota/2 \, ; \hspace{0.5cm} \Phi_{s1}^0 = 0 \, , \hspace{0.50cm} \iota = \pm 1 \, ,
\nonumber \\
{\hat{g}}_{\iota} (k) & = & {\bar{f}}^{\dag}_{\bar{q}(\iota\, 2k_F),c}\,{\bar{f}}^{\dag}_{\bar{q}(-\iota k_{F\downarrow}),s1}  
\sum_{q = -2k_F}^{2k_F} 
\Theta (k_{F\downarrow} -\vert k - \iota\,2k_F + q\vert)\, 
{\bar{f}}_{\bar{q}(q),c}\,{\bar{f}}_{\bar{q}(-k + \iota\,2k_F-q),s1} \, .
\label{downElremo}
\end{eqnarray}
The operator ${\bar{f}}_{\iota\,2k_F,c}$ in ${\hat{c}}_{\odot}$ leads to a momentum $-\iota 2k_F$ 
that exactly cancels the momentum $\iota 2k_F$
stemming from the overall $c$ band momentum shift associated with $\Phi_{c}^0 = \iota/2$
whereas the operator ${\bar{f}}^{\dag}_{\bar{q}(\iota\, 2k_F),c}$
in ${\hat{g}}_{\iota} (k)$ leads to a momentum contribution that restores such a momentum $\iota 2k_F$.

The $\omega <0$ spectrum generated by application of the $\downarrow$ one-electron removal
leading-order generator, Eq. (\ref{downElremo}), onto the ground state reads $-\omega = -\varepsilon_c (q) -\varepsilon_{s1} (k - \iota\,2k_F+q)$
and has two branches corresponding to $\iota = \pm 1$, 
\begin{eqnarray}
\omega (k) & = & -\varepsilon_c (q) - \varepsilon_{s1} (q') \, ; \hspace{0.75cm} k = \iota\,2k_F - q - q' 
\, ; \hspace{0.5cm} q \in [-2k_F,2k_F]
 \, ; \hspace{0.50cm} q' \in [-k_{F\downarrow},k_{F\downarrow}] \, ,
\nonumber \\
k & = & 2k_F - q - q' \in [-k_{F\downarrow}, (4k_F + k_{F\uparrow})] \, , \hspace{0.4cm}{\rm branch}\hspace{0.1cm}A \, ,
\nonumber \\
k & = & - 2k_F - q - q' \in [-(4k_F + k_{F\uparrow}), k_{F\downarrow}] \, , \hspace{0.4cm}{\rm branch}\hspace{0.1cm}B  \, .
\label{SpdownElremo}
\end{eqnarray}

In the case of the $\downarrow$ one-electron LHB addition processes (5) the leading-order operator reads,
\begin{eqnarray}
c^{\dag}_{k,\downarrow} & \approx & {\hat{g}}_{\iota} (k)\,{\hat{c}}_{\odot} \, ,
\nonumber \\
{\hat{c}}_{\odot} & = & {\bar{f}}^{\dag}_{-\iota\,2k_F,c}\,{\bar{f}}^{\dag}_{\iota k_{F\downarrow},s1} \, ; \hspace{0.5cm}
\Phi_{c}^0 = \iota/2 \, ; \hspace{0.5cm} \Phi_{s1}^0 = 0 \, , \hspace{0.50cm} \iota = \pm 1\, ,
\nonumber \\
{\hat{g}}_{\iota} (k) & = & {\bar{f}}_{\bar{q}(-\iota\, 2k_F),c}\,{\bar{f}}_{\bar{q}(\iota k_{F\downarrow}),s1}  
\nonumber \\
& \times & (\sum_{q = -\pi}^{-2k_F} + \sum_{q = 2k_F}^{\pi})\,\delta_{-\iota,{\rm sgn}\{k - \iota\,2k_F -q\}}
\Theta (k_{F\uparrow} -\vert k -\iota\,2k_F - q\vert)\Theta (\vert k -\iota\,2k_F - q\vert - k_{F\downarrow})
\nonumber \\
& \times &  
{\bar{f}}^{\dag}_{\bar{q}(q),c}\,{\bar{f}}^{\dag}_{\bar{q}(k - \iota\,2k_F -q),s1} \, .
\label{downElLHBadd}
\end{eqnarray}
Here and throughout this paper one has that ${\rm sgn}\{x\}=1$ for $x>0$, ${\rm sgn}\{x\}=-1$ for $x<0$,
and ${\rm sgn}\{x\}=0$ for $x=0$. The operator ${\bar{f}}^{\dag}_{-\iota\,2k_F,c}$
in the operator ${\hat{c}}_{\odot}$ leads to a momentum $-\iota 2k_F$ that exactly cancels the momentum $\iota 2k_F$
stemming from the $c$ band overall momentum shift whereas the operator ${\bar{f}}_{\bar{q}(-\iota\, 2k_F),c}$
in ${\hat{g}}_{\iota} (k)$ leads to a momentum contribution that restores such a momentum $\iota 2k_F$.

Within a $k$ extended zone scheme the $\omega >0$ spectrum generated by application of the $\downarrow$ one-electron LHB addition
leading-order generator, Eq. (\ref{downElLHBadd}), onto the ground state reads $\omega = \varepsilon_c (q) + \varepsilon_{s1} (k - \iota\,2k_F -q)$
and has four branches, 
\begin{eqnarray}
\omega (k) & = & \varepsilon_c (q) + \varepsilon_{s1} (q') \, ; \hspace{0.75cm} k = \iota\,2k_F + q + q' 
\, ; \hspace{0.50cm} {\rm sgn}\{q'\} = -\iota\hspace{0.10cm}{\rm for}\hspace{0.10cm}q'\neq 0 \, ,
\nonumber \\
k & = & 2k_F + q + q' \in [(4k_F +k_{F\uparrow}), (\pi + 2k_F + k_{F\uparrow})] \, , \hspace{0.4cm}{\rm branch}\hspace{0.1cm}A \, ,
\nonumber \\
& & q \in [2k_F,\pi] \, ; \hspace{0.50cm} q' \in [k_{F\downarrow},k_{F\uparrow}] \, , 
\nonumber \\
k & = & 2k_F + q + q' \in [ - (\pi - 2k_F - k_{F\downarrow}),k_{F\uparrow}] \, , \hspace{0.4cm}{\rm branch}\hspace{0.1cm}B \, ,
\nonumber \\
& & q \in [-\pi,-2k_F] \, ; \hspace{0.50cm} q' \in [k_{F\downarrow},k_{F\uparrow}] \, , 
\nonumber \\ 
k & = & - 2k_F + q + q' \in [-(\pi + 2k_F + k_{F\uparrow}), -(4k_F +k_{F\uparrow})] \, , \hspace{0.4cm}{\rm branch}\hspace{0.1cm}A' \, ,
\nonumber \\
& & q \in [-\pi,-2k_F]
 \, ; \hspace{0.50cm} q' \in [-k_{F\uparrow},-k_{F\downarrow}] \, , 
\nonumber \\
k & = & - 2k_F + q + q' \in [-k_{F\uparrow}, (\pi - 2k_F - k_{F\downarrow})] \, , \hspace{0.4cm}{\rm branch}\hspace{0.1cm}B'  \, ,
\nonumber \\
& & q \in [2k_F,\pi]
 \, ; \hspace{0.50cm} q' \in [-k_{F\uparrow},-k_{F\downarrow}] \, .
\label{SpdownElLHBadd}
\end{eqnarray}

In the case of the UHB addition of one $\downarrow$ electron processes (6) the leading-order operator is given by,
\begin{eqnarray}
{\hat{c}}^{\dag}_{k,\downarrow} & \approx & {\hat{g}} (k)\,{\hat{c}}_{\odot} \, ,
\nonumber \\
{\hat{c}}_{\odot} & = & {\bar{f}}_{\iota\,2k_F,c}\,{\bar{f}}^{\dag}_{-\iota (\pi -2k_F),\eta 1} 
 \, ; \hspace{0.5cm}
\Phi_{c}^0 = \iota/2 \, ; \hspace{0.5cm} \Phi_{s1}^0 = \pm 1/2 \, , \hspace{0.50cm} \iota = \pm 1\, ,
\nonumber \\
{\hat{g}} (k) & = & {\bar{f}}^{\dag}_{\bar{q}(\iota\,2k_F),c}\,{\bar{f}}_{\bar{q}(\pm k_{F\downarrow}),s1}  
\sum_{q = -2k_F}^{2k_F} \Theta (k_{F\downarrow} - \vert k - \iota\,\pi + q\vert)\, 
{\bar{f}}_{\bar{q}(q),c}\,{\bar{f}}^{\dag}_{\bar{q}(k - \iota\,\pi + q),s1} \, .
\label{downElUHBadd}
\end{eqnarray}
The operator ${\bar{f}}_{\iota\,2k_F,c}$ in ${\hat{c}}_{\odot}$ leads to a momentum 
$-\iota 2k_F$ that exactly cancels the momentum $\iota 2k_F$
stemming from the $c$ band overall momentum shift whereas the operator ${\bar{f}}^{\dag}_{\bar{q}(\iota\, 2k_F),c}$
in ${\hat{g}}_{\iota} (k)$ leads to a momentum contribution that restores such a momentum $\iota 2k_F$. The
latter momentum is finally cancelled by the momentum $-\iota 2k_F$ from the second term of
the momentum $\iota (\pi - 2k_F)$ stemming from ${\bar{f}}^{\dag}_{-\iota(\pi -2k_F),\eta 1}$.
Indeed, as in the case of the $\uparrow$ one-electron UHB addition processes (3), Eq. (\ref{upElUHBadd}),
one has $N_{\eta 1} (q_j)=1$ where $q_j = -\iota (\pi - 2k_F)$ and $M_{\eta,-1/2}=1$ for the excited energy 
eigenstates in the general momentum expression, Eq. (\ref{P}),
so that the momentum $\pi\,M_{\eta,-1/2}=\pi$ combines with $(\pi -q_j)\, N_{\eta 1} (q_j) 
= \pi - q_j$ to give $2\pi - q_j = -q_j =\iota (\pi - 2k_F)$. Moreover, the momentum $\mp k_{F\downarrow}$
resulting from the $s1$ pseudofermion annihilation at $\bar{q}(\pm k_{F\downarrow})$ exactly
cancels the momentum $\pm k_{F\downarrow}$ stemming from the $s1$ band overall 
momentum shift. 

The $\omega >0$ spectrum generated by application of the $\downarrow$ one-electron UHB addition
leading-order generator, Eq. (\ref{downElUHBadd}), onto the ground state reads $\omega = 2\mu -\varepsilon_c (q) + \varepsilon_{s1} (k - \iota\,\pi +q)$
and has within a $k$ extended zone scheme the following two branches, 
\begin{eqnarray}
\omega (k) & = & 2\mu -\varepsilon_c (q) +\varepsilon_{s1} (q') \, ; \hspace{0.75cm} k = \iota\,\pi - q + q' = \pi - q + q' \, ,
\nonumber \\
k & \in & [(\pi-k_{F\uparrow}), (\pi + 2k_F + k_{F\uparrow})] \, ; \hspace{0.5cm} q \in [-2k_F,2k_F]
 \, ; \hspace{0.50cm} q' \in [k_{F\downarrow},k_{F\uparrow}] \, , \hspace{0.4cm}{\rm branch}\hspace{0.1cm}A \, ,
\nonumber \\
k & \in & [(\pi - 2k_F - k_{F\uparrow}),(\pi + k_{F\uparrow})] \, ; \hspace{0.5cm} q \in [-2k_F,2k_F]
 \, ; \hspace{0.50cm} q' \in [-k_{F\uparrow},-k_{F\downarrow}] \, , \hspace{0.4cm}{\rm branch}\hspace{0.1cm}B \, .
\label{SpdownElUHBadd}
\end{eqnarray}

In the above expressions, the $c$ and/or $s1$ pseudofermion momentum values $\pm 2k_{F}$
and $\pm k_{F\downarrow}$, respectively, appearing in the
operators ${\hat{c}}_{\odot}$ belong to the initial ground state $\beta = c,s1$ band 
whereas the $\beta$ pseudofermion momentum values ${\bar{q}} (q) = q + 2\pi\,\Phi_{\beta} (q)/L$
in the operators ${\hat{g}} (k)$ expressions belong to the excited energy eigenstates $\beta =c,s1$ bands.

\subsection{The $\sigma$ one-electron operators matrix elements between the
ground state and the excited energy eigenstates and corresponding spectral functions
in terms of $\beta =c,s1$ pseudofermion spectral functions}
\label{matrixOnel} 

The $\sigma$ one-electron spectral functions, Eq. (\ref{Bkomega}), can be written in the pseudofermion representation
as follows,
\begin{equation}
B (k,\omega)  = \sum_{i' =0}^{\infty}\sum_{\nu}\, \vert\langle\nu\vert\,
{\hat{g}}_{i'} (k)\,{\hat{c}}_{\odot}\vert GS\rangle\vert^2\,
\delta\Bigl(\omega - \gamma (E_{\nu} - E_{GS})\Bigr) \, , \hspace{0.50cm} \gamma\,\omega > 0 \, , 
\label{ABON-odotGEN}
\end{equation}
where for simplicity we have omitted from $B (k,\omega)$ the labels $\sigma $ and $\gamma =\pm 1$
and denoted the excited-state indices $\nu^-$ and $\nu^+$ generally by $\nu$.

Following the above properties, one approximates the general spectral function,
Eq. (\ref{ABON-odotGEN}), by its pseudofermion leading-order term
involving the operators given in Eqs. (\ref{upElremo}), (\ref{upElLHBadd}), (\ref{upElUHBadd}), (\ref{downElremo}), 
(\ref{downElLHBadd}), and (\ref{downElUHBadd}),
\begin{equation}
B (k,\omega) \approx B^{\odot} (k,\omega) = \sum_{\nu}\, \vert\langle\nu\vert\,
{\hat{g}} (k)\,{\hat{c}}_{\odot}\vert GS\rangle\vert^2\,\delta\Bigl(
\omega - \gamma (E_{\nu} - E_{GS})\Bigr) \, , \hspace{0.50cm} \gamma\,\omega > 0 \, . 
\label{ABON-odot}
\end{equation}

Both the generator onto the electron vacuum of the initial ground state in Eq. (\ref{GS}) and the
operator ${\hat{c}}_{\odot}$ in ${\hat{c}}_{\odot}\vert GS\rangle$ are written in terms 
of $c$ and $s1$ pseudofermion creation and/or annihilation operators, Eqs. (\ref{f-f-Q}) and (\ref{f-f-FT-Q}),
whose discrete canonical momentum values equal the corresponding momentum 
values $q_j$, Eqs. (\ref{q-j}) and (\ref{Ic-an}), of that initial ground state. In the case of
the $\sigma$ one-electron UHB addition operators in Eqs. (\ref{upElUHBadd}) and (\ref{downElUHBadd}), 
the expression of the operator ${\hat{c}}_{\odot}$ includes as well a $\eta 1$ pseudofermion creation operator
of canonical momentum $\pm (\pi - 2k_F)$.

On the other hand, both the operator ${\hat{g}} (k)$ and the generators onto the electron vacuum of the
excited energy eigenstates $\vert\nu\rangle$ are written in terms of $c$ and $s1$ pseudofermion 
operators whose discrete canonical momentum values ${\bar{q}}_j$, Eq. (\ref{barqan}),
are those of these excited energy eigenstates. Interestingly, there is always an exact excited energy 
eigenstate $\vert f_G \rangle$ of the final $N_{\sigma}\pm 1$ ground state $\vert GS_f\rangle$ such that,
\begin{equation}
\vert f_G \rangle = {\hat{g}} (k)\vert GS_f\rangle \, .
\label{statef}
\end{equation}

In the case of the $c$ and $s1$ bands, the two types of 
discrete canonical momentum values that correspond to the initial ground state and excited energy
eigenstates, respectively, account for the Anderson orthogonality catastrophe \cite{Karlo-97,Anderson} 
occurring in these bands under the transitions to the excited energy eigenstates $\vert\nu\rangle$.
Such an Anderson orthogonality catastrophe is behind the exotic character of the
quantum overlaps that control the one-electron spectral functions. On the other hand, since the initial ground 
state is not populated by $\eta 1$ pseudofermions and in the case of $\sigma$ one-electron UHB addition
the $\eta 1$ band limiting canonical momentum values $\pm (\pi - 2k_F)$ of the created
$\eta 1$ pseudofermion are unchanged relative to the corresponding $\eta 1$ pseudoparticle momentum values,
the $\sigma$ one-electron operators matrix elements overlaps involving such a $\eta 1$ pseudofermion 
are straightforwardly computed. 

The excitation ${\hat{g}} (k)\,{\hat{c}}_{\odot} \vert GS\rangle$ in the matrix elements
of the spectral function expression, Eq. (\ref{ABON-odot}), has finite overlap
with the corresponding specific energy eigenstate, Eq. (\ref{statef}), which gives,
\begin{eqnarray}
\langle f_G \vert\,{\hat{g}} (k)\,{\hat{c}}_{\odot} \vert GS\rangle & = & \langle GS_f^{\rm ex}\vert {\hat{c}}_{\odot} \vert GS\rangle
\nonumber \\
& = & \langle 0\vert \prod_{\beta =c,s1}
{\bar{f}}_{{{\bar{q}}_{N_{\beta}^{\odot}},\,\beta}}...{\bar{f}}_{{{\bar{q}}_2},\,\beta}\,{\bar{f}}_{{{\bar{q}}_1},\,\beta}
{\bar{f}}^{\dag }_{{{q'}_1},\,\beta}\,{\bar{f}}^{\dag }_{{{q'}_2},\,\beta}...{\bar{f}}^{\dag}_{{{q'}_{N_{\beta}^{\odot}}},\,\beta}\vert 0\rangle 
\nonumber \\
& = & \langle 0\vert \prod_{\beta =c,s1}{\bar{f}}_{{{q'}_{N_{\beta}^{\odot}},\,\beta}}...{\bar{f}}_{{{q'}_2},\,\beta}\,{\bar{f}}_{{{q'}_1},\,\beta}
{\bar{f}}^{\dag }_{{{\bar{q}}_1},\,\beta}\,{\bar{f}}^{\dag }_{{{\bar{q}}_2},\,\beta}...{\bar{f}}^{\dag}_{{{\bar{q}}_{N_{\beta}^{\odot}}},\,\beta}\vert 0\rangle^* \, ,
\label{matrixel}
\end{eqnarray}
where $\vert GS_f^{\rm ex}\rangle$ is a state with the same $c$ and $s1$ pseudofermion occupancy as $\vert GS_f\rangle$
but whose $\beta = c,s1$ band discrete momentum values are those of its excited energy eigenstate $\vert f_G \rangle = {\hat{g}} (k) \vert GS_f\rangle$
and $N_{\beta}^{\odot}$ is the number of $\beta =c$ and $\beta =s1$ pseudofermions of the states 
${\hat{c}}_{\odot}\vert GS\rangle$ and $\vert GS_f\rangle$. 

The $\beta = c,s1$ bands discrete canonical momentum values ${q'}_1$, ${q'}_2$, ...,${q'}_{N_{\beta}^{\odot}}$ in Eq. (\ref{matrixel})
equal the corresponding initial ground state discrete momentum values whereas
${\bar{q}}_1$, ${\bar{q}}_2$, ...,${\bar{q}}_{N_{\beta}^{\odot}}$ are the discrete canonical
momentum values of the excited energy eigenstate $\vert f_G \rangle$, Eq. (\ref{statef}).
Since these two sets of discrete momenta have different values, an Anderson orthogonality catastrophe 
occurs such that the excited energy eigenstates of general form,
\begin{eqnarray}
\vert f_{G_C} \rangle & = & \prod_{\beta =c,s1}{\hat{g}}_C (m_{\beta,+1},m_{\beta,-1})\,{\hat{g}} (k) \vert GS_f\rangle 
\nonumber \\
& = & \prod_{\beta =c,s1}{\hat{g}}_C (m_{\beta,+1},m_{\beta,-1})\,\vert f_G \rangle
\, , \hspace{0.5cm} \beta = c, s1 \, , \hspace{0.50cm} \iota = \pm 1 \, ,
\label{statefGC}
\end{eqnarray}
which result from application onto the state $\vert f_G \rangle$, Eq. (\ref{statef}), of the
$\beta =c,s1$ generators ${\hat{g}}_C (m_{\beta,+1},m_{\beta,-1})$ of the low-energy and small-momentum processes (C),
also have overlap with the excitation ${\hat{g}} (k)\,{\hat{c}}_{\odot} \vert GS\rangle$.

One then finds that,
\begin{eqnarray}
& & \langle f_G\vert\prod_{\beta =c,s1}{\hat{g}}^{\dag}_C (m_{\beta,+1},m_{\beta,-1})
{\hat{g}} (k)\,{\hat{c}}_{\odot} \vert GS\rangle = 
\langle GS_f^{\rm ex}\vert \prod_{\beta =c,s1}{\hat{g}}^{\dag}_C (m_{\beta,+1},m_{\beta,-1}){\hat{c}}_{\odot} \vert GS\rangle
\nonumber \\
& = & \langle 0\vert\prod_{\beta =c,s1}{\bar{f}}_{{{\bar{q}}_{N_{\beta}^{\odot}},\,\beta}}...{\bar{f}}_{{{\bar{q}}_2},\,\beta}\,{\bar{f}}_{{{\bar{q}}_1},\,\beta}\,
{\hat{g}}^{\dag}_C (m_{\beta,+1},m_{\beta,-1})
{\bar{f}}^{\dag }_{{{q'}_1},\,\beta}\,{\bar{f}}^{\dag }_{{{q'}_2},\,\beta}...{\bar{f}}^{\dag}_{{{q'}_{N_{\beta}^{\odot}}},\,\beta}\vert 0\rangle 
\nonumber \\
& = & \langle 0\vert\prod_{\beta =c,s1}{\bar{f}}_{{{q'}_{N_{\beta}^{\odot}},\,\beta}}...{\bar{f}}_{{{q'}_2},\,\beta}\,{\bar{f}}_{{{q'}_1},\,\beta}\,
{\hat{g}}^{\dag}_C (m_{\beta,+1},m_{\beta,-1})\,
{\bar{f}}^{\dag }_{{{\bar{q}}_1},\,\beta}\,{\bar{f}}^{\dag }_{{{\bar{q}}_2},\,\beta}...{\bar{f}}^{\dag}_{{{\bar{q}}_{N_{\beta}^{\odot}}},\,s1}\vert 0\rangle^* \, .
\label{mtrixelC}
\end{eqnarray}
The number of elementary $\beta =c,s1$ pseudofermion - pseudofermion-hole processes (C)
of momentum $\pm 2\pi/L$ in the vicinity of the $\beta;\iota=\pm 1$ Fermi points of $\vert GS_f\rangle$ is
denoted here and in the following by $m_{\beta,\iota}=1,2,3,...$. Such processes conserve the number
$N_{\beta}^{\odot}$ of $\beta =c,s1$ pseudofermions, so that the matrix elements, Eq. (\ref{mtrixelC}), have the same form
as that in Eq. (\ref{matrixel}) but with the excited-state occupied discrete canonical momentum
values ${\bar{q}}_1$, ${\bar{q}}_2$, ...,${\bar{q}}_{N_{\beta}^{\odot}}$ in the vicinity of the $\beta =c,s1$ bands
Fermi points being slightly different from those in that equation. 

The function $B^{\odot} (k,\omega)$, Eq. (\ref{ABON-odot}), is below expressed in terms of a sum of terms each
of which is a convolution of $c$ and $s1$ pseudofermion spectral functions. The expression of such pseudofermion spectral functions
involves sums that run over the processes (C) numbers $m_{\beta,\iota}=1,2,3,...$. It reads,
\begin{eqnarray}
B_{Q_{\beta}} (k',\omega') & = & {L\over 2\pi}\sum_{m_{\beta,\,+1};m_{\beta,\,-1}}\,A^{(0,0)}_{\beta}\,a_{\beta} (m_{\beta,\,+1},\,m_{\beta,\,-1})
\nonumber \\
& \times & \delta \Bigl(\omega' -{2\pi\over L}\,v_{\beta}\sum_{\iota =\pm1} (m_{\beta,\iota}+\Delta_{\beta}^{\iota})\Bigr)\,
\delta \Bigl(k' -{2\pi\over L}\,\sum_{\iota =\pm1}\iota\,(m_{\beta,\iota}+\Delta_{\beta}^{\iota})\Bigr) 
\, , \hspace{0.50cm} \beta = c,s1 \, ,
\label{BQ-gen}
\end{eqnarray}
where the $\beta =c,s1$ {\it lowest peak weights} $A^{(0,0)}_{\beta}$ are associated with a transition from the ground state
to a PS excited energy eigenstate generated by processes (A) and (B), the relative weights 
$a_{\beta}=a_{\beta} (m_{\beta,\,+1},\,m_{\beta,\,-1})$ are generated by additional processes (C) whose
$\beta =c,s1$ generators ${\hat{g}}_C (m_{\beta,+1},m_{\beta,-1})$ are those in Eq. (\ref{statefGC}), 
and $\Delta_{\beta}^{\iota}$ refers to the functional $2\Delta_{\beta}^{\iota}= 
(\iota\delta N^{F}_{\beta,\iota} + \Phi_{\beta}(\iota q_{F\beta}))^2$ associated
with the $\beta =c,s1$ pseudofermion number deviation $\delta N^{F}_{\beta,\iota}$ 
at the $\iota =\pm 1$ Fermi points and corresponding phase shift $2\pi\,\Phi_{\beta}(\iota q_{F\beta})$, 
Eq. (\ref{Phibetaq}), in units of $2\pi$ acquired by the $\beta =c,s1$ pseudofermions with
momenta $\iota q_{F\beta}=\pm q_{F\beta}$ under the above transition. This functional plays a key 
role in the PDT and is found below to emerge naturally from the $\beta =c,s1$ pseudofermion 
spectral weights.

In the case of $\sigma$ one-electron UHB addition, the $\beta =c,s1$ weights $A^{(0,0)}_{\beta}\,a_{\beta} (m_{\beta,\,+1},\,m_{\beta,\,-1})$ 
in Eq. (\ref{BQ-gen}) are reached after the quantum overlap stemming from creation of the $\eta 1$ pseudofermion 
is trivially computed. For all the $\sigma $ one-electron removal, LHB addition, and UHB addition processes that contribute to the
spectral functions in the vicinity of their singular features the $\beta =c,s1$ weights $A^{(0,0)}_{\beta}\,a_{\beta} (m_{\beta,\,+1},\,m_{\beta,\,-1})$ 
have the general form,
\begin{equation}
\vert\langle 0\vert {\bar{f}}_{{{q'}_{N_{\beta}^{\odot}},\,\beta}}...{\bar{f}}_{{{q'}_2},\,\beta}\,{\bar{f}}_{{{q'}_1},\,\beta}
{\bar{f}}^{\dag }_{{{\bar{q}}_1},\,\beta}\,{\bar{f}}^{\dag }_{{{\bar{q}}_2},\,\beta}...{\bar{f}}^{\dag}_{{{\bar{q}}_{N_{\beta}^{\odot}}},\,\beta}\vert 0\rangle\vert^2 \, , \hspace{0.5cm} \beta = c,s1 \, ,
\label{Aa}
\end{equation}
where $N_{\beta}^{\odot}$  stands for the number of $\beta =c,s1$ pseudofermions of the excited energy eigenstate 
generated by the processes (A) and (B). Such matrix element square can be expressed in terms of a Slater determinant 
of $\beta =c,s1$ pseudofermion operators, Eqs. (\ref{f-f-Q}) and (\ref{f-f-FT-Q}), as follows,
\begin{equation}
\left|\left|
\begin{array}{llcl} \{{\bar{f}}^{\dag }_{{{\bar{q}}_1},\,\beta}\, ,{\bar{f}}_{{{q'}_1},\,\beta}\} & \{{\bar{f}}^{\dag }_{{{\bar{q}}_1},\,\beta}\, ,{\bar{f}}_{{{q'}_2},\,\beta}\} 
& \cdots & 
\{{\bar{f}}^{\dag }_{{{\bar{q}}_1},\,\beta}\, ,{\bar{f}}_{{{q'}_{N_{\beta}^{\odot}},\,\beta}}\} \\
\{{\bar{f}}^{\dag }_{{{\bar{q}}_2},\,\beta}\, ,{\bar{f}}_{{{q'}_1},\,\beta}\} & \{{\bar{f}}^{\dag}_{{{\bar{q}}_2},\,\beta}\, ,{\bar{f}}_{{{q'}_2},\,\beta}\} 
& \cdots & \{{\bar{f}}^{\dag}_{{{\bar{q}}_2},\,\beta}\, ,{\bar{f}}_{{{q'}_{N_{\beta}^{\odot}},\,\beta}}\} \\
\multicolumn{4}{c}\dotfill\\
\{{\bar{f}}^{\dag }_{{{\bar{q}}_{N_{\beta}^{\odot}}},\,\beta}\, ,{\bar{f}}_{{{q'}_1},\,\beta}\} 
& \{{\bar{f}}^{\dag }_{{{\bar{q}}_{N_{\beta}^{\odot}}},\,\beta}\, ,{\bar{f}}_{{{q'}_2},\,\beta}\} 
& \cdots & 
\{{\bar{f}}^{\dag}_{{{\bar{q}}_{N_{\beta}^{\odot}}},\,\beta}\, ,{\bar{f}}_{{{q'}_{N_{s1}^{\odot}}},\,\beta}\}
\end{array} \right| \right|^2 \, , \hspace{0.5cm} \beta = c,s1 \, .
\label{det1}
\end{equation}
The $\beta =c,s1$ pseudofermion operators matrix elements 
$\langle 0\vert {\bar{f}}_{{{q'}_{N_{\beta}^{\odot}},\,\beta}}...{\bar{f}}_{{{q'}_2},\,\beta}\,{\bar{f}}_{{{q'}_1},\,\beta}{\bar{f}}^{\dag }_{{{\bar{q}}_1},\,\beta}\,{\bar{f}}^{\dag }_{{{\bar{q}}_2},\,\beta}...{\bar{f}}^{\dag}_{{{\bar{q}}_{N_{\beta}^{\odot}}},\,\beta}\vert 0\rangle$ in Eq. (\ref{Aa})
are associated with the two factors of the product $\prod_{\beta =c,s1}$ in the matrix elements, Eq. (\ref{matrixel}). 

The function $B^{\odot} (k,\omega)$, Eq. (\ref{ABON-odot}), can be written as follows,
\begin{equation}
B^{\odot} (k,\omega) = \sum_{\nu}
\Theta\Bigl(\Omega -\delta\omega_{\nu}\Bigr)\,\Theta\Bigl(\delta\omega_{\nu}\Bigr)\,
\Theta\left(\vert v_{\nu}\vert -v_{{\bar{\beta}}} \right)
{\breve{B}}^{\odot}_{\nu} (\delta\omega_{\nu},v_{\nu}) \, .
\label{B-PAR-J-CPHS-sum-0}
\end{equation}
The summation $\sum_{\nu}$ runs here over excited energy eigenstates 
generated by processes (A), (B), and (C) of the general form, Eq. (\ref{statefGC}), at fixed values 
of $k$ and $\omega$. Such states have excitation energy and momentum, Eq. (\ref{dE-dP}), in the ranges $\delta E_{\nu}^{\odot} \in [\omega - \Omega,\omega]$ 
and $\delta P_{\nu}^{\odot}\in [k - \Omega/v_{\nu},k]$ where,
\begin{eqnarray}
\delta\omega_{\nu} & = & (\omega-\gamma\,\delta E_{\nu}^{\odot} ) = (\omega - \gamma\,E_{\nu}^{\odot} + \gamma\,E_{GS})  \, ; \hspace{0.50cm}
\delta k_{\nu} = k -\delta P_{\nu}^{\odot} \, ,
\nonumber \\
\delta E_{\nu} & = & \gamma\,\delta E_{\nu}^{\odot} + \delta\omega_{\nu} = \omega \, ; \hspace{0.50cm}
P_{\nu} = \delta P_{\nu}^{\odot} + \delta k_{\nu} = k \, .
\label{dEdPvv*}
\end{eqnarray}
Here the energy and momentum spectra,
\begin{equation}
\delta E_{\nu}^{\odot} = E_{\nu}^{\odot} - E_{GS} \, ; \hspace{0.75cm} \delta P^{\odot}_{\nu} = P^{\odot}_{\nu} - P_{GS} \, ,
\label{dE-dP}
\end{equation}
are those of the excited energy eigenstates $\vert f_G \rangle$, Eq. (\ref{statef}), generated by the processes (A) and (B), 
which have finite quantum overlap with the excitation ${\hat{g}} (k)\,{\hat{c}}_{\odot}\vert GS\rangle$.
The velocities in Eq. (\ref{B-PAR-J-CPHS-sum-0}) read,
\begin{equation}
v_{\nu} = \delta\omega_{\nu}/\delta k_{\nu} 
\, ; \hspace{0.50cm}
v_{{\bar{\beta}}} = {\rm min}\{v_c,v_{s1}\} 
\, ; \hspace{0.50cm}
v_{\beta} = {\rm max}\{v_c,v_{s1}\}  \, ,
\label{v-bb-b}
\end{equation}
where $v_c$ and $v_{s1}$ are the $\beta =c,s1$ Fermi velocities, Eq. (\ref{vel-beta}). 
The energy deviation $\delta E_{\nu} =\omega$ and momentum deviation $\delta P_{\nu} =k$ 
in Eq. (\ref{dEdPvv*}) denote the excitation energy and momentum of the excited energy eigenstates,
respectively. $\Omega$ is the processes (C) energy range. It is self-consistently determined as that 
for which the velocity $v_{\nu}$, Eq. (\ref{dEdPvv*}), remains nearly unchanged. 

The lack of $c$ and $s1$ pseudofermion interaction terms in the PS finite-$u$ energy spectrum, Eq. (\ref{DE}), 
enables the function ${\breve{B}}^{\odot}_{\nu} (\delta\omega_{\nu},v_{\nu})$ in Eq. (\ref{B-PAR-J-CPHS-sum-0})
being expressed as the following convolution of $c$ and $s1$ peudofermion spectral functions, Eq. (\ref{BQ-gen}),
\begin{equation}
{\breve{B}}^{\odot}_{\nu} (\delta\omega_{\nu},v_{\nu}) = {{\rm sgn} (v_{\nu})\over 2\pi}\int_{0}^{\delta\omega_{\nu}}d\omega'\int_{-{\rm sgn}
(v_{\nu})\delta\omega_{\nu}/v_{\beta}}^{+{\rm sgn} (v_{\nu})\delta\omega_{\nu}/v_{\beta}}dk'
\,B_{Q_{{\bar{\beta}}}} (\delta\omega_{\nu}/v_{\nu} -k',\delta\omega_{\nu}-\omega')\,B_{Q_{\beta}} (k',\omega') \, . 
\label{B-l-i-breve}
\end{equation}
Here ${\bar{\beta}}=c,s1$ and $\beta =s1,c$, respectively, 
are chosen according to the criterion, Eq. (\ref{v-bb-b}),
concerning the relative magnitudes of the two $c$ and $s1$ Fermi velocities, Eq. (\ref{vel-beta}).

In addition to leading to a non-interacting like spectral-function matrix-element overlap,
the $\sigma$ one-electron UHB addition processes involving the creation of one $\eta 1$ pseudofermion 
of momentum $\pm (\pi-2k_F)$ are accounted for by their contributions 
$2\mu$ and $\mp (\pi-2k_F)$ to the excitation energy and momentum spectra 
$\delta E^{\odot}$ and $\delta P^{\odot}$, Eq. (\ref{dE-dP}), respectively. On the other
hand and as mentioned above, under transitions from the present $n_e\in [0,1[$ and
$m\in [0,n_e]$ initial ground states, the zero-momentum $q_{\eta,+1/2} = 0$ and $q_{s,+1/2} = 0$, Eq. (\ref{q-eta-s}), 
and zero-energy $\varepsilon_{\eta,+1/2} = 0$ and $\varepsilon_{s,+1/2} = 0$, Eq. (\ref{energy-eta}),
unpaired $+1/2$ rotated $\eta$-spin and unpaired $+1/2$ rotated spin processes 
are accounted for by the $c$ and $s1$ pseudofermion holes, respectively. This follows
from they playing the role of unoccupied sites of the $c$ and $s1$ effective lattices,
respectively.    

The Slater determinant of $\beta =c,s1$ pseudofermion operators, Eq. (\ref{det1}), 
involves the pseudofermion anti-commutators. The apparent simplicity of such a Slater determinant 
masks the complexity of the main technical problem of the PDT, which lays in performing
the state summations in the spectral functions Lehmann representation, Eq. (\ref{Bkomega}). As reported in the following,
it results from the involved form of such anti-commutators and thus of the corresponding Slater determinants
of $\beta =c,s1$ pseudofermion operators. 

The unitarity of the pseudoparticle - pseudofermion transformation implies that the local $\beta =c,s1$ 
pseudofermion operators ${\bar{f}}^{\dag}_{j',\beta}$ and ${\bar{f}}_{j',\beta}$ in Eq. (\ref{f-f-FT-Q}) 
obey the following fermionic algebra similar to that in Eqs. (\ref{albegra-cf}) and (\ref{ffs1}) for the 
corresponding local $\beta =c,s1$ pseudoparticle operators,
\begin{equation}
\{{\bar{f}}^{\dag }_{j,\beta},{\bar{f}}_{j',\beta}\} = \delta_{j,j'} \, , \hspace{0.50cm} \beta = c,s1 \, . 
\label{pfalocal-Q}
\end{equation}

Consider two $\beta =c,s1$ pseudofermions of canonical momentum 
${\bar{q}}_j$ and ${\bar{q}_{j'}}$, respectively. Here ${\bar{q}}_j$ and ${\bar{q}}_{j'}=q_{j'}$ 
correspond to the $\beta =c,s1$ bands of a PS excited energy eigenstate and the corresponding ground state, respectively. 
Due to the $\beta =c,s1$ pseudofermion phase-shift functional $2\pi\,\Phi_{\beta} (q_j)$, Eq. (\ref{Phibetaq}), 
being incorporated in the canonical momentum, Eq. (\ref{barqan}), one straightforwardly finds from the use of  
Eqs. (\ref{f-f-FT-Q}) and (\ref{pfalocal-Q}) that the anti-commutator of ${\bar{f}}^{\dag}_{j',\beta}$ and 
${\bar{f}}_{j',\beta}$ reads,
\begin{equation}
\{{\bar{f}}^{\dag }_{{\bar{q}}_j,\beta},{\bar{f}}_{{\bar{q}}_{j'},\beta}\} =
{1\over L_{\beta}}\,e^{-i({\bar{q}}_j-{\bar{q}}_{j'})/
2}\,e^{i\,2\pi\,\Phi_{\beta}^T(q_j)/2}\,{\sin\Bigl(2\pi\,\Phi_{\beta}^T (q_j)/
2\Bigr)\over\sin ([{\bar{q}}_j-{\bar{q}}_{j'}]/2)} \, ; \hspace{0.50cm}
\Phi_{\beta}^T (q_j) = \Phi_{\beta}^0 + \Phi_{\beta} (q_j) \, , \hspace{0.50cm} \beta = c,s1 \, , 
\label{pfacrGS}
\end{equation}
whereas $\{{\bar{f}}^{\dag}_{{\bar{q}}_{j},\beta},{\bar{f}}^{\dag}_{{\bar{q}}_{j'},\beta}\} = \{{\bar{f}}_{{\bar{q}}_{j},\beta},{\bar{f}}_{{\bar{q}}_{j'},\beta}\}=0$.
Here $2\pi\,\Phi_{\beta}^T(q_j)$ is the overall phase shift acquired by a $\beta =c,s1$ pseudofermion of momentum $q_j$ under
the transition from the ground state to the PS excited energy eigenstate, $2\pi\,\Phi_{\beta}^0$, Eq. (\ref{pican}), is the corresponding 
non-scattering part of that phase shift, and $2\pi\,\Phi_{\beta} (q_j)$, Eq. (\ref{Phibetaq}), is its scattering part.

For $2\pi\,\Phi_{\beta}^T(q_j)\rightarrow 0$ the anti-commutator relation, Eq. (\ref{pfacrGS}), would be the usual one,
$\{f^{\dag }_{{\bar{q}}_j,\beta},f_{q_{j'},\beta}\} = \delta_{{\bar{q}}_j,{\bar{q}}_j'}$. That such an anti-commutator relation
has not that simple form is the price to pay to render the $\beta =c,s1$ pseudofermions without interaction terms
in their energy spectrum, which is of the form, Eq. (\ref{DE}). Indeed this is achieved by incorporating the 
$\beta$ pseudofermion scattering phase shift $2\pi\,\Phi_{\beta} (q_j)$, Eq. (\ref{Phibetaq}),
in the $\beta =c,s1$ band canonical momentum, Eq. (\ref{barqan}). The unusual form, Eq. (\ref{pfacrGS}),
of that anti-commutator relation is behind such a scattering phase shift controlling the spectral weight
distributions of the $\sigma$ one-electron spectral functions, Eq. (\ref{Bkomega}), as confirmed below. 

The unitarity of the pseudoparticle - pseudofermion transformation would preserve the
pseudoparticle operator algebra provided that the sets of $\beta =c,s1$ band $j=1,...,L_{\beta}$ and $j'=1,...,L_{\beta}$ 
canonical momentum values $\{{\bar{q}}_j\}$ and $\{{\bar{q}}_{j'}\}$, respectively, were the same. The exotic form
of the anti-commutator, Eq. (\ref{pfacrGS}), follows from ${\bar{q}}_j$ and ${\bar{q}}_{j'}$
corresponding to different sets of slightly shifted canonical momentum values. This is due to the shakeup effects
introduced by the state-dependent $\beta =c,s1$ pseudofermion scattering phase-shift functional $2\pi\,\Phi_{\beta} (q_j)$.

The derivation of the spectral weights in the $\beta =c,s1$ pseudofermion spectral functions, Eq. (\ref{BQ-gen}),
which include the $\beta =c,s1$ lowest peak weights $A^{(0,0)}_{\beta}$ generated by processes (A) and (B) and 
the relative weights $a_{\beta}=a_{\beta} (m_{\beta,\,+1},\,m_{\beta,\,-1})$ generated by processes (C) resulting
from the application of the $\beta =c,s1$ operators ${\hat{g}}_C (m_{\beta,+1},m_{\beta,-1})$, Eq. (\ref{statefGC}), 
onto the energy eigenstates generated by the processes (A) and (B), proceeds much as for the
corresponding $u\rightarrow\infty$ spinless fermion spectral function in Ref. \cite{Karlo-97}. 
Following the procedures of such a reference, after some algebra that involves the use of the 
pseudofermion anti-commutators, Eq. (\ref{pfacrGS}), in Eq. (\ref{det1}) one arrives
to the expressions given in Eqs. (\ref{A00}) of Appendix \ref{Ele2PsPhaShi} for the lowest peak weights $A^{(0,0)}_{\beta}$
and in Eqs. (\ref{aNNDP}) and (\ref{aNDP}) of that Appendix for the relative weights 
$a_{\beta}=a_{\beta} (m_{\beta,\,+1},\,m_{\beta,\,-1})$.

Also the corresponding computation of the one-electron spectral-weight $(k,\omega)$-plane distributions follows
steps similar to those used in Ref. \cite{Karlo-97}. The PDT is indeed an extension to finite $u$ of the method used 
in that reference for $u\rightarrow\infty$ \cite{V-1}. Note though that the
mapping to a Heisenberg chain used in that reference to deal with the spin part of the problem is valid only
at $m=0$ and $u\gg 1$. In our case for which $u$ is finite and $m\in [0,n_e]$ the alternative use of the $s1$ 
pseudofermion representation renders the treatment of the corresponding rotated spins $1/2$ formally similar 
to that of the related $c$ pseudofermion representation. 

For $m_{\beta,\iota}=1$ the relative weights given in Eq. (\ref{aNDP}) of Appendix \ref{Ele2PsPhaShi} read,
\begin{equation}
2\Delta_{\beta}^{\iota} \equiv a_{\beta,\iota}(1) = \left({\delta {\bar{q}}_{F\beta}^{\iota}\over (2\pi/L)}\right)^2 
= \left(\iota\delta N^{F}_{\beta,\iota} + \Phi_{\beta}(\iota q_{F\beta})\right)^2 \, ,
\hspace{0.50cm} \beta = c, s1 \, , \hspace{0.50cm} \iota =\pm 1 \, . 
\label{a10DP-iota}
\end{equation}
These four $\beta =c,s1$ and $\iota =\pm 1$ relative weights $2\Delta_{\beta}^{\iota} \equiv a_{\beta,\iota}(1)$,
which appear in the $c$ and $s1$ pseudofermion spectral function expression, Eq. (\ref{BQ-gen}),
are controlled by the $\beta =c,s1$ and $\iota =\pm 1$ Fermi-points pseudofermion
scattering phase shifts $2\pi\,\Phi_{\beta} (\iota q_{F\beta})$, Eq. (\ref{Phibetaq}),
and corresponding excited energy eigenstate canonical momentum deviations 
$\delta {\bar{q}}_{F\beta}^{\iota} = (\iota\,2\pi\,\delta N^F_{\beta,\iota}+
2\pi\,\Phi_{\beta} (\iota q_{F\beta}))/L$. Here $\delta N^F_{\beta,\iota}=\delta N^{0,F}_{\beta,\iota}+\iota\,\Phi_{\beta}^0$
so that $\delta {\bar{q}}_{F\beta}^{\iota} = (\iota\,2\pi\,\delta N^{0,F}_{\beta,\iota}+
2\pi\,\Phi_{\beta}^T (\iota q_{F\beta}))/L$. The bare deviation $\delta N^{0,F}_{\beta,\iota}$ accounts for the 
number of $\beta =c,s1$ pseudofermions created or annihilated at the right ($\iota =+1$) and left 
($\iota =+1$) $\beta =c,s1$ Fermi points. The overall deviation $\delta N^{F}_{\beta,\iota}$ accounts
in addition to the non-scattering phase shifts $\Phi_{\beta}^0$.

For general PS excited energy eigenstates populated by $c$ pseudofermions and 
composite $\alpha n$ pseudofermions with arbitrary numbers
$n\geq 1$ of pairs such that $(\delta N_c +\delta N_{\rm ps}^{SU(2)})/L\rightarrow 0$ as $L\rightarrow\infty$ where
the deviations from the initial ground state refer to the number $N_c$ of $c$ pseudofermions and 
$N_{\rm ps}^{SU(2)}$ of $\alpha n$ pseudofermions of the different $\alpha\nu$ branches, Eq. (\ref{F-beta}), 
the four $\beta =c,s1$ and $\iota =\pm 1$ functionals, Eq. (\ref{a10DP-iota}), can be written as,
\begin{eqnarray}
2\Delta^{\iota}_{\beta} & = & \left(\sum_{\beta'=c,s1}\left(\iota\, \xi^0_{\beta\,\beta'}\,{\delta N^F_{\beta'}\over 2} 
+ \xi^1_{\beta\,\beta'}\,\delta J^F_{\beta'}\right)
+ \sum_{\beta''=c,\alpha n}\sum_{j'=1}^{L_{\beta''}}\Phi_{\beta,\beta''}(\iota q_{F\beta},q_{j'})\delta N^{NF}_{\beta''} (q_{j'})\right)^2 \, .
\label{functional}
\end{eqnarray}
In this expression $\xi^{0}_{\beta\,\beta'}$ and $\xi^{1}_{\beta\,\beta'}$ are the 
$\beta =c,s1$ pseudofermion phase-shift parameters, Eq. (\ref{x-aa}),
$\delta N^F_{\beta'}=\sum_{\iota=\pm 1}\delta N_{\beta',\iota}$, and $\delta J_{\beta'}^F={1\over 2}\sum_{\iota=\pm 1}(\iota)\,\delta N_{\beta',\iota}$.
The deviations $\delta N^{NF}_{\beta''} (q_{j'})$ refer to $\beta''=c,\alpha n$ band momentum values $q_{j'}$,
which for the $\beta''=c,s1$ branches are away from the $\beta''=c,s1$ Fermi points. (The $c$ and $s1$
pseudofermion creation or annihilation at and in the vicinity of such points is rather accounted for by the deviations $\delta N^F_{\beta'}$
and $\delta J^F_{\beta'}$ in Eq. (\ref{functional}).)

A property that in the present TL plays a key role in our derivation of the $\sigma$ one-electron spectral weights is that the $\delta$-functions 
in the $\beta =c,s1$ pseudofermion spectral function expression, Eq. (\ref{BQ-gen}),
impose that,
\begin{equation}
{L\over 4\pi\,v_{\beta}}(\omega' +\iota\,v_{\beta}\,k')-\Delta_{\beta}^{\iota})=m_{\beta,\iota} \, ,
\hspace{0.50cm} \beta = c, s1 \, , \hspace{0.50cm} \iota =\pm 1 \, .
\label{relaTL}
\end{equation}
That the quantity $((L/4\pi\,v_{\beta})(\omega' +\iota\,v_{\beta}\,k')-\Delta_{\beta}^{\iota})$ on the left-hand
side of this equation is proportional to $L$ implies that 
for any arbitrarily small $k'$ and $\omega'$ values for which $0<(\omega' +\iota\,v\,k')/(4\pi v)\ll 1$ the corresponding 
values of the $\iota=\pm 1$ integer numbers $m_{\beta,\iota}=((L/4\pi\,v_{\beta})(\omega' +\iota\,v_{\beta}\,k')-\Delta_{\beta}^{\iota})$
are in the TL such that $m_{\beta,\iota}\gg 1$.
Hence the following asymptotic behavior of the $\beta,\iota$ relative weight, 
Eq. (\ref{aNDP}) of Appendix \ref{Ele2PsPhaShi}, is {\it exact}
within the TL and is thus used in the derivation of the spectral-function expressions given below,
\begin{equation}
a_{\beta,\iota} (m_{\beta,\iota}) \approx \frac{1}{\Gamma (2\Delta_{\beta}^{\iota})}
\Bigl(m_{\beta,\iota}+\Delta_{\beta}^{\iota}\Bigr)^{2\Delta_{\beta}^{\iota}-1}
\, ; \hspace{0.75cm} 2\Delta_{\beta}^{\iota}\neq 0 \, ,
\hspace{0.50cm} \beta = c, s1 \, , \hspace{0.50cm} \iota = \pm 1 \, . 
\label{f}
\end{equation}

A relation also useful for such a derivation involves the $\beta =c,s1$ lowest peak weight $A^{(0,0)}_{\beta}$, 
Eq. (\ref{A00}) of Appendix \ref{Ele2PsPhaShi}, 
in the $\beta =c,s1$ pseudofermion spectral function $B_{Q_{\beta}} (k',\omega')$, Eq. (\ref{BQ-gen}). It reads,
\begin{equation}
A^{(0,0)}_{\beta} = {F^{(0,0)}_{\beta}\over (L\,S_{\beta})^{-1+2\Delta_{\beta}^{+1} +2\Delta_{\beta}^{-1}}}
\, , \hspace{0.50cm} \beta = c,\,s1 \, .
\label{F00}
\end{equation}
Here $F^{(0,0)}_{\beta}$ and $S_{\beta}$ are in the $L\rightarrow\infty$ limit independent of $L$ 
and $2\Delta_{c}^{+1}$, $2\Delta_{c}^{-1}$, $2\Delta_{s1}^{+1}$, and $2\Delta_{s1}^{-1}$
are the four functionals, Eq. (\ref{functional}). (The product $S_{c}\times S_{s1}\approx 1$ 
is given by $1$ both in the $u\rightarrow 0$ and $u\rightarrow\infty$ limits.)

In the general case in which the four $\beta =c,s1$ and $\iota =\pm 1$ parameters $2\Delta_{\beta}^{\iota}$ are finite, 
one finds that the $\beta =c,s1$ pseudofermion spectral function $B_{Q_{\beta}} (k',\omega')$, Eq. (\ref{BQ-gen}), 
reads in the TL,
\begin{eqnarray}
& & B_{Q_{\beta}} (k',\omega') = {L\over 4\pi v_{\beta}}\,
A^{(0,0)}_{\beta}\,\prod_{\iota =\pm 1}\,a_{\beta,\iota}
\Bigl({\omega' +\iota\,v_{\beta}\,k'\over 4\pi v_{\beta}/L}\Bigr) 
\nonumber \\
& \approx &   
{F^{(0,0)}_{\beta}\over 4\pi\,v_{\beta}\,S_{\beta}}\,
\prod_{\iota =\pm 1}\,{\Theta (\omega' +\iota\,v_{\beta}\,k')\over 
\Gamma (2\Delta_{\beta}^{\iota})}\,
\Bigl({\omega' +\iota\,v_{\beta}\,k'\over 4\pi \,v_{\beta}\,S_{\beta}}\Bigr)^{-1 +2\Delta_{\beta}^{\iota}} 
\, , \hspace{0.50cm} \beta = c,s1 \, .
\label{B-J-i-sum-GG}
\end{eqnarray}
To reach the second expression, which in the TL is exact, Eqs. (\ref{f}) and (\ref{F00}) were used.
The $\beta =c,s1$ pseudofermion spectral functions, Eq. (\ref{BQ-gen}), have a different form when 
$2\Delta_{\beta}^{\iota}>0$ and $2\Delta_{\beta}^{-\iota}=0$, as given in Eq. (\ref{B-J-i-sum-GG2}) 
of Appendix \ref{Ele2PsPhaShi}. When $2\Delta_{\beta}^{\iota} =2\Delta_{\beta}^{-\iota}=0$ it
is $\delta$-function like, Eq. (\ref{B-J-i-sum-GG3}) of that Appendix.

\subsection{The small higher-order pseudofermion contributions to the $\sigma$ one-electron spectral weight}
\label{hocontribu} 

The pseudofermion representation spectral functions expression, Eq. (\ref{ABON-odotGEN}), 
includes all higher-order processes that generate little $\sigma$ one-electron spectral weight and
do not contribute to the line shape near singular spectral features studied 
in this paper. The PDT also accounts for the corresponding contributions of ground-state transitions 
to excited energy eigenstates of general form,
\begin{equation}
\vert f_G (i')\rangle = {\hat{g}}_{i'} (k)\vert GS_f\rangle \, , \hspace{0.50cm} i' = 0,1,...,\infty \, .
\label{statefGEN}
\end{equation}
Those may be populated by $\alpha n$ pseudofermions of branches with $n>1$ pairs.
For finite values of the spin density, the small weight contribution from such transitions 
higher-order pseudofermion processes appear at high excitation energy scales, which 
for each created $n>1$ $\alpha n$ pseudofermion is around $n\,2\mu_{\alpha}$, Eq. (\ref{2mu-eta-s}). 

The contribution to the $\sigma$ electron operators matrix elements of the creation of such composite 
$\alpha n$ pseudofermions is simpler to compute than that of the $c$ and $s1$ pseudofermions. 
As above for the $i'=0$ operator ${\hat{g}} (k)$, the $\alpha n$ pseudofermion operators in the expression of any
$i'\geq 0$ operator ${\hat{g}}_{i'} (k)$ in the spectral function expression, Eq. (\ref{ABON-odotGEN}),
and energy eigenstate, Eq. (\ref{statefGEN}), have discrete canonical momentum
values that belong to the excited energy eigenstate $\alpha n$ band. One then finds that, 
\begin{equation}
\langle f_G \vert{\hat{g}}_{i'} (k)\,{\hat{c}}_{\odot}\vert GS\rangle
= \langle GS_f\vert {\hat{g}}_{i'}^{\dag} (k)\,{\hat{g}}_{i'} (k)\,{\hat{c}}_{\odot}\vert GS\rangle
= \langle GS_f^{\rm ex(i')}\vert {\hat{c}}_{\odot} \vert GS\rangle \, ,
\label{melemsn}
\end{equation}
where $\vert GS_f^{\rm ex(i')}\rangle$ is a state with the same $c$ and $s1$ pseudofermion occupancy as $\vert GS_f\rangle$
but whose $c$ and $s1$ band discrete momentum values are those of its excited energy eigenstate 
$\vert f_G (i')\rangle = {\hat{g}}_{i'} (k) \vert GS_f\rangle$. 

That the $\sigma$ one-electron matrix elements quantum overlaps resulting from the creation of $n>1$ $\alpha n$ pseudofermios 
by the operators ${\hat{g}}_{i'}^{\dag} (k)\,{\hat{g}}_{i'} (k)$ in Eq. (\ref{melemsn}) are 
Fermi-liquid like is due to the lack of such occupancies in the ground states $\vert GS_f\rangle$ and $\vert GS\rangle$.
Their creation is thus not associated with Anderson orthogonality catastrophes. This is why
after computing such trivial quantum overlaps, one is left with matrix elements 
$\langle GS_f^{\rm ex(i')}\vert {\hat{c}}_{\odot} \vert GS\rangle$, Eq. (\ref{melemsn}), that
only involve $c$ and $s1$ pseudofermion operators and have the same general form as that in Eq. (\ref{matrixel}).
The same applies to higher-order additional $\beta =c,s1$ pseudofermion particle-hole
processes of type (A) also generated by the operators ${\hat{g}}_{i'}^{\dag} (k)\,{\hat{g}}_{i'} (k)$.

However, $\vert\langle GS_f^{\rm ex(i')}\vert {\hat{c}}_{\odot} \vert GS\rangle\vert$ strongly
decreases upon increasing the index $i'=0,1,...,\infty$, most of the spectral weight being
associated with the $i'=0$ matrix element $\langle GS_f^{\rm ex(0)}\vert {\hat{c}}_{\odot} \vert GS\rangle
= \langle GS_f^{\rm ex}\vert {\hat{c}}_{\odot} \vert GS\rangle$, Eq. (\ref{matrixel}). 
As a result, the corresponding higher-order pseudofermion processes lead to very
small $\sigma$ one-electron spectral weight contributions. Moreover, the transitions to the 
excited energy eigenstates, Eq. (\ref{statefGEN}), generated from the ground state by
such higher-order pseudofermion processes do not contribute to the $\sigma$ one-electron spectral weight 
in the vicinity of the singular features, which is the issue studied in this paper. 

\subsection{The involved state summations problem and
analytical expressions obtainable near singular spectral features}
\label{statesumm2} 

The numerical computation of the momentum and state summations in Eqs. (\ref{ABON-odotGEN})
and (\ref{ABON-odot}) needed to access the corresponding finite-$u$ spectral-weight distributions over 
the whole $(k,\omega)$ plane is a very involved technical problem. This is a procedure that
enormously simplifies in the $u\rightarrow\infty$ limit. The reason is that within it the $c$ pseudofermion phase-shift functional 
$2\pi\,\Phi^T_{c}(q_j)$ defined by Eqs. (\ref{Phibetaq}) and (\ref{pfacrGS}) becomes independent of $q_j$, being the 
quantity called $Q'-Q$ in Ref. \cite{Karlo-97}. This enables, in the case of the $u\rightarrow\infty$ and $m=0$ 
one-electron removal and LHB addition spectral functions, the numerical computation of all state summations. The authors 
of Refs. \cite{Karlo,Karlo-97} have performed that exercise. They obtained the beautiful one-electron spectral-weight 
distributions plotted in Fig. 1 of Ref. \cite{Karlo} for the whole $(k,\omega)$ plane, $u\gg 1$, $n_e=0.5$, and $m=0$.

On the other hand, for finite $u$ values the $\beta =c,s1$ pseudofermion phase-shift functionals $\Phi^T_{\beta}(q_j)$
are both momentum $q_j$ and densities $n_e$ and $m$ dependent and have different values for 
each excited energy eigenstate. Hence the numerical computation of the momentum and state summations needed to access 
the corresponding finite-$u$ spectral-weight distributions over the whole $(k,\omega)$ plane
becomes an extremely difficult technical task. 

Fortunately, though, the use of Eq. (\ref{B-J-i-sum-GG}) and Eqs. (\ref{B-J-i-sum-GG2})
and (\ref{B-J-i-sum-GG3}) of Appendix \ref{Ele2PsPhaShi} 
for the $\beta$ pseudofermion spectral function $B_{Q_{\beta}} (k',\omega')$, Eq. (\ref{BQ-gen}), in 
the function ${\breve{B}}^{\odot}_{\nu} (\delta\omega_{\nu},v_{\nu})$, Eq. (\ref{B-l-i-breve}),
that appears in the expression of the spectral function leading-order term
$B^{\odot} (k,\omega)$, Eq. (\ref{B-PAR-J-CPHS-sum-0}), enables partially performing the 
summations in the latter equation for the $(k,\omega)$-plane vicinity of 
most $\sigma $ one-electron singular spectral features. 

An important such a feature is a {\it branch line}. In the present case of the $\sigma $ one-electron
spectral functions, Eq. (\ref{Bkomega}), the one-parametric branch lines that at least for some momentum interval
correspond to a singular feature are all contained in the two-parametric spectra given in
Eqs. (\ref{SpupElremo}), (\ref{SpupElLHBadd}), (\ref{SpupElUHBadd}), (\ref{SpdownElremo}), 
(\ref{SpdownElLHBadd}), and (\ref{SpdownElUHBadd}). Those correspond
to excited energy eigenstates generated by the leading-order pseudofermion processes. 

Such a branch line results from transitions to a well-defined subclass of these excited energy eigenstates.
Specifically, a particle and hole branch line is generated by creation
of one $\beta =c,s1$ pseudofermion and one $\beta =c,s1$ pseudofermion hole, respectively, 
of canonical momentum ${\bar{q}} = {\bar{q}} (q)$ corresponding to a
well-defined $\beta$ band momentum value $q$ as defined by Eq. (\ref{barqan}).
The set of such transitions scans the whole corresponding $\beta$ band
momentum range. Specifically, for a $\beta=c$ branch line the $c$ band momentum $q$ runs in 
the intervals $q \in [-\pi,-2k_F]$ and $q \in [2k_F,\pi]$ for a particle branch line and 
in the range $q \in [-2k_F,2k_F]$ for a hole branch line. In the case of a $\beta =s1$ branch line, the
$s1$ band momentum $q$ runs in the ranges $q \in [-k_{F\uparrow},-k_{F\downarrow}]$ and $q \in [k_{F\downarrow},k_{F\uparrow}]$ 
for a particle branch line and in the interval $q \in [-k_{F\downarrow},k_{F\downarrow}]$ for a hole branch line.

For a $c$ and $s1$ branch line, the $s1$ and $c$, respectively, pseudofermion or pseudofermion
hole created under the transitions to the excited energy eigenstates whose two-parametric spectra are given in
Eqs. (\ref{SpupElremo}), (\ref{SpupElLHBadd}), (\ref{SpupElUHBadd}), (\ref{SpdownElremo}), 
(\ref{SpdownElLHBadd}), and (\ref{SpdownElUHBadd}) is added to one of the $\iota =\pm 1$ corresponding 
Fermi points. As given in Eqs. (\ref{upElUHBadd}) and (\ref{downElUHBadd}), in the case of $\sigma$ 
one-electron UHB addition the corresponding $\eta 1$ pseudofermion is created at one of the $\eta 1$ 
band limiting momentum values, $q = \pm (\pi -2k_F)$. 

The PS excited energy eigenstates generated from the ground state by the types of processes 
described above have a one-parametric $(k,\omega)$-plane $\beta =c,s1$ branch line spectrum,
\begin{equation}
\omega_{\beta}^{\sigma} (k) = \omega_0 + \varepsilon_{\beta} (q)\,\delta N_{\beta} (q) \, ; \hspace{0.75cm} 
k = k_0 + q\,\delta N_{\beta} (q) \, , \hspace{0.50cm} \beta = c,s1 \, , 
\label{dE-dP-bl}
\end{equation}
where $\sigma =\uparrow,\downarrow$ refers to the one-electron spectral function under
consideration, $\varepsilon_{\beta} (q)$ is the $\beta =c,s1$ band energy dispersion, Eq. (\ref{epsilon-q}), 
$\delta N_{\beta} (q) = +1$ and $\delta N_{\beta} (q) = -1$ for a particle and hole branch 
line, respectively, and the energy scale $\omega_0$ and momentum $k_0$ are given by,
\begin{eqnarray}
\omega_0 & = & 2\mu\,\delta N_{\eta 1} \, , \hspace{0.5cm} \delta N_{\eta 1} = 0,1 \, ,
\nonumber \\
k_0 & = & 4k_F\,\delta J_{c}^F + 2k_{F\downarrow}\,\delta J_{s1}^F + 2(\pi -2k_F)\,\delta J_{\eta 1} \, ,
\label{1el-omega0}
\end{eqnarray}
respectively. Here the $\beta=c,s1$ current number deviations $\delta J_{\beta}^F$ are those in Eq. (\ref{functional}),
$\delta N_{\eta 1} =\delta J_{\eta 1}=0$ for both $\sigma$ electron removal and $\sigma$ electron LHB addition, 
$\delta N_{\eta 1} =1$ and $\delta J_{\eta 1}=-{1\over 2}\sum_{\iota=\pm 1}(\iota)\,\delta N_{\eta 1,\iota} = \mp 1/2$
for $\sigma$ electron UHB addition, and $\delta N_{\eta 1,\iota} = 1$ for creation of the $\eta 1$ pseudofermion
at the $\iota =\pm 1$ limiting $\eta 1$ band momentum $\iota (\pi -2k_F)$. 

In the case of the $(k,\omega)$-plane region in the vicinity of a $\beta =c,s1$ branch line,
the summation $\sum_{\nu}$ in Eq. (\ref{B-PAR-J-CPHS-sum-0}) runs over excited energy eigenstates 
with the specific $k$ and $\omega$ values that appear in the argument of the corresponding function $B^{\odot} (k,\omega)$. 
At such fixed values, the two corresponding $\beta =c,s1$ lowest peak weights $A^{(0,0)}_{\beta}$, 
Eq. (\ref{A00}) of Appendix \ref{Ele2PsPhaShi}, have nearly 
the same magnitude for all such states. The state summations
can then be partially performed. The technical details of such summations are provided in Appendix B of Ref. \cite{V-1}.
They lead to the following general behavior in the vicinity of a $\sigma$ one-electron $\beta = c,s1$ branch line,
\begin{eqnarray}
B_{\sigma,\gamma} (k,\omega) & = & C_{\sigma,\gamma,\beta}\,
\Bigl(\gamma\,\omega - \omega_{\beta}^{\sigma} (k)\Bigr)^{\xi_{\beta}^{\sigma} (k)}  \, ; \hspace{0.50cm} 
(\gamma\,\omega - \omega_{\beta}^{\sigma} (k)) \geq 0 \, , \hspace{0.50cm} \gamma = \pm 1 \, ,
\nonumber \\
\xi_{\beta}^{\sigma} (k) & = & -1+\sum_{\beta' = c,s1}\sum_{\iota =\pm 1}2\Delta_{\beta'}^{\iota} (q)\vert_{q=(k-k_0)\delta N_{\beta} (q)}  \, .
\label{branch-l}
\end{eqnarray}
Here $C_{\sigma,\gamma,\beta}$ is a $n_e$, $m$, and $u$ dependent constant that is independent of $k$ and $\omega$,
$\omega\geq 0$ and $\omega\leq 0$ for $\gamma=1$ and $\gamma=-1$, respectively, and $2\Delta_{\beta'}^{\iota} (q)$
refers to the following specific form that the functionals, Eq. (\ref{functional}), have for the excited energy
eigenstates that control the $\sigma$ one-electron spectral weight distribution near the $\beta =c,s1$ branch line,
\begin{eqnarray}
2\Delta^{\iota}_{c} (q) & = & \left(\sum_{\beta'=c,s1}\left(\iota\, \xi^0_{c\,\beta'}\,{\delta N^F_{\beta'}\over 2} 
+ \xi^1_{c\,\beta'}\,\delta J^F_{\beta'}\right) + \xi^1_{c\,c}\,\delta J_{\eta 1}
+ \Phi_{c\,\beta}(\iota 2k_F,q)\,\delta N^{NF}_{\beta} (q)\right)^2 \, ,
\nonumber \\
2\Delta^{\iota}_{s1} (q) & = & \left(\sum_{\beta'=c,s1}\left(\iota\, \xi^0_{s1\,\beta'}\,{\delta N^F_{\beta'}\over 2} 
+ \xi^1_{s1\,\beta'}\,\delta J^F_{\beta'}\right) + \xi^1_{s1\,c}\,\delta J_{\eta 1}
+ \Phi_{s1\,\beta}(\iota k_{F\downarrow},q)\,\delta N^{NF}_{\beta} (q)\right)^2 \, .
\label{OESFfunctional}
\end{eqnarray}
In these expressions one has that $\delta N^{NF}_{\beta} (q) = +1$ and $ \delta N^{NF}_{\beta} (q) = -1$ for a 
particle and hole $\beta =c,s1$ branch line, respectively, and $q$ is not at the $\beta =c,s1$ Fermi points. 
For the $\sigma$ one-electron UHB addition
energy eigenstates for which $\delta J_{\eta 1} =\mp 1/2$ the relation
$\Phi_{\beta'',\eta 1}(\iota q_{F\beta''},\pm (\pi -2k_F)) = \pm \xi^{1}_{\beta''\,c}/2$, Eq. (\ref{Phibetaeta1Fbeta}),
was used to express the phase shift acquired by the $\beta'' =c,s1$ pseudofermions of $\iota =\pm 1$ Fermi momenta
$\iota q_{F\beta''}$ due to the creation of the $\eta 1$ pseudofermion of 
$\eta 1$ band momentum $\pm(\pi -2k_F)$.

In addition to the parameter,
\begin{eqnarray}
\gamma & = & -1 \hspace{0.5cm}{\rm for}\hspace{0.1cm}{\rm electron}\hspace{0.1cm}{\rm removal} \, ,
\nonumber \\
& = & + 1 \hspace{0.5cm}{\rm for}\hspace{0.1cm}{\rm electron}\hspace{0.1cm}{\rm addition} \, ,
\label{c0RA}
\end{eqnarray}
the one $\sigma$ one-electron spectra associated with the singular spectral features 
considered in Sec. \ref{DSGzzxx} involve a second parameter $\gamma_{\sigma}$ and the use of the symbol $\bar{\sigma}$ that are
given by,
\begin{eqnarray}
\gamma_{\uparrow} & = & + 1 \, ; \hspace{0.75cm} {\bar{\uparrow}} = \downarrow \, ,
\nonumber \\
\gamma_{\downarrow} & = & - 1 \, ; \hspace{0.75cm} {\bar{\downarrow}} = \uparrow \, .
\label{cssigma}
\end{eqnarray}
That in Eq. (\ref{branch-l}) the $\beta =c,s1$ branch line spectrum $\omega_{\beta}^{\sigma} (k)$ is not multiplied by $\gamma$ is 
justified by it being according to Eq. (\ref{dE-dP-bl}) always such that $\omega_{\beta}^{\sigma} (k)\geq 0$. 

The $\sigma$ one-electron spectral function line shapes near the branch lines, Eq. (\ref{branch-l}), 
are beyond the reach of the techniques associated with the low-energy Tomonaga-Luttinger-liquid. In the
limit of low-energy, the PDT describes the well-known behaviors predicted by
such techniques. This refers specifically to the vicinity of $(k,\omega)$-plane points $(k_0,0)$
of which $(k_0,\omega_0)$ is a generalization for $\omega_0>0$.
Near them, the $\sigma =\uparrow, \downarrow$ one-electron spectral function $B_{\sigma,\gamma} (k,\omega)$, Eq. (\ref{Bkomega}),
behavior rather is \cite{LE},
\begin{eqnarray}
B_{\sigma,\gamma} (k,\omega) & \propto & \Bigl(\gamma\,\omega -\omega_0\Bigr)^{\zeta^{\sigma}} \, , \hspace{0.50cm} 
(\gamma\,\omega - \omega_0) \geq 0 \, ,
\nonumber \\
\zeta^{\sigma} & = & -2+\sum_{\beta' = c,s1}\sum_{\iota =\pm 1}2\Delta_{\beta'}^{\iota} \, , \hspace{0.50cm} 
(\gamma\,\omega - \omega_0) \neq \pm v_{\beta}\,(k-k_0) \, , \hspace{0.50cm} \beta =c,s1 \, ,
\nonumber \\
B_{\sigma,\gamma} (k,\omega) & \propto & \Bigl(\gamma\,\omega - \omega_0 \mp v_{\beta}\,(k-k_0)\Bigr)^{\zeta_{\pm}^{\sigma}}  \, , \hspace{0.50cm} 
(\gamma\,\omega - \omega_0 \mp v_{\beta}\,(k-k_0)) \geq 0 \, ,
\nonumber \\
\zeta_{\pm}^{\sigma} & = & -1- 2\Delta_{\beta}^{\mp 1} +\sum_{\beta' = c,s1}\sum_{\iota =\pm 1}2\Delta_{\beta'}^{\iota} \, , \hspace{0.50cm} 
(\gamma\,\omega - \omega_0) \approx \pm v_{\beta}\,(k-k_0) \, , \hspace{0.50cm} \beta =c,s1 \, ,
\label{point}
\end{eqnarray}
where the form of the $\beta' =c,s1$ functionals $2\Delta_{\beta'}^{\iota}$ , Eq. (\ref{functional}), simplifies to,
\begin{eqnarray}
2\Delta^{\iota}_{c} & = & \left(\sum_{\beta'=c,s1}\left(\iota\, \xi^0_{c\,\beta'}\,{\delta N^F_{\beta'}\over 2} 
+ \xi^1_{c\,\beta'}\,\delta J^F_{\beta'}\right) + \xi^1_{c\,c}\,\delta J_{\eta 1}\right)^2 \, ,
\nonumber \\
2\Delta^{\iota}_{s1} (q) & = & \left(\sum_{\beta'=c,s1}\left(\iota\, \xi^0_{s1\,\beta'}\,{\delta N^F_{\beta'}\over 2} 
+ \xi^1_{s1\,\beta'}\,\delta J^F_{\beta'}\right) + \xi^1_{s1\,c}\,\delta J_{\eta 1}\right)^2 \, .
\label{pointfunctional}
\end{eqnarray}
The $\sigma$ spectral function expressions, Eq. (\ref{point}), apply to the small finite-weight region 
very near and above ($\gamma=1$) or below ($\gamma=-1$) the $(k,\omega)$-plane point $(k_0,\omega_0)$. 

There is a third type of $\sigma$ one-electron spectral feature in the vicinity of which the PDT provides an analytical expression. 
It is generated by processes where one $c$ pseudofermion or $c$ pseudofermion hole is created at a momentum value $q$
and one $s1$ pseudofermion or one $s1$ pseudofermion hole is created at a momentum value $q'$, such that their group velocities,
Eq. (\ref{vel-beta}), obey the equality $v_{c}(q) = v_{s1}(q')$.
It corresponds to a $c-s1$ border line whose $(k,\omega)$-plane spectrum is,
\begin{equation}
\omega_{c-s1}^{\sigma} (k)= \left(\omega_0 + \vert\epsilon_{c}(q)\vert+\vert\epsilon_{s1}(q')\vert\right)\,\delta_{v_{c}(q) ,\,v_{s1}(q')}
\, ; \hspace{0.50cm} 
k = k_0 + q\,\delta N_{c} (q) + q'\,\delta N_{s1} (q') \,  
\label{dE-dP-c-s1}
\end{equation}
Whether each of the deviations $\delta N_{c} (q)$ and $\delta N_{s1} (q')$ reads $+1$ or $-1$ is unrelated and is specific to 
border line under consideration.

The following $\sigma$ one-electron spectral function behavior in the vicinity of such a $c-s1$ border line,
\begin{equation}
B_{\sigma,\gamma} (k,\omega) \propto \Bigl(\gamma\,\omega -\omega_{c-s1}^{\sigma} (k)\Bigr)^{-1/2}  \, , \hspace{0.50cm} 
(\gamma\,\omega -\omega_{c-s1}^{\sigma} (k)) \geq 0 \, ,
\label{B-bol}
\end{equation}
is determined by the density of the two-parametric states generated upon varying $q$ and $q'$
within the corresponding $c$ and $s1$ band values, respectively. A $\sigma$ one-electron border line is part of the boundary line of the
two-parametric spectra, Eqs. (\ref{SpupElremo}), (\ref{SpupElLHBadd}), (\ref{SpupElUHBadd}), (\ref{SpdownElremo}), 
(\ref{SpdownElLHBadd}), and (\ref{SpdownElUHBadd}), $(k,\omega)$-plane regions. 

\subsection{Validity of the expressions for the line shape near the singular spectral features}
\label{validity}

The general behavior $B_{\sigma,\gamma} (k,\omega) = C_{\sigma,\gamma,\beta}\,
(\gamma\,\omega - \omega_{\beta}^{\sigma} (k))^{\xi_{\beta}^{\sigma} (k)}$
for small $(\gamma\,\omega - \omega_{\beta}^{\sigma} (k))>0$ in the vicinity of $\beta =c,s1$ branch lines, Eq. (\ref{branch-l}), 
also occurs in the case of two-particle dynamical correlation functions $B (k,\omega)$ for which 
the convention is $\gamma =1$ and $\omega \geq 0$. However, such a $B (k,\omega)$ expression
near a $\beta =c,s1$ branch line is in that case exact provided that the branch line
coincides with a lower threshold of the $(k,\omega)$-plane finite spectral-weight region \cite{CarCadez},
{\it i.e.} for which $B (k,\omega) =0$ for $(\omega - \omega_{\beta} (k))<0$. 

The $(k,\omega)$-plane spectral weight distribution of two-particle dynamical correlation functions is in general plateau-like.
It then follows that for $k$ ranges of a branch line for which $B (k,\omega) >0$ for $(\omega - \omega_{\beta} (k))<0$
there is a sufficient amount of two-particle spectral weight just below the line for the coupling to that generated
by the processes that contribute to the weight distribution as given in Eq. (\ref{branch-l}) changing the type
of $k$ and $\omega$ dependence for $(\omega - \omega_{\beta} (k))>0$. The microscopic processes behind
such a coupling are accounted for by the PDT yet performing the corresponding state summations needed to reach a simple 
analytical expression for $B (k,\omega)$ at small $(\omega - \omega_{\beta} (k))>0$ turns out to be
a complex technical problem.

In the present case of the $\sigma$ one-electron spectral functions $B_{\sigma,\gamma} (k,\omega)$, Eq. (\ref{Bkomega}), 
the behavior, Eq. (\ref{branch-l}), in the vicinity of a $\beta =c,s1$ 
branch line is exact for $k$ ranges for which such a line coincides with a lower threshold ($\gamma=1$) or a upper 
threshold ($\gamma=-1$) of the $(k,\omega)$-plane finite spectral-weight regions associated with
the corresponding two-parametric spectra. This requires that $B_{\sigma,\gamma} (k,\omega)=0$
for $\gamma\,\omega < \omega_{\beta}^{\sigma} (k)$.

The physically more important $\beta =c,s1$ branch line $k$ ranges are those for which the exponent $\xi_{\beta}^{\sigma} (k)$, Eq. (\ref{branch-l}),
is negative and that line corresponds to a singular feature. Fortunately and in contrast to two-particle dynamical correlation functions, 
along the line $k$ ranges for which $\xi_{\beta}^{\sigma} (k)<0$ in Eq. (\ref{branch-l})
and $B_{\sigma,\gamma} (k,\omega)>0$ for small $(\gamma\,\omega - \omega_{\beta}^{\sigma} (k))<0$ the corresponding
spectral weight at $\gamma\,\omega < \omega_{\beta}^{\sigma} (k)$ is much smaller than that at 
$\gamma\,\omega > \omega_{\beta}^{\sigma} (k)$. As a result, the coupling of the small weight at
$\gamma\,\omega < \omega_{\beta}^{\sigma} (k)$ to that at $\gamma\,\omega > \omega_{\beta}^{\sigma} (k)$
changes the distribution near the singular feature, Eq. (\ref{branch-l}), very little. The processes 
behind the small weight at $\gamma\,\omega < \omega_{\beta}^{\sigma} (k)$ are generated as well by
the pseudofermion leading-order operator term that depending on the $\sigma$ one-electron spectral function is 
one of the operators given in Eqs. (\ref{upElremo}), (\ref{upElLHBadd}), (\ref{upElUHBadd}), (\ref{downElremo}), 
(\ref{downElLHBadd}), and (\ref{downElUHBadd}). Indeed, the subclass of one-parametric processes that generate 
the line shape, Eq. (\ref{branch-l}), just above ($\gamma =+1$) or below ($\gamma =-1$) the branch line
refers to a particular case of such more general two-parametric processes.

For the $k$ ranges for which $B_{\sigma,\gamma} (k,\omega)>0$ at
$\gamma\,\omega < \omega_{\beta}^{\sigma} (k)$, the spectral function $B_{\sigma,\gamma} (k,\omega)$ remains 
having the power-law like behavior, Eq. (\ref{branch-l}),  in the vicinity of the line for $\gamma\,\omega > \omega_{\beta}^{\sigma} (k)$.
Specifically, the line spectrum $\omega_{\beta}^{\sigma} (k)$, Eq. (\ref{dE-dP-bl}), remains insensitive to the 
coupling, which only slightly affects the value of the exponent $\xi_{\beta}^{\sigma} (k)$. 
Such an effect is small and very small when $0<\xi_{\beta}^{\sigma} (k)<1$ and $\xi_{\beta}^{\sigma} (k)<0$,
respectively, in Eq. (\ref{branch-l}). The theory includes actually a small $k$ dependent parameter,
\begin{equation}
\gamma_{\sigma,\gamma} (k) = \left({\int_{\omega_{\beta}^{\sigma} (k)-\Omega}^{\omega_{\beta}^{\sigma} (k)}
B_{\sigma,\gamma} (k,\omega)\,d\omega
\over \int_{\omega_{\beta}^{\sigma} (k)}^{\omega_{\beta}^{\sigma} (k)+\Omega}
B_{\sigma,\gamma} (k,\omega)\,d\omega}\right)^{\gamma} \, , \hspace{0.5cm} \gamma = \pm 1 \, .
\label{tausigk}
\end{equation}
Here $\Omega$ stands for the processes (C) energy range for $\omega>\gamma\,\omega_{\beta}^{\sigma} (k)$. 
It is self-consistently determined as that for which the velocity $v_{\nu}$, Eq. (\ref{dEdPvv*}), remains nearly unchanged.
One can then expand the exponent expression in that small parameter, the zeroth order
leading term being $\xi_{\beta}^{\sigma} (k)$, as given in Eq. (\ref{branch-l}). 

In the vicinity of the line $k$ ranges for which $\xi_{\beta}^{\sigma} (k)$, Eq. (\ref{branch-l}), 
is negative there is a much larger amount of spectral weight for $\omega>\gamma\,\omega_{\beta}^{\sigma} (k)$
than for $\omega<\gamma\,\omega_{\beta}^{\sigma} (k)$. The $k$ dependent parameter, Eq. (\ref{tausigk}),
is thus extremely small for such $k$ intervals, {\it i.e.} $\gamma_{\sigma,\gamma} (k)\ll 1$. Since the corresponding exponent 
corrections are also extremely small and do not change the physics, for simplicity in the studies
of Sec. \ref{DSGzzxx} we use the leading-order exponent expression $\xi_{\beta}^{\sigma} (k)$, Eq. (\ref{branch-l}).
The otherwise very small exponent corrections vanish in a $\beta =c,s1$ branch line $k$ ranges for which 
it coincides with the a lower threshold ($\gamma=1$) or upper threshold ($\gamma=-1$) of the $(k,\omega)$-plane 
finite spectral-weight region.

Moreover, the $\sigma $ one-electron spectral function expression
near a $\beta =c,s1$ branch line, Eq. (\ref{branch-l}), is valid provided that the exponent 
in it obeys the inequality $\xi_{\beta}^{\sigma} (k)>-1$.
When for a given $\beta =c,s1$ branch line $k$ range one finds that $\xi_{\beta}^{\sigma} (k)=-1$, the exact expression
of the spectral function is not that given in Eq. (\ref{branch-l}). For these $k$ ranges 
one has that the four functionals $2\Delta^{\iota}_{\beta}$, Eq. (\ref{pointfunctional}) for $\beta =c,s1$ and
$\iota =\pm 1$, vanish. This corresponds to the $\beta =c,s1$ pseudofermion spectral function form,
Eq. (\ref{B-J-i-sum-GG3}) of Appendix \ref{Ele2PsPhaShi}. One then finds that the corresponding 
$\sigma $ one-electron spectral function behavior is also $\delta$-function-like and given by,
\begin{equation}
B_{\sigma,\gamma} (k,\omega) = \delta \Bigl(\gamma\,\omega - \omega_{\beta}^{\sigma} (k)\Bigr) \, .
\label{branch-lexp-1}
\end{equation}

As expected, it is confirmed in the ensuing section that only as $u\rightarrow 0$ some $\beta =c,s1$ branch line 
exponents read $\xi_{\beta}^{\sigma} (k)=-1$. For the corresponding $k$ momentum ranges one recovers
parts of the exact $u=0$ $\sigma$ one-electron spectrum, with $\omega_{\beta}^{\sigma} (k)$ on the
right-hand side of Eq. (\ref{branch-lexp-1}) becoming the
corresponding non-interacting electronic spectrum. This is confirmed by accounting for the
$u\rightarrow 0$ limiting behaviors of the $\beta =c,s1$ energy dispersions $\varepsilon_{\beta} (q)$ appearing
in the spectrum $\omega_{\beta}^{\sigma} (k)$, Eq. (\ref{dE-dP-bl}). Such limiting behaviors are reported 
in Eqs. (\ref{varepsiloncu0}) and (\ref{varepsilonsu0}) of Appendix \ref{LimitBV}.

Furthermore, the branch-line exponent expression, Eq. (\ref{branch-l}), is not valid in its 
limiting $k$ points when they coincide with boundary points $(k_0,\omega_0)$ in the vicinity 
of which the line shape has rather the form given in Eqs. (\ref{point}) and (\ref{pointfunctional}). 
The PDT naturally accesses such an alternative behavior. For $\sigma$ electron removal and LHB addition 
it corresponds to the known low-energy behavior of the spectral function in the vicinity of $(k,\omega)$-plane points $(k_0,0)$. Since for
the densities ranges $n_e \in [0,1[$ and $m\in [0,n_e]$ considered here the latter low-energy behavior is known and 
coincides with that reported in Eq. (5.7) of Ref. \cite{Frahm-91}, we restrict our study of 
Section \ref{DSGzzxx} to the high-energy spectral features.
The previous studies of the high-energy spectral features of the 1D Hubbard model by means of
the PDT \cite{TTF,spectral0,spectral,spectral-06} and most of those relying on other methods 
\cite{Essler,Essler-14,Kohno-10,Benthien-04} have been limited to zero spin density. Hence the analysis 
of Sec. \ref{DSGzzxx} is mainly focused on finite spin densities $m \in ]0,n_e]$. 

Concerning the behavior of the spectral functions near the border lines, Eq. (\ref{B-bol}), 
in the related cases of charge-charge and spin-spin two-electron dynamical correlation 
functions the boundary line exponent $-1/2$ that results from the density of the two-parametric states
is changed to $1/2$ by the two-electron matrix elements between the ground state and the excited energy eigenstate. 
This always occurs when the two values $q$ and $q'$ and corresponding 
group velocities $v_{\beta}(q)$ and $v_{\beta}(q')$ such that $v_{\beta}(q)=v_{\beta}(q')$
belong to the same $\beta =c,s1$ band.
\begin{figure}
\includegraphics[scale=1.00]{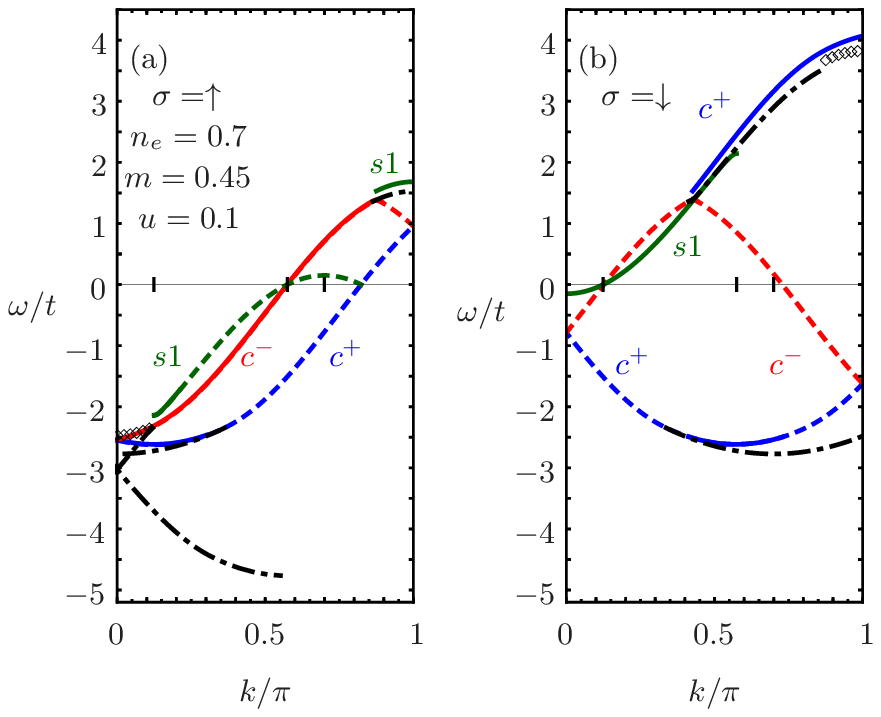}
\includegraphics[scale=1.00]{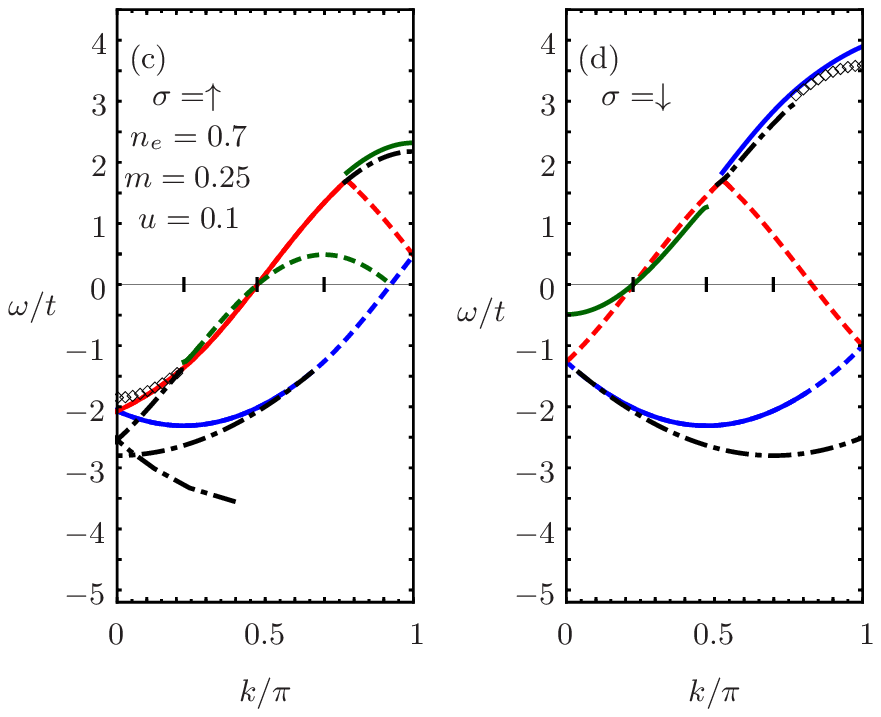}
\caption{\label{s4525u01} The $(k,\omega)$-plane singular branch lines $k$ ranges (solid lines) and other branch lines
$k$ ranges (dashed lines) for which the corresponding exponent $\xi_{\beta}^{\sigma} (k)$, Eq. (\ref{branch-l}), is negative and positive,
respectively, and singular boundary lines (dashed-dotted lines) of the weight distribution associated with the $\uparrow$ and 
$\downarrow$ one-electron spectral function for $u=0.1$, electronic density $n_e =0.7$, and (a)-(b) spin densities $m=0.45$ and 
(c)-(d) $m=0.25$. The branch line spectra plotted here are defined in Section \ref{DSGzzxx}. The $c^+$, $c^-$, and $s1$ branch-line
labels appearing here in panels (a) for $\sigma =\uparrow$ and (b) for $\sigma =\downarrow$
apply to the branch lines with similar topology in the panels (c) and (d), respectively.
(Online the $c^+$, $c^-$, and $s1$ branch lines plotted here as defined in Section \ref{DSGzzxx} 
are blue, red, and green, respectively.)
The lines represented by sets of diamond symbols contribute to the $u\rightarrow 0$ one-electron spectrum yet are
not branch lines. For $\sigma$ one-electron UHB addition only the branch lines that contribute to the $u\rightarrow 0$ spectral
weight are represented.}
\end{figure}

\begin{figure}
\includegraphics[scale=1.00]{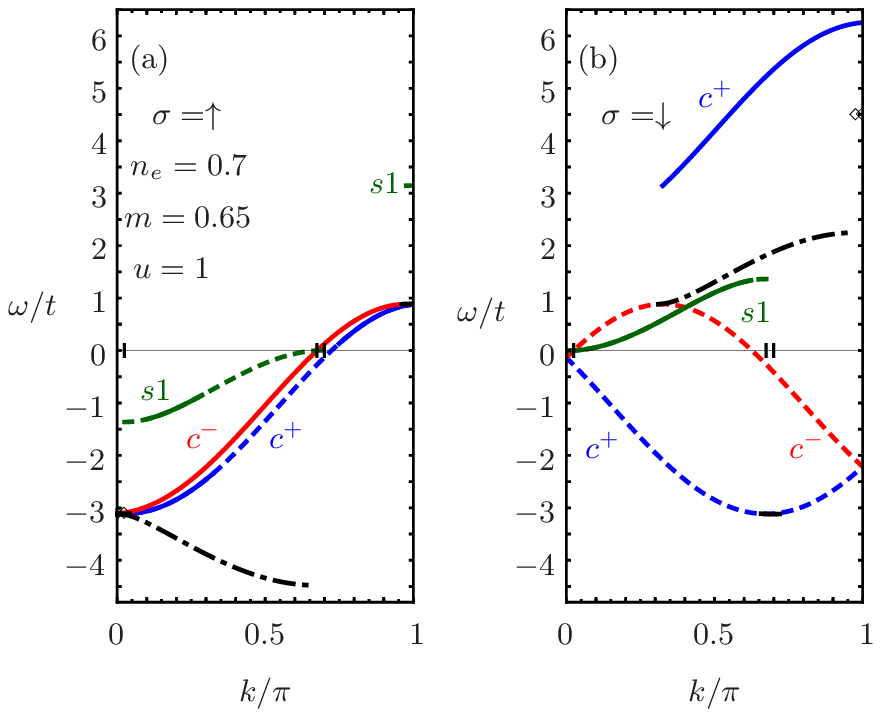}
\includegraphics[scale=1.00]{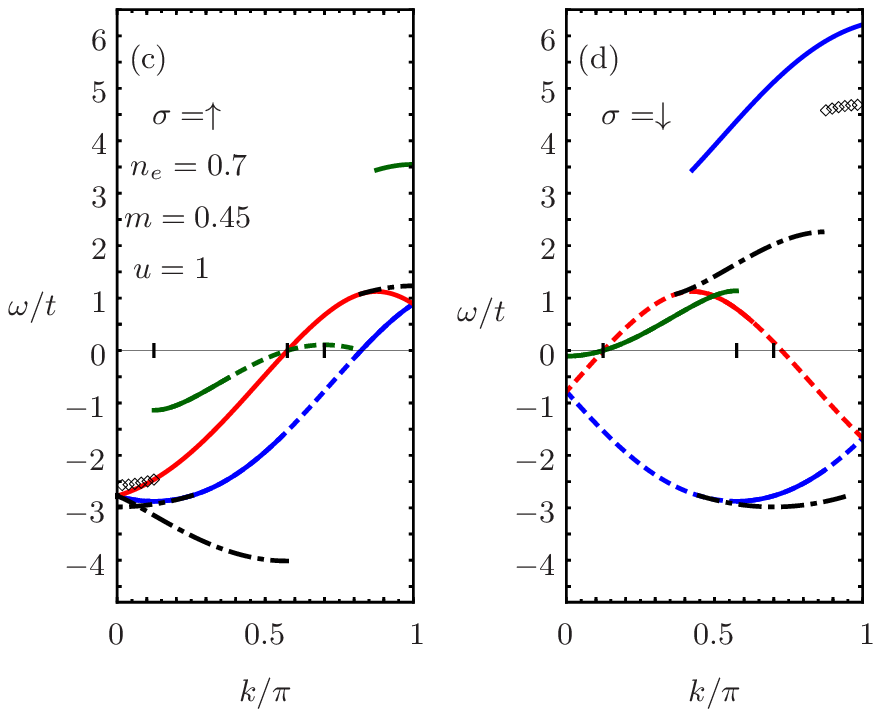}
\caption{\label{s6545u1} The same $(k,\omega)$-plane lines as in Fig. \ref{s4525u01}
for $u=1$, electronic density $n_e =0.7$, and spin densities (a)-(b) $m=0.65$ and (c)-(d) $m=0.45$.
(Online the $c^+$, $c^-$, and $s1$ branch lines are blue, red, and green, respectively.)}
\end{figure}

In the present case of the $\sigma$ one-electron spectral functions
the border lines are generated by pairs of values $q$ and $q'$ belonging to the $c$ and
$s1$ bands, respectively, such that $v_{c}(q)=v_{s1}(q')$. The $\sigma$ one-electron matrix elements 
between the ground state and the excited energy eigenstates do not change the exponent $-1/2$ 
resulting from the density of the two-parametric states, so that the border-line
singularities, Eq. (\ref{B-bol}), prevail. The border lines of the $\sigma $ one-electron 
removal and LHB addition spectral functions are plotted in Figs. \ref{s4525u01}-\ref{s032505u1} yet for 
simplicity their specific analytical form is not given in this paper. 

In what the $\sigma$ one-electron LHB and UHB addition spectral functions as defined in 
Eq. (\ref{LUHB}) for $u>0$, $n_e \in [0,1[$, and $m\in [0,n_e]$ is concerned, we have a few comments. 
At $n_e=1$ there is no $\sigma$ one-electron LHB. That eletronic
density refers to the Mott-Hubbard insulator phase at which there is a gap $2\mu_u$, Eq. (\ref{2mu0}),
between the top of the $\sigma$ one-electron removal band and the addition UHB. On the other hand, for the metallic phase electronic
density range $n_e \in [0,1[$ considered here, the spectral weight associated with the $\sigma$ one-electron 
addition LHB has not an exact top, yet such a weight becomes very small above some $u>0$, 
$n_e \in [0,1[$, and $m\in [0,n_e]$ dependent finite energy scale. Hence for intermediate
and large $u$ values there emerges a pseudogap between that region of the $\sigma$ one-electron 
addition LHB and the well-defined bottom of the UHB. Our study focuses on singular spectral features, such a pseudogap
being clearly visible in Figs. \ref{s6545u1}-\ref{s032505u1} for intermediate and large $u$ values, where
as discussed below the $(k,\omega)$-plane solid lines and dashed-dotted lines refer to negative-exponent
singular branch lines $k$ ranges and singular border lines, respectively.

An interesting property is that, when expressed as function of the $\beta =c,s1$ band momentum $q$, the 
$\sigma=\uparrow,\downarrow$ one-electron UHB addition $\beta =c,s1$ branch lines 
spectrum and exponent are exactly the same as for the $\beta =c,s1$ branch lines
of the $\bar{\sigma}=\downarrow,\uparrow$ one-electron removal spectral function. That relation is also preserved 
in terms of the momentum $k$ and energy $\omega$ provided that they are replaced by $\pi -k$
and $2\mu -\omega$, respectively.

Such a relation follows from model symmetries whose consequences are fully explicit at
$n_e=1$ for chemical potential $\mu =0$ at the middle of the Mott-Hubbard gap. Then there is no $\sigma$
one-electron LHB addition spectral function and the following exact relation holds,
\begin{equation}
B_{\sigma,+1}^{\rm UHB} (k,\omega) =
B_{\bar{\sigma},-1} (\pi-k,-\omega) \, , \hspace{0.5cm} n_e = 1
\, , \hspace{0.5cm} \mu = 0 \, .
\label{ne1SFsymm}
\end{equation}
For $n_e\rightarrow 1$ and thus chemical potential $\mu\rightarrow \mu_u$ this relation 
is also valid yet reads $B_{\sigma,+1}^{\rm UHB} (k,\omega) =
B_{\bar{\sigma},-1} (\pi-k,2\mu-\omega)$.

At $n_e = 1$ the rotated-electron doubly occupied site of the excited energy eigenstates associated
with the $\sigma$ one-electron UHB addition spectral function corresponds to a $\eta$-spin doublet
configuration of a single unpaired rotated spin of projection $-1/2$. On the other hand, for electronic
densities $n_e \in [0,1[$ such states are rather populated by one $\eta 1$ pseudofermion that
corresponds to a $\eta$-spin singlet configuration of two paired rotated $\eta$-spins of opposite projection. 

That the $\sigma$ one-electron UHB addition $s1$ and $c^{\pm}$ branch lines $(k,\omega)$-plane 
spectrum and exponent momentum dependence studied below in Section \ref{DSGzzxx}
are for electronic densities in the range $n_e \in [0,1[$ and under the transformations 
$k\rightarrow \pi -k$ and $\omega\rightarrow 2\mu -\omega$ exactly the same as for the $\bar{\sigma}$ 
one-electron removal $s1$ and $c^{\pm}$ branch lines, respectively, is a weaker consequence of 
the same symmetry. It follows from a $\eta 1$ pseudofermion of momentum at the
$\eta 1$ band limting values $\bar{q}=q = \pm (\pi -2k_F)$, Eq. (\ref{qcanGS}),  
being invariant under the pseudoparticle - pseudofermion unitary transformation. Indeed, for such $\sigma$ 
one-electron UHB addition singular features the $\eta 1$ pseudofermion is created at one of such two $\eta 1$ band limiting values.
Hence the corresponding $\eta 1$ pseudofermion energy, Eq. (\ref{epsilon-q}) for $\beta =\eta 1$, reads 
$\varepsilon_{\eta 1} (\pm (\pi -2k_F)) = 2\mu_{\eta}=2\mu$. It thus equals that of two unpaired rotated 
$\eta$-spins of opposite projection, Eq. (\ref{varep}) with $2\mu_{\alpha}$ given by Eq. (\ref{2mu-eta-s}) for $\alpha = \eta$.
The invariance under the pseudoparticle - pseudofermion unitary transformation of the
$\eta 1$ pseudofermion created at the momentum $\bar{q}=q = \pm (\pi -2k_F)$
is behind this property by implying that the corresponding anti-bounding energy $\varepsilon_{\eta 1}^0 (q)\geq 0$ on the right-hand side
of Eq. (\ref{e-0-bands}) vanishes, $\varepsilon_{\eta 1}^0 (\pm (\pi -2k_F))=0$. This means that at these momentum
values the two rotated $\eta$-spins within the composite $\eta 1$ pseudofermion
are in a $\eta$-spin singlet configuration yet are unpaired, similarly to the unpaired rotated $\eta$-spins in 
the multiplet configurations and specifically to the projection $-1/2$ unpaired and single rotated $\eta$-spin of the $n_e=1$ $\eta$-spin 
doublet $\sigma $ one-electron UHB addition spectral function $B_{\sigma,+1}^{\rm UHB} (k,\omega)$,
Eq. (\ref{ne1SFsymm}).

Although the $\sigma$ one-electron UHB addition spectral weight generated
by transitions to excited energy eigenstates for which the $\eta 1$ pseudofermion emerges at
a $\eta 1$ band canonical momentum $\bar{q} = \bar{q} (q)$ corresponding to a bare
momentum value $-(\pi -2k_F)<q < (\pi -2k_F)$ is small, such processes imply that
the relation $B_{\sigma,+1}^{\rm UHB} (k,\omega) =B_{\bar{\sigma},-1} (\pi-k,2\mu-\omega)$ 
is not exact for $n_e<1$. It becomes exact only in the $n_e\rightarrow 1$ limit
and thus for chemical potential $\mu\rightarrow \mu_u$ when $(\pi -2k_F)\rightarrow 0$. 
In a weaker way it nevertheless survives for $n_e \in [0,1[$ in what the $\sigma$ one-electron 
UHB addition singular $\beta =c,s1$ branch lines 
$(k,\omega)$-plane spectrum and exponent momentum dependence are concerned for
the reasons reported above.

In Figs. \ref{s4525u01}-\ref{s032505u1} the $\uparrow$ and $\downarrow$ one-electron removal and LHB addition 
$\beta$ branch lines whose exponent $\xi_{\beta}^{\sigma} (k)$, Eq. (\ref{branch-l}), is negative for at least some 
$k$ interval and $u$, $n_e$, and $m$ ranges and the boundary lines considered in the ensuing section are shown 
in the $(k,\omega)$-plane for several values of $u$, electronic densities $n_e=0.3$ and $n_e=0.7$, and sets 
of spin density values $m<n_e$. For $\uparrow$ and $\downarrow$ one-electron UHB addition only the main 
branch lines that in the $u\rightarrow 0$ limit contribute to the $u=0$ $\sigma $ one-electron addition spectrum
are shown. (Online the $c^+$, $c^-$, and $s1$ branch lines defined in Section \ref{DSGzzxx} and
plotted in these figures are blue, red, and green, respectively.)

Indeed, since the behavior of the $\downarrow$ and $\uparrow$ one-electron removal spectral functions
near their $\beta =c,s1$ branch lines is studied in some detail, for simplicity in the following the study of
the related $\uparrow$ and $\downarrow$, respectively, one-electron UHB addition branch lines is limited to those
that in the $u\rightarrow 0$ limit contribute to the $u=0$ $\sigma $ one-electron addition 
$\delta$-function-like spectrum.

The $\sigma$ one-electron $\beta$ branch lines are in Figs. \ref{s4525u01}-\ref{s032505u1}
represented by solid lines and dashed lines for the $k$ ranges for which the 
corresponding exponent $\xi_{\beta}^{\sigma} (k)$, Eq. (\ref{branch-l}), is negative and positive,
respectively. The $\sigma$ one-electron removal and LHB addition boundary lines are represented by dashed-dotted lines. 
Most of the $u=0$ $\delta$-function like $\sigma$ one-electron spectrum $k$ ranges 
are obtained from branch lines in the $u\rightarrow 0$ limit. The two exceptions are the $u=0$ $\uparrow$ one-electron removal spectrum 
for the momentum range $k \in [-k_{F\downarrow},k_{F\downarrow}]$ and the $u=0$ $\downarrow$ one-electron addition spectrum 
for the $k$ interval $\vert k\vert \in [\pi-k_{F\downarrow},\pi]$, which emerge in the $u\rightarrow 0$ limit from the non-branch lines
that are represented in Figs. \ref{s4525u01}-\ref{s032505u1} by sets of diamond symbols. 

\section{The singular $\sigma$ one-electron spectral features}
\label{DSGzzxx}

\begin{figure}
\includegraphics[scale=1.00]{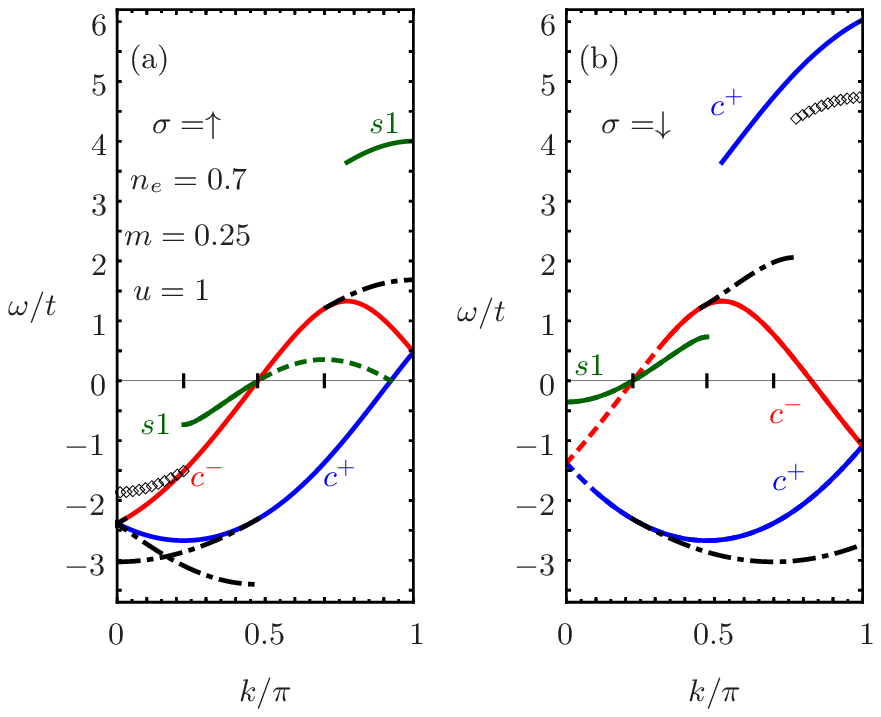}
\includegraphics[scale=1.00]{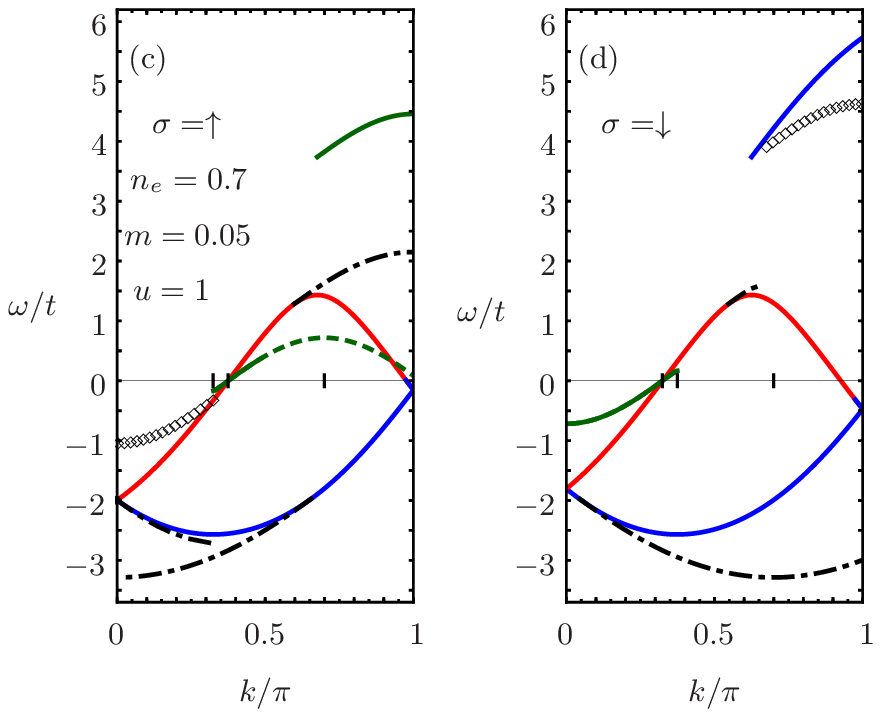}
\caption{\label{s2505u1} The same $(k,\omega)$-plane lines as in Fig. \ref{s4525u01}
for $u=1$, electronic density $n_e =0.7$, and spin densities (a)-(b) $m=0.25$ and (c)-(d) $m=0.05$.
(Online the $c^+$, $c^-$, and $s1$ branch lines are blue, red, and green, respectively.)}
\end{figure}

In this section we study the line shape behavior of the $\sigma$ one-electron spectral 
functions, Eq. (\ref{Bkomega}), in the vicinity of the branch lines shown in Figs. \ref{s4525u01}-\ref{s032505u1}.
For the $k$ ranges for which the exponents controlling the line shape near these
lines are negative, there are singularity cusps in the corresponding 
$\sigma$ one-electron spectral functions. 

The $\sigma $ one-electron removal and LHB addition $c^{\pm}$ and $s1$ branch lines are the topics of
Sections \ref{upRcc} and \ref{upRs}, respectively. Section \ref{upUHBs} addresses
the issue of the $\sigma $ one-electron UHB addition branch lines.
Finally, the $\uparrow$ one-electron removal and $\downarrow$ one-electron UHB addition 
$s1'$ non-branch lines that for $m\neq 0$ contribute to the $u\rightarrow 0$ one-electron 
spectrum is the subject of Section \ref{upRsl}.

\subsection{The $\sigma $ one-electron removal and LHB addition $c^{\pm}$ branch lines}
\label{upRcc}

\begin{figure}
\includegraphics[scale=0.98]{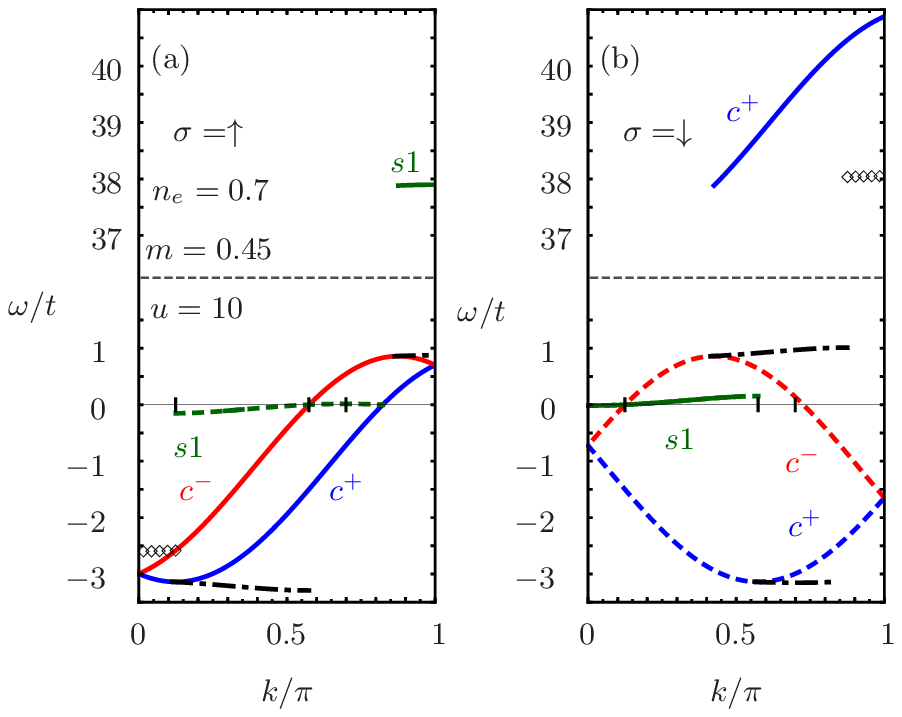}
\includegraphics[scale=0.98]{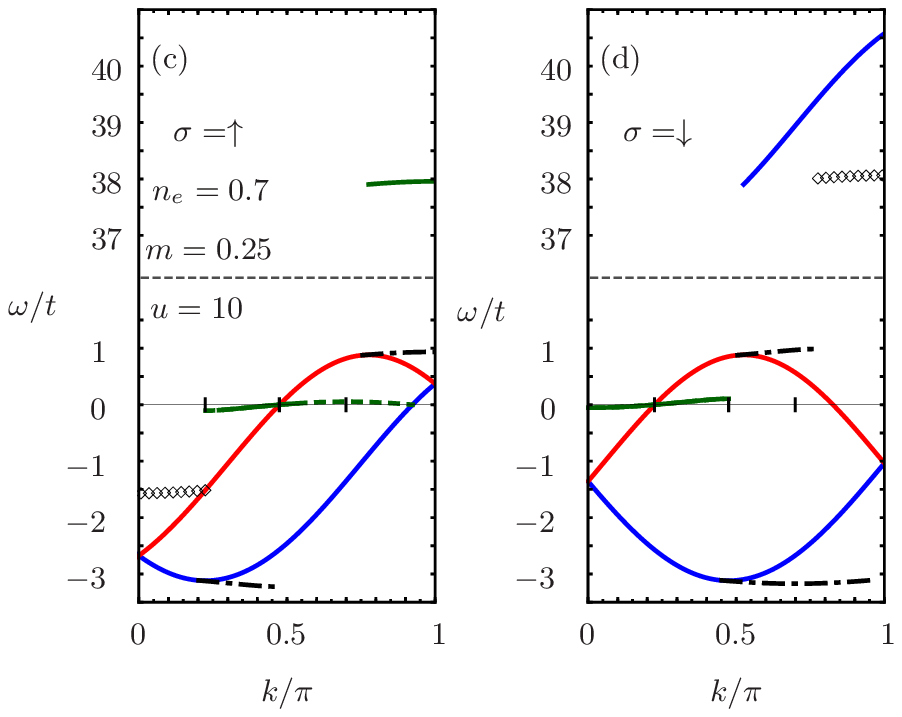}
\caption{\label{s4525u10} The same $(k,\omega)$-plane lines as in Fig. \ref{s4525u01}
for $u=10$, electronic density $n_e =0.7$, and spin densities (a)-(b) $m=0.45$ and (c)-(d) $m=0.25$.
Note the different $\omega$ intervals separated by a horizontal dashed line used for the removal
and LHB addition spectral features and the UHB addition branch line, respectively.
(Online the $c^+$, $c^-$, and $s1$ branch lines are blue, red, and green, respectively.)}
\end{figure}

\begin{figure}
\includegraphics[scale=0.98]{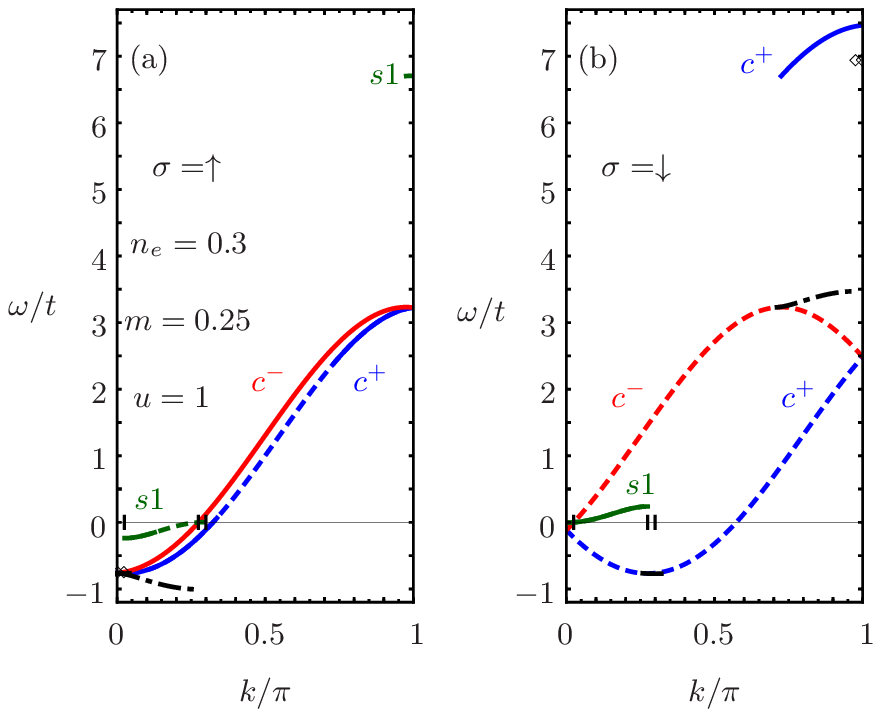}
\includegraphics[scale=0.98]{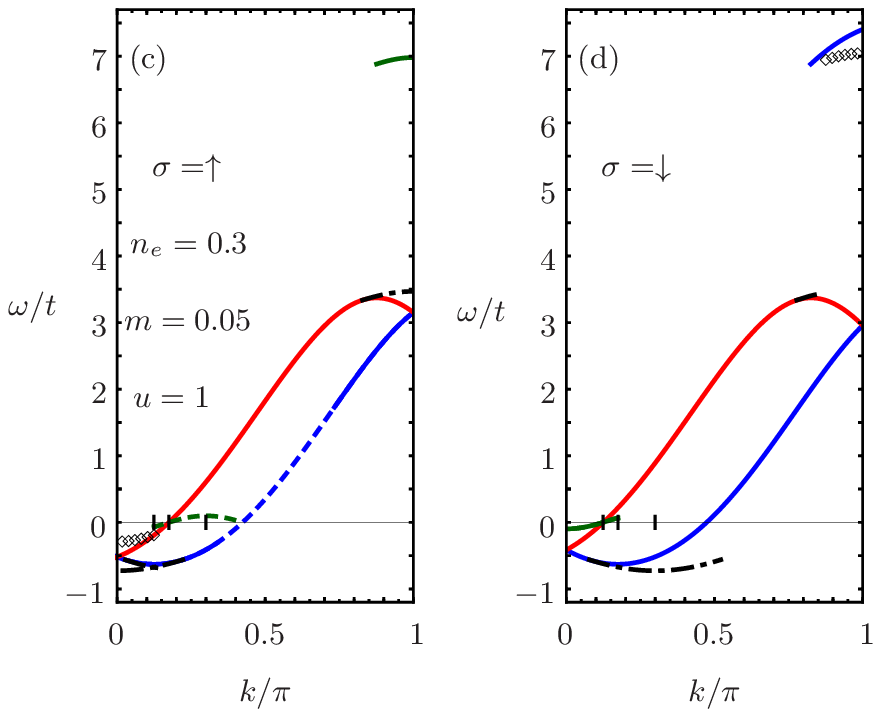}
\caption{\label{s032505u1} The same $(k,\omega)$-plane lines as in Fig. \ref{s4525u01}
for $u=1$, electronic density $n_e =0.3$, and spin densities (a)-(b) $m=0.25$ and (c)-(d) $m=0.05$.
(Online the $c^+$, $c^-$, and $s1$ branch lines are blue, red, and green, respectively.)}
\end{figure}

The $\sigma $ electron removal and LHB addition $c^{\pm}$ branch lines
are generated by processes that correspond to particular cases of those generated by
the leading-order operators, Eqs. (\ref{upElremo}), (\ref{upElLHBadd}), (\ref{downElremo}),
and (\ref{downElLHBadd}) that are behind the $\uparrow$ one-electron removal spectrum, Eq. (\ref{SpupElremo}),
$\uparrow$ one-electron LHB addition spectrum, Eq. (\ref{SpupElLHBadd}),
$\downarrow$ one-electron removal spectrum, Eq. (\ref{SpdownElremo}), and
$\downarrow$ one-electron LHB addition spectrum, Eq. (\ref{SpdownElLHBadd}). Hence
these lines one-parametric spectra plotted in Figs. \ref{s4525u01}-\ref{s032505u1}
are contained within such two-parametric spectra that 
occupy well defined regions in the $(k,\omega)$ plane.
(Online the $c^+$ and $c^-$ branch lines are blue and red, respectively, in these figures.)

These one-parametric spectra $\omega_{c^{\pm}}^{\sigma} (k)$ and the exponents $\xi_{c^{\pm}}^{\sigma} (k)$
associated with these branch lines are such that,
\begin{equation}
\omega_{c^{+}}^{\sigma} (k) = \omega_{c^{-}}^{\sigma} (-k) \, ; \hspace{0.75cm}
\xi_{c^{+}}^{\sigma} (k) = \xi_{c^{-}}^{\sigma} (-k) \, , \hspace{0.50cm} \sigma =\uparrow,\downarrow \, .
\label{c+-rela}
\end{equation}
Within a reduced first-Brillouin zone scheme, considering both the $c^{+}$ and $c^{-}$ branch lines for
$k \in [0,\pi]$ or only the $c^{+}$ branch line for $k \in [-\pi,\pi]$ contains exactly the same information.
Here we chose the latter option. 

The $\sigma$ one-electron removal and LHB addition $c^{+}$ branch line refers to excited energy 
eigenstates with the following number deviations relative to those of the initial ground state,
\begin{equation}
\delta N_c^F = 0 \, ; \hspace{0.5cm} \delta J_c^F = \delta_{\sigma,\downarrow}/2 \, ; \hspace{0.5cm} \delta N_c^{NF} = \gamma \, ; \hspace{0.5cm} 
\delta N_{s1}^F = \delta_{\sigma,\downarrow}\,\gamma \, ; \hspace{0.5cm} \delta J_{s1}^F = \gamma_{\sigma}/2 \, .
\label{NudRcc}
\end{equation}

The spectrum of general form, Eq. (\ref{dE-dP-bl}), that defines the $(k,\omega)$-plane shape of the 
$\sigma $ one-electron removal and LHB addition $c^+$ branch line reads,
\begin{eqnarray}
\omega_{c^{+}}^{\sigma} (k) & = & \gamma\,\varepsilon_c (q) \, ,  \hspace{0.6cm} \gamma = \pm 1 \, ,
\nonumber \\
q & \in & [-2k_F,2k_F] \hspace{0.5cm}{\rm for}\hspace{0.1cm}\sigma\hspace{0.1cm}{\rm electron}\hspace{0.1cm}{\rm removal} \, ,
\nonumber \\
q & \in & [-\pi,-2k_F] \hspace{0.1cm}{\rm and}\hspace{0.1cm}q \in [2k_F,\pi]
\hspace{0.5cm}{\rm for}\hspace{0.1cm}\sigma\hspace{0.1cm}{\rm electron}\hspace{0.1cm}{\rm LHB}\hspace{0.1cm}{\rm addition} \, ,
\label{OkudRLAcc}
\end{eqnarray}
where $\varepsilon_c (q)$ is the $c$ band energy dispersion, Eq. (\ref{epsilon-q}) for $\beta =c$.
The relation of the $c$ band momentum $q$ to the excitation momentum $k$ is within an extended-zone scheme given by,
\begin{eqnarray}
k & = & \gamma\,q + k_{F\bar{\sigma}} \, ,
\nonumber \\
k &\in & [-k_{F\sigma},(2k_F+k_{F\bar{\sigma}})] \hspace{0.5cm}{\rm for}\hspace{0.1cm}\gamma = -1
\nonumber \\
k & \in & [-(\pi -k_{F\bar{\sigma}}),-k_{F\sigma}] \hspace{0.1cm}{\rm and}\hspace{0.1cm}
k \in [(2k_F+k_{F\bar{\sigma}}),(\pi +k_{F\bar{\sigma}})] \hspace{0.5cm}{\rm for}\hspace{0.1cm}\gamma = +1 \, .
\label{OkudLAcc}
\end{eqnarray}

As mentioned above, we consider a reduced first Brillouin-zone scheme for $k \in [-\pi,\pi]$ within which the $c^{+}$ branch line separates 
into several subbranches. One finds that these subbranches refer to the following momentum $k$ intervals,
\begin{eqnarray}
k & = & \gamma\,q + k_{F\bar{\sigma}} \hspace{0.2cm}{\rm subbranch} \, ,
\nonumber \\
k &\in &  [-k_{F\sigma},(2k_F+k_{F\bar{\sigma}})] \hspace{0.5cm}{\rm for}\hspace{0.1cm}\gamma = -1
\nonumber \\
k & \in & [-(\pi -k_{F\bar{\sigma}}),-k_{F\sigma}] \hspace{0.1cm}{\rm and}\hspace{0.1cm}
k \in [(2k_F+k_{F\bar{\sigma}}),\pi] \hspace{0.5cm}{\rm for}\hspace{0.1cm}\gamma = +1 \, ,
\nonumber \\
k & = & q + k_{F\bar{\sigma}} - 2\pi  \hspace{0.2cm}{\rm subbranch} \, ,
\nonumber \\
k & \in & [-\pi,-(\pi -k_{F\bar{\sigma}})]  \hspace{0.5cm}{\rm for}\hspace{0.1cm}\gamma = +1 \, ,
\label{kqsigc+1}
\end{eqnarray}
that are valid for the densities ranges,
\begin{eqnarray}
& & \uparrow\hspace{0.1cm}{\rm electron:}\hspace{0.1cm}(i)\hspace{0.1cm} n_e \in [0,2/3]\hspace{0.1cm}{\rm and}\hspace{0.1cm}m\in [0,n_e]
\hspace{0.1cm}{\rm and}\hspace{0.1cm}(ii)\hspace{0.1cm} n_e \in [2/3,1]\hspace{0.1cm}{\rm and}\hspace{0.1cm}m\in [(3n_e-2),n_e] \, ,
\nonumber \\
& & \downarrow\hspace{0.1cm}{\rm electron:}\hspace{0.1cm}(i)\hspace{0.1cm} n_e \in [0,1/2]\hspace{0.1cm}{\rm and}\hspace{0.1cm}m\in [0,n_e]
\hspace{0.1cm}{\rm and}\hspace{0.1cm}(ii)\hspace{0.1cm} n_e \in [1/2,2/3]\hspace{0.1cm}{\rm and}\hspace{0.1cm}m\in [0,(2-3n_e)] \, .
\nonumber
\end{eqnarray}
On the other hand, the momentum $k$ intervals,
\begin{eqnarray}
k & = & \gamma\,q + k_{F\bar{\sigma}}  \hspace{0.2cm}{\rm subbranch} \, ,
\nonumber \\
k &\in & [-k_{F\sigma},\pi] \hspace{0.5cm}{\rm for}\hspace{0.1cm}\gamma = -1
\nonumber \\
k & \in &[-(\pi -k_{F\bar{\sigma}}),-k_{F\sigma}]\hspace{0.5cm}{\rm for}\hspace{0.1cm}\gamma = +1 \, ,
\nonumber \\
k & = & q + k_{F\bar{\sigma}} - 2\pi  \hspace{0.2cm}{\rm subbranch} \, ,
\nonumber \\
k &\in & [-\pi,-(2\pi - 2k_F - k_{F\bar{\sigma}})] \hspace{0.5cm}{\rm for}\hspace{0.1cm}\gamma = -1
\nonumber \\
k & \in & [-(2\pi - 2k_F - k_{F\bar{\sigma}}),-(\pi -k_{F\bar{\sigma}})]  \hspace{0.5cm}{\rm for}\hspace{0.1cm}\gamma = +1 \, ,
\label{kqsigc+2}
\end{eqnarray}
are valid for the densities ranges,
\begin{eqnarray}
& & \uparrow\hspace{0.1cm}{\rm electron:}\hspace{0.1cm}n_e \in [2/3,1]\hspace{0.1cm}{\rm and}\hspace{0.1cm}m\in [0,(3n_e-2)] \, ,
\nonumber \\
& & \downarrow\hspace{0.1cm}{\rm electron:}\hspace{0.1cm}(i)\hspace{0.1cm} n_e \in [1/2,2/3]\hspace{0.1cm}{\rm and}\hspace{0.1cm}m\in [(2-3n_e),n_e]
\hspace{0.1cm}{\rm and}\hspace{0.1cm}(ii)\hspace{0.1cm} n_e \in [2/3,1]\hspace{0.1cm}{\rm and}\hspace{0.1cm}m\in [0,n_e] \, .
\nonumber
\end{eqnarray}
The corresponding $k$ intervals of the $c^-$ branch line subbranches are obtained from those provided here upon
exchanging $k$ by $-k$.

The one-parametric spectrum $\omega_{c^{+}}^{\sigma} (k)$ of each $c^+$ branch line subbranch is given by Eq. (\ref{OkudRLAcc}) with the relation 
between the excitation momentum $k$ and the $c$ band momentum $q$ provided in the corresponding $k$ interval, Eqs. (\ref{kqsigc+1}) 
and (\ref{kqsigc+2}). Combining the analysis of such momentum $k$ intervals with the relation 
$\omega_{c^{+}}^{\sigma} (k) = \omega_{c^{-}}^{\sigma} (-k)$, Eq. (\ref{c+-rela}), reveals that the $\sigma$ one-electron LHB 
addition $c^{\pm}$ branch lines are the natural continuation of the $\sigma$ one-electron removal $c^{\pm}$ branch lines. 

\begin{figure}
\includegraphics[scale=1.00]{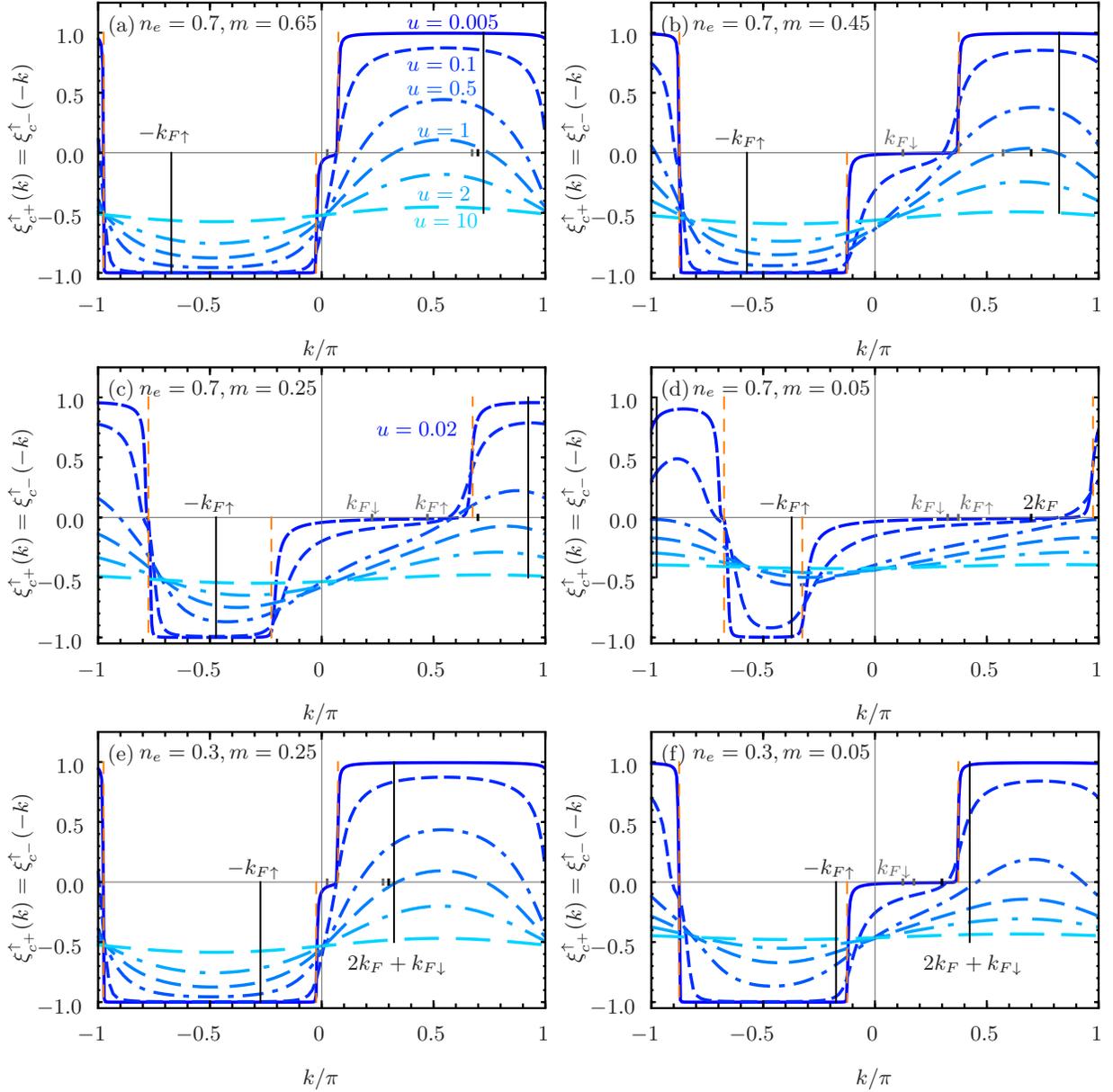}
\caption{\label{fc+up} The exponent $\xi_{c^{+}}^{\uparrow} (k)=\xi_{c^{-}}^{\uparrow} (-k)$, Eq. (\ref{xiupRLAcc}),
that controls the singularities in the vicinity of the $c^+$ branch line whose $(k,\omega)$-plane
one-parametric spectrum is defined by Eqs. (\ref{OkudRLAcc}), (\ref{kqsigc+1}), and (\ref{kqsigc+2}) for the $\sigma =\uparrow$
one-electron removal and LHB addition spectral function, Eq. (\ref{cpm-branch}), as a function of the momentum
$k/\pi\in ]-1,1[$ for several $u$ values, electronic density $n_e =0.7$, and spin densities
(a) $m=0.65$, (b) $m=0.45$, (c) $m=0.25$, and (d)  $m=0.05$, and for electronic density $n_e =0.3$ 
and spin densities (e) $m=0.25$ and (f) $m=0.05$. The type of exponent line associated with
each $u$ value is for all figures the same. Full and dashed vertical lines denote specific momentum
values between different subbranches and momenta where the $u\rightarrow 0$ limiting value
of the exponent changes, respectively.}
\end{figure}

\begin{figure}
\includegraphics[scale=1.00]{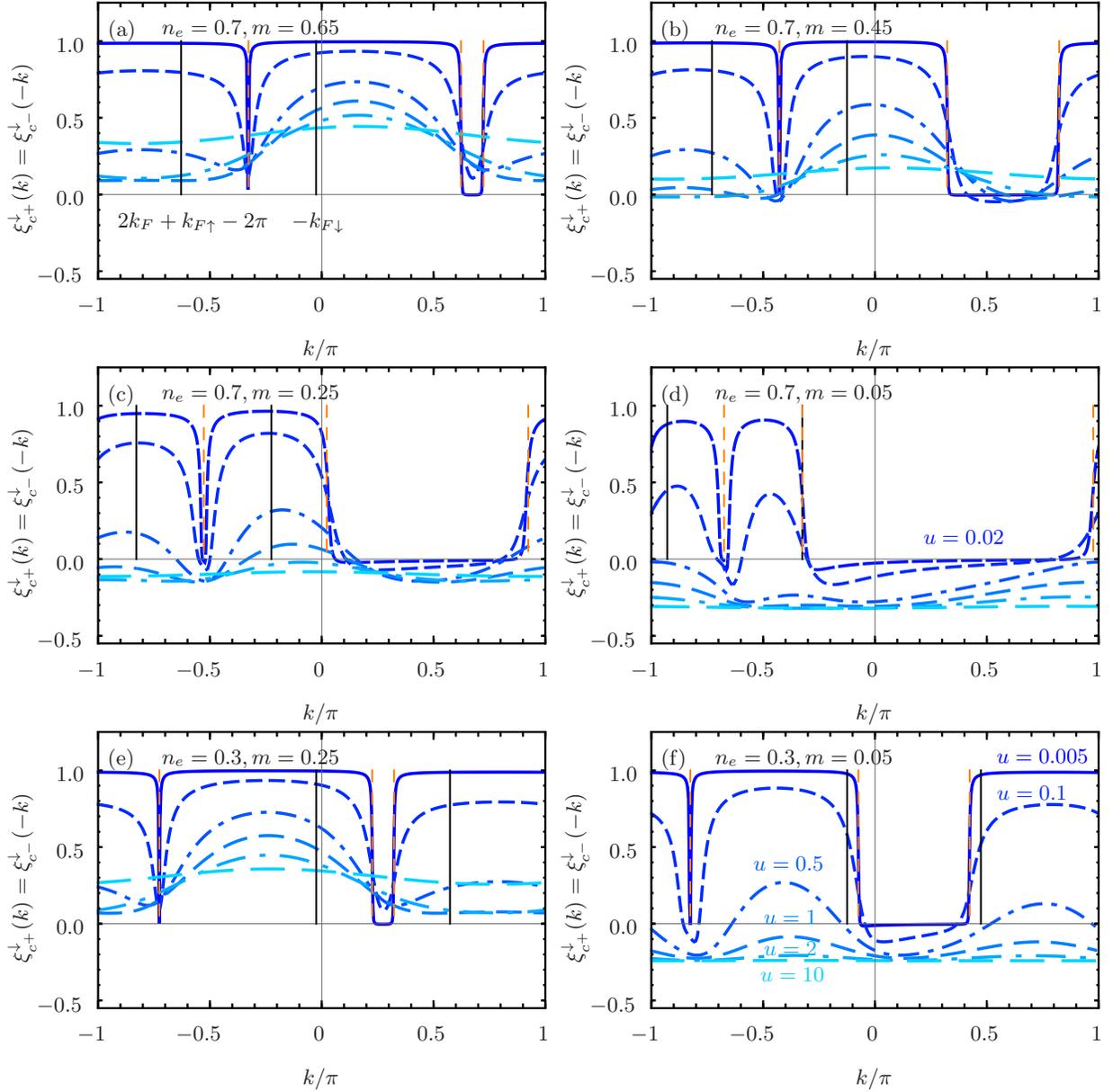}
\caption{\label{fc+down} The exponent $\xi_{c^{+}}^{\downarrow} (k)=\xi_{c^{-}}^{\downarrow} (-k)$, Eq. (\ref{xidownRLAcc}),
that controls the singularities in the vicinity of the $c^+$ branch line whose $(k,\omega)$-plane
shape is defined by Eqs. (\ref{OkudRLAcc}), (\ref{kqsigc+1}), and (\ref{kqsigc+2}) for the $\sigma =\downarrow$
one-electron removal and LHB addition spectral function, Eq. (\ref{cpm-branch}), as a function of the momentum
$k/\pi\in ]-1,1[$ for the same values of $u$, electronic density $n_e$, and spin density $m$ as in 
Fig. \ref{fc+up}.}
\end{figure}

From the use of the values of the functional, Eq. (\ref{OESFfunctional}), specific to the excited energy eigenstates that
determine spectral weight distribution near the $c^{\pm}$ branch lines, one finds that the momentum dependent exponents of 
general form, Eq. (\ref{branch-l}), that control such a line shape read,
\begin{equation}
\xi_{c^{+}}^{\uparrow} (k) = \xi_{c^{-}}^{\uparrow} (-k) = -1 + \sum_{\iota=\pm1}\left({\xi_{c\,s1}^1\over 2} + \gamma\,\Phi_{c,c}(\iota 2k_{F},q)\right)^2 
+ \sum_{\iota=\pm1}\left({\xi_{s1\,s1}^1\over 2} + \gamma\,\Phi_{s1,c}(\iota k_{F\downarrow},q)\right)^2 \, ,
\label{xiupRLAcc}
\end{equation}
for the $\sigma = \uparrow$ one-electron $c^{\pm}$ branch lines and,
\begin{eqnarray}
\xi_{c^{+}}^{\downarrow} (k) & = & \xi_{c^{-}}^{\downarrow} (-k) = -1 + \sum_{\iota=\pm1}\left({\iota\,\gamma\,\xi_{c\,s1}^0\over 2} + {(\xi_{c\,c}^1-\xi_{c\,s1}^1)\over 2} 
+ \gamma\,\Phi_{c,c}(\iota 2k_{F},q)\right)^2 
\nonumber \\
& + & \sum_{\iota=\pm1}\left({\iota\,\gamma\,\xi_{s1\,s1}^0\over 2} + {(\xi_{s1\,c}^1-\xi_{s1\,s1}^1)\over 2} 
+ \gamma\,\Phi_{s1,c}(\iota k_{F\downarrow},q)\right)^2 \, ,
\label{xidownRLAcc}
\end{eqnarray}
for the $\sigma = \downarrow$ one-electron $c^{\pm}$ branch lines. These $\uparrow$ 
and $\downarrow$ one-electron exponents are plotted in Figs. \ref{fc+up} and \ref{fc+down}, respectively,
as a function of the momentum $k/\pi\in ]-1,1[$ for several $u$ values, electronic densities $n_e =0.3$
and $n_e =0.7$, and a set of spin density values $m<n_e$.

The specific form of the general expression, Eq. (\ref{branch-l}), of the $\sigma$ one-electron spectral function
$B_{\sigma,\gamma} (k,\omega)$, Eq. (\ref{Bkomega}), in the vicinity of the present $c^{\pm}$ branch lines is,
\begin{equation}
B_{\sigma,\gamma} (k,\omega) = C_{\sigma,\gamma,c^{\pm}} \Bigl(\gamma\omega - \omega_{c^{\pm}}^{\sigma} (k)\Bigr)^{\xi_{c^{\pm}}^{\sigma} (k)}  
\, ; \hspace{0.50cm} (\gamma\,\omega - \omega_{c^{\pm}}^{\sigma} (k)) \geq 0 \, , \hspace{0.50cm} \gamma = \pm 1 \, ,
\label{cpm-branch}
\end{equation}
where $C_{\sigma,\gamma,c^{\pm}}$ are constants independent of $k$ and $\omega$, the spectra 
$\omega_{c^{+}}^{\sigma} (k) = \omega_{c^{-}}^{\sigma} (-k)$ 
of the several subbranches are given in Eqs. (\ref{OkudRLAcc}), (\ref{kqsigc+1}), and (\ref{kqsigc+2}),
and the exponents $\xi_{c^{+}}^{\sigma} (k)=\xi_{c^{-}}^{\sigma} (-k)$ are defined in Eqs. (\ref{xiupRLAcc}) 
and (\ref{xidownRLAcc}) for $\sigma =\uparrow$ and $\sigma =\downarrow$, respectively.

The following exponents behaviors reached in the $u\rightarrow 0$ limit are derived from
the use in Eqs. (\ref{xiupRLAcc}) and (\ref{xidownRLAcc}) of the values corresponding to
that limit of the phase-shift parameters $\xi^{j}_{\beta\,\beta'}$ and $\beta =c,s1$ pseudofermion phase 
shifts in units of $2\pi$, $\Phi_{\beta,\beta'} (\iota q_{F\beta},q)$, given in Eqs. (\ref{ZZ-gen-u0}) and 
(\ref{Phis-all-qFqu0}) of Appendix \ref{LimitBV}, respectively. The found behaviors in the 
$u\rightarrow 0$ limit of the $c^{+}$ branch line subbranches exponents for $\sigma = \uparrow$ 
one-electron removal ($\gamma=-1$) are,
\begin{eqnarray}
& & \lim_{u\rightarrow 0}\xi_{c^{+}}^{\uparrow} (k) = -1 \, , \hspace{0.5cm} k \in [-k_{F\uparrow},-k_{F\downarrow}] 
\hspace{0.1cm}{\rm for}\hspace{0.1cm}\gamma = -1
\nonumber \\
& & \hspace{0.75cm}{\rm for}\hspace{0.1cm}n_e \in [0,1]\hspace{0.1cm}{\rm and}\hspace{0.1cm}m\in [0,n_e] \, ,
\label{xiupRc+U0-0}
\end{eqnarray}
\begin{eqnarray}
& & \lim_{u\rightarrow 0}\xi_{c^{+}}^{\uparrow} (k) = 0 \, , 
\nonumber \\
& & \hspace{0.5cm}k \in [-k_{F\downarrow},3k_{F\downarrow}]\hspace{0.1cm}{\rm for}\hspace{0.1cm}\gamma = -1
\nonumber \\
& & \hspace{0.75cm}{\rm for}\hspace{0.1cm}n_e \in [0,2/3]\hspace{0.1cm}{\rm and}\hspace{0.1cm}m\in [0,n_e] 
\nonumber \\
& & \hspace{0.75cm}{\rm for}\hspace{0.1cm}n_e \in [2/3,1]\hspace{0.1cm}{\rm and}\hspace{0.1cm}m\in [(n_e-2/3),n_e] 
\nonumber \\
& & \hspace{0.5cm}k \in [-k_{F\downarrow},\pi]\hspace{0.1cm}{\rm and}\hspace{0.1cm}k \in [-\pi,-(2\pi -3k_{F\downarrow})]
\hspace{0.1cm}{\rm for}\hspace{0.1cm}\gamma = -1
\nonumber \\
& & \hspace{0.75cm}{\rm for}\hspace{0.1cm}n_e \in [2/3,1]\hspace{0.1cm}{\rm and}\hspace{0.1cm}m\in [0,(n_e-2/3)] \, ,
\label{xiupRc+U0-1}
\end{eqnarray}
\begin{eqnarray}
& & \lim_{u\rightarrow 0}\xi_{c^{+}}^{\uparrow} (k) = 1 \, , 
\nonumber \\
& & \hspace{0.5cm} k \in [3k_{F\downarrow},(2k_F+k_{F\downarrow})] \hspace{0.1cm}{\rm for}\hspace{0.1cm}\gamma = -1
\nonumber \\
& & \hspace{0.75cm}{\rm for}\hspace{0.1cm}n_e \in [0,2/3]\hspace{0.1cm}{\rm and}\hspace{0.1cm}m\in [0,n_e] 
\nonumber \\
& & \hspace{0.75cm}{\rm for}\hspace{0.1cm}n_e \in [2/3,1]\hspace{0.1cm}{\rm and}\hspace{0.1cm}m\in [(3n_e-2),n_e] 
\nonumber \\
& & \hspace{0.5cm} k \in [3k_{F\downarrow},\pi]\hspace{0.1cm}{\rm and}\hspace{0.1cm}
k \in [-\pi,-(2\pi -2k_F-k_{F\downarrow})]\hspace{0.1cm}{\rm for}\hspace{0.1cm}\gamma = -1
\nonumber \\
& & \hspace{0.75cm}{\rm for}\hspace{0.1cm}n_e \in [2/3,1]\hspace{0.1cm}{\rm and}\hspace{0.1cm}m\in [(n_e-2/3),(3n_e-2)] 
\nonumber \\
& & \hspace{0.5cm} k \in [-(2\pi -3k_{F\downarrow}),-(2\pi -2k_F-k_{F\downarrow})]
\nonumber \\
& & \hspace{0.75cm}{\rm for}\hspace{0.1cm}n_e \in [2/3,1]\hspace{0.1cm}{\rm and}\hspace{0.1cm}m\in [0,(n_e-2/3)] \, .
\label{xiupRc+U0--1}
\end{eqnarray}

For LHB addition ($\gamma=+1$), one finds,
\begin{eqnarray}
& & \lim_{u\rightarrow 0}\xi_{c^{+}}^{\uparrow} (k) = -1 \, , \hspace{0.5cm} k \in [-(\pi -k_{F\downarrow}),-k_{F\uparrow}] 
\hspace{0.1cm}{\rm for}\hspace{0.1cm}\gamma = +1
\nonumber \\
& & \hspace{0.75cm}{\rm for}\hspace{0.1cm}n_e \in [0,1]\hspace{0.1cm}{\rm and}\hspace{0.1cm}m\in [0,n_e] \, ,
\nonumber \\
& & \lim_{u\rightarrow 0}\xi_{c^{+}}^{\uparrow} (k) = 1 \hspace{0.1cm}{\rm for}\hspace{0.1cm}\gamma = +1
\nonumber \\
& & \hspace{0.75cm}{\rm for}\hspace{0.1cm}{\rm the}\hspace{0.1cm}{\rm other}\hspace{0.1cm}
k\hspace{0.1cm}{\rm ranges}\hspace{0.1cm}{\rm in}\hspace{0.1cm}{\rm Eqs.}\hspace{0.1cm}(\ref{kqsigc+1})
\hspace{0.1cm}{\rm and}\hspace{0.1cm}(\ref{kqsigc+2})
\hspace{0.1cm}{\rm with}\hspace{0.1cm}\sigma=\uparrow\hspace{0.1cm}{\rm and}\hspace{0.1cm}\bar{\sigma}=\downarrow \, .
\label{xiupLAccU0}
\end{eqnarray}

Similar values for the exponent $\xi_{c^{-}}^{\downarrow} (k)$ are obtained upon exchanging $k$ by $-k$. Important
$c^{-}$ branch line subbranches are those for which $\lim_{u\rightarrow 0}\xi_{c^{-}}^{\uparrow} (k) = -1$. They
refer to the $k$ ranges,
\begin{eqnarray}
& & \lim_{u\rightarrow 0}\xi_{c^{-}}^{\uparrow} (k) = -1 \, ,
\nonumber \\
& & \hspace{0.50cm}k \in [k_{F\downarrow},k_{F\uparrow}] 
\hspace{0.1cm}{\rm for}\hspace{0.1cm}\gamma = -1\hspace{0.1cm}{\rm and}\hspace{0.1cm}k \in [k_{F\uparrow},(\pi -k_{F\downarrow})] 
\hspace{0.1cm}{\rm for}\hspace{0.1cm}\gamma = +1\, ,
\label{xiupRc+U0-0-1}
\end{eqnarray}
that are valid for $n_e \in [0,1[$ and $m\in [0,n_e]$.

For the $k$ ranges for which $\lim_{u\rightarrow 0}\xi_{c^{\pm}}^{\uparrow} (k) = -1$ the line shape 
has not the form given in Eq. (\ref{cpm-branch}) and rather becomes $\delta$-function like,
Eq. (\ref{branch-lexp-1}). In the present case this gives,
\begin{eqnarray}
\lim_{u\rightarrow 0} B_{\uparrow,-1} (k,\omega) & = & \delta\Bigl(\omega + \omega_{c^{+}}^{\uparrow} (k)\Bigr) 
= \delta\Bigl(\omega - 2t(\cos k - \cos k_{F\uparrow})\Bigr) \, , \hspace{0.50cm} k \in [-k_{F\uparrow},-k_{F\downarrow}] \, ,
\nonumber \\
\lim_{u\rightarrow 0} B_{\uparrow,+1} (k,\omega) & = & \delta\Bigl(\omega - \omega_{c^{+}}^{\uparrow} (k)\Bigr) 
= \delta\Bigl(\omega + 2t(\cos k - \cos k_{F\uparrow})\Bigr) 
\, , \hspace{0.50cm} k \in [-(\pi -k_{F\downarrow}),-k_{F\uparrow}] \, ,
\nonumber \\
\lim_{u\rightarrow 0} B_{\uparrow,-1} (k,\omega) & = & \delta\Bigl(\omega + \omega_{c^{-}}^{\uparrow} (k)\Bigr) 
= \delta\Bigl(\omega - 2t(\cos k - \cos k_{F\uparrow})\Bigr)
\, , \hspace{0.50cm} k \in [k_{F\downarrow},k_{F\uparrow}] \, ,
\nonumber \\
\lim_{u\rightarrow 0} B_{\uparrow,+1} (k,\omega) & = & \delta\Bigl(\omega - \omega_{c^{-}}^{\uparrow} (k)\Bigr) 
= \delta\Bigl(\omega + 2t(\cos k - \cos k_{F\uparrow})\Bigr)
\, , \hspace{0.50cm} k \in [k_{F\uparrow},(\pi -k_{F\downarrow})] \, .
\label{cpm-branch-delta}
\end{eqnarray}
The behaviors reported here thus recover parts of the exact $u=0$ $\sigma$ one-electron spectrum.
That the spectra $\omega_{c^{\pm}}^{\sigma} (k)$ become in the $u\rightarrow 0$ limit the corresponding 
non-interacting electronic spectra is confirmed by accounting for the limiting behavior of the $c$ energy 
dispersion $\varepsilon_{c} (q)$ appearing in these $u>0$ general spectra expression, Eq. (\ref{OkudRLAcc}). 
Such a limiting behavior is reported in Eq. (\ref{varepsiloncu0}) of Appendix \ref{LimitBV}.

On the other hand, for the $k$ ranges for which the exponents are for $u\rightarrow 0$ given by $0$ and/or $1$ the 
$\uparrow$ one-electron spectral weight at and near the corresponding branch lines vanishes in the $u\rightarrow 0$ limit.

One finds that in the $u\rightarrow 0$ limit the $\sigma = \downarrow$ one-electron removal exponent, Eq. (\ref{xidownRLAcc}), 
has the following behaviors,
\begin{eqnarray}
& & \lim_{u\rightarrow 0}\xi_{c^{+}}^{\downarrow} (k) = 1 \, , 
\nonumber \\
& & \hspace{0.50cm}k \in [-k_{F\downarrow},(k_{F\uparrow}-2k_{F\downarrow})] 
\hspace{0.1cm}{\rm for}\hspace{0.1cm}\gamma = -1
\nonumber \\
& & \hspace{0.75cm}{\rm for}\hspace{0.1cm}n_e \in [0,1]\hspace{0.1cm}{\rm and}\hspace{0.1cm}m\in [0,n_e] 
\nonumber \\
& & \hspace{0.50cm}k \in [(2k_F+k_{F\downarrow}),(2k_F+k_{F\uparrow})]\hspace{0.1cm}{\rm for}\hspace{0.1cm}\gamma = -1
\nonumber \\
& & \hspace{0.75cm}{\rm for}\hspace{0.1cm}n_e \in [0,1/2]\hspace{0.1cm}{\rm and}\hspace{0.1cm}m\in [0,n_e] 
\nonumber \\
& & \hspace{0.75cm}{\rm for}\hspace{0.1cm}n_e \in [1/2,2/3]\hspace{0.1cm}{\rm and}\hspace{0.1cm}m\in [0,(2-3n_e)] 
\nonumber \\
& & \hspace{0.50cm}k \in [(2k_F+k_{F\downarrow}),\pi]\hspace{0.1cm}{\rm and}\hspace{0.1cm}
k \in [-\pi,-(2\pi -2k_F-k_{F\uparrow})]\hspace{0.1cm}{\rm for}\hspace{0.1cm}\gamma = -1
\nonumber \\
& & \hspace{0.75cm}{\rm for}\hspace{0.1cm}n_e \in [1/2,2/3]\hspace{0.1cm}{\rm and}\hspace{0.1cm}m\in [(2-3n_e),n_e] 
\nonumber \\
& & \hspace{0.75cm}{\rm for}\hspace{0.1cm}n_e \in [2/3,1]\hspace{0.1cm}{\rm and}\hspace{0.1cm}m\in [(3n_e-2),n_e] 
\nonumber \\
& & \hspace{0.50cm}k \in [-(2\pi -2k_F-k_{F\downarrow}),-(2\pi -2k_F-k_{F\uparrow})]\hspace{0.1cm}{\rm for}\hspace{0.1cm}\gamma = -1
\nonumber \\
& & \hspace{0.75cm}{\rm for}\hspace{0.1cm}n_e \in [2/3,1]\hspace{0.1cm}{\rm and}\hspace{0.1cm}m\in [0,(3n_e-2)] 
\nonumber \\
\label{xidownRc+U0-1}
\end{eqnarray}
and
\begin{eqnarray}
&& \lim_{u\rightarrow 0}\xi_{c^{+}}^{\downarrow} (k) = 0 \, , 
\nonumber \\
& & \hspace{0.50cm}k \in [(k_{F\uparrow}-2k_{F\downarrow}),(2k_F+k_{F\downarrow})]\hspace{0.1cm}{\rm for}\hspace{0.1cm}\gamma = -1 
\nonumber \\ 
& & \hspace{0.75cm}{\rm for}\hspace{0.1cm}n_e \in [0,2/3]\hspace{0.1cm}{\rm and}\hspace{0.1cm}m\in [0,n_e] 
\nonumber \\
& & \hspace{0.75cm}{\rm for}\hspace{0.1cm}n_e \in [2/3,1]\hspace{0.1cm}{\rm and}\hspace{0.1cm}m\in [(3n_e-2),n_e] 
\nonumber \\
& & \hspace{0.50cm}k \in [(k_{F\uparrow}-2k_{F\downarrow}),\pi]
\hspace{0.1cm}{\rm and}\hspace{0.1cm}k \in [-\pi,-(2\pi -2k_F-k_{F\downarrow})]
\hspace{0.1cm}{\rm for}\hspace{0.1cm}\gamma = -1 
\nonumber \\
& & \hspace{0.75cm}{\rm for}\hspace{0.1cm}n_e \in [2/3,1]\hspace{0.1cm}{\rm and}\hspace{0.1cm}m\in [0,(3n_e-2)] \, .
\label{xidownRc+U0-0}
\end{eqnarray}

On the other hand, the $\sigma = \downarrow$ one-electron LHB exponent is found to behave as,
\begin{equation}
\lim_{u\rightarrow 0}\xi_{c^{+}}^{\downarrow} (k) = 1 \hspace{0.1cm}{\rm for}\hspace{0.1cm}\gamma = +1
\hspace{0.1cm}{\rm and}\hspace{0.1cm}{\rm the}\hspace{0.1cm}
k\hspace{0.1cm}{\rm ranges}\hspace{0.1cm}{\rm in}\hspace{0.1cm}{\rm Eqs.}\hspace{0.1cm}(\ref{kqsigc+1})
\hspace{0.1cm}{\rm and}\hspace{0.1cm}(\ref{kqsigc+2})
\hspace{0.1cm}{\rm with}\hspace{0.1cm}\sigma=\downarrow\hspace{0.1cm}{\rm and}\hspace{0.1cm}\bar{\sigma}=\uparrow \, .
\label{xidownRccU0}
\end{equation}
Hence the $\downarrow$ one-electron spectral weight at and near these branch lines vanishes in the $u\rightarrow 0$ limit both
for electron removal and LHB addition. Similar values for the exponent $\xi_{c^{-}}^{\downarrow} (k)$ are obtained upon exchanging $k$ by $-k$. 

Analytical expressions for the above exponents can be derived for $u\gg 1$. 
These expressions are continuous functions of the spin density $m$ whose
limiting behaviors for $m\rightarrow 0$ and $m\rightarrow n_e$ we provide here.
For $u\gg 1$ and spin density $m\rightarrow 0$ such expressions are derived
from the use in Eqs. (\ref{xiupRLAcc}) and (\ref{xidownRLAcc}) 
of the parameters $\xi^{j}_{\beta\,\beta'}$ expressions obtained by combining
Eqs. (\ref{ZZ-gen-m0}) and (\ref{x0limits}) of Appendix \ref{LimitBV}
for $u\gg 1$ and of those of the $\beta =c,s1$ pseudofermion phase shifts 
provided in Eq. (\ref{PhiallFULm0}) of that Appendix. 
One then finds the following $c^{+}$ branch line exponent expression
that applies to all its above subbranches $k$ intervals whereas for the 
twin $c^{-}$ branch line it refers to subbranches $k$ intervals generated from those of the
$c^{+}$ branch line upon exchanging $k$ by $-k$,
\begin{equation}
\xi_{c^{\pm}}^{\sigma} (k) = -{3\over 8} + {\ln 2\over 4\pi\,u}
\left(\sin (\pi n_e) \pm 2\sin \left(k \mp {\pi\over 2}n_e\right)\right) \, , \hspace{0.5cm} 
\sigma = \uparrow, \downarrow \, .
\label{xiudRLAccUim0}
\end{equation}

On the other hand, for $u\gg 1$ and spin density $m\rightarrow n_e$ one
uses in Eqs. (\ref{xiupRLAcc}) and (\ref{xidownRLAcc}) 
the parameters $\xi^{j}_{\beta\,\beta'}$ expressions obtained by combining
Eqs. (\ref{ZZ-gen-m1}) and (\ref{eta0lim}) of Appendix \ref{LimitBV}
and those of the $\beta =c,s1$ pseudofermion phase shifts
provided in Eq. (\ref{PhiallFULmne}) of that Appendix. One then finds 
that the $c^{\pm}$ branch line exponents have different expressions for
the $\uparrow$ one-electron and $\downarrow$ one-electron spectral functions that read,
\begin{eqnarray}
\xi_{c^{\pm}}^{\uparrow} (k) & = & - {1\over 2} \pm {2\over \pi\,u}\sin k \, , 
\nonumber \\
\xi_{c^{\pm}}^{\downarrow} (k) & = & {1\over 2} - {2\over \pi\,u}
\left(\sin (\pi n_e) \pm \sin (k \mp \pi n_e)\right) \, , 
\label{xiudRLAccUim1}
\end{eqnarray}
respectively.

As shown in Fig. \ref{fc+up}, the main effect on the $k$ dependence of the 
$\uparrow$ one-electron removal and LHB addition exponent 
$\xi_{c^{+}}^{\uparrow} (k) = \xi_{c^{-}}^{\uparrow} (-k)$, Eq. (\ref{xiupRLAcc}), of increasing the on-site repulsion
$u$ from $u\ll 1$ to $u\gg 1$ is to continuously changing its $u\rightarrow 0$ values $-1$,
$0$, and $1$ for the $k$ ranges given in Eqs. (\ref{xiupRc+U0-0})-(\ref{xiupLAccU0})
to a $k$ independent value for $k\in [-\pi,\pi]$ as $u\rightarrow\infty$, which smoothly changes 
from $-3/8$ for $m\rightarrow 0$ to $-1/2$ for for $m\rightarrow n_e$.
The general trend of such an exponent $u$ dependence is thus that for the momentum
$k$ ranges for which it reads $0$ and $1$ in the $u\rightarrow 0$ limit it decreases
upon increasing $u$ whereas for the $k$ intervals for which it is given by $-1$
in that limit it rather increases for increasing $u$ values.

On other hand, the exponent $\xi_{c^{+}}^{\downarrow} (k)=\xi_{c^{-}}^{\downarrow} (k)$, Eq. (\ref{xidownRLAcc}),
plotted in Fig. \ref{fc+down} becomes negative only for large $u$ and small spin density values.
For $u\rightarrow 0$ it reads $0$ and $1$ for the $k$ intervals provided in Eqs. (\ref{xidownRc+U0-1})-(\ref{xidownRccU0})
whereas as $u\rightarrow\infty$ it continuously evolves to a $k$ independent value for $k\in [-\pi,\pi]$
that smoothly changes from $-3/8$ for $m\rightarrow 0$ to $1/2$ for for $m\rightarrow n_e$.
The general trend of that exponent $u$ dependence is different upon changing the densities.
As shown in Fig. \ref{fc+down}, for some densities it always decreases upon increasing $u$
whereas for other densities it first decreases upon increasing $u$ until reaching some minimum
at a finite $u$ value above which it increases upon further increasing $u$.

\subsection{The $\sigma$ one-electron removal and LHB addition $s1$ branch line}
\label{upRs}

The $\sigma $ electron removal and LHB addition $s1$ branch line
is generated by processes that correspond again to a particular case of those generated by
the leading-order operators, Eqs. (\ref{upElremo}), (\ref{upElLHBadd}), (\ref{downElremo}),
and (\ref{downElLHBadd}). Hence for the $\uparrow$ and $\downarrow$ one-electron spectral 
functions its one-parametric spectrum plotted in Figs. \ref{s4525u01}-\ref{s032505u1}
is contained within the $(k,\omega)$-plane region occupied by the two-parametric spectra 
corresponding to such more general processes.
(Online the $s1$ branch lines are green in these figures.)

The one-parametric spectrum of this branch line is such that
$\omega_{s1}^{\sigma} (k) = \omega_{s1}^{\sigma} (-k)$ and the corresponding
exponent given below is also such that $\xi_{s1}^{\sigma} (k) = \xi_{s1}^{\sigma} (-k)$.
Hence for simplicity we restrict our following analysis to $k \geq 0$. For
such a momentum range the $\sigma $ electron removal and LHB addition parts
of the $s1$ branch line refer to excited energy eigenstates with the following number deviations relative 
to those of the initial ground state,
\begin{equation}
\delta N_c^F = \gamma \, ; \hspace{0.5cm} \delta J_c^F = \delta_{\sigma,\uparrow}/2 \, ; \hspace{0.5cm} 
\delta N_{s1}^F = \delta_{\sigma,\uparrow}\,\gamma \, ; \hspace{0.5cm} \delta J_{s1}^F = 0 \, ; \hspace{0.5cm} 
\delta N_{s1}^{NF} = - \gamma_{\sigma}\,\gamma  \, .
\label{NudRLAs}
\end{equation}

The spectrum $\omega_{s1}^{\sigma} (k)$ of general form, Eq. (\ref{dE-dP-bl}), is for the present branch line given by,
\begin{eqnarray}
\omega_{s1}^{\sigma} (k) & = & 
- \gamma_{\sigma}\,\gamma\, \varepsilon_{s1} (q) \, ,
\nonumber \\
q & \in & [-k_{F\uparrow},-k_{F\downarrow}] \hspace{0.5cm}{\rm for}\hspace{0.1cm}\uparrow\hspace{0.1cm}{\rm electron}\hspace{0.1cm}{\rm removal} \, ,
\nonumber \\
q & \in & [-k_{F\downarrow},k_{F\downarrow}] \hspace{0.5cm}{\rm for}\hspace{0.1cm}\uparrow\hspace{0.1cm}{\rm electron}\hspace{0.1cm}{\rm LHB}\hspace{0.1cm}{\rm addition} \, ,
\nonumber \\
q & \in & [-k_{F\downarrow},0] \hspace{0.5cm}{\rm for}\hspace{0.1cm}\downarrow\hspace{0.1cm}{\rm electron}\hspace{0.1cm}{\rm removal} \, ,
\nonumber \\
q & \in & [k_{F\downarrow},k_{F\uparrow},] \hspace{0.5cm}{\rm for}\hspace{0.1cm}\downarrow\hspace{0.1cm}{\rm electron}\hspace{0.1cm}{\rm LHB}\hspace{0.1cm}{\rm addition} \, ,
\label{OkudRs}
\end{eqnarray}
where $\varepsilon_{s1} (q)$ is the $s1$ band energy dispersion, Eq. (\ref{epsilon-q}) for $\beta =s1$.

The relation of the $s1$ band momentum $q$ to the excitation momentum $k$ is,
\begin{equation}
k = \delta_{\sigma,\uparrow}\,2k_{F} - \gamma_{\sigma}\,\gamma\,q \geq 0 \, ,
\label{kqs1up}
\end{equation}
which gives,
\begin{eqnarray}
k & \in & [k_{F\downarrow},k_{F\uparrow}]  
\hspace{0.5cm}{\rm for}\hspace{0.1cm}\uparrow\hspace{0.05cm}{\rm electron}\hspace{0.1cm}{\rm removal} \, ,
\nonumber \\
k & \in &  [k_{F\uparrow},(2k_F+k_{F\downarrow})] 
\hspace{0.5cm}{\rm for}\hspace{0.1cm}\uparrow\hspace{0.05cm}{\rm electron}\hspace{0.1cm}{\rm LHB}\hspace{0.1cm}{\rm addition} \, ,
\label{kupRLAs}
\end{eqnarray}
and
\begin{eqnarray}
k & \in & [0,k_{F\downarrow}]  
\hspace{0.5cm}{\rm for}\hspace{0.1cm}\downarrow\hspace{0.05cm}{\rm electron}\hspace{0.1cm}{\rm removal} \, ,
\nonumber \\
k & \in & [k_{F\downarrow},k_{F\uparrow}] 
\hspace{0.5cm}{\rm for}\hspace{0.1cm}\downarrow\hspace{0.05cm}{\rm electron}\hspace{0.1cm}{\rm LHB}\hspace{0.1cm}{\rm addition} \, ,
\label{OkdownRLAsoth}
\end{eqnarray}
respectively. 

Except for $\uparrow$ one-electron LHB addition, the above $s1$ branch-line $k$ ranges are within the first Brillouin-zone. 
In that specific case it refers for some densities to an extended-zone scheme. Here we consider a reduced first Brillouin-zone scheme 
for $k \in [0,\pi]$ within which the $s1$ branch line separates for $\uparrow$ one-electron LHB addition into
two subbranches. Actually, one of such subbranches stems for $k>0$ from $k$ momentum values
that within an extended-zone scheme  arise from second Brillouin-zone $k<0$ momentum values.
(For such processes one has in Eq. (\ref{NudRLAs}) that $\delta J_c^F = -1/2$ rather than $\delta J_c^F = 1/2$.) This gives,
\begin{eqnarray}
k & = & 2k_{F} - q\hspace{0.2cm}{\rm subbranch} \, ,
\nonumber \\
k & \in &  [k_{F\uparrow},(2k_F+k_{F\downarrow})] 
\hspace{0.5cm}{\rm for}\hspace{0.1cm}\gamma = 1 \, ,
\nonumber \\
& & \uparrow {\rm electron}\hspace{0.1cm}{\rm addition}\hspace{0.1cm}(i)\hspace{0.1cm} n_e \in [0,2/3]\hspace{0.1cm}{\rm and}\hspace{0.1cm}m\in [0,n_e]
\hspace{0.1cm}
\nonumber \\
& & {\rm and}\hspace{0.1cm}(ii)\hspace{0.1cm} n_e \in [2/3,1]\hspace{0.1cm}{\rm and}\hspace{0.1cm}m\in [(3n_e-2),n_e] \, ,
\nonumber \\
k & = & 2k_{F} - q\hspace{0.2cm}{\rm subbranch} \, ,
\nonumber \\
k & \in & [k_{F\uparrow},\pi] 
\hspace{0.5cm}{\rm for}\hspace{0.1cm}\gamma = 1 \, ,
\nonumber \\
k & = & - 2k_{F} - q + 2\pi \hspace{0.2cm}{\rm subbranch} \, ,
\nonumber \\
k & \in & [(2\pi - 2k_F - k_{F\downarrow}),\pi] \hspace{0.5cm}{\rm for}\hspace{0.1cm}\gamma = 1 \, ,
\nonumber \\
& & \uparrow {\rm electron}\hspace{0.1cm}{\rm addition}\hspace{0.1cm}n_e \in [2/3,1]\hspace{0.1cm}{\rm and}\hspace{0.1cm}m\in [0,(3n_e-2)] \, .
\label{kupRLAs1BZ}
\end{eqnarray}

\begin{figure}
\includegraphics[scale=1.00]{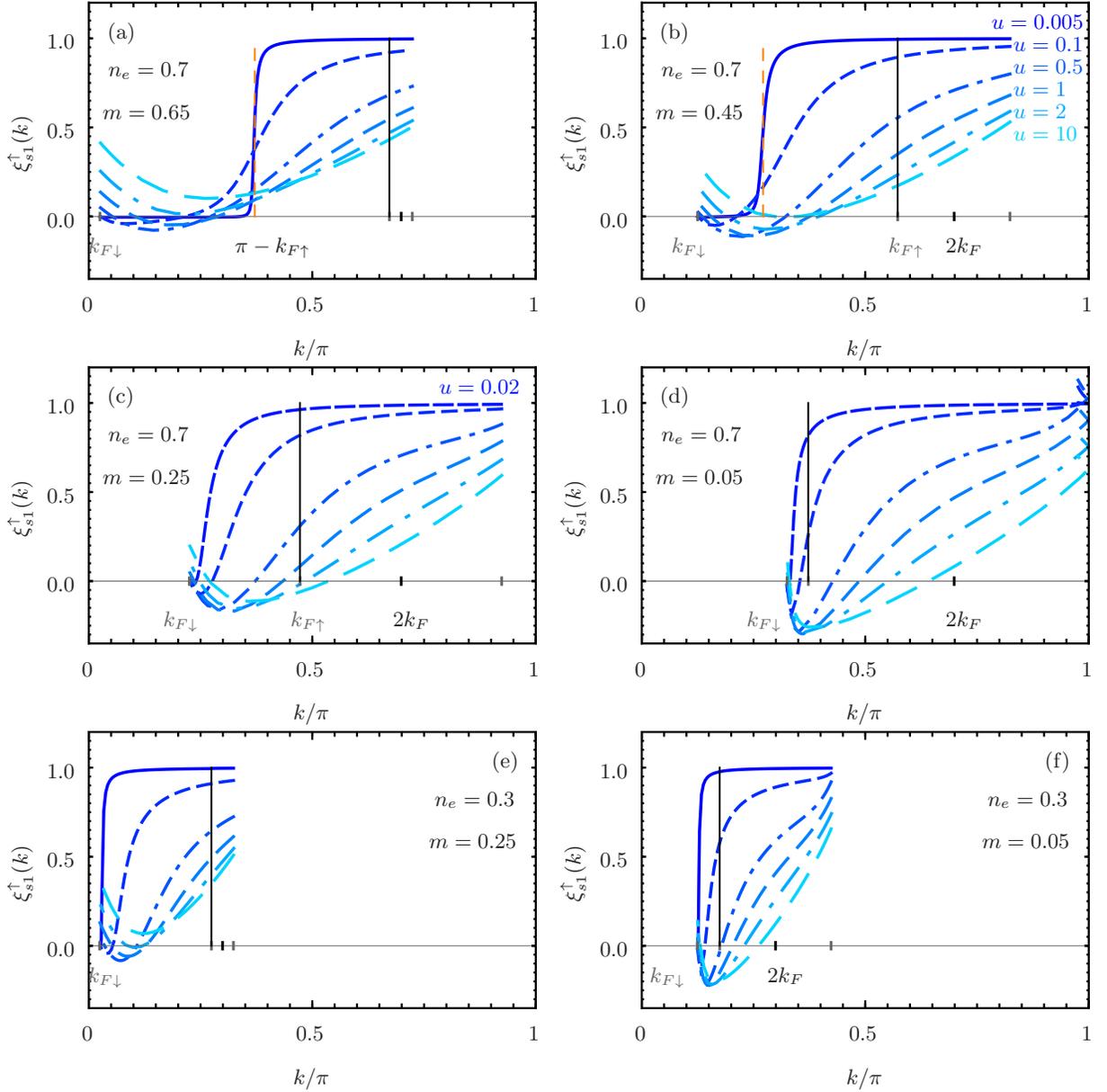}
\caption{\label{fs1up} The exponent $\xi_{s1}^{\uparrow} (k)$, Eq. (\ref{xiupRLAs}),
that controls the singularities in the vicinity of the $s1$ branch line whose $(k,\omega)$-plane
shape is defined by Eqs. (\ref{OkudRs}), (\ref{kupRLAs}), and (\ref{kupRLAs1BZ}) for the $\sigma =\uparrow$
one-electron removal and LHB addition spectral function, Eq. (\ref{s1-branch}), as a function of the momentum
$k/\pi\in ]0,1[$ for the same values of $u$, electronic density $n_e$, and spin density $m$ as in 
Fig. \ref{fc+up}. (For $k/\pi\in ]-1,0[$ the exponent $\xi_{s1}^{\uparrow} (k)$
is given by $\xi_{s1}^{\uparrow} (k)=\xi_{s1}^{\uparrow} (-k)$ with $-k/\pi\in ]0,1[$ as plotted here.)}
\end{figure}

\begin{figure}
\includegraphics[scale=1.00]{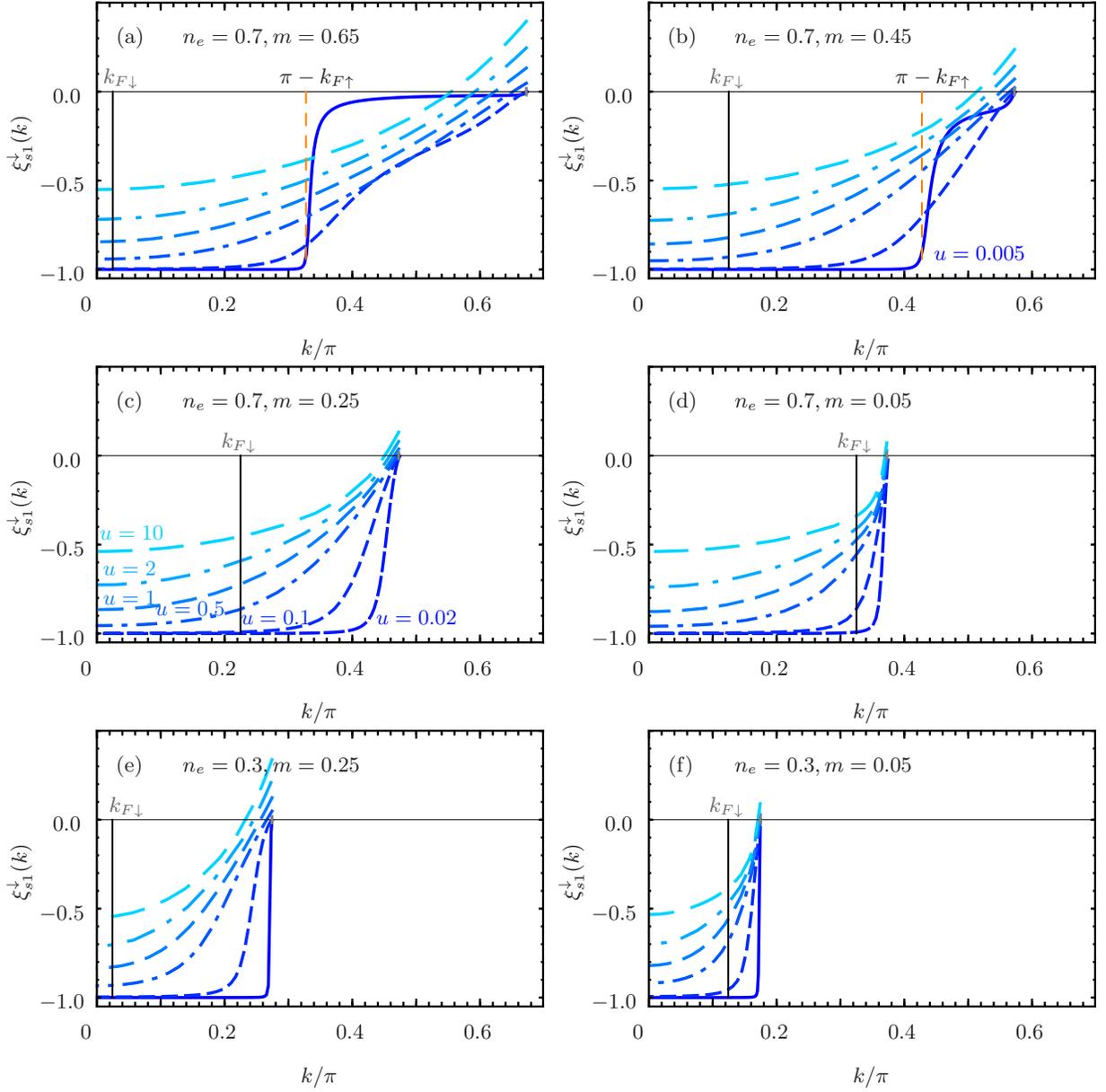}
\caption{\label{fs1down} The exponent $\xi_{s1}^{\downarrow} (k)$, Eq. (\ref{xidownRLAsoth}),
that controls the singularities in the vicinity of the $s1$ branch line whose $(k,\omega)$-plane
one-parametric spectrum is defined by Eqs. (\ref{OkudRs}) and (\ref{OkdownRLAsoth}) for the $\sigma =\downarrow$
one-electron removal and LHB addition spectral function, Eq. (\ref{s1-branch}), as a function of the momentum
$k/\pi\in ]0,1[$ for the same values of $u$, electronic density $n_e$, and spin density $m$ as in 
Fig. \ref{fc+up}. (For $k/\pi\in ]-1,0[$ the exponent $\xi_{s1}^{\downarrow} (k)$
is again given by $\xi_{s1}^{\downarrow} (k)=\xi_{s1}^{\downarrow} (-k)$ with $-k/\pi\in ]0,1[$ as plotted here.)}
\end{figure}

Analysis of the momentum $k$ intervals in Eqs. (\ref{OkdownRLAsoth}) and (\ref{kupRLAs1BZ}) reveals that the 
$\sigma$ one-electron LHB addition $s1$ branch line is the natural continuation of the $\sigma$ 
one-electron removal $s1$ branch line. The momentum dependent exponent of general form, Eq. (\ref{branch-l}),
that controls the line shape near the $\sigma =\uparrow$ one-electron removal and LHB addition $s1$ branch line is given by,
\begin{eqnarray}
\xi_{s1}^{\uparrow} (k) & = & -1 + \sum_{\iota=\pm1}\left({\iota \,\gamma(\xi_{c\,c}^0+\xi_{c\,s1}^0)\over 2} 
+ {\xi_{c\,c}^1\over 2} - \gamma\,\Phi_{c,s1}(\iota 2k_{F},q)\right)^2 
\nonumber \\
& + & \sum_{\iota=\pm1}\left({\iota \,\gamma(\xi_{s1\,c}^0+\xi_{s1\,s1}^0)\over 2} + 
{\xi_{s1\,c}^1\over 2} - \gamma\,\Phi_{s1,s1}(\iota k_{F\downarrow},q)\right)^2  \, ,
\label{xiupRLAs}
\end{eqnarray}
whereas that that controls it in the vicinity of the $\sigma =\downarrow$ one-electron removal and LHB addition $s1$ branch line reads,
\begin{equation}
\xi_{s1}^{\downarrow} (k) = -1 + \sum_{\iota=\pm1}\left({\iota\,\xi_{c\,c}^0\over 2} + \Phi_{c,s1}(\iota 2k_{F},q)\right)^2 
+ \sum_{\iota=\pm1}\left({\iota\,\xi_{s1\,c}^0\over 2} + \Phi_{s1,s1}(\iota k_{F\downarrow},q)\right)^2  \, .
\label{xidownRLAsoth}
\end{equation}
This latter exponent has the same formal expression for $\gamma =-1$ and $\gamma =+1$ the corresponding $q$
ranges being though different, as given in Eq. (\ref{OkudRs}).
These $\uparrow$ and $\downarrow$ one-electron exponents are plotted in Figs. \ref{fs1up} and \ref{fs1down}, respectively,
as a function of the momentum $k/\pi\in ]0,1[$ for several $u$ values, electronic densities $n_e =0.3$
and $n_e =0.7$, and a set of spin density values $m<n_e$.

The general expression, Eq. (\ref{branch-l}), of the $\sigma$ one-electron spectral function
$B_{\sigma,\gamma} (k,\omega)$, Eq. (\ref{Bkomega}), near the $s1$ branch lines is 
in the present case given by, 
\begin{equation}
B_{\sigma,\gamma} (k,\omega) = C_{\sigma,\gamma,s1} \Bigl(\gamma\omega - \omega_{s1}^{\sigma} (k)\Bigr)^{\xi_{s1}^{\sigma} (k)}  
\, ; \hspace{0.50cm} (\gamma\,\omega - \omega_{s1}^{\sigma} (k)) \geq 0 \, , \hspace{0.50cm} \gamma = \pm 1 \, ,
\label{s1-branch}
\end{equation}
where $C_{\sigma,\gamma,s1}$ is a constant independent of $k$ and $\omega$,
the spectrum $\omega_{s1}^{\sigma} (k)$ is that in Eq. (\ref{OkudRs}), and the exponent $\xi_{s1}^{\sigma} (k)$
is given in Eqs. (\ref{xiupRLAs}) and (\ref{xidownRLAsoth}).

The behaviors reached in the $u\rightarrow 0$ limit by the exponents, Eqs. (\ref{xiupRLAs}) and (\ref{xidownRLAsoth}),
can be found by use in these equations of the parameters $\xi^{j}_{\beta\,\beta'}$ values given in Eq. (\ref{ZZ-gen-u0}) of Appendix \ref{LimitBV}
and of the $\beta =c,s1$ pseudofermion phase shifts $\Phi_{\beta,\beta'} (\iota q_{F\beta},q)$ expressions provided in 
Eq. (\ref{Phis-all-qFqu0}) of that  Appendix. One then finds that the $\sigma =\uparrow$ one-electron removal exponent and 
the $\sigma =\downarrow$ one-electron LHB addition exponent have the following related behaviors,
\begin{eqnarray}
\lim_{u\rightarrow 0}\xi_{s1}^{\sigma} (k) & = & \gamma_{\sigma} \, , \hspace{0.5cm} k \in [k_{F\downarrow},k_{F\uparrow}] 
\nonumber \\
& & {\rm for}\hspace{0.50cm} m \in [0,n_e] \hspace{0.50cm}{\rm and}\hspace{0.50cm} n_e \in [0,1/2] 
\hspace{0.50cm}{\rm and}\hspace{0.50cm}{\rm for}\hspace{0.50cm} m \in [0,1-n_e] \hspace{0.50cm}{\rm and}\hspace{0.50cm} n_e \in [1/2,1]
\nonumber \\
\lim_{u\rightarrow 0}\xi_{s1}^{\sigma} (k) & = & \gamma_{\sigma} \, , \hspace{0.5cm} k \in [k_{F\downarrow},\pi - k_{F\uparrow}] 
\nonumber \\
& = & 0 \, , \hspace{0.5cm} k \in [\pi - k_{F\uparrow},k_{F\uparrow}] 
\nonumber \\
& & {\rm for}\hspace{0.50cm} m \in [1-n_e,n_e] \hspace{0.50cm}{\rm and}\hspace{0.50cm} n_e \in [1/2,1] \, .
\label{xiupRsU0}
\end{eqnarray}
Furthermore, one finds that the $\sigma =\uparrow$ electron LHB addition and
$\sigma =\downarrow$ electron removal exponents have also related behaviors that read,
\begin{equation}
\lim_{u\rightarrow 0}\xi_{s1}^{\sigma} (k) = \gamma_{\sigma} \hspace{0.3cm}{\rm (for}\hspace{0.1cm}{\rm the}\hspace{0.1cm}
{\rm whole}\hspace{0.1cm}{\rm branch}\hspace{0.1cm}{\rm lines}\hspace{0.1cm}k\hspace{0.1cm}{\rm range)} \, .
\label{xiupLAsU0}
\end{equation}
Hence the $\sigma =\uparrow$ one-electron spectral weight at and near these $s1$ branch lines vanishes in the $u\rightarrow 0$ limit
both for $\sigma =\uparrow$ electron removal and LHB addition. 

As given generally in Eq. (\ref{branch-lexp-1}), for the $n_e$, $m$, and $k$ ranges for which 
$\lim_{u\rightarrow 0}\xi_{s1}^{\downarrow} (k) = -1$ the line shape
near the branch line is not of the power-law form, Eq. (\ref{s1-branch}). In that limit it rather corresponds to
the following $\delta$-function-like $\sigma =\downarrow$ one-electron spectral weight distribution along it,
\begin{eqnarray}
& & \lim_{u\rightarrow 0} B_{\downarrow,-1} (k,\omega) = \delta\Bigl(\omega + \omega_{s1}^{\downarrow} (k)\Bigr) 
= \delta\Bigl(\omega - 2t(\cos k - \cos k_{F\downarrow})\Bigr)
\, , \hspace{0.20cm} k \in [-k_{F\downarrow},k_{F\downarrow}]  \, ,
\nonumber \\
& & \lim_{u\rightarrow 0} B_{\downarrow,+1} (k,\omega) = \delta\Bigl(\omega - \omega_{s1}^{\downarrow} (k)\Bigr) 
= \delta\Bigl(\omega + 2t(\cos k - \cos k_{F\downarrow})\Bigr) \, ,
\nonumber \\
& & \hspace{0.75cm} k \in [k_{F\downarrow},k_{F\uparrow}] 
\hspace{0.20cm} {\rm for}\hspace{0.10cm} m \in [0,n_e] \hspace{0.10cm}{\rm and}\hspace{0.10cm} n_e \in [0,1/2] 
\hspace{0.10cm}{\rm and}\hspace{0.10cm}{\rm for}\hspace{0.10cm} m \in [0,1-n_e] \hspace{0.10cm}{\rm and}\hspace{0.10cm} n_e \in [1/2,1]
\nonumber \\
& & \hspace{0.75cm} k \in [k_{F\downarrow},\pi - k_{F\uparrow}] 
\hspace{0.20cm} {\rm for}\hspace{0.10cm} m \in [1-n_e,n_e] \hspace{0.10cm}{\rm and}\hspace{0.10cm} n_e \in [1/2,1] \, .
\label{s1-branch-delta}
\end{eqnarray}
The $u\rightarrow 0$ limiting behavior reported in Eq. (\ref{varepsilonsu0}) of Appendix \ref{LimitBV}
for the $s1$ energy dispersion $\varepsilon_{s1} (q)$ appearing in the spectrum $\omega_{s1}^{\downarrow} (k)$, 
Eq. (\ref{OkudRs}), confirms that the latter spectrum becomes in the $u\rightarrow 0$ limit the corresponding $u=0$ 
non-interacting electronic spectrum, as given in Eq. (\ref{s1-branch-delta}).

On the other hand, for the $k$ range for which $\lim_{u\rightarrow 0}\xi_{s1}^{\downarrow} (k) = 0$ the $\downarrow$ one-electron addition spectral 
weight at and near the present $s1$ branch line vanishes in the $u\rightarrow 0$ limit.

For $u\gg 1$ the $s1$ branch line exponent expression is a continuous function of the spin density $m$.
We have derived the corresponding exponent analytical expressions valid for $u\gg 1$ in 
the $m\rightarrow 0$ and $m\rightarrow n_e$ limits.
The $s1$ branch line momentum width vanishes in the $m\rightarrow 0$ limit both for $\downarrow$ one-electron
LHB addition and $\uparrow$ one-electron removal. On the other hand, in that limit the $s1$ branch line for
$\uparrow$ one-electron LHB addition and $\downarrow$ one-electron removal becomes the 
$s1$ branch line for one-electron LHB addition and removal, respectively. By using 
in Eqs. (\ref{xiupRLAs}) and (\ref{xidownRLAsoth}) the values of the parameters $\xi^{j}_{\beta\,\beta'}$ obtained by combining
Eqs. (\ref{ZZ-gen-m0}) and Eq. (\ref{x0limits}) of Appendix \ref{LimitBV} for $u\gg 1$ and of the 
expressions of the $\beta =c,s1$ pseudofermion phase shifts provided in Eq. (\ref{PhiallFULm0}) of that Appendix,
which refer to $u\gg 1$ and spin density $m\rightarrow 0$, one finds that  
the exponent in the spectral function expression, Eq. (\ref{s1-branch}), that controls
the line shape near the $\downarrow$ one-electron removal and $\uparrow$ one-electron LHB addition
$s1$ branch line reads in these limits,
\begin{eqnarray}
\xi_{s1}^{\sigma} (k) & = & - {1\over 2}\left(1- \left({k\over \pi n_e}\right)^2\right)
\left(1 + {2\ln 2\over \pi\,u}\sin (\pi n_e)\right)
- {1\over 2u}\cos\left({k\over n_e}\right)\sin (\pi n_e) \, ,
\nonumber \\
& & \sigma = \uparrow {\rm electron}\hspace{0.1cm}{\rm addition}\hspace{0.1cm}{\rm for}\hspace{0.1cm}k\in [k_F,3k_F]
\hspace{0.1cm}{\rm and}\hspace{0.1cm} n_e \in [0,2/3] 
\nonumber \\
& & \sigma = \uparrow {\rm electron}\hspace{0.1cm}{\rm addition}\hspace{0.1cm}{\rm for}\hspace{0.1cm}k\in [k_F,\pi]
\hspace{0.1cm}{\rm and}\hspace{0.1cm} n_e \in [2/3,1] 
\nonumber \\
& & \sigma = \downarrow {\rm electron}\hspace{0.1cm}{\rm removal}\hspace{0.1cm}{\rm for}\hspace{0.1cm}k\in [0,k_F]
\hspace{0.1cm}{\rm and}\hspace{0.1cm} n_e \in [0,1]
\nonumber \\
\xi_{s1}^{\uparrow} (k) & = & - {1\over 2}\left(1- \left({(k-2\pi)\over \pi n_e}\right)^2\right)
\left(1 + {2\ln 2\over \pi\,u}\sin (\pi n_e)\right)
- {1\over 2u}\cos\left({k-2\pi\over n_e}\right)\sin (\pi n_e) \, ,
\nonumber \\
& & \uparrow {\rm electron}\hspace{0.1cm}{\rm addition}\hspace{0.1cm}{\rm for}\hspace{0.1cm}k\in [(2\pi - 3k_F),\pi]
\hspace{0.1cm}{\rm and}\hspace{0.1cm} n_e \in [2/3,1]  \, ,
\label{xiupRsUim0}
\end{eqnarray}
so that,
\begin{eqnarray}
\lim_{k\rightarrow 0}\xi_{s1}^{\downarrow} (k) & = & - {1\over 2}
- {1\over 2u}\left(1 + {2\ln 2\over \pi}\right)\sin (\pi n_e) \, ,
\nonumber \\
\lim_{k\rightarrow k_F}\xi_{s1}^{\sigma} (k) & = & - {3\over 8}
- {3\ln 2\over 4\pi\,u}\sin (\pi n_e) \, ; \hspace{0.75cm}
\lim_{k\rightarrow 2k_F}\xi_{s1}^{\sigma} (k) = {1\over 2u}\sin (\pi n_e) \, ,
\nonumber \\
\lim_{k\rightarrow 3k_F}\xi_{s1}^{\sigma} (3k_F) & = & \lim_{k\rightarrow 2\pi -3k_F}\xi_{s1}^{\uparrow} (k) = {5\over 8} + {5\ln 2\over 4\pi\,u}\sin (\pi n_e) \, . 
\label{xiupRsUim0-aux}
\end{eqnarray}
To reach the second exponent expression given in Eq. (\ref{xiupRsUim0}) one can either (i) use a new general exponent expression
obtained upon replacing $\delta J_c^F = 1/2$ by $\delta J_c^F = -1/2$, which changes the terms
$\xi_{c\,c}^1/2$ and $\xi_{s1\,c}^1/2$ in Eq. (\ref{xiupRLAs}) to $-\xi_{c\,c}^1/2$ and $-\xi_{s1\,c}^1/2$, respectively, 
or (ii) use the present exponent expression, Eq. (\ref{xiupRLAs}),
upon bringing a $k>0$ second Brillouin zone contribution to $k\in [-\pi,-(2\pi - 3k_F)]$ and then
relying on the $\xi_{s1}^{\uparrow} (k)=\xi_{s1}^{\uparrow} (-k)$ symmetry to reach the expression valid
for $k\in [(2\pi - 3k_F),\pi]$. For $u\gg 1$ and $m\rightarrow 0$ the $\uparrow$ one-electron LHB addition 
exponent $\xi_{s1}^{\uparrow} (k)$ continuously changes from $\xi_{s1}^{\uparrow} (k) = - 3/8$ for $k\rightarrow k_F$
to $\xi_{s1}^{\uparrow} (k) = 0$ for $k\rightarrow 2k_F$. For its other $k$ ranges it is
positive. In these limits the $\downarrow$ one-electron removal exponent $\xi_{s1}^{\downarrow} (k)$
continuously changes from $\xi_{s1}^{\downarrow} (k) = - 1/2$ for $k\rightarrow 0$
to $\xi_{s1}^{\sigma} (k) = - 3/8$ for $k\rightarrow k_F$.

On the other hand, in the $m\rightarrow n_e$ limit the situation is the opposite relative to that for $m\rightarrow 0$, 
as the $s1$ branch line momentum width vanishes in the former limit both for $\uparrow$ one-electron
LHB addition and $\downarrow$ one-electron removal. The use in the exponent 
expressions, Eqs. (\ref{xiupRLAs}) and (\ref{xidownRLAsoth}), of the values 
for $u\gg 1$ and spin density $m\rightarrow n_e$ of the parameters $\xi^{j}_{\beta\,\beta'}$ obtained 
by combining Eq. (\ref{ZZ-gen-m1}) and (\ref{eta0lim}) of Appendix \ref{LimitBV}
for $u\gg 1$ and of the expressions of the $\beta =c,s1$ pseudofermion phase shifts
provided in Eq. (\ref{PhiallFULmne}) of that Appendix we find the following
exponent expressions for the $\uparrow$ one-electron removal
and $\downarrow$ one-electron LHB addition $s1$ branch line,
\begin{eqnarray}
\xi_{s1}^{\uparrow} (k) & = & {1\over 2}\left({k\over \pi n_e}\right)^2 
+ {2\over\pi^2}\left[\arctan\left({1\over 2}\cot \left({k\over 2n_e}\right)\right)\right]^2 
\nonumber \\
& - & {2\over\pi\,u}\left[\cos^2 \left({k\over 2n_e}\right) 
- {k\over \pi n_e}{2\over\pi}\arctan\left({1\over 2}\cot \left({k\over 2n_e}\right)\right)\right]\sin (\pi n_e) \, ,
\nonumber \\
& & \uparrow {\rm electron}\hspace{0.1cm}{\rm removal}\hspace{0.1cm}{\rm for}\hspace{0.1cm}k\in [0,2k_F] \, ,
\nonumber \\
\xi_{s1}^{\downarrow} (k) & = & - {1\over 2}\left(1-\left({k\over \pi n_e}\right)^2 \right)
+ {2\over\pi^2}\left[\arctan\left({1\over 2}\tan \left({k\over 2n_e}\right)\right)\right]^2 
\nonumber \\
& - & {2\over\pi\,u}\left[\cos^2 \left({k\over 2n_e}\right) 
+ {k\over \pi n_e}{2\over\pi}\arctan\left({1\over 2}\tan \left({k\over 2n_e}\right)\right)\right]\sin (\pi n_e) \, ,
\nonumber \\
& & \downarrow {\rm electron}\hspace{0.1cm}{\rm addition}\hspace{0.1cm}{\rm for}\hspace{0.1cm}k\in [0,2k_F] \, ,
\label{xiupLAsUim0}
\end{eqnarray}
so that,
\begin{eqnarray}
\lim_{k\rightarrow 0}\xi_{s1}^{\uparrow} (k) & = & {1\over 2} - {2\over\pi\,u}\sin (\pi n_e) \, ,
\nonumber \\
\lim_{k\rightarrow k_F}\xi_{s1}^{\uparrow} (k) & = & {1\over 8} + 2\left({1\over\pi}\arctan\left({1\over 2}\right)\right)^2 
- {1\over\pi\,u}\left(1 - {2\over\pi}\arctan\left({1\over 2}\right)\right)\sin (\pi n_e)  \, ,
\nonumber \\
& \approx & 0.16856 - {0.22436\over u}\sin (\pi n_e) \, ,
\nonumber \\
\lim_{k\rightarrow 2k_F}\xi_{s1}^{\uparrow} (k) & = & {1\over 2} \, ,
\nonumber \\
\lim_{k\rightarrow 0}\xi_{s1}^{\downarrow} (k) & = & - {1\over 2} - {2\over\pi\,u}\sin (\pi n_e) \, ,
\nonumber \\
\lim_{k\rightarrow k_F}\xi_{s1}^{\downarrow} (k) & = &  - {3\over 8}
+ 2\left({1\over\pi}\arctan\left({1\over 2}\right)\right)^2 - {1\over\pi\,u}\left(1
+ {2\over\pi}\arctan\left({1\over 2}\right)\right)\sin (\pi n_e) \, ,
\nonumber \\
& \approx & - 0.33144 - {0.41226\over u}\sin (\pi n_e)
\nonumber \\
\lim_{k\rightarrow 2k_F}\xi_{s1}^{\downarrow} (k) & = & {1\over 2} - {2\over\pi\,u}\sin (\pi n_e) \, .
\label{xiupLAsUim0-aux}
\end{eqnarray}
Analysis of these expressions and values reveals that in the $u\gg 1$ limit and $m\rightarrow n_e$ the $\uparrow$ 
one-electron removal exponent $\xi_{s1}^{\uparrow} (k)$
smoothly decreases from $\xi_{s1}^{\uparrow} (k) = 1/2$ for $k\rightarrow 0$
until it reaches a minimum value at $k=k_F$. For $k>k_F$ it continuously increases
to $\xi_{s1}^{\uparrow} (k) = 1/2$ as $k\rightarrow 2k_F$.
In the same limits the $\downarrow$ 
one-electron LHB addition exponent $\xi_{s1}^{\downarrow} (k)$
smoothly varies from $\xi_{s1}^{\downarrow} (k) = - 1/2$ for $k\rightarrow 0$
to $\xi_{s1}^{\downarrow} (k) = 1/2$ for $k\rightarrow 2k_F$. 

Moreover, analysis of Fig. \ref{fs1up} shows that the exponent $\xi_{s1}^{\uparrow} (k)$ 
only becomes negative for a part of the $s1$ branch line $k$ interval that
starts at $k=k_{F\downarrow}$ and ends at a $k$ momentum that 
for smaller and larger spin density values refers to one-electron
LHB addition and removal, respectively. The $u$ values for which it
is negative are dependent of the densities.
For the densities ranges $n_e \in [0,1/2] $ and $m \in [0,1-n_e]$ 
and also for $n_e \in [1/2,1]$ and $m \in [0,1-n_e]$ the exponent $\xi_{s1}^{\uparrow} (k)$ decreases
upon increasing $u$ from $1$ for $u\rightarrow 0$ to its $u\gg 1$ values.
In addition, according to Fig. \ref{fs1up} its $u$ dependence is more involved
for the densities intervals $n_e \in [1/2,1]$ and $m \in [1-n_e,n_e]$ for which it is
given by $0$ and $1$ in the $u\rightarrow 0$ limit for different $k$ ranges, respectively. 
For the $k$ ranges for which it reads $1$ for $u\rightarrow 0$ it remains being an
increasing function of $u$ for the whole $u$ interval. For the $k$ 
intervals for which it is given by $0$ in the $u\rightarrow 0$ limit, upon increasing
$u$ it first decreases, goes through a minimum value, and then becomes an
increasing function of $u$ until reaching its $u\rightarrow\infty$ $k$ dependent values.

On the other hand, for $u>0$ the exponent $\xi_{s1}^{\downarrow} (k)$ whose $k$
dependence is plotted in Fig. \ref{fs1down} is in general negative except
for a small $k$ region that corresponds to the larger $k$ values of its range.
Both for the densities ranges $n_e \in [0,1/2] $ and $m \in [0,1-n_e]$ and for 
$n_e \in [1/2,1]$ and $m \in [0,1-n_e]$ the exponent $\xi_{s1}^{\downarrow} (k)$ increases
upon increasing $u$ from $-1$ for $u\rightarrow 0$ to its $u\gg 1$ $k$ dependent values.
As also shown in that figure, its $u$ dependence is more complex
for the densities intervals $n_e \in [1/2,1]$ and $m \in [1-n_e,n_e]$ for which it is
given by $-1$ and $0$ in the $u\rightarrow 0$ limit for different $k$ ranges, respectively. 
For the $k$ ranges for which it reads $-1$ for $u\rightarrow 0$ it remains being an
increasing function of $u$ for the whole $u$ interval. However, for the $k$ 
domains for which it is given by $0$ in the $u\rightarrow 0$ limit, upon increasing
$u$ it first decreases, goes through a minimum value, and then becomes an
increasing function of $u$ until reaching its $u\rightarrow\infty$ $k$ dependent values.

\subsection{The $\sigma$ one-electron UHB addition branch lines}
\label{upUHBs}

The $\sigma$ one-electron UHB addition branch lines are generated by 
processes that correspond to particular cases of those generated by the leading-order operators, Eqs. (\ref{upElUHBadd})
and (\ref{downElUHBadd}), that are behind the $\uparrow$ one-electron UHB addition spectrum, Eq. (\ref{SpupElUHBadd}),
and $\downarrow$ one-electron UHB addition spectrum, Eq. (\ref{SpdownElUHBadd}). Hence
they are contained within such two-parametric spectra that occupy well defined regions in the $(k,\omega)$ plane.

As discussed in Sec. \ref{validity}, following the direct relation of the $\sigma $ one-electron UHB addition branch lines
spectra and exponents to those of the $\bar{\sigma}$ one-electron removal branch lines, for simplicity here
we limit our study to the $\sigma$ one-electron UHB addition branch lines
that in the $u\rightarrow 0$ limit contribute to the $u=0$ $\sigma $ one-electron addition spectrum.
In the case of the $\uparrow$ and $\downarrow$ one-electron UHB addition spectral functions
those are the $s1$ branch line and one of the subbranches of the $c^{\pm}$ branch lines, respectively.

As for the $\downarrow$ one-electron removal $s1$ branch line, the spectrum that defines the 
$(k,\omega)$-plane spectrum of the $\uparrow$ one-electron UHB addition $s1$ branch line is such that
$\omega_{s1}^{\sigma} (k) = \omega_{s1}^{\sigma} (-k)$ for $k \leq 0$ and the corresponding
exponent given below is also such that $\xi_{s1}^{\sigma} (k) = \xi_{s1}^{\sigma} (-k)$
for $k \leq 0$. Hence for simplicity we restrict our following analysis to a reduced first
Brillouin-zone scheme for positive momentum values $k \in [0,\pi]$.

This $s1$ branch line refers to excited energy eigenstates with the following number deviations relative to those of the
initial ground state,
\begin{equation}
\delta N_c^F = -1  \, ; \hspace{0.5cm} \delta J_c^F = 1/2 \, ; \hspace{0.5cm} 
\delta N_{s1}^F = \delta J_{s1}^F = 0 \, ; \hspace{0.5cm} \delta N_{s1}^{NF} = -1 
\, ; \hspace{0.5cm} \delta N_{\eta 1} = 1 \, ; \hspace{0.5cm} \delta J_{\eta 1} = -1/2 \, .
\label{NupUHBs}
\end{equation}
Its $(k,\omega)$-plane one-parametric spectrum reads,
\begin{equation}
\omega_{s1}^{\uparrow} (k) = 2\mu - \varepsilon_{s1} (q) \, ;  \hspace{0.5cm} q \in [-k_{F\downarrow},k_{F\downarrow}]  \, .
\label{OkupUHBs}
\end{equation}
Here $\varepsilon_{s1} (q)$ is the $s1$ band energy dispersion, Eq. (\ref{epsilon-q}) for $\beta =s1$,
and $2\mu$ stands for the energy scale defined in Eq. (\ref{mu-muBH}). Within an extended zone scheme 
the general relation of the $k>0$ excitation momentum to the $s1$ band momentum $q$ in Eq. (\ref{OkupUHBs}) is,
\begin{equation}
k = \pi - q \in [(\pi - k_{F\downarrow}),(\pi+k_{F\downarrow})]  \, .
\label{kupUHBs}
\end{equation}

Bringing this spectrum to the first Brillouin zone leads to two subbranches that refer to 
excitation momentum ranges $k\in [(\pi - k_{F\downarrow}),\pi]$ and 
$k =\in [-\pi,-(\pi - k_{F\downarrow})]$, respectively. On the other hand, a contribution from
$k<0$ extended zone scheme second Brillouin zone interval also leads to 
the $k\in [(\pi - k_{F\downarrow}),\pi]$ range. We checked that the two corresponding
spectral-function contributions to the momentum range $k\in [(\pi - k_{F\downarrow}),\pi]$ lead to the 
same power-law type of spectral-weight distributions in the vicinity of
the $s1$ branch line. The corresponding reduced first-Brillouin-zone scheme used 
here for $k\in [0,\pi]$ excitation momentum relates to the $s1$ band momentum as,
\begin{equation}
k = \pi - q = [(\pi - k_{F\downarrow}),\pi] \, ,
\label{OkupUHBs1BZ}
\end{equation}
for $q \in [0,k_{F\downarrow}]$. (Online the $\uparrow$ one-electron UHB addition $s1$ branch line 
is green in Figs. \ref{s4525u01}-\ref{s032505u1}; This branch line lays above the UHB pseudogap in 
Figs. \ref{s6545u1}-\ref{s032505u1}, which refer to intermediate and large $u$ values.)

\begin{figure}
\includegraphics[scale=1.00]{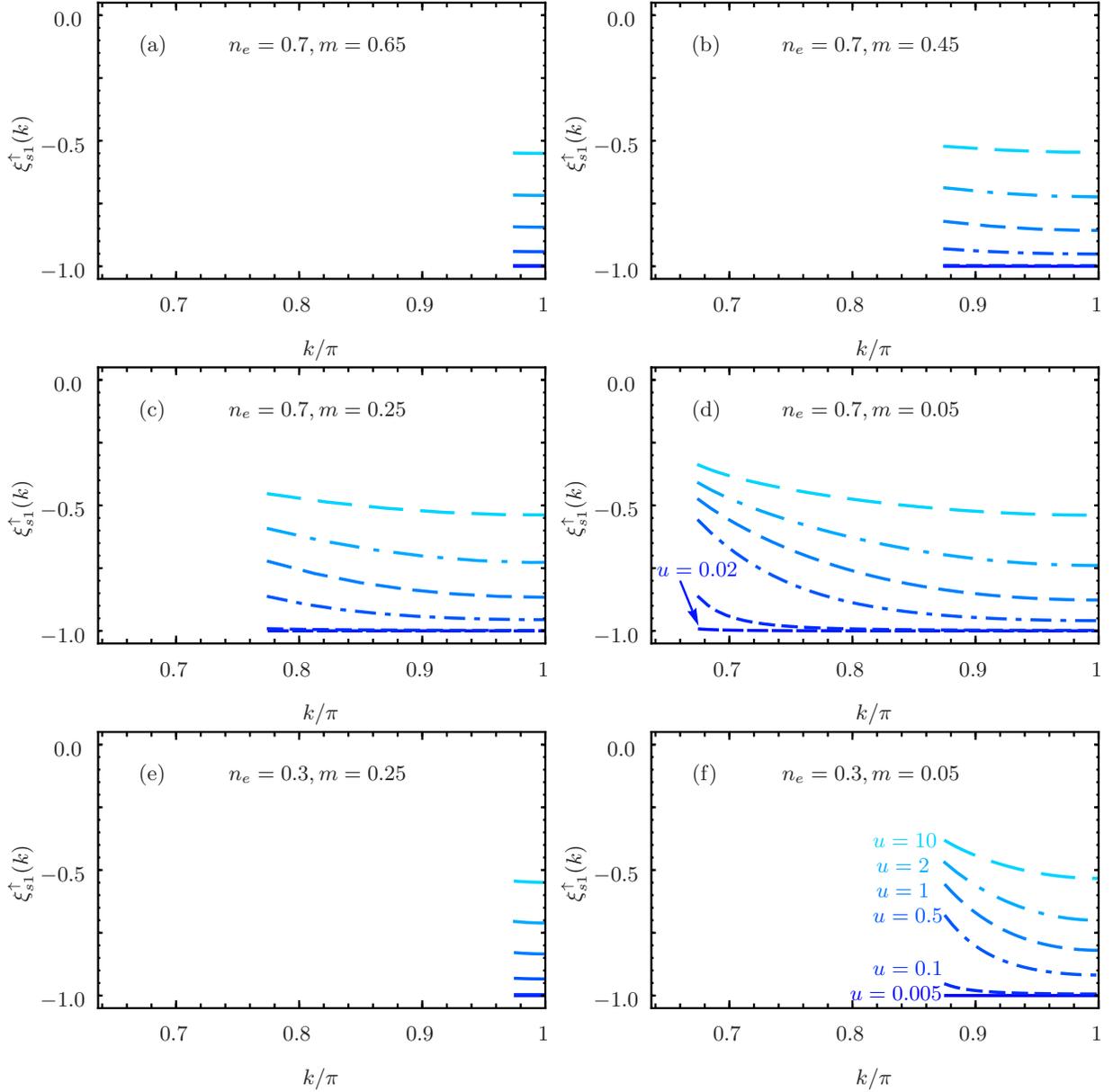}
\caption{\label{fs1upUHB} The exponent $\xi_{s1}^{\uparrow} (k)$, Eq. (\ref{xiupUHBs}),
that controls the singularities in the vicinity of the $s1$ branch line whose $(k,\omega)$-plane
one-parametric spectrum is defined by Eq. (\ref{OkupUHBs}) for the $\sigma =\uparrow$
one-electron UHB addition spectral function, Eq. (\ref{s1-branch-UHB}), as a function of the momentum
$k/\pi\in ]k_0/\pi,1[$ where $]k_0/\pi,1[$ with $0<k_0<\pi$ is a $k$ interval that contains
the branch line for the same values of $u$, electronic density $n_e$, and spin density $m$ as in 
Fig. \ref{fc+up}. (For $k/\pi\in ]-1,-k_0/\pi[$ the exponent $\xi_{s1}^{\uparrow} (k)$
is given by $\xi_{s1}^{\uparrow} (k)=\xi_{s1}^{\uparrow} (-k)$ with $-k/\pi\in ]k_0/\pi,1[$ as plotted here.)}
\end{figure}

The momentum dependent exponent of general form, Eq. (\ref{branch-l}), that controls the line shape near the branch line is given by,
\begin{equation}
\xi_{s1}^ {\uparrow} (k) = -1 + \sum_{\iota=\pm1}\left(-{\iota\,\xi_{c\,c}^0\over 2} - \Phi_{c,s1}(\iota 2k_{F},q)\right)^2 
+ \sum_{\iota=\pm1}\left(-{\iota\,\xi_{s1\,c}^0\over 2} - \Phi_{s1,s1}(\iota k_{F\downarrow},q)\right)^2  \, .
\label{xiupUHBs}
\end{equation}
This exponent is plotted in Fig. \ref{fs1upUHB} as a function of the momentum $k/\pi\in ]0,1[$ for 
several $u$ values, electronic densities $n_e =0.3$ and $n_e =0.7$, and a set of spin density 
values $m<n_e$.

Near the present $s1$ branch line the $\sigma =\uparrow$ one-electron addition spectral function 
$B_{\uparrow,+1} (k,\omega)$, Eq. (\ref{Bkomega}), corresponds to the UHB and has
the following power-law behavior, 
\begin{equation}
B^{\rm UHB}_{\uparrow,+1} (k,\omega) = C_{\uparrow,s1}^{UHB} \Bigl(\omega - \omega_{s1}^{\uparrow} (k)\Bigr)^{\xi_{s1}^{\uparrow} (k)}  
\, ; \hspace{0.50cm} (\omega - \omega_{s1}^{\uparrow} (k)) \geq 0 \, ,
\label{s1-branch-UHB}
\end{equation}
where $C_{\uparrow,s1}^{UHB}$ is a constant independent of $k$ and $\omega$,
the spectrum $\omega_{s1}^{\uparrow} (k)$ is that in Eq. (\ref{OkupUHBs}), and the exponent $\xi_{s1}^{\uparrow} (k)$
is given in Eq. (\ref{xiupUHBs}).

The direct relation of the exponent, Eq. (\ref{xiupUHBs}), to that of the $\downarrow$ one-electron removal
$s1$ branch line enables deriving its behaviors for both $u\rightarrow 0$ and $u\gg 1$ from those of that
other exponent. In the $u\rightarrow 0$ limit one finds the following value,
\begin{equation}
\lim_{u\rightarrow 0}\xi_{s1}^{\uparrow} (k) = -1 \hspace{0.3cm}{\rm (for}\hspace{0.1cm}{\rm the}\hspace{0.1cm}
{\rm whole}\hspace{0.1cm}{\rm above}\hspace{0.1cm}{\rm branch}\hspace{0.1cm}{\rm line}\hspace{0.1cm}k\hspace{0.1cm}{\rm range)} \, .
\label{xiupUHBsU0}
\end{equation}
Hence, consistently with Eq. (\ref{branch-lexp-1}), for $u\rightarrow 0$ this branch line acquires the following $\delta$-function-like 
one-electron spectral weight distribution along it,
\begin{equation}
\lim_{u\rightarrow 0} B^{\rm UHB}_{\uparrow,+1} (k,\omega) = \delta\Bigl(\omega - \omega_{s1}^{\uparrow} (k)\Bigr) 
= \delta\Bigl(\omega + 2t(\cos k - \cos k_{F\uparrow})\Bigr) \, , \hspace{0.50cm} \vert k\vert \in  [(\pi - k_{F\downarrow}),\pi] \, .
\label{BUHBs1up}
\end{equation}
The $u\rightarrow 0$ limiting behavior reported in Eq. (\ref{varepsilonsu0}) of Appendix \ref{LimitBV}
for the $s1$ energy dispersion $\varepsilon_{s1} (q)$ appearing in the spectrum $\omega_{s1}^{\uparrow} (k)$, 
Eq. (\ref{OkupUHBs}), confirms that the latter spectrum becomes in the $u\rightarrow 0$ limit the corresponding $u=0$ 
non-interacting electronic spectrum, as given in Eq. (\ref{BUHBs1up}).

The expression found for $u\gg 1$ and $m\rightarrow 0$ for the exponent, Eq. (\ref{xiupUHBs}),
is given by, 
\begin{equation}
\xi_{s1}^{\uparrow} (k) = - {1\over 2}\left(1- \left({\pi - k\over \pi n_e}\right)^2\right)
\left(1 + {2\ln 2\over \pi\,u}\sin (\pi n_e)\right)
- {1\over 2u}\cos\left({\pi -k\over n_e}\right)\sin (\pi n_e)  \, ,
\label{xiupUHBsUim0}
\end{equation}
so that,
\begin{eqnarray}
\lim_{k\rightarrow \pi-k_F}\xi_{s1}^{\sigma} (k) & = & - {3\over 8}
- {3\ln 2\over 4\pi\,u}\sin (\pi n_e) \, ,
\nonumber \\
\lim_{k\rightarrow \pi}\xi_{s1}^{\downarrow} (k) & = & - {1\over 2}
- {1\over 2u}\left(1 + {2\ln 2\over \pi}\right)\sin (\pi n_e) \, . 
\label{xiupUHBsUim0-aux}
\end{eqnarray}
In the $m\rightarrow n_e$ limit the present $s1$ branch line momentum width 
vanishes so that it does not exist. 

Analysis of Fig. \ref{fs1upUHB} reveals that for $m<n_e$ the
$s1$ branch-line exponent, Eq. (\ref{xiupUHBs}), is a decreasing function of the momentum $k$.
Moreover, it increases upon increasing $u$ and remains negative for all momentum $k$ and
$m<n_e$ densities ranges.

Next, concerning the $\downarrow$ one-electron UHB addition spectral function,
the spectra $\omega_{c^{\pm}}^{\sigma} (k)$ that define the $(k,\omega)$-plane shape of the $c^{+}$ branch line
and its twin $c^{-}$ branch line and the corresponding exponents $\xi_{c^{\pm}}^{\sigma} (k)$ are related as given in Eq. (\ref{c+-rela})
for $\uparrow$ electron removal. Considering the $c^{+}$ branch line in a reduced first Brillouin-zone 
scheme for which $k\in [-\pi,\pi]$ contains the same information as considering both the $c^{+}$ and
$c^{-}$ branch lines for the positive excitation momentum range $k\in [0,\pi]$. Below we only consider
the $k$ range associated with the subbranches for which the exponent $\xi_{c^{+}}^{\downarrow} (k) = \xi_{c^{-}}^{\downarrow} (-k)$
contributes to the $\downarrow$ one-electron spectral weight as $u\rightarrow 0$. It turns out that
for the exponent $\xi_{c^{+}}^{\downarrow} (k)$ such a subbranch is contained in the positive 
excitation momentum range $k\in [0,\pi]$.

The one $\sigma$ one-electron UHB addition $c^{+}$ branch line is associated with 
excited energy eigenstates with the following number deviations relative to those of the initial ground state,
\begin{equation}
\delta N_c^F = 0 \, ; \hspace{0.5cm} \delta J_c^F = \mp 1/2 \, ; \hspace{0.5cm} \delta N_c^{NF} = -1 \, ; \hspace{0.5cm} 
\delta N_{s1}^F = 0 \, ; \hspace{0.5cm} \delta J_{s1}^F = 1/2 \, ; \hspace{0.5cm} \delta N_{\eta 1} = 1 
\, ; \hspace{0.5cm} \delta J_{\eta 1} = \pm 1/2 \, .
\label{NdownHAcc1BZ}
\end{equation}

The one-parametric spectrum of general form, Eq. (\ref{dE-dP-bl}), that defines the $(k,\omega)$-plane shape of this line reads,
\begin{equation}
\omega_{c^{+}}^{\downarrow} (k) = 2\mu - \varepsilon_c (q) \, ;  \hspace{0.5cm} q \in [-2k_F,2k_F] \, ,
\label{OkdownHAcc}
\end{equation}
where $\varepsilon_c (q)$ is the $c$ band energy dispersion, Eq. (\ref{epsilon-q}) for $\beta =c$, and
the corresponding $c$ band momentum $q$ is within an extended zone scheme related to the excitation momentum $k$ as,
\begin{equation}
k = \pi + k_{F\downarrow} - q \in [(\pi -k_{F\uparrow}),(\pi + 2k_F+k_{F\downarrow})] \, .
\label{kdownHAcc}
\end{equation}

Bringing this spectrum to the $k\in [-\pi,\pi]$ reduced first Brillouin-zone 
leads to two $(k,\omega)$-plane $c^{+}$ branch line subbranches whose $k$ intervals are given by
$k = - \pi + k_{F\downarrow} - q \in [-\pi,-(\pi - 2k_F -k_{F\downarrow})]$ and
$k= \pi + k_{F\downarrow} - q \in [(\pi -k_{F\uparrow}),\pi]$, respectively. 
As mentioned above, in the following we only consider the second of such momentum ranges, 
\begin{equation}
k = \pi + k_{F\downarrow} - q \in [(\pi -k_{F\uparrow}),\pi] \, .
\label{O2kdownHAcc}
\end{equation}
Indeed, it is that for which the exponent 
$\xi_{c^{+}}^{\downarrow} (k) = \xi_{c^{-}}^{\downarrow} (-k)$ reads $-1$ in the
$u\rightarrow 0$ limit and thus the branch line contributes to the 
$\delta$-function-like $\downarrow$ one-electron spectrum in that limit.
(Online the $\downarrow$ one-electron UHB addition $c^+$ branch line 
is is blue in Figs. \ref{s4525u01}-\ref{s032505u1}; This branch line lays above the UHB pseudogap 
in Figs. \ref{s6545u1}-\ref{s032505u1}, which refer to intermediate and large $u$ values.)

\begin{figure}
\includegraphics[scale=1.00]{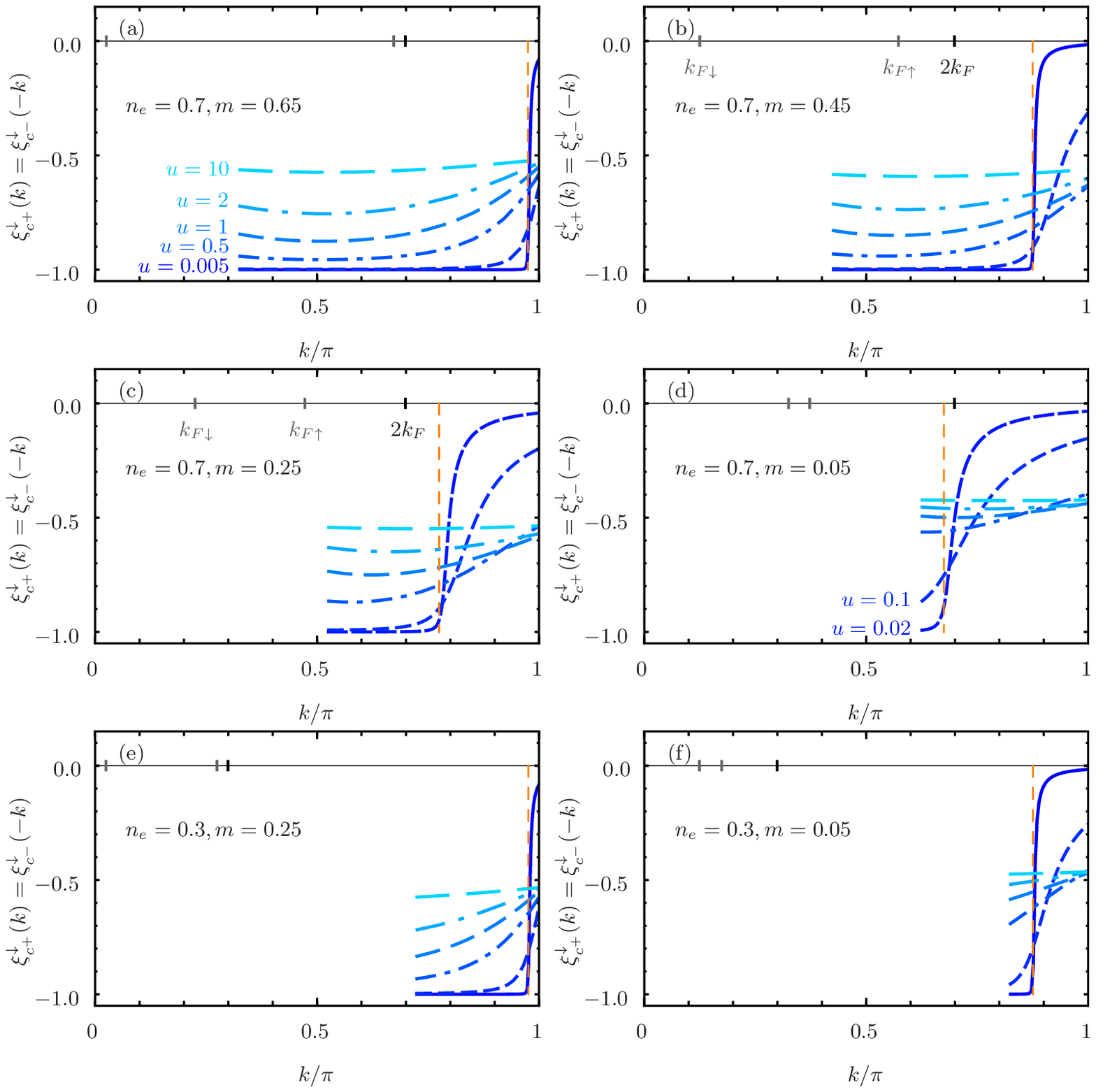}
\caption{\label{fc+downUHB} The exponent $\xi_{c^{+}}^{\downarrow} (k)=\xi_{c^{-}}^{\downarrow} (-k)$, Eq. (\ref{xidownHAcc}),
that controls the singularities in the vicinity of the $c^+$ branch line whose $(k,\omega)$-plane
one-parametric spectrum is defined by Eq. (\ref{OkdownHAcc}) for the $\sigma =\downarrow$
one-electron UHB addition spectral function, Eq. (\ref{cpm-branch-UHB}), as a function of the momentum
$k/\pi\in ]0,1[$ for the same values of $u$, electronic density $n_e$, and spin density $m$ as in 
Fig. \ref{fc+up}.}
\end{figure}

The momentum dependent exponent of general form, Eq. (\ref{branch-l}), that controls the line shape near the branch line 
is in the present case given by,
\begin{equation}
\xi_{c^{+}}^{\downarrow} (k) = \xi_{c^{-}}^{\downarrow} (-k) =
-1 + \sum_{\iota=\pm1}\left({\xi_{c\,s1}^1\over 2} - \Phi_{c,c}(\iota 2k_{F},q)\right)^2 
+ \sum_{\iota=\pm1}\left({\xi_{s1\,s1}^1\over 2} - \Phi_{s1,c}(\iota k_{F\downarrow},q)\right)^2 \, .
\label{xidownHAcc}
\end{equation}
It is plotted in Fig. \ref{fc+downUHB} as a function of the momentum $k/\pi\in ]0,1[$ for 
several $u$ values, electronic densities $n_e =0.3$ and $n_e =0.7$, and a set of spin density 
values $m<n_e$.

In the vicinity of the present $c^{\pm}$ branch lines the $\sigma =\downarrow$ one-electron addition spectral 
function $B_{\downarrow,+1} (k,\omega)$, Eq. (\ref{Bkomega}), refers to the UHB and
has the following power-law behavior, 
\begin{equation}
B^{\rm UHB}_{\downarrow,+1} (k,\omega) = C_{\downarrow,c^{\pm}}^{UHB} 
\Bigl(\omega - \omega_{c^{\pm}}^{\downarrow} (k)\Bigr)^{\xi_{c^{\pm}}^{\downarrow} (k)}  
\, ; \hspace{0.50cm} (\omega - \omega_{c^{\pm}}^{\downarrow} (k)) \geq 0 \, ,
\label{cpm-branch-UHB}
\end{equation}
where $C_{\downarrow,c^{\pm}}^{UHB}$ is a constant independent of $k$ and $\omega$,
the spectrum $\omega_{c^{+}}^{\downarrow} (k)$ is that in Eqs. (\ref{OkdownHAcc}) and
(\ref{O2kdownHAcc}), and the exponent $\xi_{c^{+}}^{\downarrow} (k)$
is given in Eq. (\ref{xidownHAcc}). Furthermore, $\omega_{c^{-}}^{\downarrow} (k)=\omega_{c^{+}}^{\downarrow} (-k)$
and $\xi_{c^{-}}^{\downarrow} (k)=\xi_{c^{+}}^{\downarrow} (-k)$.

The direct relation of the exponent, Eq. (\ref{xidownHAcc}), to that of the corresponding $\uparrow$ one-electron removal
$c^{\pm}$ branch lines subbranches enables deriving its behaviors for both $u\rightarrow 0$ and $u\gg 1$ from those of these
other exponents. In the $u\rightarrow 0$ limit one finds the following values in the $k$ range, Eq. (\ref{O2kdownHAcc}),
\begin{eqnarray}
\lim_{u\rightarrow 0}\xi_{c^+}^{\downarrow} (k) & = & 0 \, , \hspace{0.5cm} k \in [(\pi - k_{F\downarrow}),\pi] \, ,
\nonumber \\
& = & -1 \, , \hspace{0.5cm} k \in [(\pi -k_{F\uparrow}),(\pi - k_{F\downarrow})] 
\nonumber \\
\lim_{u\rightarrow 0}\xi_{c^-}^{\downarrow} (k) & = & -1 \, , \hspace{0.5cm} k \in [-(\pi - k_{F\downarrow}),-(\pi -k_{F\uparrow})] 
\nonumber \\
& = & 0 \, , \hspace{0.5cm} k \in [-\pi,-(\pi - k_{F\downarrow})] \, .
\label{xidownHA+U0}
\end{eqnarray}
For the $k$ ranges for which such exponents read $-1$ the line shape becomes $\delta$-function-like for
$u\rightarrow 0$, as given in Eq. (\ref{branch-lexp-1}). In the present cases we find,
\begin{eqnarray}
\lim_{u\rightarrow 0} B^{\rm UHB}_{\downarrow,+1} (k,\omega) & = & \delta\Bigl(\omega - \omega_{c^{-}}^{\downarrow} (k)\Bigr) 
= \delta\Bigl(\omega + 2t(\cos k - \cos k_{F\downarrow})\Bigr) \, , \hspace{0.50cm} k \in [-(\pi - k_{F\downarrow}),-(\pi -k_{F\uparrow})] \, ,
\nonumber \\
& = & \delta\Bigl(\omega - \omega_{c^{+}}^{\downarrow} (k)\Bigr) 
= \delta\Bigl(\omega + 2t(\cos k - \cos k_{F\downarrow})\Bigr) \, , \hspace{0.50cm} k \in [(\pi -k_{F\uparrow}),(\pi - k_{F\downarrow})] \, .
\label{BUHBcpmdown}
\end{eqnarray}
That the spectrum $\omega_{c^{+}}^{\downarrow} (k) = \omega_{c^{-}}^{\downarrow} (-k)$, Eq. (\ref{OkdownHAcc}),
becomes in the $u\rightarrow 0$ limit the corresponding $u=0$ non-interacting electronic spectrum
is confirmed by the $u\rightarrow 0$ limiting behavior reported in Eq. (\ref{varepsiloncu0}) of Appendix \ref{LimitBV}
for the $c$ band energy dispersion $\varepsilon_{c} (q)$ appearing in the $u>0$ spectrum 
general expression, Eq. (\ref{OkdownHAcc}). On the other hand, for the $k$ ranges for which the exponent is 
given by $0$ for $u\rightarrow 0$ the one-electron spectral weight at and near the corresponding branch lines 
vanishes in the $u\rightarrow 0$ limit.

For $u\gg 1$ and $m\rightarrow 0$ one finds the following expressions,
\begin{equation}
\xi_{c^{\pm}}^{\downarrow} (k) = -{3\over 8} + {\ln 2\over 4\pi\,u}
\left(\sin (\pi n_e) \mp 2\sin \left(k \mp {\pi\over 2}n_e\right)\right) \, .
\label{xidownHAUim0}
\end{equation}
In the $m\rightarrow n_e$ limit the exponents expressions are found to read,
\begin{equation}
\xi_{c^{\pm}}^{\downarrow} (k) = - {1\over 2} \mp {2\over \pi\,u}\sin k \, . 
\label{xidownHAUim1}
\end{equation}

As it follows from analysis of Fig. \ref{fc+downUHB}, the main effect on the $k$ dependence of the 
$\downarrow$ one-electron UHB addition exponent 
$\xi_{c^{+}}^{\downarrow} (k) = \xi_{c^{-}}^{\downarrow} (-k)$, Eq. (\ref{xidownHAcc}), of increasing the on-site repulsion
$u$ from $u\ll 1$ to $u\gg 1$ is to continuously changing its $u\rightarrow 0$ values $-1$ and
$0$ for the $k$ ranges given in Eq. (\ref{xidownHA+U0})
to a $k$ independent value for $k\in [0,\pi]$ as $u\rightarrow\infty$, which smoothly changes 
from $-3/8$ for $m\rightarrow 0$ to $-1/2$ for for $m\rightarrow n_e$.

\subsection{The $\uparrow$ one-electron removal and $\downarrow$ one-electron UHB addition 
$s1'$ non-branch lines for $0<m<n_e$}
\label{upRsl}

The importance of the branch lines is confirmed by in the $u\rightarrow 0$ limit they recovering
most of the $u=0$ $\delta$-function-like $\sigma$ one-electron spectrum $k$ ranges, as confirmed
by combining Eqs. (\ref{cpm-branch-delta}), (\ref{s1-branch-delta}), (\ref{BUHBs1up}), (\ref{BUHBcpmdown}).
Interestingly, part of that spectral weight stems from the $u>0$ UHB.

The $k$ subrange of the $u=0$ $\sigma$ one-electron spectrum that does not stem from
branch lines refers for $0<m<n_e$ to the momentum interval $k\in [-k_{F\downarrow},k_{F\downarrow}]$
for $\uparrow$ one-electron removal and $\vert k\vert\in [\pi-k_{F\downarrow},\pi]$
for $\downarrow$ one-electron addition. That spectral weight stems from well-defined $u>0$
spectral features whose line-shape expressions involve state summations difficult to compute.

Specifically, the $u=0$ $\uparrow$ one-electron removal spectral weight missing 
for $k\in [-k_{F\downarrow},k_{F\downarrow}]$ and $0<m<n_e$ stems from a $u>0$ $s1'$ non-branch line that 
is generated by transitions to excited energy eigenstates with the following number deviations relative to those of the
initial ground state,
\begin{equation}
\delta N_c^F = \delta J_c^F = 0 \, ; \hspace{0.5cm} \delta N_c^{NF} = - 1  \, ; \hspace{0.5cm} 
\delta N_{s1}^F = 1 \, ; \hspace{0.5cm} \delta J_{s1}^F = \pm 1 \, ; \hspace{0.5cm} \delta N_{s1}^{NF} = - 1 \, .
\label{NupRsl}
\end{equation}

The one-parametric spectrum of this line is given by,
\begin{eqnarray}
\omega_{s1'}^{\uparrow} (k) & = & - \varepsilon_{s1} (-k) - \varepsilon_c (\pm 2k_{F\downarrow}) = - \varepsilon_{s1} (q) 
- \varepsilon_c (\pm 2k_{F\downarrow}) \, ,  \hspace{0.5cm} q \in [-k_{F\downarrow},k_{F\downarrow}] \, ,
\nonumber \\
k & = & - q \in [-k_{F\downarrow},k_{F\downarrow}] \, ,
\label{OkupRsl}
\end{eqnarray}
where $\varepsilon_{s1} (q)$ is the $s1$ band energy dispersion, Eq. (\ref{epsilon-q}) for $\beta =s1$.

While the line shape analytical expression near this $s1'$ non-branch line remains an unsolved problem for $u>0$, in the $u\rightarrow 0$ limit
it becomes $\delta$-function-like,
\begin{equation}
\lim_{u\rightarrow 0} B_{\uparrow,-1} (k,\omega) = \delta\Bigl(\omega + \omega_{s1'}^{\uparrow} (k)\Bigr) 
= \delta\Bigl(\omega - 2t(\cos k - \cos k_{F\uparrow})\Bigr) \, , \hspace{0.50cm} k \in [-k_{F\downarrow},k_{F\downarrow}] \, .
\label{Bremovslineup}
\end{equation}

On the other hand, the $u=0$ $\downarrow$ one-electron addition
spectral weight missing for $\vert k\vert\in [\pi-k_{F\downarrow},\pi]$ and $0<m<n_e$ stems from a $u>0$ UHB
$s1'$ non-branch line that is generated by transitions to excited energy eigenstates with the following number 
deviations relative to those of the initial ground state,
\begin{equation}
\delta N_c^F = -1 \, ; \hspace{0.5cm} \delta J_c^F = 0 \, ; \hspace{0.5cm} 
\delta N_{s1}^F = 0 \, ; \hspace{0.5cm} \delta J_{s1}^F = 1/2 
\, ; \hspace{0.5cm} \delta N_{\eta 1} = 1  \, ; \hspace{0.5cm} \delta J_{\eta 1} = -1/2 \, .
\label{NdownUAsl}
\end{equation}
There is another such a $s1'$ non-branch line for $k<0$.

The one-parametric spectrum that defines the $(k,\omega)$-plane form of this line reads,
\begin{eqnarray}
\omega_{s1'}^{\downarrow} (k) & = & 2\mu - \varepsilon_{s1} (\pi -k) + \varepsilon_{s1} (k_{F\uparrow}) = 2\mu - \varepsilon_{s1} (q) 
+ \varepsilon_{s1} (k_{F\uparrow}) \, , \hspace{0.5cm} q \in [0,k_{F\downarrow}] \, .
\nonumber \\
k & = & \pi - q \in [\pi -k_{F\downarrow},\pi] \, .
\label{OkdownUAsl}
\end{eqnarray}

The line shape analytical expression near this $s1'$ non-branch line remains again an open problem for $u>0$
except in the $u\rightarrow 0$ limit in which it is $\delta$-function-like,
\begin{equation}
\lim_{u\rightarrow 0} B^{\rm UHB}_{\downarrow,+1} (k,\omega) = \delta\Bigl(\omega - \omega_{s1'}^{\downarrow} (k)\Bigr) 
= \delta\Bigl(\omega + 2t(\cos k - \cos k_{F\downarrow})\Bigr)
\, , \hspace{0.5cm} \vert k\vert \in [\pi -k_{F\downarrow},\pi] \, .
\label{BUHBs1pdown}
\end{equation}

The $\uparrow$ one-electron removal and $\downarrow$ one-electron UHB addition 
$s1'$ non-branch lines are represented in Figs. \ref{s4525u01}-\ref{s032505u1} by 
sets of diamond symbols. 

\section{Concluding remarks}
\label{concluding}

In this paper we have studied the momentum and energy dependence of the 
$\sigma $ one-electron spectral functions, Eq. (\ref{Bkomega}),
of the 1D Hubbard model at finite magnetic field in the vicinity of two types
of singular features: The branch lines and border lines whose $(k,\omega)$-plane spectra
general form is given in Eqs. (\ref{dE-dP-bl}) and 
(\ref{dE-dP-c-s1}), respectively. The branch lines are represented in Figs. \ref{s4525u01}-\ref{s032505u1}
by solid lines and dashed lines for the $k$ ranges for which the 
corresponding exponent $\xi_{\beta}^{\sigma} (k)$, Eq. (\ref{branch-l}), is negative and positive,
respectively. The one-electron removal and LWS addition boundary lines are 
in these figures represented by dashed-dotted lines. 

To access the line shapes near these singular features we have used the PDT introduced in Refs. \cite{V-1,LE} whose applications to the 
study of the 1D Hubbard model one-electron spectral functions have been limited to zero magnetic
field \cite{TTF,spectral0,spectral,spectral-06}. The momentum dependence of the exponents that in the TL control the 
line shapes in the vicinity of the $\sigma $ one-electron spectral functions branch lines was derived. 
For the $k$ ranges for which such exponents $\xi_{\beta}^{\sigma} (k)$ (which are plotted in 
Figs. (\ref{fc+up})-(\ref{fc+downUHB})) are negative, there are singularity cusps in the corresponding 
$\sigma$ one-electron spectral functions, Eq. (\ref{Bkomega}). The same occurs in the $(k,\omega)$-plane vicinity of the border lines.

The important role played by the branch lines singularity cusps is confirmed by in the $u\rightarrow 0$ limit they recovering
the $u=0$ $\delta$-function-like $\sigma$ one-electron spectrum for most of its momentum $k$ range, 
as confirmed by combining Eqs. (\ref{cpm-branch-delta}), (\ref{s1-branch-delta}), (\ref{BUHBs1up}), (\ref{BUHBcpmdown}).
The low-energy behavior of the correlation functions of the 1D Hubbard model at finite magnetic field has been
the subject of several previous studies \cite{Woy-89,Frahm,Frahm-91,Ogata-91}. To our knowledge, no previous 
investigations accessed for finite magnetic fields the repulsion $u$, electronic density $n_e$, spin density
$m$, and momentum dependence of the exponents that in the TL control at high-energy the
$\sigma$ one-electron spectral functions in the vicinity of such branch lines singularity cusps.

The momentum subrange for which the $u=0$ $\delta$-function-like $\sigma$ one-electron spectrum
does not stem from branch lines is $k\in [0,k_{F\downarrow}]$ for $\uparrow $ one-electron removal 
and $k\in [\pi-k_{F\downarrow},\pi]$ for $\downarrow $ one-electron addition. The PDT also accounts for
the non-branch-line processes that give rise in the $u\rightarrow 0$ limit to the $u=0$ 
one-electron spectrum at such a $k$ interval yet the line shape of the corresponding
spectral features remains for $u>0$ an involved unsolved technical problem. 
(These $u>0$ non-branch lines are represented in Figs. \ref{s4525u01}-\ref{s032505u1}
by sets of diamond symbols.) 

Complementarily, we have clarified beyond the results of Refs. \cite{V-1,LE} how the
$\sigma$ one-electron creation and annihilation operators matrix elements between the ground state and
excited energy eigenstates are accounted for by the PDT. Specifically, we have shown that the corresponding 
microscopic processes involve the rotated electrons as a needed link of the non-perturbative relation between
the electrons and the pseudofermions. Moreover, in this paper the $\sigma$ one-electron addition 
LHB and UHB were defined in terms of the occupancy configurations of such rotated electrons
for the whole $u>0$ range and all electronic densities and spin densities. 

Concerning the relation of our theoretical results to actual condensed-matter systems, angle-resolved photoemission spectroscopy 
at finite magnetic field is not possible, since the field would severely deflect the photoelectrons. However, it is possible to measure the local 
spectral function on quasi-1D metals by (scanning) tunneling spectroscopy at finite magnetic field. Such experiments
would provide some partial information on the spectral features theoretically studied in this paper by means
of the 1D Hubbard model at finite magnetic field.

On the other hand, such a model has been implemented with ultra-cold atoms on optical lattices 
\cite{Campo-07,Greif-15} and the related antiferromagnetic Heisenberg spin chain has been prepared to characterize its 
spin configurations \cite{Campo-15}. An interesting program would be the observation of the 
one-atom spectral weight distributions over the $(k,\omega)$ plane associated with the 
spectral functions studied in this paper in systems of spin $1/2$ ultra-cold atoms on optical lattices.  

%%%%%%%%%%%%%%%%%%%%%%%%%%%%%%%%%%%%%%%%%%%%%%%%%%%%%%%%%%%%%%%%%%%%%%%%%
\acknowledgements
We thank Ralph Claessen, Henrik Johannesson, Alexander Moreno, and Pedro D. Sacramento for illuminating discussions 
and the support by the Beijing CSRC and the FEDER through the COMPETE Program and the Portuguese FCT in the framework 
of the Strategic Projects PEST-C/FIS/UI0607/2013 and UID/CTM/04540/2013. J. M. P. C. acknowledges the hospitality of the 
Department of Physics at the University of Gothenburg, where the final part of this work was conducted.
%%%%%%%%%%%%%%%%%%%%%%%%%%%%%%%%%%%%%%%%%%%%%%%%%%%%%%%%%%%%%%%%%%%%%%%%%

%%%%%%%%%%%%%%%%%%%%%%%%%%%%%%%%%%%%%%%%%%%%%%%%%%%%%%%%%%%%%%%%%%%%%%%%%%
\appendix

\section{The Bethe-ansatz equations within the $\beta$ pseudoparticle representation and
related quantities needed for the studies of this paper}
\label{Ele2PsPhaShi}

Here we provide the pseudoparticle momentum distribution functional notation used in this paper
for the 1D Hubbard model BA equations introduced in Ref. \cite{Takahashi} for the TL, 
express the energy eigenvalues in terms of the rapidities that are the solutions of
such equations, and provide useful information on the specific solutions of these equations 
for the excited energy eigenstates belonging to a PS as defined in Section \ref{quantum-liquid}.

Moreover, the integral equations that define the rapidity dressed phase shifts 
$2\pi\,\bar{\Phi }_{\beta,\beta'} (r,r')$ in the expression, Eq. (\ref{Phi-barPhi}), of the
related $\beta$ pseudofermion phase shifts $2\pi\,\Phi_{\beta,\beta'} (q_j,q_{j'})$
are introduced, the $f$ functions in the second-order terms of the 
energy functional, Eq. (\ref{DE-fermions}), are expressed in
terms of such $\beta$ pseudofermion phase shifts, and the $\beta =c,s1$ lowest peak weights $A^{(0,0)}_{\beta}$ and
relative weights $a_{\beta}=a_{\beta}(m_{\beta,\,+1},\,m_{\beta,\,-1})$ in the $\beta$ pseudofermion 
spectral functions, Eq. (\ref{BQ-gen}), are written in terms of the related $\beta$ pseudofermion phase-shift 
functional $\Phi^T_{\beta}({q}_j)$, Eq. (\ref{pfacrGS}), which is a well-defined superposition of  
$\beta$ pseudofermion phase shifts $2\pi\,\Phi_{\beta,\beta'} (q_j,q_{j'})$. Two different forms
that the $\beta =c,s1$ pseudofermion spectral function $B_{Q_{\beta}} (k',\omega')$
whose general expression, Eq. (\ref{BQ-gen}), involves these lowest peak weights and relative weights
acquires in the TL as a result of the specific values of four functionals controlled by $\Phi^T_{\beta}({q}_j)$ 
are also provided.

Within the pseudoparticle momentum distribution functional notation used in this 
paper the BA equations considered in Ref. \cite{Takahashi} read,
\begin{eqnarray}
q_j & = & k^c (q_j) + {2\over L}\sum_{n =1}^{\infty}
\sum_{j'=1}^{L_{s n}}\,N_{sn}(q_{j'})\arctan\left({\sin
k^c (q_j)-\Lambda^{sn}(q_{j'}) \over n u}\right)
\nonumber \\
& + & {2\over L}\sum_{n =1}^{\infty}
\sum_{j'=1}^{L_{\eta n}}\, N_{\eta n}(q_{j'}) \arctan\left({\sin
k^c (q_j)-\Lambda^{\eta n}(q_{j'}) \over n u}\right) 
\, , \hspace{0.50cm} j = 1,...,L \, , 
\label{Tapco1}
\end{eqnarray}
and
\begin{eqnarray}
q_j & = & \delta_{\alpha,\eta}
\sum_{\iota =\pm1}\arcsin (\Lambda^{\alpha n} (q_{j}) - i\,\iota\,u)
+ {2\,(-1)^{\delta_{\alpha,\eta}}\over L} \sum_{j'=1}^{L}\,
N_{c}(q_{j'})\arctan\left({\Lambda^{\alpha n}(q_j)-\sin k^c (q_{j'})\over n u}\right)
\nonumber \\
& - & {1\over L}\sum_{n' =1}^{\infty}\sum_{j'=1}^{L_{\alpha n'}}\, N_{\alpha n'}(q_{j'})\Theta_{n\,n'}
\left({\Lambda^{\alpha n}(q_j)-\Lambda^{\alpha n'}(q_{j'})\over u}\right) \, , \hspace{0.35cm} 
j = 1,...,L_{\alpha n} \, , \hspace{0.35cm} \alpha = \eta, s \, , \hspace{0.35cm} n =1,...,\infty \, .
\label{Tapco2}
\end{eqnarray}
The sets of $j = 1,...,L$ and $j = 1,...,L_{\alpha n}$ quantum numbers $q_j$ in Eqs. (\ref{Tapco1}) and (\ref{Tapco2}), 
respectively, which are defined in Eqs. (\ref{q-j}) and (\ref{Ic-an}), play the role of microscopic momentum values of 
different BA excitation branches. The corresponding $\beta$-band momentum distribution functions $N_{\beta} (q_j)$ 
read $N_{\beta} (q_j)=1$ and $N_{\beta} (q_j)=0$ for occupied and unoccupied discrete momentum values, respectively,
the rapidity function $\Lambda^{\alpha n}(q_{j})$ is the real part of the complex rapidity, Eq. (\ref{complex-rap}),
and $\Theta_{n\,n'} (x)$ is the function,
\begin{eqnarray}
\Theta_{n\,n'}(x) & = & \delta_{n,n'}\Bigl\{2\arctan\Bigl({x\over 2n}\Bigl) 
+ \sum_{l=1}^{n -1}4\arctan\Bigl({x\over 2l}\Bigl)\Bigr\} 
\nonumber \\
& + & (1-\delta_{n,n'})\Bigl\{ 2\arctan\Bigl({x\over \vert\,n-n'\vert}\Bigl)
+ 2\arctan\Bigl({x\over n+n'}\Bigl) 
+ \sum_{l=1}^{{n+n'-\vert\,n-n'\vert\over 2} -1}4\arctan\Bigl({x\over \vert\, n-n'\vert +2l}\Bigl)\Bigr\} \, ,
\label{Theta}
\end{eqnarray}
where $n, n' = 1,...,\infty$. The indices $\alpha =\eta,s$ and numbers $n =1,...,\infty$ refer to different 
BA excitation branches that are associated with the composite $\alpha n$ pseudoparticles as
defined in this paper.

The corresponding energy eigenvalues have for densities ranges $n_e\in [0,1[$ and $m\in [0,n_e]$ the following form,
\begin{equation}
E = \sum_{j=1}^{L}\left(N_{c} (q_j)\,E_c (q_j) + U/4 - \mu_{\eta}\right)
+ \sum_{\alpha=\eta,s}\sum_{n=1}^{\infty}\sum_{j=1}^{L_{\alpha n}}\,N_{\alpha n} (q_j)\,E_{\alpha n} (q_j) 
+ \sum_{\alpha=\eta,s}2\mu_{\alpha}\,(S_{\alpha}+S_{\alpha}^z) \, ,
\label{E}
\end{equation}
where the $\alpha = \eta,s$ energy scales $2\mu_{\alpha}$ are given in Eq. (\ref{2mu-eta-s})
and the spectra $E_c (q_j)$ and $E_{\alpha n} (q_j)$ read,
\begin{eqnarray}
E_c (q_j) & = & - 2t\cos k^c (q_j) - U/2 + \mu_{\eta} - \mu_s \, ,
\nonumber \\
E_{\alpha n} (q_j) & = & n\,2\mu_{\alpha} + 
\delta_{\alpha,\eta}\left(4t\,{\rm Re}\Bigl\{\sqrt{1-(\Lambda^{\eta n} (q_j) -i\,n u)^2}\Bigr\} - n\,U\right)
\, , \hspace{0.50cm} \alpha = \eta,s \, , \hspace{0.50cm} n =1,...,\infty \, ,
\label{spectra-E-an-c-0}
\end{eqnarray}
respectively. (The corresponding momentum eigenvalues of general $u>0$ energy and momentum 
eigenstates are provided in Eq. (\ref{P}).)

Useful solutions for our studies of the BA equations, Eqs. (\ref{Tapco1}) and (\ref{Tapco2}), are those for 
a ground state and its excited energy eigenstates that span a PS, as defined in Section \ref{quantum-liquid}. 
We denote the $c$ and $s1$ band PS ground-state rapidity functions by $\Lambda_0^{c}(q_j) = \sin k_0^c (q_j)$ 
and $\Lambda_0^{s1}(q_j)$, respectively. They are the solutions of the 
BA equations, Eq. (\ref{Tapco1}) and Eq. (\ref{Tapco2}) for $\alpha n=s1$, respectively, with the $\beta =c,\alpha n$ band 
momentum distribution functions as given in Eq. (\ref{N0q1DHm}). Hence they read,
\begin{eqnarray}
q_j & = & k_0^c (q_j) + {2\over L}\sum_{q'=-k_{F\downarrow}}^{k_{F\downarrow}}
\arctan\left({\sin k_0^c (q_j)-\Lambda_0^{s1}(q') \over u}\right)
\, , \hspace{0.50cm} j = 1,...,L \, , 
\nonumber \\
q_j & = & {2\over L} \sum_{q'=-2k_F}^{2k_F}\arctan\left({\Lambda_0^{s1}(q_j)-\sin k_0^c (q')\over u}\right)
\nonumber \\
& - & {2\over L}\sum_{q'=-k_{F\downarrow}}^{k_{F\downarrow}}
\arctan\left({\Lambda_0^{s1}(q_j)-\Lambda_0^{s1}(q')\over 2u}\right) \, , \hspace{0.35cm} 
j = 1,...,N_{\uparrow} \, .
\label{TapcoGS}
\end{eqnarray}
In the TL the ground state momentum rapidity function $k_0^c (q)$ and rapidity function $\Lambda_0^{s1}(q)$
have well-defined inverse functions $q^c = q^c (k)$ where $k\in [-\pi,\pi]$ and $q^{s1} = q^{s1} (\Lambda)$
where $\Lambda\in [-\infty,\infty]$, respectively. One can then derive coupled integral equations 
from the coupled algebraic equations, Eq. (\ref{TapcoGS}), whose solutions are the distributions 
$2\pi\rho (k) =\partial q^c (k)/\partial k$ and $2\pi\sigma (\Lambda) =\partial q^{s1} (\Lambda)/\partial \Lambda$. 
From such solutions one can then access the TL ground-state momentum rapidity function $k_0^c (q)$ and 
rapidity function $\Lambda_0^{s1}(q)$, respectively.

A result that plays a key role in the pseudoparticle - pseudofermion unitary transformation studied in 
Section \ref{matrixelem} is that the $c$ and $s1$ band rapidity functions $\Lambda^{c}(q_j) = \sin k^c (q_j)$ 
and $\Lambda^{s1}(q_j)$ of a PS excited energy eigenstates can be expressed in terms of those of the corresponding initial
ground state. From straightforward yet lengthly manipulations of the BA equations, Eqs. (\ref{Tapco1}) and (\ref{Tapco2}), 
that involve expansions up to arbitrary order in the deviations $\delta N_{\beta} (q_j)$, Eq. (\ref{DNq}), one finds that,
\begin{eqnarray}
\Lambda^{c}(q_j) & = & \Lambda_0^{c}\Bigl({\bar{q}} (q_j)\Bigr) =
\sin k_0^c\Bigl({\bar{q}} (q_j)\Bigr) \, , \hspace{0.50cm} j = 1,...,L_c \, ,
\nonumber \\
\Lambda^{s1}(q_j) & = & \Lambda_0^{s1}\Bigl({\bar{q}} (q_j)\Bigr) \, , \hspace{0.50cm} j = 1,...,L_{s1} \, ,
\label{FL}
\end{eqnarray}
where ${\bar{q}}_j = {\bar{q}} (q_j)$ with $j=1,...,L_{\beta}$ are the discrete $\beta =c,s1$ band canonical momentum 
values given in Eq. (\ref{barqan}).

The integral equations that define the rapidity dressed phase shifts $2\pi\,\bar{\Phi }_{\beta,\beta'} (r,r')$ 
in Eq. (\ref{Phi-barPhi}) are for densities in the ranges $n_e \in [0,1]$ and $m \in [0,n_e]$ derived by
solving such BA equations up to first order in the deviations $\delta N_{\beta} (q_j)$. 
In the following we write the rapidity dressed phase shifts in units of $2\pi$.
A first set of rapidity dressed phase shifts obey integral equations by their
own. These equations read,
\begin{equation}
\bar{\Phi }_{s1,c}\left(r,r'\right) = -{1\over\pi}\arctan (r-r') + \int_{-r^0_s}^{r^0_s}
dr''\,G(r,r'')\,{\bar{\Phi }}_{s1,c}\left(r'',r'\right) \, ,
\label{Phis1c-m}
\end{equation}
\begin{equation}
\bar{\Phi }_{s1,\eta n}\left(r,r'\right) =  -{1\over{\pi^2}}\int_{-r^0_c}^{r^0_c} dr''{\arctan
\Bigl({r''-r'\over n}\Bigr)\over{1+(r-r'')^2}} +
\int_{-r^0_s}^{r^0_s} dr''\,G(r,r'')\,{\bar{\Phi}}_{s1,\eta n}\left(r'',r'\right) \, , 
\label{Phis1cn-m}
\end{equation}
and
\begin{eqnarray}
\bar{\Phi }_{s1,sn}\left(r,r'\right) & = & \delta_{1
,n}\,{1\over\pi}\arctan\Bigl({r-r'\over 2}\Bigl) + (1-\delta_{1
,n}){1\over\pi}\Bigl\{ \arctan\Bigl({r-r'\over n-1}\Bigl) +
\arctan\Bigl({r-r'\over
n+1}\Bigl)\Bigr\} \nonumber \\
& - &  {1\over{\pi^2}}\int_{-r^0_c}^{r^0_c} dr''{\arctan \Bigl({r''-r'\over n}\Bigr)\over{1+(r-r'')^2}} +
\int_{-r^0_s}^{r^0_s} dr''\,G(r,r'')\,{\bar{\Phi
}}_{s1,s1}\left(r'',r'\right) \, . 
\label{Phis1sn-m}
\end{eqnarray}
The parameters $r_c^0$ and $r_s^0$ appearing in these equations are
defined in Eq. (\ref{QB-r0rs}) and the kernel $G(r,r')$ is given by,
\begin{equation}
G(r,r') = - {1\over{2\pi}}\left[{1\over{1+((r-r')/2)^2}}\right]
\left[1 - {1\over 2}
\left(t(r)+t(r')+{{l(r)-l(r')}\over{r-r'}}\right)\right] \, .
\label{G}
\end{equation}
Here
\begin{equation}
t(r) = {1\over{\pi}}\left[\arctan(r + r^0_c) - {\rm
arc}{\rm tan}(r -r^0_c)\right] \, , \label{t}
\end{equation}
and
\begin{equation}
l(r) = {1\over{\pi}}\left[ \ln (1+(r + r^0_c)^2) - \ln (1+(r
-r^0_c)^2)\right] \, . 
\label{l}
\end{equation}

A second set of rapidity dressed phase shifts are expressed in terms of those in Eqs.
(\ref{Phis1c-m})-(\ref{Phis1sn-m}) as follows,
\begin{equation}
\bar{\Phi }_{c,c}\left(r,r'\right) =
{1\over{\pi}}\int_{-r^0_s}^{r^0_s} dr''{\bar{\Phi}_{s1,c}\left(r'',r'\right) \over {1+(r-r'')^2}} \, ,
\label{Phicc-m}
\end{equation}
\begin{equation}
\bar{\Phi }_{c,\eta n}\left(r,r'\right) = -{1\over\pi}\arctan\Bigl({r-r'\over n}\Bigr) +
{1\over{\pi}}\int_{-r^0_s}^{r^0_s} dr''{\bar{\Phi
}_{s1,\eta n}\left(r'',r'\right) \over {1+(r-r'')^2}} \, ,
\label{Phiccn-m}
\end{equation}
and
\begin{equation}
\bar{\Phi }_{c,sn}\left(r,r'\right) = -{1\over\pi}\arctan\Bigl({r-r'\over n}\Bigr) + {1\over{\pi}}\int_{-r^0_s}^{r^0_s} dr''
{\bar{\Phi}_{s1,sn}\left(r'',r'\right) \over {1+(r-r'')^2}} \, .
\label{Phicsn-m}
\end{equation}

Finally, the remaining rapidity dressed  phase shifts can be
expressed either in terms of those in Eqs. (\ref{Phicc-m})-(\ref{Phicsn-m}) only,
\begin{equation}
{\bar{\Phi }}_{\eta n,c}\left(r,r'\right) = {1\over\pi}\arctan\Bigl({r-r'\over {n}}\Bigr) 
- {1\over{\pi}}\int_{-r_c^0}^{+r_c^0} dr''{{\bar{\Phi}}_{c,c}\left(r'',r'\right) \over {n[1+({r-r''\over {n}})^2]}} \, , 
\label{Phicnc-m}
\end{equation}
\begin{equation}
\bar{\Phi }_{\eta n,\eta n'}\left(r,r'\right) = {\Theta_{n,n'}(r-r')\over{2\pi}} -
{1\over{\pi}}\int_{-r_c^0}^{+r_c^0} dr''{\bar{\Phi }_{c,\eta n'}\left(r'',r'\right) \over
{n[1+({r-r''\over n})^2]}} \, , 
\label{Phicncn-m}
\end{equation}
\begin{equation}
\bar{\Phi }_{\eta n,sn'}\left(r,r'\right) = - {1\over{\pi}}\,\int_{-r_c^0}^{+r_c^0}
dr''{\bar{\Phi }_{c,sn'}\left(r'',r'\right) \over {n[1+({r-r''\over n})^2]}} \, , 
\label{Phicnsn-m}
\end{equation}
or in terms of both those in Eqs. (\ref{Phis1c-m})-(\ref{Phis1sn-m}) and in Eqs. (\ref{Phicc-m})-(\ref{Phicsn-m}),
\begin{equation}
{\bar{\Phi }}_{sn,c}\left(r,r'\right) = - {1\over\pi}\arctan\Bigl({r-r'\over {n}}\Bigr) +
{1\over{\pi}}\int_{-r^0_c}^{r^0_c} dr''{{\bar{\Phi}}_{c,c}\left(r'',r'\right) \over {n[1+({r-r''\over
n})^2]}} - \int_{-r^0_s}^{r^0_s} dr''{\bar{\Phi}}_{s1,c}\left(r'',r'\right){\Theta^{[1]}_{n,1}(r-r'')\over{2\pi}} 
\, ; \hspace{0.5cm} n > 1 \, , 
\label{Phisnc-m}
\end{equation}
\begin{equation}
{\bar{\Phi }}_{sn ,\eta n'}\left(r,r'\right) = {1\over{\pi}}\int_{-r^0_c}^{r^0_c} dr''{{\bar{\Phi}}_{c,\eta n'}\left(r'',r'\right) \over {n[1+({r-r''\over n})^2]}} 
- \int_{-r^0_s}^{r^0_s} dr''{\bar{\Phi}}_{s1,\eta n'}\left(r'',r'\right) {\Theta^{[1]}_{n,1}(r-r'')\over {2\pi}} 
\, ; \hspace{0.5cm} n > 1 \, , 
\label{Phisncn-m}
\end{equation}
\begin{equation}
{\bar{\Phi }}_{sn ,sn'}\left(r,r'\right) = {\Theta_{n,n'}(r-r')\over{2\pi}} +
{1\over{\pi}}\int_{-r^0_c}^{r^0_c} dr''{{\bar{\Phi}}_{c,sn'}\left(r'',r'\right) \over {n[1+({r-r''\over n})^2]}} 
- \int_{-r^0_s}^{r^0_s} dr''{\bar{\Phi}}_{s1,sn'}\left(r'',r'\right){\Theta^{[1]}_{n,1}(r-r'')\over{2\pi}} \, . 
\label{Phisnsn-m}
\end{equation}

In the above equations, $\Theta_{n\,n'}(x)$ is the function given in Eq. (\ref{Theta}) and $\Theta^{[1]}_{n\,n'}(x)$ is its derivative,
\begin{eqnarray}
\Theta^{[1]}_{n,n'}(x) & = & {\partial\Theta_{n,n'}(x)\over
\partial x} = \delta_{n ,n'}\Bigl\{{1\over n[1+({x\over 2n})^2]}+
\sum_{l=1}^{n -1}{2\over l[1+({x\over 2l})^2]}\Bigr\} +
(1-\delta_{n ,n'})\Bigl\{ {2\over |n-n'|[1+({x\over
|n-n'|})^2]} \nonumber \\
& + & {2\over (n+n')[1+({x\over n+n'})^2]} +
\sum_{l=1}^{{n+n'-|n-n'|\over 2} -1}{4\over
(|n-n'|+2l)[1+({x\over |n-n'|+2l})^2]}\Bigr\} \, .
\label{The1}
\end{eqnarray}

The $f$ functions in the second-order terms of the energy functional, Eq. (\ref{DE-fermions}),
can be expressed in terms of the related $\beta$ pseudofermion phase shifts $2\pi\,\Phi_{\beta,\beta'} (q_j,q_{j'})$,
Eq. (\ref{Phi-barPhi}), as follows \cite{Carmelo-91-92},
\begin{eqnarray}
f_{\beta\,\beta'}(q_j,q_{j'}) & = & v_{\beta}(q_{j})\,2\pi \,\Phi_{\beta,\beta'}(q_{j},q_{j'})+
v_{\beta'}(q_{j'})\,2\pi \,\Phi_{\beta',\beta}(q_{j'},q_{j}) 
\nonumber \\
& + & {1\over 2\pi}\sum_{\beta''=c,s1} \sum_{\iota =\pm 1} v_{\beta''}\,
2\pi\,\Phi_{\beta'',\beta}(\iota q_{F\beta''},q_{j})\,2\pi\,\Phi_{\beta'',\beta'} (\iota q_{F\beta''},q_{j'}) \, ,
\label{ff}
\end{eqnarray}
where the group velocities are defined in Eq. (\ref{vel-beta}).

Other important quantities controlled by $\beta$ pseudofermion phase shifts are the $\beta =c,s1$ lowest peak weights 
$A^{(0,0)}_{\beta}$ and relative weights $a_{\beta}=a_{\beta}(m_{\beta,\,+1},\,m_{\beta,\,-1})$ in the $\beta$ 
pseudofermion spectral functions, Eq. (\ref{BQ-gen}). These weights are
derived by the use of the pseudofermion anti-commutators, Eq. (\ref{pfacrGS}), in Eq. (\ref{det1}). After 
some suitable algebra one finds,
\begin{eqnarray}
A^{(0,0)}_{\beta} & = & \Big({1\over
L}\Bigr)^{2N^{\odot}_{\beta}}\, \prod_{j=1}^{L_{\beta}}\,
\sin^2\Bigl({\pi\over 2}\left(1- (1-2\Phi^T_{\beta}(q_j))N_{\beta}^{\odot}(q_j)\right)\Big)\, \prod_{j=1}^{L_{\beta}-1}\,
\Bigl(\sin\Bigl({\pi j\over L}\Bigr)\Bigr)^{2(L_{\beta} -j)} 
\nonumber \\
& \times &
\prod_{i=1}^{L_{\beta}}\prod_{j=1}^{L_{\beta}}\,\theta (j-i)\,
\sin^2\left({\pi\over 2}\left(1 - \left(1 - {(2(j-i) + 2\Phi^T_{\beta}({q}_j) - 2\Phi^T_{\beta}({q}_i))
\over L}\right)N_{\beta}^{\odot}({q}_j)N_{\beta}^{\odot}({q}_i)\right)\right) 
\nonumber \\
& \times &
\prod_{i=1}^{L_{\beta}}\prod_{j=1}^{L_{\beta}}\,{1\over
\sin^2\left({\pi\over 2}\left(1 - \left(1 - {2(j-i) + 2\Phi^T_{\beta}({q}_j)\over L}\right)
N_{\beta}^{\odot}({q}_i)N_{\beta}^{\odot}({q}_j)\right)\right)} \, , \hspace{0.50cm} \beta = c, s1 \, ,
\label{A00}
\end{eqnarray}
and
\begin{equation}
a_{\beta}(m_{\beta,\,+1},m_{\beta,\,-1})=\Bigl(\prod_{\iota =\pm 1}
a_{\beta,\iota}(m_{\beta,\iota})\Bigr)
\Bigl(1+{\cal{O}}\Bigl(\ln L/L\Bigr)\Bigr) \, ,
\hspace{0.50cm} \beta = c, s1 \, ,
\label{aNNDP}
\end{equation}
respectively, where,
\begin{equation}
a_{\beta,\iota}(m_{\beta,\iota}) = \prod_{j=1}^{m_{\beta,\iota}}
{(2\Delta_{\beta}^{\iota} + j -1)\over j} = \frac{\Gamma (m_{\beta,\iota} +
2\Delta_{\beta}^{\iota})}{\Gamma (m_{\beta,\iota}+1)\,
\Gamma (2\Delta_{\beta}^{\iota})} \, , \hspace{0.50cm}
\beta = c, s1 \, , \hspace{0.50cm} \iota =\pm 1 \, .
\label{aNDP}
\end{equation}
In these expressions, $N_{\beta}^{\odot}=\sum_{j=1}^{L_{\beta}}N_{\beta}^{\odot}({q}_j)$ and
$N_{\beta}^{\odot}({q}_j)$ are the number of $\beta =c,s1$ pseudofermions
and the $\beta$ band momentum distribution function, respectively, of the excited energy eigenstate generated 
by the PDT processes (A) and (B) defined in Section \ref{leading}, $L_{\beta}$ is the number of $\beta =c,s1$ band discrete momentum values
given by $L_c = L$ and $L_{s1}$ by Eq. (\ref{N-h-an}) for $\alpha n=s1$,
$\Phi^T_{\beta}({q}_j)$ is the $\beta =c,s1$ pseudofermion phase-shift functional, Eq. (\ref{pfacrGS}),
$\Gamma (x)$ is the usual gamma function, and the functionals $2\Delta^{\iota}_{\beta}$ are defined in Eqs.
(\ref{a10DP-iota}) and (\ref{functional}). 

When such functionals are such that $2\Delta_{\beta}^{\iota}>0$ and $2\Delta_{\beta}^{-\iota}=0$, the 
$\beta =c,s1$ pseudofermion spectral function $B_{Q_{\beta}} (k',\omega')$, Eq. (\ref{BQ-gen}), has in 
the TL the following form,
\begin{eqnarray}
& & B_{Q_{\beta}} (k',\omega') = {A^{(0,0)}_{\beta}\over v_{\beta}}\,a_{\beta,\iota}
\left({L\over 2\pi\,v_{\beta}}\,\omega' - \Delta_{\beta}^{\iota}\right)\,\delta \Bigl(k' - {\iota\,\omega'\over v_{\beta}}\Bigr)
\nonumber \\
& \approx &   
{F^{(0,0)}_{\beta}\over v_{\beta}\,\Gamma (2\Delta_{\beta}^{\iota})}\,\Theta (\iota\,\omega')\,\Bigl({\omega'\over 2\pi\,S_{\beta}\,v_{\beta}}\Bigr)^{-1 +2\Delta_{\beta}^{\iota}}
\,\delta \Bigl(k' - {\iota\,\omega'\over v_{\beta}}\Bigr) \, , \hspace{0.50cm} \beta = c,s1 \, .
\label{B-J-i-sum-GG2}
\end{eqnarray}
The second expression provided here is obtained from the use of Eqs. (\ref{f}) and (\ref{F00}). 

On the other hand, when $2\Delta_{\beta}^{\iota} =2\Delta_{\beta}^{-\iota}=0$ one finds that 
in the TL such a function reads,
\begin{equation}
B_{Q_{\beta}} (k',\omega') = {2\pi\over L}\,A^{(0,0)}_{\beta}\,\delta (k')\,\delta (\omega')
\approx 2\pi\,F^{(0,0)}_{\beta}\,S_{\beta}\,\delta (k')\,\delta (\omega') \, , \hspace{0.50cm} \beta = c,s1 \, .
\label{B-J-i-sum-GG3}
\end{equation}

\section{Limiting behaviors of the $\beta =c,s1$ band energy dispersions, group velocities, and
pseudofermion phase shifts}
\label{LimitBV}

The one-parametric spectra of the $\sigma $ one-electron spectral functions branch lines and border lines 
given in Eqs. (\ref{dE-dP-bl}) and  (\ref{dE-dP-c-s1}), respectively, are expressed in terms of the 
$c$ and $s1$ band energy dispersions, Eq. (\ref{epsilon-q}) for $\beta=c,s1$. 
The corresponding $\sigma $ one-electron spectral weight distribution in the
vicinity of the branch lines is controlled by the exponent 
$\xi_{\beta}^{\sigma} (k)$, Eq. (\ref{branch-l}), whose expression is
linear in the functionals, Eq. (\ref{OESFfunctional}), that involve the $\beta$ pseudofermion 
phase shifts $2\pi\,\Phi_{\beta,\beta'} (q_j,q_{j'})$.

Here we provide limiting behaviors of such $c$ and $s1$ band energy dispersions,
corresponding $c$ and $s1$ band group velocities, Eq. (\ref{vel-beta}) for $\beta=c,s1$,
and $\beta$ pseudofermion phase shifts $2\pi\,\Phi_{\beta,\beta'} (q_j,q_{j'})$, Eq. (\ref{Phi-barPhi}).
Except if otherwise stated, the expressions given in the following refer to electronic densities 
and spin densities in the ranges $n_e \in [0,1[$ and $m \in ]0,n_e]$, respectively.

In the $u\rightarrow 0$ limit the $c$ and $s1$ energy dispersions, Eq. (\ref{epsilon-q}) for $\beta=c,s1$,
have the following behaviors,
\begin{eqnarray}
\varepsilon_c (q) & = & -2t\left(2\cos \left({q\over 2}\right) - \cos k_{F\uparrow} - \cos k_{F\downarrow}\right)
\, , \hspace{0.5cm} \vert q\vert \leq 2k_{F\downarrow} \, ,
\nonumber \\
& = & -2t\left(\cos (\vert q\vert - k_{F\downarrow}) - \cos k_{F\uparrow}\right)
\, , \hspace{0.5cm} 2k_{F\downarrow} \leq \vert q\vert < \pi \, ,
\label{varepsiloncu0}
\end{eqnarray}
and
\begin{eqnarray}
\varepsilon_{s1} (q) & = & -2t\left(\cos q - \cos k_{F\downarrow}\right) \, , \hspace{0.5cm}
q \in [-k_{F\uparrow},k_{F\uparrow}] \, ,
\label{varepsilonsu0}
\end{eqnarray}
respectively. 

On the other hand, for $u\gg 1$ and $m\rightarrow 0$ the behavior of these energy dispersions is,
\begin{eqnarray}
\varepsilon_c (q) & = & -2t\left(\cos q - \cos 2k_{F} + {n\ln 2\over u}(\sin^2 q - \sin^2 2k_F)\right) 
\, , \hspace{0.5cm} q \in [-\pi,\pi]  \, ,
\nonumber \\
\varepsilon_{s1} (q) & = & - {\pi n_e\,t\over 2u}\left(1 - {\sin 2\pi n_e\over 2\pi n_e}\right)
\cos\left({q\over n_e}\right) 
\, , \hspace{0.5cm} q \in [-k_{F},k_{F}]  \, ,
\label{varepsiloncsulm0}
\end{eqnarray}
whereas for $u\gg 1$ and $m\rightarrow n_e$ they read,
\begin{eqnarray}
\varepsilon_c (q) & = & -2t\left(\cos q - \cos 2k_{F}\right)
\, , \hspace{0.5cm} q \in [-\pi,\pi]  \, ,
\nonumber \\
\varepsilon_{s1} (q) & = & - {n_e\,t\over u}\left(1 - {\sin 2\pi n_e\over 2\pi n_e}\right)
\left(\cos\left({q\over n_e}\right) - 1\right)
\, , \hspace{0.5cm} q \in [-2k_{F},2k_{F}]  \, .
\label{varepsiloncsulm1}
\end{eqnarray}

In the $u\rightarrow 0$ limit the corresponding $c$ and $s1$ group velocities, Eq. (\ref{vel-beta}) for $\beta=c,s1$, 
have the following behaviors,
\begin{eqnarray}
v_c (q) & = & 2t\sin\left({q\over 2}\right) 
\, , \hspace{0.5cm} \vert q\vert \leq 2k_{F\downarrow} \, ,
\nonumber \\
& = & {\rm sgn}\{q\}\,2t\sin (\vert q\vert - k_{F\downarrow}) 
\, , \hspace{0.5cm} 2k_{F\downarrow} \leq \vert q\vert < \pi \, ,
\label{vc0}
\end{eqnarray}
and
\begin{eqnarray}
v_{s1} (q) & = & 2t\sin q \, , \hspace{0.5cm}
q \in [-k_{F\uparrow},k_{F\uparrow}] \, ,
\label{vs0}
\end{eqnarray}
respectively. Moreover, for $u\gg 1$ and $m\rightarrow 0$ the group
velocities behavior is,
\begin{eqnarray}
v_c (q) & = & 2t\left(\sin q - {n_e\ln 2\over u}\sin 2q)\right) 
\, , \hspace{0.5cm} q \in [-\pi,\pi]  \, ,
\nonumber \\
v_{s1} (q) & = & {\pi\,t\over 2u}\left(1 - {\sin 2\pi n_e\over 2\pi n_e}\right)\sin\left({q\over n_e}\right) 
\, , \hspace{0.5cm} q \in [-k_{F},k_{F}]  \, ,
\label{vcsm0}
\end{eqnarray}
whereas for $u\gg 1$ and $m\rightarrow n_e$ they are given by,
\begin{eqnarray}
v_c (q) & = & 2t\sin q \, , \hspace{0.5cm} q \in [-\pi,\pi]  \, ,
\nonumber \\
v_{s1} (q) & = & {t\over u}\left(1 - {\sin 2\pi n_e\over 2\pi n_e}\right)
\sin\left({q\over n_e}\right)
\, , \hspace{0.5cm} q \in [-2k_{F},2k_{F}]  \, .
\label{vcsm1}
\end{eqnarray}

In the $u\rightarrow 0$ limit the phase shifts $2\pi\,\Phi_{\beta,\beta'} (q_j,q_{j'})$, Eq. (\ref{Phi-barPhi}), 
acquired by $\beta =c,s1$ pseudofermions due to the creation or annihilation under
transitions to excited energy eigenstates of other $\beta' =c,s1$ pseudofermions
have the following limiting behaviors,
\begin{eqnarray}
\Phi_{s1,s1} (q,q') & = & 0 \, , 
\nonumber \\
\Phi_{s1,c} (q,q') & = & - {1\over 2}{\rm sgn}\left\{\sin q - \sin\left({q'\over 2}\right)\right\} \, ,
\hspace{0.5cm} \vert q'\vert \leq 2k_{F\downarrow} 
\nonumber \\
& = & - {1\over 2}{\rm sgn}\left\{\sin q - {\rm sgn}\{q'\}\sin (\vert q'\vert - k_{F\downarrow})\right\} \, ,
\hspace{0.5cm} 2k_{F\downarrow} \leq \vert q'\vert < \pi \, ,
\nonumber \\
\Phi_{c,c} (q,q') & = & - {1\over 2}{\rm sgn}\{q-q'\} \, ,
\hspace{0.5cm} \vert q\vert, \vert q'\vert \leq 2k_{F\downarrow} 
\nonumber \\
& = & {1\over 2}{\rm sgn}\{q'\} \, ,
\hspace{0.5cm} \vert q\vert \leq 2k_{F\downarrow} \, ,
\hspace{0.5cm} 2k_{F\downarrow} \leq \vert q'\vert < \pi \, ,
\nonumber \\
& = & 0 \, ,
\hspace{0.5cm} 2k_{F\downarrow} < \vert q\vert < \pi  \, ,
\nonumber \\
\Phi_{c,s1} (q,q') & = & - {1\over 2}{\rm sgn}\left\{\sin\left({q\over 2}\right) - \sin q'\right\} \, ,
\hspace{0.5cm} \vert q\vert \leq 2k_{F\downarrow} 
\nonumber \\
& = & - {1\over 2}{\rm sgn}\left\{{\rm sgn}\{q\}\sin (\vert q'\vert - k_{F\downarrow})-\sin q'\right\} \, ,
\hspace{0.5cm} 2k_{F\downarrow} \leq \vert q\vert < \pi \, .
\label{Phis-all-qqu0}
\end{eqnarray}

Particular cases of these $\beta =c,s1$ pseudofermion phase shifts are those involved
in the functionals, Eq. (\ref{OESFfunctional}), which in the $u\rightarrow 0$ limit
are then given by,
\begin{eqnarray}
\Phi_{s1,s1}\left(\iota k_{F\downarrow},q\right) & = & \Phi_{c,c}\left(\iota 2k_{F},q\right) = 0 \, , 
\nonumber \\
\Phi_{s1,c}\left(\iota k_{F\downarrow},q\right) & = & - {\iota\over 2} \, ,
\hspace{0.5cm} \vert q\vert < 2k_{F\downarrow} \, ,
\hspace{0.5cm} q = - \iota 2k_{F\downarrow} \, , \hspace{0.5cm} \iota = \pm 1
\nonumber \\
& = & 0 \, , \hspace{0.5cm} q = \iota 2k_{F\downarrow} \, , \hspace{0.5cm} \iota = \pm 1
\nonumber \\
& = & - {1\over 2}{\rm sgn}\left\{\iota\sin  k_{F\downarrow} - {\rm sgn}\{q\}\sin (\vert q\vert - k_{F\downarrow})\right\} \, ,
\hspace{0.5cm} 2k_{F\downarrow} \leq \vert q\vert < \pi \, , \hspace{0.5cm} \iota = \pm 1 \, ,
\nonumber \\
\Phi_{c,s1}\left(\iota 2k_{F},q\right) & = & - {\iota\over 2} \, ,
\hspace{0.5cm} \vert q\vert < k_{F\uparrow} \, , \hspace{0.5cm} \iota = \pm 1
\nonumber \\
& = & {1\over 2}{\rm sgn}\{q\} \, ,
\hspace{0.5cm} \vert q\vert = k_{F\uparrow}  \, .
\label{Phis-all-qFqu0}
\end{eqnarray}

On the other hand, for $u\gg 1$ and spin density $m\rightarrow 0$ the above $\beta =c,s1$ pseudofermion  
phase shifts behave as,
\begin{eqnarray}
\Phi_{s1,s1}(q,q') & = & {1\over\pi}\int_{0}^{\infty}
d\omega{\sin\left(\omega\,{2\over\pi}\left[\arcsinh\left(\tan\left({q\over n_e}\right)\right)
- \arcsinh\left(\tan\left({q'\over n_e}\right)\right)\right]\right)\over \omega \left(1+e^{2\omega}\right)} 
\nonumber \\
& + & {q'\over 4u}{\sin (\pi n_e)\over \pi n_e}\cos\left({q\over n_e}\right) \, ,
\hspace{0.5cm} \vert q\vert \neq k_F 
\nonumber \\
& = & {\iota\over 2\sqrt{2}} \, , \hspace{0.5cm} q = \iota k_F
\, , \hspace{0.5cm} q' \neq \iota k_F \, , \hspace{0.5cm} \iota = \pm 1
\nonumber \\
& = & {\iota\over 2\sqrt{2}}(3-2\sqrt{2}) \, , \hspace{0.5cm} q = q' = \iota k_F \, , \hspace{0.5cm} \iota = \pm 1 \, , 
\nonumber \\
\Phi_{s1,c}(q,q') & = & - {q\over 2\pi n_e} + {1\over 4u}\cos\left({q\over n_e}\right)\,\sin q' \, , \hspace{0.5cm} 
\vert q\vert \neq k_F 
\nonumber \\
& = & - {\iota\over 2\sqrt{2}} \, , \hspace{0.5cm} q = \iota k_F \, , \hspace{0.5cm} \iota = \pm 1 \, ,
\nonumber \\
\Phi_{c,c}(q,q') & = & - {\ln 2\over 2\pi u}(\sin q -\sin q') \, ,
\nonumber \\
\Phi_{c,s1}(q,q') & = & {q'\over 2\pi n_e} - {1\over 4u}\sin q\,\cos\left({q'\over n_e}\right)
+ q'\,{\ln 2\over 2\pi u}{\sin (\pi n_e)\over \pi n_e} \, . 
\label{PhiallUlm0}
\end{eqnarray}

Those involved in the functionals, Eq. (\ref{OESFfunctional}), are in that limit and for the same
densities then given by,
\begin{eqnarray}
\Phi_{s1,s1}(\iota k_F,q) & = & {\iota\over 2\sqrt{2}} \, , \hspace{0.5cm} q \neq \iota k_F \, , \hspace{0.5cm} \iota = \pm 1
\nonumber \\
& = & {\iota\over 2\sqrt{2}}(3-2\sqrt{2}) \, , \hspace{0.5cm} q = \iota k_F \, , \hspace{0.5cm} \iota = \pm 1 \, , 
\nonumber \\
\Phi_{s1,c}(\iota k_F,q) & = & - {\iota\over 2\sqrt{2}} \, , \hspace{0.5cm} \iota = \pm 1 \, ,
\nonumber \\
\Phi_{c,c}(\iota 2k_F,q) & = & - {\ln 2\over 2\pi u}(\iota\sin (\pi n_e) -\sin q) \, ,
\nonumber \\
\Phi_{c,s1}(\iota 2k_F,q) & = & {q\over 2\pi n_e} - {\iota\over 4u}\sin 2k_F\,\cos\left({q\over n_e}\right)
+ q\,{\ln 2\over 2\pi u}{\sin (\pi n_e)\over \pi n_e} \, . 
\label{PhiallFULm0}
\end{eqnarray}

For $u\gg 1$ and $m\rightarrow n_e$ the $\beta =c,s1$ pseudofermion phase shifts under consideration behave as,
\begin{eqnarray}
\Phi_{s1,s1}(q,q') & = & {1\over\pi}\arctan\left({\tan\left({q\over 2n_e}\right) -\tan\left({q'\over 2n_e}\right)\over 2}\right) 
+ {q'\over \pi u}{\sin (\pi n_e)\over \pi n_e}\cos^2\left({q\over 2n_e}\right) \, ,
\nonumber \\
\Phi_{s1,c}(q,q') & = & - {q\over 2\pi n_e} + {1\over \pi u}\cos^2\left({q\over 2n_e}\right)\,\sin q' \, ,
\nonumber \\
\Phi_{c,c}(q,q') & = & 0 \, ,
\nonumber \\
\Phi_{c,s1}(q,q') & = & {q'\over 2\pi n_e} - {1\over \pi u}\sin q\,\cos^2\left({q'\over 2n_e}\right) \, . 
\label{PhiallULmne}
\end{eqnarray}

As a result, in that limit in which $k_{F\downarrow}=0$ the $\beta =c,s1$ pseudofermion phase shifts 
involved in the functionals, Eq. (\ref{OESFfunctional}), read,
\begin{eqnarray}
\Phi_{s1,s1}(0,q) & = & - {1\over\pi}\arctan\left({1\over 2}\tan\left({q\over 2n_e}\right)\right) 
+ {q\over \pi u}{\sin (\pi n_e)\over \pi n_e} \, ,
\nonumber \\
\Phi_{s1,c}(0,q) & = & {\sin q\over \pi u} \, ; \hspace{0.75cm}
\Phi_{c,c}(\iota 2k_F,q) = 0 \, ,
\nonumber \\
\Phi_{c,s1}(\iota 2k_F,q) & = & {q\over 2\pi n_e} - {\iota\over \pi u}\sin (\pi n_e)\,\cos^2\left({q\over 2n_e}\right)
\, , \hspace{0.5cm} \iota = \pm 1 \, . 
\label{PhiallFULmne}
\end{eqnarray}

The limiting behaviors of the related $\beta =c,s1$ pseudofermion phase-shift parameters, Eq. (\ref{x-aa}), which are the entries of the
matrices, Eq. (\ref{ZZ-gen}), are given in the following. In the $u\rightarrow 0$ limit such matrices read,
\begin{equation}
\lim_{u\rightarrow 0}\,Z^1 =
\lim_{u\rightarrow 0}\,\left[\begin{array}{cc}
\xi^{1}_{c\,c} & \xi^{1}_{c\,s1}  \\
\xi^{1}_{s1\,c} & \xi^{1}_{s1\,s1}  
\end{array}\right]
= \left[\begin{array}{cc}
1 & 0 \\
1 & 1 
\end{array}\right]
\, ; \hspace{0.75cm}
\lim_{u\rightarrow 0}\,Z^0  =
\lim_{u\rightarrow 0}\,\left[\begin{array}{cc}
\xi^{0}_{c\,c} & \xi^{0}_{c\,s1}  \\
\xi^{0}_{s1\,c} & \xi^{0}_{s1\,s1}  
\end{array}\right]
= \left[\begin{array}{cc}
1 & -1 \\
0 & 1
\end{array}\right] \, .  
\label{ZZ-gen-u0}
\end{equation}
These values apply to the limit $\lim_{u\rightarrow 0}\lim_{m\rightarrow 0}$.
However, if one takes the limit $\lim_{m\rightarrow 0}$ before $\lim_{u\rightarrow 0}$ one
finds instead,
\begin{equation}
\lim_{u\rightarrow 0}\lim_{m\rightarrow 0}\,Z^1 = \left[\begin{array}{cc}
\sqrt{2} & 1/\sqrt{2} \\
0 & 1/\sqrt{2} 
\end{array}\right]
\, ; \hspace{1.5cm}
\lim_{u\rightarrow 0}\lim_{m\rightarrow 0}\,Z^0 = \left[\begin{array}{cc}
1/\sqrt{2} & 0 \\
-1/\sqrt{2} & \sqrt{2} 
\end{array}\right] \, .
\label{ZZ-gen-m0u0}
\end{equation}
This singular behavior means that at $m=0$ and for $m\rightarrow 0$
the matrices, Eq. (\ref{ZZ-gen}), have different values at $u=0$ and in the $u\rightarrow 0$ limit. 
Interestingly, this singular behavior does nor show up in the physical quantities whose
expressions involve the $\beta =c,s1$ pseudofermion phase-shift parameters, Eq. (\ref{x-aa}),
which are the entries of the matrices under consideration.

For $m\rightarrow 0$ and all $u$ values the matrices in Eq. (\ref{ZZ-gen}) are given by,
\begin{equation}
\lim_{m\rightarrow 0}\,Z^1 = \left[\begin{array}{cc}
\xi_{0} & \xi_{0}/2 \\
0 & 1/\sqrt{2} 
\end{array}\right]
\, ; \hspace{1.5cm}
\lim_{m\rightarrow 0}\,Z^0 = \left[\begin{array}{cc}
1/\xi_{0} & 0 \\
-1/\sqrt{2} & \sqrt{2} 
\end{array}\right] \, ,  
\label{ZZ-gen-m0}
\end{equation}
where the $m\rightarrow 0$ parameter $\xi_0$ has the following limiting behaviors,
\begin{eqnarray}
\xi_0 & = & \sqrt{2} \, , \hspace{0.5cm} u\rightarrow 0 \, ,
\nonumber \\
& = & 1 + {\ln 2\over \pi u}\sin (\pi n_e)  \, , \hspace{0.5cm} u\gg 1 \, .
\label{x0limits}
\end{eqnarray}

In the $m\rightarrow n_e$ limit the matrices in Eq. (\ref{ZZ-gen}) simplify to,
\begin{equation}
\lim_{m\rightarrow n_e}\,Z^1 = \left[\begin{array}{cc}
1 & 0 \\
\eta_0 & 1 
\end{array}\right]
\, ; \hspace{1.5cm}
\lim_{m\rightarrow n_e}\,Z^0 = \left[\begin{array}{cc}
1 & -\eta_0 \\
0 & 1
\end{array}\right] \, ,  
\label{ZZ-gen-m1}
\end{equation}
where the parameter $\eta_0$ reads $\eta_0 = {2\over\pi}\arctan\left({\sin (\pi n_e)\over u}\right)$ and
thus has limiting behaviors, 
\begin{eqnarray}
\eta_0 & = & 1 \, , \hspace{0.5cm} u\rightarrow 0 \, ,
\nonumber \\
& = & {2\over\pi\,u}\sin (\pi n_e) \, , \hspace{0.5cm} u \gg 1 \, .
\label{eta0lim}
\end{eqnarray}

\end{document}